%% file: example_paper.tex
\documentclass{article}

\usepackage{microtype}
\usepackage{graphicx}
\usepackage{subcaption}
\usepackage{booktabs} 

\usepackage{hyperref}

\usepackage[accepted]{icml2023}

\usepackage{amsmath}
\usepackage{amssymb}
\usepackage{mathtools}
\usepackage{amsthm}

\usepackage[capitalize,noabbrev]{cleveref}

\theoremstyle{plain}
\newtheorem{theorem}{Theorem}[section]
\newtheorem*{theorem31}{Theorem 3.1}

\theoremstyle{definition}

\theoremstyle{remark}

\newcommand\norm[1]{\lVert#1\rVert}
\DeclareMathOperator*{\argmax}{arg\,max}

\usepackage[disable,textsize=tiny]{todonotes}

\icmltitlerunning{Reasons for the Superiority of Stochastic Estimators over Deterministic Ones: Robustness, Consistency and Perceptual Quality}

\begin{document}

\twocolumn[
\icmltitle{Reasons for the Superiority of Stochastic Estimators over Deterministic Ones: Robustness, Consistency and Perceptual Quality}




\begin{icmlauthorlist}
\icmlauthor{Guy Ohayon}{techcs}
\icmlauthor{Theo Adrai}{techcs}
\icmlauthor{Michael Elad}{techcs}
\icmlauthor{Tomer Michaeli}{techee}
\end{icmlauthorlist}

\icmlaffiliation{techcs}{Faculty of Computer Science, Technion, Haifa, Israel}
\icmlaffiliation{techee}{Faculty of Electrical and Computer Engineering, Technion, Haifa, Israel}

\icmlcorrespondingauthor{Guy Ohayon}{ohayonguy@campus.technion.ac.il}
\icmlkeywords{Machine Learning, ICML}

\vskip 0.3in
]



\printAffiliationsAndNotice{}  

\input{latex/8pages.tex} 

\bibliography{latex/egbib}
\bibliographystyle{icml2023}

\newpage
\onecolumn
\input{latex/supp.tex}

\end{document}

%% file: latex/8pages.tex
\begin{figure*}[hbt!]
\centering
\setlength{\tabcolsep}{1pt}
\newcommand{\Miki}[1]{{\color{red} [Miki: #1]}}

\renewcommand{\arraystretch}{0.5}
\begin{tabular}{ccccc}
 & \multicolumn{2}{c}{\textbf{Stochastic}} & \multicolumn{2}{c}{\textbf{Deterministic}} \\
\cmidrule(rl){2-3}\cmidrule(rl){4-5}
Input & Erratic & Robust & Erratic & Robust \\
\includegraphics[scale=0.65]{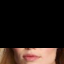}&\includegraphics[scale=0.65]{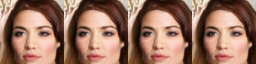}&\includegraphics[scale=0.65]{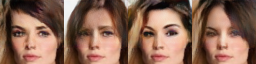}&\includegraphics[scale=0.65]{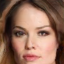}&\includegraphics[scale=0.65]{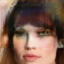}\\
\includegraphics[scale=0.65]{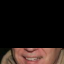}&\includegraphics[scale=0.65]{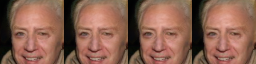}&\includegraphics[scale=0.65]{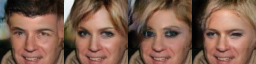}&\includegraphics[scale=0.65]{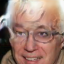}&\includegraphics[scale=0.65]{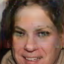}\\
\includegraphics[scale=0.65]{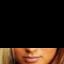}&\includegraphics[scale=0.65]{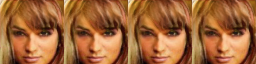}&\includegraphics[scale=0.65]{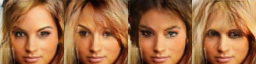}&\includegraphics[scale=0.65]{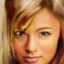}&\includegraphics[scale=0.65]{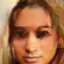}\\
\includegraphics[scale=0.65]{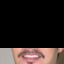}&\includegraphics[scale=0.65]{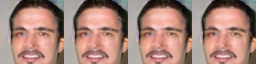}&\includegraphics[scale=0.65]{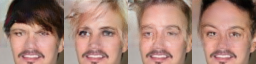}&\includegraphics[scale=0.65]{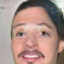}&\includegraphics[scale=0.65]{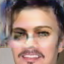}\\
\includegraphics[scale=0.65]{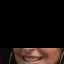}&\includegraphics[scale=0.65]{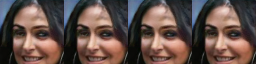}&\includegraphics[scale=0.65]{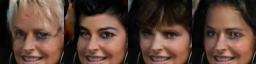}&\includegraphics[scale=0.65]{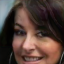}&\includegraphics[scale=0.65]{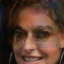}\\
\end{tabular}
\caption{Output samples from several consistent restoration algorithms that solve the image inpainting task on the CelebA data set. The erratic stochastic and deterministic algorithms are trained solely with a GAN loss. As can be seen, the erratic stochastic estimator barely produces output variability per input, which reveals a tendency of mode collapse of CGANs~\cite{gan,pix2pix2017,mathieu2016deep,dsganICLR2019,pscgan}. The robust algorithms are trained to also defend against adversarial attacks by adding~\cref{eq:stochastic-robustness-practical} to the GAN objective. Robustifying the deterministic algorithm deteriorates its perceptual quality, while doing so for the stochastic algorithm preserves this quality and significantly improves its output variability. Refer to~\cref{tab:inpainting-sto-vs-det} for quantitative evaluation.}
\label{fig:inpainting-sto-vs-det}
\end{figure*}
\begin{abstract}
Stochastic restoration algorithms allow to explore the space of solutions that correspond to the degraded input.
In this paper we reveal additional fundamental advantages of stochastic methods over deterministic ones, which further motivate their use.
First, we prove that any restoration algorithm that attains perfect perceptual quality and whose outputs are consistent with the input must be a posterior sampler, and is thus required to be stochastic.
Second, we illustrate that while deterministic restoration algorithms may attain high perceptual quality, this can be achieved only by filling up the space of all possible source images using an extremely sensitive mapping, which makes them highly vulnerable to adversarial attacks.
Indeed, we show that enforcing deterministic models to be robust to such attacks profoundly hinders their perceptual quality, 
while robustifying stochastic models hardly influences their perceptual quality, and improves their output variability.
These findings provide a motivation to foster progress in stochastic restoration methods, paving the way to better recovery algorithms.
\end{abstract}
\section{Introduction}
Image restoration has a central role in imaging sciences, enabling much of the imaging revolution we see today. Modern restoration methods allow to squeeze out ever fantastic quality from ever smaller sensors (e.g. on mobile devices), which are prone to severe noise, blur, limited resolution, and color degradations.
Traditionally, most image restoration algorithms were deterministic: providing one restored image for a given degraded input.
Stochastic restoration algorithms, which started to receive notable attention in recent years~\cite{ntire2021challenge,ntire2022challenge,explorable_sr,pscgan,kawar2022denoising,kawar2021snips,bahjat-posterior,palette,Kadkhodaie,song-generative}, instead sample from a distribution conditioned on the degraded input.
Yet, beyond their core ability to explore the space of possible solutions, the advantages of stochastic methods over deterministic ones are still unclear.
For instance, a fundamental mystery is whether stochastic restoration algorithms are theoretically superior to deterministic ones in terms of the achievable perceptual quality, the robustness to adversarial attacks, and the faithfulness of the results to the measurements (a property we refer to as consistency).

In this paper we provide novel justifications for using stochastic restoration methods.
First, we show that any estimator that generates consistent restorations and attains perfect perceptual quality must be sampling from the posterior distribution.
An immediate, yet significant, consequence is that if a deterministic restoration algorithm is consistent then it cannot attain perfect perceptual quality\footnote{Unless the posterior assigns nonzero probability to only one reconstruction for every input, in which case the inverse problem is non-ambiguous and it is possible to restore images with zero error.}.
Somewhat strangely, despite this theoretical limitation, deterministic restoration algorithms are known to be able to attain high perceptual quality, or at least produce images with high precision~\cite{esrgan,Blau_2018,srgan,deepfillv2,deepfillv2,zhao2021comodgan,Yi_2020_CVPR}.
While this may seem as a contradiction, we argue that it is not.
We show empirically that, in order to cope with the lack of output diversity, such deterministic estimators adopt an erratic\footnote{By erratic we mean that the restoration algorithm is not robust to an input adversarial attack: a small, unnoticeable change in the input degraded image may lead to an unreasonably large change in the output estimated image.
I.e., the value of~\cref{eq:stochastic-robustness} is too high.} output behavior in an attempt to ``fill up'' the space of all possible natural images, so as to minimize the distance between the distribution of their outputs and the distribution of natural images. 
As a result, such estimators are highly sensitive to small changes in their input.
Indeed, it has been experimentally observed that a visually unnoticeable input perturbation leads to unreasonable changes in the output in this kind of estimators~\cite{estimators_robustness,vulnerability-image-to-image,instabilites-deep-learning-image-reconstruction,ijcai2022p211,pmlr-v119-raj20a,adv-attacks-image-deblurring,yu2022towards,Choi_2020_ACCV,Robust-Real-World-Image-Super-Resolution}. Furthermore, we illustrate that making deterministic methods more robust to adversarial attacks causes the quality of their outputs to deteriorate (see~\cref{fig:inpainting-sto-vs-det}).
Hence, our findings provide a novel explanation for  the increased vulnerability of high perceptual quality deterministic estimators to adversarial attacks.

Since robustness for deterministic estimators comes at the expense of perceptual quality, it is appealing to resort to stochastic estimators. Such estimators can produce many restorations for any given input, so that they seemingly do not need to be highly sensitive to their inputs. Indeed, we show that as opposed to deterministic methods, robustifying stochastic algorithms hardly impairs their perceptual quality. Interestingly, some stochastic methods do exhibit a relatively high sensitivity to their inputs, but this is indicative of mode-collapse (they effectively behave like deterministic methods). As we illustrate in~\cref{fig:inpainting-sto-vs-det}, in such cases robustification improves the output diversity without hampering perceptual quality.
All at all, unlike deterministic algorithms, stochastic methods allow to attain consistent, high perceptual quality restorations, while remaining robust to adversarial attacks.

\begin{figure*}
\centering
\includegraphics{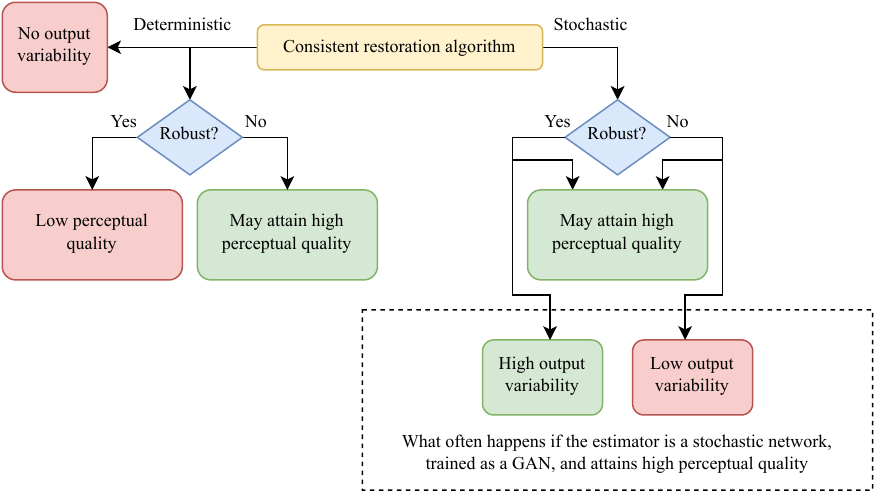}
    \caption{The main conclusions of this paper, summarized in a flow chart. We only consider perfectly consistent restoration algorithms. A deterministic restoration algorithm has, by definition, no output variability. Such an algorithm attains low perceptual quality if it is robust, and may attain high perceptual quality otherwise. A stochastic restoration algorithm may always attain high perceptual quality, whether it is robust or not. In the case where the stochastic algorithm is a neural network trained as a GAN, robustifying such an algorithm improves (increases) its output variation per input. Without robustification, such an algorithm often attains low output variability, and essentially behaves as a non-robust (erratic), deterministic algorithm.}
    \label{figure:main_contrib}
\end{figure*}

The contributions of this paper are the following: 1)~We prove that a posterior sampler is the only consistent restoration algorithm that attains perfect perceptual quality; 2) While deterministic restoration algorithms could still be consistent and attain very high perceptual quality, we reveal that they behave erratically in order to do so, i.e., we empirically show that for deterministic estimators, high perceptual quality comes with the cost of vulnerability to input adversarial attacks; 3)~We develop a novel notion of robustness for stochastic estimators; and 4)~We show that robustifying deterministic algorithms deteriorates their perceptual quality, while doing so for stochastic algorithms only improves their output diversity. Hence, we empirically confirm that stochastic restoration algorithms can be robust and attain high perceptual quality at the same time. We thus suggest robustness as an effective regularization for promoting meaningful diversity in stochastic restoration methods.
Please refer to~\cref{figure:main_contrib} for a flow chart that clarifies the main conclusions of this paper.
\section{Problem setting and preliminaries}\label{section:problem-setting}
We consider a natural image $x\in\mathbb{R}^{n_{x}}$ as a realization of a multivariate random variable $X$ with probability density function $p_{X}$. A degraded version $y\in\mathbb{R}^{n_{y}}$ is also a realization of a random vector $Y$ which is related to $X$ via some conditional distribution $p_{Y|X}$.
In this paper we assume that the degradation is deterministic, i.e., $y=D(x)$, which implies that $p_{Y|X}(y|x)=\delta(y-D(x))$, where $D:\mathbb{R}^{n_{x}}\rightarrow\mathbb{R}^{n_{y}}$ is a non-invertible function.
Problems such as image colorization, inpainting, demosaicing, single-image super-resolution, JPEG-deblocking, and more all follow this assumed structure. Here we focus on ill-posed inverse problems in which $X$ cannot be retrieved from $Y$ with zero error, i.e., $p_{X|Y}(\cdot|y)$ is not a delta function for almost every $y$.

Given a particular input $Y=y$, an image restoration algorithm produces an estimate $\hat{X}$ according to some distribution $\smash{p_{\hat{X}|Y}(\cdot|y)}$ such that the estimate $\smash{\hat{X}}$ is statistically independent of $X$ given~$Y$. For deterministic algorithms, $\smash{p_{\hat{X}|Y}(\cdot|y)}$ is a delta function for every $y$, while for stochastic algorithms it is a non-degenerate distribution. 
\subsection{Perceptual quality}
Conceptually, the perceptual quality of a restoration algorithm is a quantification of its ability to produce images that appear natural.
While there are several notions of perceptual quality, we measure it as the deviation of the restorations from natural image statistics~\cite{Blau_2018}.
    Formally, we quantify the perceptual quality of an estimator $\hat{X}$ using the \emph{perceptual index} (lower is better),
    \begin{equation}\label{eq:perceptual-index}
        d(p_{X},p_{\hat{X}}),
    \end{equation}
    where $d(p,q)$ is a divergence between distributions (e.g., Kullback-Leibler, Wasserstein).
    i.e., an estimator attains perfect perceptual quality if $p_{\hat{X}}=p_{X}$.
    As discussed in~\cite{precision_recall,improved_precision_recall}, one can measure the deviation between $p_{\hat{X}}$ and $p_{X}$ via \emph{precision} (the
probability that a random sample from $p_{\hat{X}}$ falls within the support of $p_{X}$) and \emph{recall} (the probability
that a random sample from $p_{X}$ falls within the support of $p_{\hat{X}}$).
Such an approach to measure statistical deviation is useful since it provides a meaning to the difference between $p_{\hat{X}}$ and $p_{X}$, if exists.
That is, low precision means that a sample of $\smash{\hat{X}}$ may seem unnatural, while low recall implies that part of the support of $p_{X}$ (possible natural images) can never be the output of $\smash{\hat{X}}$.
Note that
$p_{\hat{X}}=p_{X}$ if and only if both the precision and recall are perfect~\cite{precision_recall}.
Importantly, the reader should not confuse high precision with high perceptual quality.
A restoration algorithm might produce images that appear natural (with high precision), but this does not mean that the algorithm attains sufficient perceptual quality, since its recall might be compromised.
\subsection{Consistency}
Another important quality measure of a restoration algorithm is its ability to produce results that are consistent with the low quality input.
Such an ability is important to consider since only perfectly consistent restorations could potentially be the high quality source image.
In our deterministic degradation scenario, a restoration $\hat{x}$ is consistent if it satisfies $D(\hat{x})=y$ (which in turn equals $D(x)$), which can be equivalently written as $p_{Y|X}(\cdot|x)=p_{Y|\hat{X}}(\cdot|x)$. We regard an algorithm as perfectly consistent (or consistent, in short) if it satisfies this property for every $x$. 
Note that a restoration algorithm that is not perfectly consistent could still be considered as such for any practical use.
For instance,~\cite{ntire2022challenge,ntire2021challenge} consider super-resolution (SR) methods as consistent if they satisfy
    $\text{PSNR}(D(\hat{X}),D(X))\geq 45\text{dB}$.

While being an important property for restoration algorithms, apparently only recent publications provide a report of consistency~\cite{ntire2022challenge,srflow-da,lugmayr2020srflow,ntire2021challenge,explorable_sr,accurate-image-sr-using-very-deep-cnn}, while many previous ones do not~\cite{esrgan,edsr,srgan,real-time-single-image-and-video-sr-subpixel,image-sr-using-deep-nets}.
Moreover, works that do report consistency typically deal only with stochastic image restoration.
It therefore seems that consistency has been overlooked, or taken for granted by deterministic restoration algorithms.
\section{Can deterministic estimators be consistent and achieve high perceptual quality?}
While deterministic algorithms are being extensively used to produce high perceptual quality restorations (e.g.,~\cite{esrgan,srgan,deepfillv2,deepfillv2,zhao2021comodgan,Yi_2020_CVPR,deep-gen-adv-compression-artifact-remove}), the following important result must be taken into account.
\begin{theorem}\label{theorem:posterior-uniquness}
For a deterministic degradation, $y=D(x)$, an estimator $\smash{\hat{X}}$ is perfectly consistent ($p_{Y|X}=p_{Y|\hat{X}}$) and achieves perfect perceptual quality ($p_{\hat{X}}=p_{X}$) if and only if it is the posterior sampler $p_{\hat{X}|Y}=p_{X|Y}$.
\end{theorem}
The proof of the theorem makes simple use of Bayes' rule (see~\cref{appendix:thm1-proof}). But despite its simplicity, this result has several important implications. (i)~Since $p_{X|Y}(\cdot|y)$ is not a delta function for almost every $y$ (the degradation is not invertible), an immediate corollary is that \emph{a consistent deterministic algorithm can never attain perfect perceptual quality}.
(ii)~It is a known fact that there are cases in which the posterior sampler does not attain the lowest possible MSE (or any other distortion measure) among the estimators that achieve perfect perceptual quality~\cite{Blau_2018,theory-on-the-pd-tradeoff}. But from Theorem~\ref{theorem:posterior-uniquness} we see that \emph{the posterior sampler is the only perfect perceptual quality estimator that is also perfectly consistent}. Thus, for perfect perceptual quality estimators, aiming for low distortion might come on the expense of inconsistency. 
(iii) Finally, we can conclude from~\cref{theorem:posterior-uniquness} that if perfect consistency is enforced, then one can train a restoration algorithm to become a posterior sampler by \emph{solely encouraging high perceptual quality} (e.g. with a GAN loss), without any additional loss. 
\begin{figure*}[!ht]
    \centering
\input{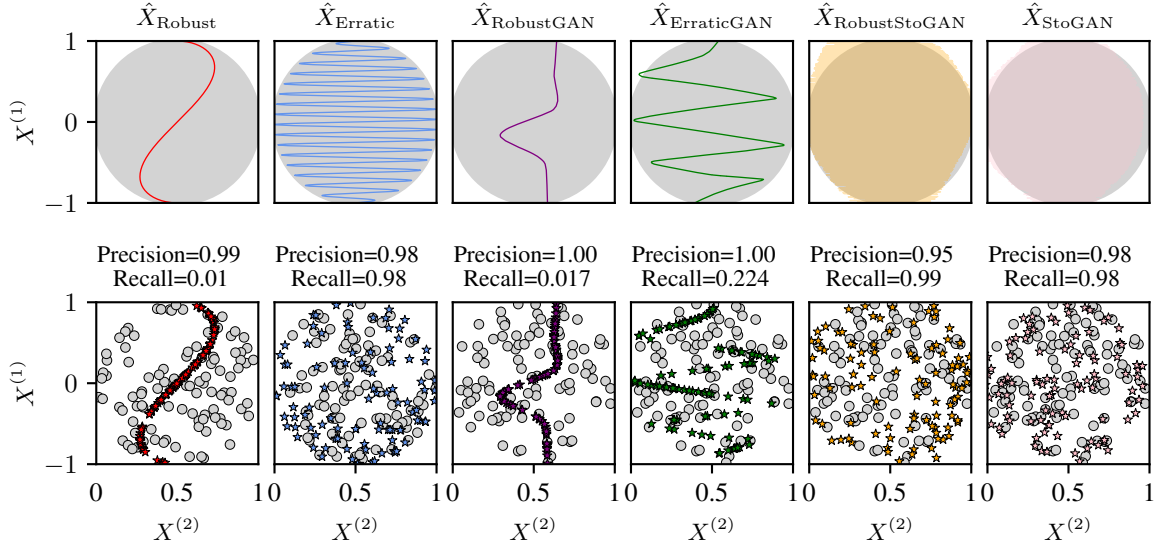}
    \caption{An illustration on a toy restoration problem of the tradeoff between robustness and perceptual quality for deterministic restoration algorithms, as well as as a demonstration that such a tradeoff does not exist for stochastic algorithms.
    Each column corresponds to a different estimator, with the estimator's name indicated in the title of the column.
    Top row: the support of the data distribution (gray disc), and the support of the output distribution of each estimator (a continuous colored line of 10,000 random samples from the output distribution of each estimator).
      Bottom row: 100 random samples from the data distribution (gray dots), and 100 random samples from the output distribution of each estimator (colored stars).
      The precision and recall~\cite{improved_precision_recall} of each estimator are computed between 1000 random output samples from the output distribution of the estimator and 1000 random samples from the data distribution. Refer to~\cref{sec:space-filling} for more details.}
    \label{fig:space-filling-curve}
\end{figure*}
\section{Erratic behavior of deterministic estimators}\label{sec:space-filling}
Earlier work~\cite{estimators_robustness,vulnerability-image-to-image} has already identified the tendency of deterministic GAN-based restoration methods to be vulnerable to adversarial attacks. In this section we shed light on this phenomenon.
The authors of both of these works hypothesize that, since GANs usually produce perceptually appealing and sharp textures, small input perturbations tend to be severely intensified in the output image.
While such a claim is valid, it does not tell the full story.
We argue that this phenomenon is indeed a result of attempting to attain high perceptual quality, but more specifically, doing so \emph{with a deterministic estimator}.
To attain high perceptual quality, such an estimator must adopt an erratic output behavior in order to fill up the space of possible source signals, and consequently be highly sensitive to small input perturbations (i.e., vulnerable to adversarial attacks).
In other words, the problem is not related to the use of GANs, but rather to the lack of randomness. Using any non-GAN-based \emph{deterministic} method with high perceptual quality would yield the same issue; and a GAN-based \emph{stochastic} method with high perceptual quality can avoid this phenomenon.
Refer to~\cref{appendix:what-else-affects-robustness} for further discussion.

Let us demonstrate this phenomenon in more detail through a toy experiment.
Suppose that $X=(X^{(1)},X^{(2)})$ is uniformly distributed on 
a disk with radius $1.0$ centered at $(0, 0)$, and let $Y=X^{(1)}$. Our goal is to estimate $X$ based on $Y$ (i.e., we need to estimate only $X^{(2)}$, as $X^{(1)}$ is known).
In~\cref{fig:space-filling-curve} (the two leftmost columns in the top row) we present two \emph{consistent} deterministic restoration algorithms of the form $\hat{X}_{\alpha}=(Y,\sqrt{1-Y^2}\sin(\alpha Y))$.
These estimators provide a mental demonstration of the anticipated behavior of robust and non-robust deterministic restoration algorithms that strive for high perceptual quality.
For an estimator of this form, notice that $\alpha$ controls the tradeoff between its perceptual quality and robustness: increasing $\alpha$ improves its perceptual quality but hinders its robustness.
For instance, $\hat{X}_{\text{Erratic}}$ (with $\alpha=50$) represents an erratic (non-robust) estimator with high perceptual quality.
It ``zigzags'' and fills the support of $p_{X}$, which intuitively leads to a smaller distance between $p_{\hat{X}}$ and $p_{X}$.
$\hat{X}_{\text{Robust}}$ (with $\alpha=1$) represents a robust estimator with low perceptual quality.
Its output mildly changes when perturbing its input, so it fills a smaller region from the support of $p_{X}$.
To numerically confirm that an increased erratic behavior results in higher perceptual quality,
we approximate the precision and recall~\cite{improved_precision_recall} of these estimators, showing that such practical statistical distance measures can be fooled by highly erratic deterministic algorithms.
For an analytical confirmation, refer to~\cref{appendix:concrete-mathematical-example}.
\subsection{Perceptual quality approximation}\label{sec:preceptual-quality-approx-deterministic}
In real world problems $p_{X}$ and $p_{\hat{X}}$ are typically unknown, and their statistical discrepancy is commonly approximated by using finite sized sets of independently drawn samples from both distributions (e.g.,~\cite{fid,improved_precision_recall,precision_recall}).
In some scenarios it would be difficult or even impossible for such a statistical distance approximation algorithm to distinguish between the output distribution of a highly erratic deterministic algorithm and the distribution of the source data.
To illustrate this, we randomly and independently sample 1000 points from the outputs of $\hat{X}_{\text{Robust}}$ and $\hat{X}_{\text{Erratic}}$ and approximate their perceptual quality via precision and recall~\cite{improved_precision_recall} (with $k=5$) and without using feature projection.
In the bottom row of~\cref{fig:space-filling-curve} we present 100 random output samples of each algorithm (colored stars), 100 random samples from $p_{X}$ (gray dots), as well as the precision and recall results.
As expected, all the algorithms attain almost perfect precision.
For the erratic estimator $\hat{X}_{\text{Erratic}}$, a human observer might be fooled to believe that its samples are actually drawn from $p_{X}$.
Indeed, it attains almost perfect recall as well, and consequently almost perfect perceptual quality according to these metrics.
However, the robust algorithm $\hat{X}_{\text{Robust}}$ attains a much lower recall, and one can easily distinguish between the distribution of its output samples and the samples from the data distribution.
This illustrates what one might expect when robustifying a deterministic estimator: its perceptual quality would be limited by a lower recall.

These toy restoration algorithms are hand-crafted to visually demonstrate our hypothesis, so let us confirm that this behavior occurs for practical estimators as well.
\subsection{Deterministic GAN estimators}\label{sec:deterministic-gan-estimators}
We trained a neural network $\hat{X}_{\text{ErraticGAN}}=G_{\theta}(Y)$ as a GAN~\cite{gan} to solve the aforementioned toy problem (we omit the use of $\theta$ from now on).
We evaluated the precision and recall in the same fashion as before, and present the results in~\cref{fig:space-filling-curve}.
As shown, $\hat{X}_{\text{ErraticGAN}}$ is indeed erratic, which aligns with our argument that this would be the behavior of a deterministic algorithm when attempting to attain high perceptual quality.
I.e., $\hat{X}_{\text{ErraticGAN}}$ becomes highly sensitive to small input perturbations.

Let us observe the effect of robustifying such an estimator.
The sensitivity of a deterministic algorithm $G(Y)$ at $Y=y$ can be measured by
\begin{gather}
    r_{G,\epsilon}(y)=\max_{\delta:\norm{\delta}_{2}\leq\epsilon}\norm{G(y)-G(y+\delta)}_{2}^{2}.
\end{gather}
That is, $r_{G,\epsilon}(y)$ is the extent to which a small input perturbation of $y$ leads to a change in the output.
From here, we can measure the \emph{robustness} of a deterministic estimator by averaging $r_{G,\epsilon}(y)$ for all $y$'s, leading to
\begin{gather}\label{eq:deterministic-robustness}
    R_{G,\epsilon}=\mathbb{E}\left[r_{G,\epsilon}(Y)\right].
\end{gather}
In~\cref{fig:space-filling-curve} we present an additional GAN based algorithm $\hat{X}_{\text{RobustGAN}}$, which was trained with $R_{G,\epsilon}$ as a regularizer (refer to~\cref{appendix:training-details-toy-gan-inpanting} for full training details of these estimators).
As expected, $\hat{X}_{\text{RobustGAN}}$ is a much more stable algorithm, but has a lower perceptual quality (lower recall). 
As hypothesized, $\hat{X}_{\text{ErraticGAN}}$ attains a higher recall than $\hat{X}_{\text{RobustGAN}}$, which can also be confirmed visually as it results in a more erratic behavior and passes through more regions in $\text{Supp}(p_{X})$.
We therefore confirm again that the robustness of a deterministic estimator should hinder its perceptual quality (and more specifically, its recall).
These experiments demonstrate again that robustness of deterministic estimators comes at the cost of lower perceptual quality.
\section{Robustness of stochastic estimators}\label{sec:robustness-of-stochastic-estimators}
Deep-learning based recovery algorithms are known to be sensitive to miniature and visually unnoticeable input perturbations, e.g. in single-image super-resolution~\cite{estimators_robustness,vulnerability-image-to-image,Choi_2020_ACCV,Robust-Real-World-Image-Super-Resolution}, deblurring~\cite{adv-attacks-image-deblurring}, rain removal~\cite{yu2022towards}, and denoising~\cite{ijcai2022p211}.
Studying the stability of image restoration methods is important and sometimes critical (e.g. in medical imaging for diagnosis), as only robust algorithms are able to provide trustworthy estimations.
Interestingly, to the best of our knowledge, all the existing publications that discuss robustness in image restoration only consider deterministic solutions.

To discuss the robustness of an arbitrary (not necessarily deterministic) estimator, we first need to generalize the notion of robustness to the stochastic case.
We extend~\cref{eq:deterministic-robustness} by thinking of robustness as the maximal deviation in $p_{\hat{X}|Y}(\cdot|y)$ that can occur due to a small change in~$y$.
Formally, we define the sensitivity of a possibly stochastic estimator $\hat{X}$ at $Y=y$ by
\begin{gather}
    r_{\hat{X},\epsilon}(y)=\max_{\delta:\norm{\delta}_{2}\leq\epsilon}W_{2}^{2}(p_{\hat{X}|Y}(\cdot|y),p_{\hat{X}|Y}(\cdot|y+\delta)),
\end{gather}
where $W_{2}$ is the Wasserstein-2 distance between distributions.
As before, we define the overall robustness of $\hat{X}$ by
\begin{gather}\label{eq:stochastic-robustness}
    R_{\hat{X},\epsilon}=\mathbb{E}\left[r_{\hat{X},\epsilon}(Y)\right].
\end{gather}
To see why~\cref{eq:stochastic-robustness} serves as a natural extension of~\cref{eq:deterministic-robustness}, we show that both of these definitions are equivalent when $\hat{X}$ is a deterministic estimator.
Indeed, if $\hat{X}=G(Y)$ for some deterministic function $G$, $p_{\hat{X}|Y}(\cdot|y)$ is a delta function for any $y$, so the $W_{2}$ distance in $r_{\hat{X}}(y)$ measures the deviation between two Dirac measures and therefore becomes a simple $L_{2}$ norm between two vectors, i.e., $ W_{2}(p_{\hat{X}|Y}(\cdot|y),p_{\hat{X}|Y}(\cdot|y+\delta))=\norm{G(y)-G(y+\delta)}_{2}$.

Practically, measuring the Wasserstein-2 distance between distributions, and not to mention minimizing it, is a highly difficult task.
We therefore offer an alternative in the case where the estimator is a neural network and the stochasticity is acquired by using an input random seed (as done in all GANs).
Let $\hat{X}=G(Y,Z)$ be a neural network, where $Y$ is the input and $Z$ is some random seed that follows any type of distribution.
We propose to measure the sensitivity of $\hat{X}$ at $Y=y$ by considering its average sensitivity over random draws of $Z$, i.e.,
\begin{gather}\label{eq:stochastic-sensitivity-practical}
    \tilde{r}_{\hat{X},\epsilon}(y)=\max_{\delta:\norm{\delta}_{2}\leq\epsilon}\mathbb{E}\left[\norm{G(y,Z)-G(y+\delta,Z)}_{2}^{2}\right].
\end{gather}
In words,~\cref{eq:stochastic-sensitivity-practical} measures the extent to which a small input perturbation changes the output with \emph{the same random seed}, averaged over many seeds.
Finally, we approximate the overall robustness of $\hat{X}$ via
\begin{gather}\label{eq:stochastic-robustness-practical}
\tilde{R}_{\hat{X},\epsilon}=\mathbb{E}\left[\tilde{r}_{\hat{X},\epsilon}(Y)\right].
\end{gather}
Observe that here as well we naturally extend the definition of robustness for deterministic estimators. As we move from a stochastic to a deterministic algorithm, the variance of the seed $Z$ drops to zero, and the two definitions coincide. Moreover, this is a rational approximation method for the stochastic case since \cref{eq:stochastic-robustness-practical} upper bounds \cref{eq:stochastic-robustness} (see proof in~\cref{appendix:practical-stochastic-robustness-upperbound}).
Hence, minimizing~\cref{eq:stochastic-robustness-practical} forces~\cref{eq:stochastic-robustness} to also minimize.


To illustrate that stochastic algorithms can simultaneously be robust and attain high perceptual quality, we trained a neural network $G(Y,Z)$ to solve the toy problem presented in~\cref{sec:space-filling}, where $Z\sim\mathcal{N}(0,I)$ is an input random seed that allows the outputs to vary for each input $Y$.
As before, we trained two consistent algorithms: $\hat{X}_{\text{StoGAN}}$ and $\hat{X}_{\text{RobustStoGAN}}$, where the former is trained solely with a GAN loss, and the latter is also regularized to be robust using~\cref{eq:stochastic-robustness-practical} (see~\cref{appendix:training-details-toy-gan-inpanting} for full training details).
Following a similar procedure to~\cref{sec:deterministic-gan-estimators}, we randomly sample 1000 outputs from each estimator (by sampling 1000 random inputs and coupling each input with one random seed) and evaluate precision and recall.
Unsurprisingly, both $\hat{X}_{\text{StoGAN}}$ and $\hat{X}_{\text{RobustStoGAN}}$ lead to precision and recall scores above $0.95$, which confirms our hypothesis that a consistent stochastic algorithm can maintain high perceptual quality even when it is robust to input adversarial attacks.

It is important to note that the whole discussion on robust restoration algorithms with high perceptual quality is relevant only when assuming that a minor perturbation in $y$ does not lead to a large change in the posterior distribution $p_{X|Y}(\cdot|y)$ when dealing with natural images.
As discussed in~\cref{sec:space-filling}, a consistent deterministic estimator with high perceptual quality attempts to become a posterior sampler by behaving erratically, and thus we propose to use stochastic estimators instead (such as a posterior sampler).
See~\cref{appendix:what-else-affects-robustness} for more details on this subject.
\section{Experiments on natural images}\label{section:experiments-on-natural}
\subsection{Demonstrations with GANs}
We train several types of GAN-based restoration algorithms to solve the image inpainting and super resolution tasks (see~\cref{appendix:why-not-diffusion} for a discussion on our choice of using GANs for the following demonstrations).
In all the experiments we use a U-Net architecture~\cite{unet} as our estimator, which we denote by $G(Y,Z)$.
For the stochastic algorithms $Z\sim\mathcal{N}(0,I)$ is a random input seed that allows their outputs to vary for each $Y$,
while for the deterministic ones we fix $Z=0$.
We use the same architecture for the stochastic and deterministic algorithms to ensure that it does not impair the evaluation.
All the algorithms are enforced to produce perfectly consistent restorations.
The optimization task in all of the experiments is a weighted sum of two objectives: $\mathcal{L}_{GAN}$ and $\mathcal{L}_{R}$.
Here, $\mathcal{L}_{GAN}$ is a non-saturating generative adversarial training loss~\cite{gan} combined with $R_{1}$ critic regularization~\cite{gan_converge}, where the critic is the same ResNet architecture as in~\cite{gan_converge}.
Optimizing such a loss would drive $p_{\hat{X}}$ to be as close as possible to $p_{X}$, and due to the consistency enforcement, a posterior sampler is the only optimal solution for this task (\cref{theorem:posterior-uniquness}).
Note that the critic's objective is to distinguish between samples from $p_{\hat{X}}$ and $p_{X}$ without taking $Y$ into account. The term
$\mathcal{L}_{R}$ is a loss that drives the restoration algorithm to be robust to input adversarial attacks.
It is equal to \cref{eq:stochastic-robustness-practical}, where $Z$ is zero for the deterministic algorithms.
In practice, we perform 5 Adam~\cite{adam} optimization steps to approximate the attack $\delta^{*}$ on each input $y$, and for the stochastic algorithms we perform the average in~\cref{eq:stochastic-sensitivity-practical} over 10 random seeds for each $y$.
The final training objective is
\begin{gather}\label{eq:loss}
\mathcal{L}_{GAN}+\lambda_{R}\mathcal{L}_{R},
\end{gather}
where $\lambda_{R}$ controls the level of the algorithm's robustness.
Complementary training details are provided in the following sections and in~\cref{appendix:training-details-natural}.
\subsubsection{Experimental setup}\label{sec:experimental-setup}
\noindent\textbf{Inpainting:} We perform several experiments on the image inpainting task, in which some pixels of a high quality image are masked and a restoration algorithm estimates their values.
We assume that the locations of the masked pixels are fixed and known.
In all of our experiments, we mask the upper $\frac{3}{4}$ part of the image, and leave the remaining bottom pixels untouched.
We use the CelebA data set~\cite{liu2015faceattributes} for the experiments in this task.
Consistency is enforced by replacing the bottom $\frac{1}{4}$ part of the output with that of the input, a typical way to enforce consistency in image inpainting~\cite{palette,deepfillv1,deepfillv2,Yi_2020_CVPR,zhao2021comodgan}.
We train several consistent restoration algorithms by optimizing~\cref{eq:loss}: two deterministic and two stochastic models, where one of each is trained with $\lambda_{R}=0$ and the other with $\lambda_{R}=500$.
We refer to the algorithms trained with $\lambda_{R}=0$ as \emph{erratic}, and to the others as \emph{robust}.
We choose $\epsilon$ in~\cref{eq:stochastic-robustness-practical} so that $\text{PSNR}(y+\delta^{*}, y)\geq 32.9\text{dB}$ for any $y$ (so that $\delta^{*}$ leads to a barely noticeable change in $y$).
We refer to this quantity as the \emph{adversarial input PSNR} (AI-PSNR).
\begin{table}[t!]
\caption{Quantitative evaluation of the CelebA image inpainting algorithms described in~\cref{sec:experimental-setup}.
Robstifying the deterministic restoration algorithm significantly deteriorates its recall and FID performance. Doing so for the stochastic algorithm only slightly hinders its performance (with the same AI-PSNR), while also improving its output variability (high per-pixel std).
Refer to~\cref{sec:anaylisis-of-results} for further analysis of these results.
}
\centering
\setlength{\tabcolsep}{4pt}
\renewcommand{\arraystretch}{1.2}
\begin{tabular}{c|c|c|c|ccc}
\multicolumn{1}{c}{} & \multicolumn{2}{c}{\textbf{Stochastic}} & \multicolumn{2}{c}{\textbf{Deterministic}} \\
\cmidrule(rl){2-3}\cmidrule(rl){4-5}
Metric & Erratic & Robust & Erratic & Robust \\
\midrule
FID$^{\downarrow}$ & 14.10 & 19.27 & 14.37 & 39.09 \\
Precision$^{\uparrow}$ & 0.670 & 0.634 & 0.670 & 0.533 \\
Recall$^{\uparrow}$ & 0.388 & 0.268 & 0.382 & 0.022 \\
Per-pixel std$^{\uparrow}$ & 0.007 & 0.105 & 0 & 0\\
Robustness$^{\uparrow}$& 6.884 & 26.19 & 3.544 & 23.69\\
AI-PSNR & 34.47 & 32.95 & 34.18 & 32.98\\
\end{tabular}
\label{tab:inpainting-sto-vs-det}
\end{table}

\noindent\textbf{Super resolution:} We train the same types of algorithms as before (erratic \& robust, deterministic \& stochastic), but this time we solve the super resolution task.
The degradation we consider is a plain averaging with scaling factors of $4\times$, $8\times$, and $16\times$.
The training objective and the used architectures remain the same.
Consistency is enforced with CEM~\cite{explorable_sr}.
Unlike in the inpainting task, this time the dimensionality of the input is different than that of the output.
Since our estimator is a neural network with input and output of the same size, we feed to it an upsampled version of the low-resolution image, using nearest-neighbor interpolation. 
Moreover, we train the models on CelebA, as well as on the $64\times64$ version of ImageNet~\cite{imagenet_small}.
\begin{figure}
    \centering
    \input{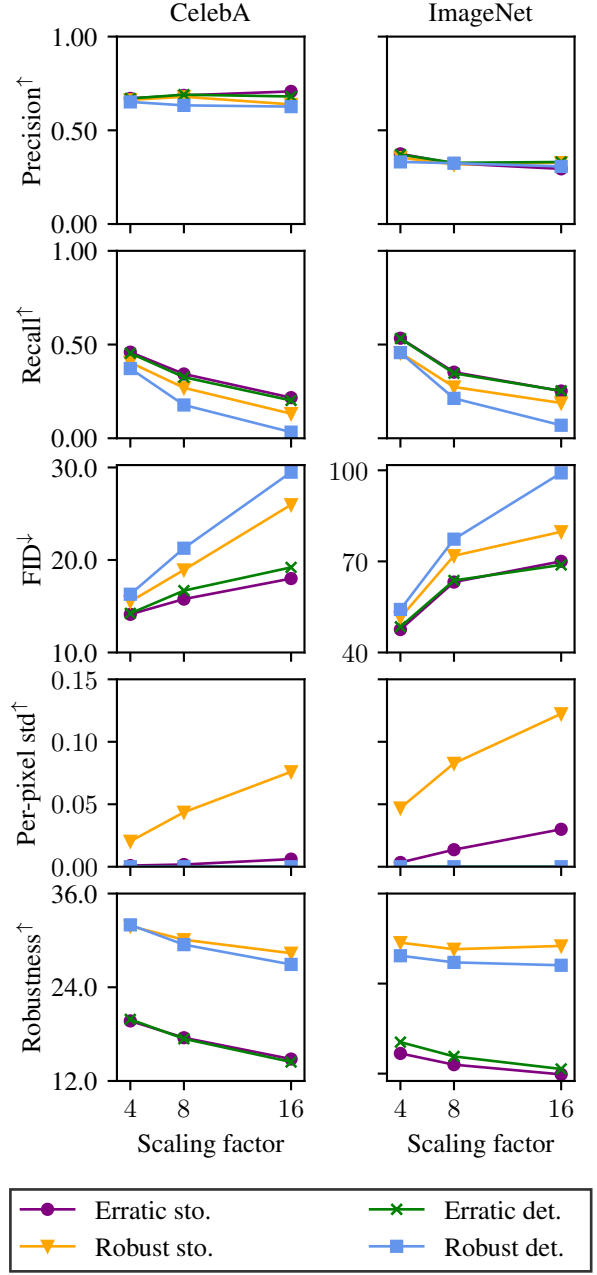}
    \caption{Quantitative evaluation of the super resolution algorithms described in~\cref{sec:experimental-setup}. AI-PSNR is equal to 34dB for all algorithms. The anticipated link between robustness and perceptual quality is revealed again: The robust deterministic models achieve the lowest perceptual quality for all scaling factors. As the scaling factor (degradation severity) increases, we observe higher output variability only for the robust stochastic models, and a larger perceptual quality gap for the robust deterministic models.}
    \label{fig:sr}
\end{figure}

\noindent\textbf{Performance metrics:}
For the inpainting algorithms, we report in~\cref{tab:inpainting-sto-vs-det} the perceptual quality performance according to FID~\cite{fid}, precision, and recall~\cite{improved_precision_recall} (with $k=3$), and present several outputs produced by the algorithms in~\cref{fig:inpainting-sto-vs-det}.
We also demonstrate the quality and the extent of the output variation of the stochastic models by showing four output samples for each input.
The perceptual quality metrics are computed by considering the training set as the real samples, and the algorithms' outputs on the validation set as the fake samples.
Moreover, we report the robustness performance of each algorithm according to
\begin{gather}
    \mathbb{E}\left[\text{PSNR}(G(Y,Z),G(Y+\delta^{*}(Y),Z)\right],
\end{gather}
where the average is computed over the entire validation set and $\delta^{*}(y)$ is the solution of~\cref{eq:stochastic-sensitivity-practical} for each $y$, and is acquired in the same fashion as in the training stage.
Lastly, we report the per-pixel standard deviation of each algorithm over 32 restored samples (for the same input),  averaged across all the output pixels and over the entire validation set.
This measures the average variation of the output pixels for all the inputs in the validation data, i.e., a higher value corresponds to a more diverse output space, per input, on average. 
In~\cref{fig:sr} we report these performance metrics for the super resolution algorithms as well.
Please refer to~\cref{appendix:training-details-natural} for complementary training details.

\subsubsection{Analysis of the results}\label{sec:anaylisis-of-results}
\noindent\textbf{Inpainting:}
The erratic deterministic algorithm is competitive with the stochastic methods in terms of FID~(\cref{tab:inpainting-sto-vs-det}).
However, while both robust algorithms attain the same level of robustness (and for the same AI-PSNR), the deterministic one suffers from a significant deterioration in FID and the stochastic one does not.
This can be confirmed visually in~\cref{fig:inpainting-sto-vs-det} as well.
As the only difference between the robust algorithms is the random noise injection (same model, same loss, etc.), this result supports our hypothesis that stochastic models can overcome the trade-off between robustness and perceptual quality of deterministic mappings.

Another interesting outcome is that the erratic stochastic algorithm produces virtually zero output diversity per input, while the robust stochastic algorithm maintains meaningful output variation.
This can be confirmed by the naked eye when observing the randomly sampled outputs in~\cref{fig:inpainting-sto-vs-det} (the robust algorithm creates several types of hair, genders, etc. for the same input), and by the higher per-pixel standard deviation.
These results yield several insights:
1)~Since the erratic stochastic algorithm barely produces output variation, it effectively behaves like a deterministic estimator.
This shows that the instability of a stochastic model is indicative of conditional mode-collapse~\cite{gan,pix2pix2017,mathieu2016deep,dsganICLR2019}.
2)~Apparently, in order to attain high perceptual quality, it is less challenging to learn an erratic output curvature rather than to map a random input seed to meaningful, high quality output variation.
3) Enforcing robustness on a stochastic model promotes meaningful output variation and effectively alleviates the conditional mode-collapse issue.
These results further support our link between the robustness and stochasticity of a restoration algorithm, as anticipated.
Indeed, we observe that the robust deterministic model suffers from a very low recall compared to the robust stochastic one.

\noindent\textbf{Super resolution:}
The first observation we make from~\cref{fig:sr} is that, as the scaling factor increases, the FID of all algorithms deteriorate.
This is supposedly a result of attempting to solve a more difficult inverse problem: higher scaling factors preserve less information (the problem is ``more'' ill-posed).
Yet, we see that for the same degradation severity, the robust deterministic algorithm always attains the worst FID performance.
Moreover, its FID gap with the other algorithms increases with the scaling factor, suggesting that the tradeoff between robustness and perceptual quality (for deterministic estimators) is more severe for a higher scaling factor.
Intuitively, a higher scaling factor is expected to increase the size of the support of $p_{X|Y}(\cdot|y)$ for any $y$, so it covers a larger portion of the support of $p_{X}$.
We therefore hypothesize that, for high scaling factors, a consistent, robust deterministic estimator would ``miss'' a larger portion in the support of $p_{X}$, and would therefore attain lower recall, just as the results show.
Interestingly, we see that the precision of all algorithms and across all scaling factors remain roughly the same, showing that a restoration algorithm can indeed output images that appear natural but still attain low perceptual quality (due to low recall).
Lastly, we again see that the erratic stochastic model barely produces output variability, while the robust one does.
\subsection{Robustness of SRFlow}\label{sec:srflow-robustness}
Previous work~\cite{estimators_robustness,vulnerability-image-to-image} hypothesize that restoration algorithms with high perceptual quality are more vulnerable to adversarial attacks.
This hypothesis stems from the low robustness levels of ESRGAN~\cite{esrgan}, a GAN-based, deterministic, high perceptual quality algorithm.
However, such a hypothesis is limited as it is based on the evaluation of a single high perceptual quality restoration algorithm.
Here we provide additional support for our claim that such an hypothesis is incomplete: it is valid only for deterministic estimators, while stochastic ones can be robust and still attain high perceptual quality.

In~\cref{appendix:srflow-robustness} we evaluate the robustness of SRFlow~\cite{lugmayr2020srflow}, a stochastic super-resolution algorithm that produces highly consistent outputs and comparable perceptual quality to ESRGAN~\cite{esrgan}.
As reported in~\cref{fig:srflow-robustness}, SRFlow exhibits substantially higher robustness than ESRGAN, and comparable robustness to distortion-optimized models.
These findings further demonstrate the potential of high perceptual quality stochastic estimators to attain superior robustness compared to deterministic estimators with comparable perceptual quality performance.

\section{Summary}
In this work we ask whether stochastic estimators are better than deterministic ones, and our short answer is \emph{yes}. 
Indeed, we proved that a posterior sampler is the only restoration algorithm with perfect consistency and perceptual quality, which shows that a consistent deterministic algorithm can never attain perfect perceptual quality.
While a deterministic estimator can still attain high perceptual quality, it must adopt an erratic output behavior in order to do so, which makes it vulnerable to adversarial attacks.
Hence, our work provides a novel explanation for the existence of adversarial attacks in high perceptual quality deterministic restoration algorithms.
We expand the notion of robustness for stochastic algorithms, and then show that such algorithms can significantly alleviate the tradeoff between robustness and perceptual quality that exists in deterministic algorithms.
Interestingly, we find that robustness can be used to promote meaningful output variability in stochastic models, which aligns well with the theory and hypothesis developed in this paper.
Our conclusion: Robust, stochastic restoration algorithms do provide better recovery.
\section{Acknowledgments}
This research was partially supported by the Israel Science Foundation (ISF) under Grants 335/18 and 2318/22 and by the Council For Higher Education - Planning \& Budgeting Committee.

%% file: latex/supp.tex
\appendix
\setcounter{theorem}{0}
\section{Proof of~\cref{theorem:posterior-uniquness}}\label{appendix:thm1-proof}
\begin{theorem31}
For a deterministic degradation, $y=D(x)$, an estimator $\smash{\hat{X}}$ is perfectly consistent ($p_{Y|X}=p_{Y|\hat{X}}$) and achieves perfect perceptual quality ($p_{\hat{X}}=p_{X}$) if and only if it is the posterior sampler $p_{\hat{X}|Y}=p_{X|Y}$.
\end{theorem31}
\begin{proof}
First, note that $D(\hat{X})=D(X)$ if and only if $p_{Y|\hat{X}}=p_{Y|X}$.
Assume that $p_{\hat{X}|Y}=p_{X|Y}$.
Then,
\begin{gather*}
    p_{\hat{X}}(x)=\int_{y}p_{\hat{X}|Y}(x|y)p_{Y}(y)dy=\int_{y}p_{X|Y}(x|y)p_{Y}(y)dy=p_{X}(x),
\end{gather*}
and
\begin{gather*}
    p_{Y|\hat{X}}(y|x)=\frac{p_{\hat{X}|Y}(x|y)p_{Y}(y)}{p_{\hat{X}}(x)}=\frac{p_{X|Y}(x|y)p_{Y}(y)}{p_{X}(x)}=p_{Y|X}(y|x).
\end{gather*}
In the other direction, assume that $p_{\hat{X}}=p_{X}$ and $p_{Y|\hat{X}}=p_{Y|X}$.
Then,
\begin{gather*}
    p_{\hat{X}|Y}(x|y)=\frac{p_{Y|\hat{X}}(y|x)p_{\hat{X}}(x)}{p_{Y}(y)}=\frac{p_{Y|X}(y|x)p_{X}(x)}{p_{Y}(y)}=p_{X|Y}(x|y),
\end{gather*}
concluding the proof.
\end{proof}
\section{Further discussion on deterministic estimators}\label{appendix:what-else-affects-robustness}
In this section we expand on the topics discussed in~\cref{sec:space-filling} and~\cref{sec:robustness-of-stochastic-estimators}.
\subsection{A tradeoff between precision and recall for continuous, consistent deterministic estimators}
\begin{figure}[h]
    \centering
    \input{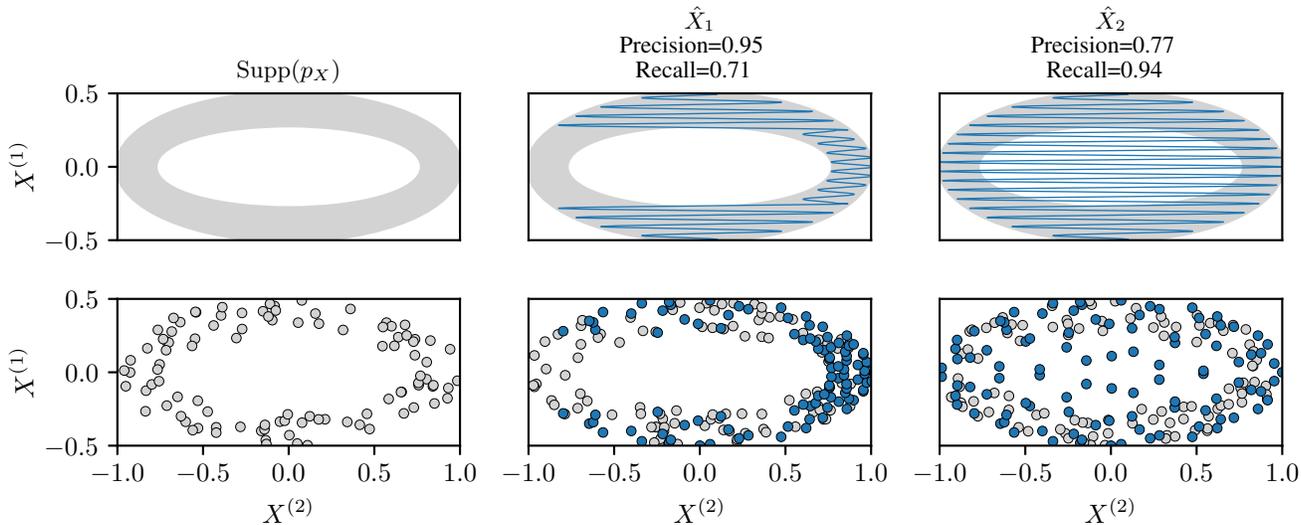}
    \caption{Mental illustration of a tradeoff between precision and recall for consistent, high perceptual quality, continuous deterministic estimators. In the top row we show the support of the data distribution and of the distribution of each estimator's outputs. In the bottom row we show 100 random samples from these distributions. $\hat{X}_{1}$ avoids the forbidden region (the middle empty ellipse which is not part of $\text{Supp}(p_{X})$), and therefore attains almost perfect precision with impaired recall. $\hat{X}_{2}$ passes through the forbidden region in order to attain higher recall, but as a result compromises on precision since it generates outputs that are not in the data distribution.}
    \label{fig:precision-recall-tradeoff}
\end{figure}
Let $X=(X^{(1)},X^{(2)})$ be a two dimensional random variable supported on the set of all points between two concentric ellipses, as shown in~\cref{fig:precision-recall-tradeoff}.
The task is to estimate $X$ given $Y=X^{(1)}$ with perfect consistency.
Notice that, different from the example in~\cref{sec:space-filling}, now there is a ``hole'' in the support of $p_{X}$.
In~\cref{fig:precision-recall-tradeoff} we present two hypothetical estimators that are consistent and continuous (these estimators are handcrafted for the purpose of this demonstration).
Observe that due to the hole in $p_{X}$, a continuous, deterministic estimator, regardless of how erratic it is, can never approach having perfect perceptual quality (unlike the two examples in~\cref{appendix:concrete-mathematical-example} and~\cref{sec:space-filling}).
Such an estimator would always have to pass through the central empty ellipse region, which is not part of $\text{Supp}(p_X)$, in order to produce outputs from both sides of $\text{Supp}(p_{X})$.
Interestingly, this example reveals a tradeoff between precision and recall for such continuous deterministic estimators. To demonstrate this, let us assume that the data distribution is uniform over its support (the set of points between the two concentric ellipses). We randomly sample 1000 points from each distribution, and compute precision and recall as in~\cref{sec:preceptual-quality-approx-deterministic}.
As the results show (\cref{fig:precision-recall-tradeoff}), $\hat{X}_{1}$, the estimator that does not pass through the forbidden region, compromises on recall in order to attain higher precision. $\hat{X}_{2}$ decides to pass through the forbidden region in order to generate samples from the left side of the ellipse, which results in lower precision and higher recall.

Notice that the situation described above has nothing to do with robustness, and occurs even for erratic, non-robust deterministic estimators.
Moreover, this tradeoff does not exist for stochastic estimators~(\cref{theorem:posterior-uniquness}).
\subsection{A note on the MMSE and MAP estimators}
In~\cref{sec:robustness-of-stochastic-estimators} we note that the discussion on robust estimators with high perceptual quality is only relevant when assuming that a posterior sampler is robust, i.e., when a small change in $y$ does not lead to an unreasonable change in the distribution $p_{X|Y}(\cdot|y)$.
In such a case, we expect that both the MMSE estimator $\mathbb{E}[X|Y]$ and the Maximum A Posteriori (MAP) estimator $\max_{x}p_{X|Y}(x|Y)$ would also be robust, since the mean and the maximum of $p_{X|Y}(\cdot|y)$ cannot significantly deviate when a small perturbation is added to $y$.
\section{Concrete mathematical example}\label{appendix:concrete-mathematical-example}
\begin{figure}[H]
    \centering
    \includegraphics{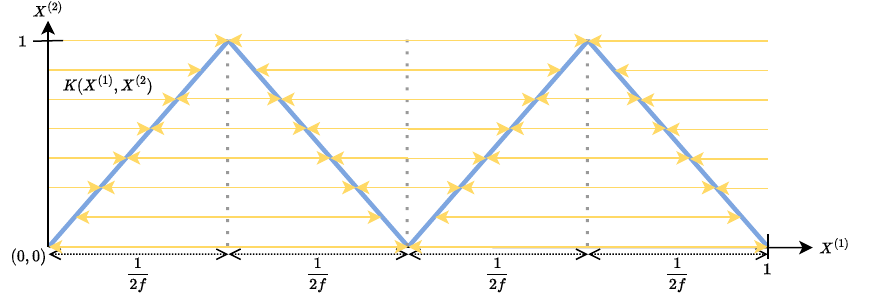}
    \caption{An illustration of $\text{Supp}(p_{\hat{X}})$ (zigzag line) and the mapping $K_f(X^{(1)},X^{(2)})$ (arrow lines), where $\hat{X}=(X^{(1)},G_{f}(X^{(1)}))$ is the estimator from~\cref{appendix:concrete-mathematical-example} and the mapping $K_f(X^{(1)},X^{(2)})$ maps each point $(X^{(1)},X^{(2)})$ onto the zigzag in a horizontal fashion, as illustrated in the figure.}
    \label{fig:concrete-expample-intuition}
\end{figure}
We provide a concrete toy example which shows that the perceptual index (according to the $W_{2}$ distance) of a consistent deterministic estimator can be arbitrarily minimized by making the estimator more erratic.
Let $X=(X^{(1)},X^{(2)})$ be a two dimensional random variable, where $X^{(1)},X^{(2)}\sim U(0,1)$, and $X^{(1)},X^{(2)}$ are statistically independent.
Let $Y=X^{(1)}$, and $\hat{X}=(Y,G_{f}(Y))$ be a consistent, deterministic estimator, where $G_{f}(y)=\frac{1}{\pi}\arccos{\left(\cos{\left(2\pi fy\right)}\right)}$ and $f\in\mathbb{N}$.
Our goal is to show that
\begin{gather}
    W_{2}(p_{X},p_{\hat{X}})\underset{f\rightarrow \infty}{\longrightarrow}0.
\end{gather}
We prove this by considering the Monge formulation of the $W_{2}$ distance, i.e.,
\begin{gather}\label{eq:monge-w2}
    W_{2}^{2}(p_{X},p_{\hat{X}})=\inf_{T}\mathbb{E}[\norm{X-T(X)}_{2}^{2}]\ s.t.\ T(X)\sim p_{\hat{X}},
\end{gather}
which is true since $p_{X}$ is absolutely continuous.
Let
\begin{gather}
    K_{f}(x_{1},x_{2})=\frac{1}{2f}\left(\lfloor 2f x_{1} \rfloor+\begin{cases}
        1-x_{2} & mod(\lfloor 2f x_{1}\rfloor,2)=1\\
        x_{2} & mod(\lfloor 2f x_{1} \rfloor,2)=0
    \end{cases}\right),\\
    T_{0}(X)=(K_{f}(X^{(1)},X^{(2)}),X^{(2)}).
\end{gather}
One can show that $T_{0}(X)\sim p_{\hat{X}}$ (see~\cref{fig:concrete-expample-intuition} for intuition), so $T_{0}$ satisfies the constraint in~\cref{eq:monge-w2} and therefore
\begin{gather}
    W_{2}^{2}(p_{X},p_{\hat{X}})\leq\mathbb{E}[\norm{X-T_{0}(X)}_{2}^{2}].
\end{gather}
Moreover, $\forall x_{1},x_{2}\in[0,1]$ we have
\begin{align}
    &\lim_{f\rightarrow\infty}(x_{1}-\frac{1}{2f}(\lfloor 2f x_{1}\rfloor-x_{2}))^{2}=0,\\
    &\lim_{f\rightarrow\infty}(x_{1}-\frac{1}{2f}(\lfloor 2f x_{1}\rfloor-1+x_{2}))^{2}=0,
\end{align}
so 
\begin{align}
    \lim_{f\rightarrow\infty}(x_{1}-K_{f}(x_{1},x_{2}))^{2}=0.
\end{align}
Since both $X^{(1)},K_{f}(X^{(1)},X^{(2)})\sim U[0,1]$, we have that $\mathbb{P}((X^{(1)}-K_{f}(X^{(1)},X^{(2)}))^{2}\leq 4)=1$.
Hence, from the dominated convergence theorem (DCT) we have
\begin{align}
    &W_{2}^{2}(p_{X},p_{\hat{X}})\leq \lim_{f\rightarrow\infty}\mathbb{E}[\norm{X-T_{0}(X)}_{2}^{2}]=\lim_{f\rightarrow\infty}\int_{[0,1]^{2}}(x_{1}-K_{f}(x_{1},x_{2}))^{2}p_{X}(x_{1},x_{2})dx_{1}dx_{2}\\
    =&\int_{[0,1]^{2}}\lim_{f\rightarrow\infty}(x_{1}-K_{f}(x_{1},x_{2}))^{2}p_{X}(x_{1},x_{2})dx_{1}dx_{2}=0.
\end{align}
We have shown that the perceptual index of the estimator $\hat{X}$ can be arbitrarily minimized (by taking larger values of $f$).
\section{Proof that~\cref{eq:stochastic-robustness-practical} upper bounds~\cref{eq:stochastic-robustness}}\label{appendix:practical-stochastic-robustness-upperbound}
\noindent Our goal is to show that
\begin{gather}
R_{\hat{X},\epsilon}\leq \tilde{R}_{\hat{X},\epsilon}.
\end{gather}
Let $\hat{X}=G(Y,Z)$ for some random variables $Y,Z$, and let
\begin{gather}
\delta^{*}=\argmax_{\delta:\norm{\delta}_{2}\leq\epsilon}W_{2}^{2}(p_{\hat{X}|Y}(\cdot|y),p_{\hat{X}|Y}(\cdot|y+\delta)).
\end{gather}
Then, for any $y$
\begin{gather}
r_{\hat{X},\epsilon}(y)=W_{2}^{2}(p_{\hat{X}|Y}(\cdot|y),p_{\hat{X}|Y}(\cdot|y+\delta^{*})).
\end{gather}
Note that the $W_{2}$ distance between $p_{\hat{X}|Y}(\cdot|y)$ and $p_{\hat{X}|Y}(\cdot|y+\delta^{*})$ is essentially the distance between the distributions of the random variables $G(y,Z)$ and $G(y+\delta^{*},Z)$.
Since the MSE between two random variables upper bounds their $W_{2}^{2}$ distance, we have
\begin{gather}
    r_{\hat{X},\epsilon}(y)\leq\mathbb{E}[\norm{G(y,Z)-G(y+\delta^{*},Z)}_{2}^{2}].
\end{gather}
By the definition of $\tilde{r}_{\hat{X},\epsilon}$ we have
\begin{gather}
    \mathbb{E}[\norm{G(y,Z)-G(y+\delta^{*},Z)}_{2}^{2}]\leq\tilde{r}_{\hat{X},\epsilon}(y),
\end{gather}
so $r_{\hat{X},\epsilon}(y)\leq\tilde{r}_{\hat{X},\epsilon}(y)$, and since this is true for any $y$, we conclude that
\begin{gather}
    R_{\hat{X},\epsilon}=\mathbb{E}\left[r_{\hat{X},\epsilon}(Y)\right]\leq\mathbb{E}\left[\tilde{r}_{\hat{X},\epsilon}(Y)\right]=\tilde{R}_{\hat{X},\epsilon}.
\end{gather}
\section{Complementary training details}
\subsection{Toy GAN experiments}\label{appendix:training-details-toy-gan-inpanting}
The sizes of the training and validation sets are 100,000 and 10,000, respectively, both of which are random, independent samples from the toy distribution described in~\cref{sec:space-filling}.
In all experiments the estimator's and critic's architecture is the simple fully connected network described in~\cref{tab:toy-archs}.
Notice that the estimator outputs only one value (the estimate of $X^{(2)}$) as the value of $X^{(1)}$ is known. The output and the input are being concatenated to result in an output of size 2, which is the final estimate. For the stochastic estimators $\hat{X}_{\text{StoGAN}}$ and $\hat{X}_{\text{ErraticStoGAN}}$ one of the inputs is $Y$ and the other is $Z\sim U[0,1]$. For the deterministic estimators we simply fix $Z=0$, like we do in the experiments on natural images. The loss we use is the exact same one we use for natural images, but with a gradient penalty coefficient of $10.0$, and $\lambda_{R}=0.1$ for the robust estimators (both the deterministic and the stochastic).
During training, we alternate between one training step for the estimator and one step for the critic, where each is trained for a total of $20,000$ steps.
To optimize the parameters of the networks we use the Adam optimizer with $(\beta_{1},\beta_{2})=(0, 0.9)$ and a learning rate of $10^{-4}$.
$\epsilon$ in the computation of~\cref{eq:stochastic-robustness-practical} is set to $10^{-3}$ in all experiments.
For the stochastic estimator, we perform the average in~\cref{eq:stochastic-sensitivity-practical} over 50 realizations of $Z$.
Finding the adversarial attack on each input in done by using the Adam optimizer for 4 steps, with $(\beta_{1},\beta_{2})=(0.9, 0.999)$ and a learning rate of $10^{-4}$.
In~\cref{tab:toy-gan-robustness-results} we report the robustness values of all toy GAN estimators.
\begin{table}[h]
    \caption{The architecture of both the estimator and critic in all toy GAN experiments.}
    \centering
        \begin{tabular}{c|c|c}
        Layer &Output size & Filter \\
        \midrule
         Fully Connected       &512 & $2\rightarrow 512$\\
         ReLU &512&-\\
         Fully Connected       &512 & $512\rightarrow 512$\\
         ReLU &512&-\\
         Fully Connected       &512 & $512\rightarrow 512$\\
         ReLU &512&-\\
         Fully Connected       &512 & $512\rightarrow 512$\\
         ReLU &512&-\\
         Fully Connected       &512 & $512\rightarrow 1$\\
    \end{tabular}
    \label{tab:toy-archs}
\end{table}
\begin{table}[h]
    \caption{Robustness of the toy GAN estimators from~\cref{sec:deterministic-gan-estimators} and~\cref{sec:robustness-of-stochastic-estimators}.}
    \centering
        \begin{tabular}{c|c|c}
        Estimator & Robustness: $\sqrt{\tilde{R}_{\hat{X},\epsilon}}(\cdot 10^{-3})$ (higher is better) & $\epsilon(\cdot 10^{-3})$ \\
        \midrule
         $\hat{X}_{\text{ErraticGAN}}$ & 3.62 & 1\\
         $\hat{X}_{\text{RobustGAN}}$ & 0.77 & 1\\
         $\hat{X}_{\text{StoErraticGAN}}$ & 0.13 & 1\\
         $\hat{X}_{\text{StoRobustGAN}}$ & 0.51 & 1
    \end{tabular}
    \label{tab:toy-gan-robustness-results}
\end{table}
\subsection{Experiments on natural images}\label{appendix:training-details-natural}
\subsubsection{Neural network architectures}
\begin{figure}
    \centering
    \includegraphics{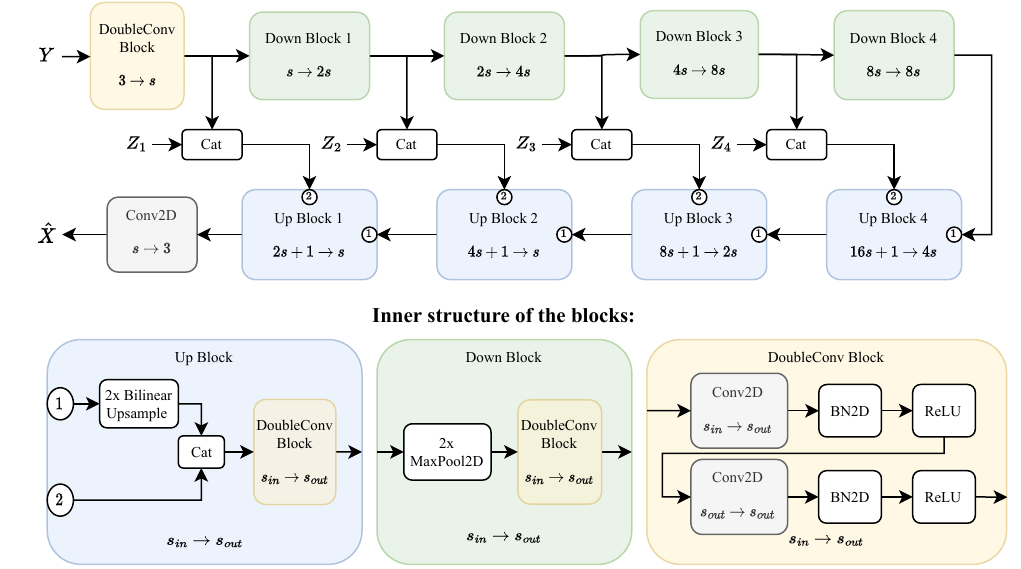}
    \caption{Description of the U-Net architecture~\cite{unet}, adopted as the estimator in all the experiments on natural images. We use $Z=(Z_{1},Z_{2},Z_{3},Z_{4})$, where for the deterministic estimators $Z_{i}=0$, and for the stochastic ones $Z_{i}$ are statistically independent random vectors, each following a normal distribution with zero mean and identity covariance matrix.
    Each $Z_{i}$ is reshaped to match the spatial size of the input to Down Block $i$.
    Cat corresponds to concatenation on the channels dimension. BN2D is a two dimensional batch normalization layer~\cite{batchnorm}. We use a ReLU activation after each BN2D layer in each DoubleConv block. There is no activation in the last Conv2D layer of the network. The only hyperparameter of the network is $s$. For the experiments on CelebA we use $s=24$ (for both the inpainting and super resolution tasks), and for the experiments on ImageNet we use $s=32$.}
    \label{fig:unet}
\end{figure}

In all the experiments the estimator is a U-Net architecture~\cite{unet} (receiving $Y$ and $Z$ as inputs), and the critic is a ResNet model~\cite{resnet} with a similar structure to the one described in~\cite{gan_converge}.
Full details of the architectures are disclosed in~\cref{fig:unet} and ~\cref{tab:critic}.
While the chosen GAN architecture for the deterministic and stochastic estimators could have been further improved, we chose the structure described herein, as it provides good quality outcomes and is relatively easy to train. We believe that, while improving the GAN training and architecture may boost performance, it would not change the interplay we exposed between deterministic and stochastic restoration methods. 
\begin{table}[h]
    \caption{Full description of the critic's architecture used in all the experiments on natural images. We use pre-activation ResNet blocks and Leaky ReLU activations with a slope of $0.2$ everywhere (except for the last layer, where there is no activation). The output of each ResNet block is multiplied by $0.1$. The only hyperparameter of the network is $f$, and the spatial extent of the input image is always $64\times 64$ in our experiments. $f=16$ in all the CelebA experiments, and $f=32$ in all ImageNet experiments. This architecture follows a similar structure to the one described in~\cite{gan_converge}.}
    \centering
    \begin{tabular}{c|c|c}
        Layer &Output size & Filter \\
        \midrule
         Conv2D       &$f\times 64\times 64$ & $3\rightarrow f$\\
         \midrule
         ResNet Block &$f\times 64\times 64$ &  $f\rightarrow f\rightarrow f$\\
         ResNet Block &$2f\times 64\times 64$ & $f\rightarrow f\rightarrow 2f$\\
         Average Pooling 2D &$2f\times 32\times 32$& -\\
         \midrule
         ResNet Block &$2f\times 32\times 32$ & $2f\rightarrow 2f\rightarrow 2f$\\
         ResNet Block &$4f\times 32\times 32$ & $2f\rightarrow 2f\rightarrow 4f$\\
         Average Pooling 2D &$4f\times 16\times 16$& -\\
         \midrule
         ResNet Block &$4f\times 16\times 16$ & $4f\rightarrow 4f\rightarrow 4f$\\
         ResNet Block &$8f\times 16\times 16$ & $4f\rightarrow 4f\rightarrow 8f$\\
         Average Pooling 2D &$8f\times 8\times 8$& -\\
         \midrule
         ResNet Block &$8f\times 8\times 8$ & $8f\rightarrow 8f\rightarrow 8f$\\
         ResNet Block &$16f\times 8\times 8$ &$8f\rightarrow 8f\rightarrow 16f$\\
         Average Pooling 2D &$16f\times 4\times 4$& -\\
         \midrule
          ResNet Block &$16f\times 4\times 4$ &$16f\rightarrow 16f\rightarrow 16f$\\
         ResNet Block &$16f\times 4\times 4$ & $16f\rightarrow 16f\rightarrow 16f$\\
         Average Pooling 2D &$16f\times 2\times 2$& -\\
         \midrule
         ResNet Block &$16f\times 2\times 2$ & $16f\rightarrow 16f\rightarrow 16f$\\
         ResNet Block &$16f\times 2\times 2$ & $16f\rightarrow 16f\rightarrow 16f$\\
         Fully Connected & 1 & $16f\cdot 2\cdot 2\rightarrow 1$\\
    \end{tabular}
    \label{tab:critic}
\end{table}
\subsubsection{Optimization settings}
We alternate between optimizing the estimator and the critic (one step for each model at a time), performing a total of 1.2M steps for each model.
At the estimator's optimization step we also perform an ``inner'' optimization procedure that finds the adversarial attack of each input, as described in~\cref{appendix:practical-comp}.
The robustness loss (\cref{eq:stochastic-robustness-practical}) is included in the estimator's objective once in every three optimization steps, i.e., we always perform two estimator training steps solely with a GAN loss, and the robustness loss is added at the third step when training a robust model.
For the optimization of the estimator' and critic's parameters, we always use the Adam optimizer with $(\beta_{1},\beta_{2})=(0.5,0.99)$ and a learning rate of $10^{-4}$.
For both models we also perform a multi-step learning rate scheduling with a decay of $\gamma=0.6$, scheduled at the steps $400k, 500k, 600k, 700k$ and $750k$.
The $R_{1}$ gradient penalty coefficient of the critic is set to 1.
\subsubsection{Computation of the robustness loss (\cref{eq:stochastic-robustness-practical})}\label{appendix:practical-comp}
Note that measuring the robustness of an estimator with~\cref{eq:stochastic-robustness-practical} involves a maximization procedure for each $y$ (\cref{eq:stochastic-sensitivity-practical}).
We approximate this maximization with 5 optimization steps using the Adam optimizer~\cite{adam} with a learning rate of $1$ and $(\beta_{1},\beta_{2})=(0.9,0.999)$.
The optimizer's parameters are re-initialized at the beginning of each training step (so its inner parameters are not transferred from step to step).
The same procedure with the same hyper-parameters is done both for training and validation.

Another thing to note is that computing~\cref{eq:stochastic-sensitivity-practical} for each $y$ requires averaging over samples of $Z$.
For the stochastic estimators we sample 10 instances of $Z$ to compute this average, and for the deterministic estimators averaging is not required since $Z=0$.

Lastly, in the inpainting task we use $\epsilon=2.5$, and in the super resolution task we use $\epsilon=0.553,0.276,0.138$ for the scaling factors $\times 4,\times 8,\times 16$, respectively. These values of $\epsilon$ ensure the same bound on the adversarial input PSNR across all scaling factors.
To ensure that the attack $\delta$ holds $\norm{\delta}_{2}\leq\epsilon$, we project $\delta$ on the $\epsilon$ ball after each update step of the attack.
\subsubsection{Data sets}\label{appendix:data-sets}
\textbf{CelebA:} We pre-process each source image by first resizing it to 102x102 via bilinear-interpolation and then cropping the center 64x64 pixels, so the source images we use are of size 64x64.
We use 80\% of the original CelebA training data to train all the algorithms (the images in the resulting subset are being chosen at random), and the remaining 20\% is added to the original CelebA validation data.
We do so to have enough images for perceptual quality evaluation using metrics such as FID~\cite{fid}.

\noindent\textbf{ImageNet:} We use the $64\times 64$ version of the ImageNet data set~\cite{imagenet_small} with no pre-processing, and with the original training and validation sets splits.
\subsubsection{Computation of perceptual quality metrics}
We compute FID~\cite{fid}, precision and recall~\cite{improved_precision_recall} (with $k=3$) using the PyTorch package~\cite{piq}, and with the default feature extraction network of that package (InceptionV3~\cite{inceptionv3}).
We randomly choose 50,000 real and fake samples to perform the evaluation (the real samples are the high quality training images, and the fake samples are the outputs of the algorithm on the validation inputs).
For the stochastic algorithms each fake sample corresponds to a different input, i.e., we sample 50,000 random seeds and each seed is combined with a different input to provide one output, effectively sampling one instance from $p_{\hat{X}|Y}(\cdot|y)$ for each $y$.
\section{What is the rationale of utilizing GANs to compare stochastic and deterministic restoration models?}\label{appendix:why-not-diffusion}
Among image restoration methods, GAN-based~\cite{esrgan,srgan,explorable_sr,pscgan}, flow-based~\cite{lugmayr2020srflow,srflow-da}, diffusion-based~\cite{song-generative, Kadkhodaie, bahjat-posterior,kawar2021snips,kawar2022denoising}, and VAE-based~\cite{Gatopoulos,Liu2021-b,Liu2021} techniques are currently the most widely adopted and effective in terms of achieving high levels of perceptual quality.
However, in order to provide sufficient evidence for our hypothesis in this paper, which posits that stochastic models can achieve high levels of perceptual quality with higher levels of robustness compared to deterministic models, it is essential to conduct a fair and unbiased comparison.
For example, a stochastic flow-based method that achieves high levels of perceptual quality may exhibit greater robustness compared to a deterministic GAN-based technique with similar levels of perceptual quality (as demonstrated in~\cref{appendix:srflow-robustness}). However, this does not necessarily imply that the superior performance of the flow-based method is solely attributed to its stochasticity.
Rather, it shows that a restoration algorithm with high levels of perceptual quality can also exhibit robustness, which refutes the hypothesis proposed in~\cite{estimators_robustness}.
Hence, to conduct a fair evaluation, it is crucial to isolate the impact of stochasticity on the model's performance.
This can be achieved by using GANs, as they naturally allow to perform a comparison ``on the same grounds'' (same data sets, same loss, same architecture, same AI-PSNR, same hyper-parameters, etc.) by manipulating the input random seed from a degenerate distribution (to obtain a deterministic model) to a non-degenerate distribution (to obtain a stochastic model).
Such an effective way to isolate the impact of output stochasticiy on the model's performance is currently much less trivial with flow-based, diffusion-based, and VAE-based methods.
We therefore leave the analysis of these methods for future work.
\section{Robustness of SRFlow: results}\label{appendix:srflow-robustness}

We assess the robustness of SRFlow~\cite{lugmayr2020srflow} by adapting the I-FGSM~\cite{estimators_robustness, kurakin2017adversarial} method to be applicable to any differentiable estimation procedure that utilizes a random seed as input.
This adaptation enables a fair comparison of stochastic methods with the I-FGSM attacks performed on other methods~\cite{estimators_robustness}.

To find the input attack $\tilde{y}$ on the SRFlow model (denoted by $\text{SRF}(Y,Z)$), we compute the loss $\mathbb{E}[\norm{\text{SRF}(y,Z)-\text{SRF}(\tilde{y},Z)}_{2}]$ for each $Y=y$, averaging over 10 realizations of $Z$ (the random seed of SRFlow), and then continue to perform 50 I-FGSM update steps in the same fashion as described in~\cite{estimators_robustness}, Sec. 3.1.
To ensure a fair comparison with other models, we adhere to the official public implementation of the I-FGSM procedure described in~\cite{estimators_robustness}, and use the same initialization for the attack.
Additionally, we utilize the official code and pre-trained model checkpoints of SRFlow as provided by the authors.
Lastly, as in~\cite{estimators_robustness}, we perform the evaluation on the Set5~\cite{set5}, Set14~\cite{set14}, and BSD100~\cite{bsd100} data sets.

The robustness evaluation of SRFlow is presented in~\cref{fig:srflow-robustness}. 
For ease of comparison, we also include the robustness levels of ESRGAN~\cite{esrgan} and EDSR~\cite{edsr} as reported in~\cite{estimators_robustness} (refer to~\cite{estimators_robustness} for the analysis of additional distortion-based models).
Our results show that the robustness levels of SRFlow are significantly higher than those of ESRGAN across all values of $\alpha$, the hyper-parameter controlling the severity of the I-FGSM attack.
This confirms that even an ``off the shelf'' stochastic restoration algorithm with high perceptual quality can exhibit much higher robustness than a deterministic algorithm with comparable output perceptual quality.
Furthermore, the results show that SRFlow exhibits comparable level of robustness to EDSR and most of the distortion-based models evaluated in~\cite{estimators_robustness}.
I.e., despite achieving high perceptual quality, SRFlow maintains a robustness level that is comparable to models that attain much lower levels of perceptual quality.
All and all, this experiment further supports our claim that the link between high perceptual quality and low robustness~\cite{estimators_robustness} is valid only for deterministic mappings.

\begin{figure}[h]
     \begin{subfigure}{0.5\textwidth}
         \centering
         \input{latex/figures/adv_set5.pgf}
         \caption{Set5 adversarial output PSNR.}
     \end{subfigure}
     \hfil
     \begin{subfigure}{0.5\textwidth}
         \centering
         \input{latex/figures/inp_set5.pgf}
         \caption{Set5 adversarial input PSNR.}
     \end{subfigure}

     \begin{subfigure}{0.5\textwidth}
         \centering
         \input{latex/figures/adv_set14.pgf}
         \caption{Set14 adversarial output PSNR.}
     \end{subfigure}
     \hfil
     \begin{subfigure}{0.5\textwidth}
         \centering
         \input{latex/figures/inp_set14.pgf}
         \caption{Set14 adversarial input PSNR.}
     \end{subfigure}

      \begin{subfigure}{0.5\textwidth}
         \centering
         \input{latex/figures/adv_bsd.pgf}
         \caption{BSD100 adversarial output PSNR.}
     \end{subfigure}
     \hfil
     \begin{subfigure}{0.5\textwidth}
         \centering
         \input{latex/figures/inp_bsd.pgf}
         \caption{BSD100 adversarial input PSNR.}
     \end{subfigure}
        \caption{Comparison of adversarial output PSNR (robustness) and adversarial input PSNR (AI-PSNR) on Set5~\cite{set5}, Set14~\cite{set14}, and BSD100~\cite{bsd100}, using the I-FGSM attack~\cite{estimators_robustness,kurakin2017adversarial} with several values of $\alpha$. The evaluated methods are SRFlow~\cite{lugmayr2020srflow}, EDSR~\cite{edsr}, and ESRGAN~\cite{esrgan}. The results of EDSR and ESRGAN are copied from~\cite{estimators_robustness}, whereas SRFlow is evaluated as described in~\cref{appendix:srflow-robustness}.}
        \label{fig:srflow-robustness}
\end{figure}

%% file: latex/figures/adv_set5.pgf
\begingroup%
\makeatletter%
\begin{pgfpicture}%
\pgfpathrectangle{\pgfpointorigin}{\pgfqpoint{3.300000in}{2.600000in}}%
\pgfusepath{use as bounding box, clip}%
\begin{pgfscope}%
\pgfsetbuttcap%
\pgfsetmiterjoin%
\definecolor{currentfill}{rgb}{1.000000,1.000000,1.000000}%
\pgfsetfillcolor{currentfill}%
\pgfsetlinewidth{0.000000pt}%
\definecolor{currentstroke}{rgb}{1.000000,1.000000,1.000000}%
\pgfsetstrokecolor{currentstroke}%
\pgfsetdash{}{0pt}%
\pgfpathmoveto{\pgfqpoint{0.000000in}{0.000000in}}%
\pgfpathlineto{\pgfqpoint{3.300000in}{0.000000in}}%
\pgfpathlineto{\pgfqpoint{3.300000in}{2.600000in}}%
\pgfpathlineto{\pgfqpoint{0.000000in}{2.600000in}}%
\pgfpathlineto{\pgfqpoint{0.000000in}{0.000000in}}%
\pgfpathclose%
\pgfusepath{fill}%
\end{pgfscope}%
\begin{pgfscope}%
\pgfsetbuttcap%
\pgfsetmiterjoin%
\definecolor{currentfill}{rgb}{1.000000,1.000000,1.000000}%
\pgfsetfillcolor{currentfill}%
\pgfsetlinewidth{0.000000pt}%
\definecolor{currentstroke}{rgb}{0.000000,0.000000,0.000000}%
\pgfsetstrokecolor{currentstroke}%
\pgfsetstrokeopacity{0.000000}%
\pgfsetdash{}{0pt}%
\pgfpathmoveto{\pgfqpoint{0.527360in}{0.541249in}}%
\pgfpathlineto{\pgfqpoint{3.165000in}{0.541249in}}%
\pgfpathlineto{\pgfqpoint{3.165000in}{2.465000in}}%
\pgfpathlineto{\pgfqpoint{0.527360in}{2.465000in}}%
\pgfpathlineto{\pgfqpoint{0.527360in}{0.541249in}}%
\pgfpathclose%
\pgfusepath{fill}%
\end{pgfscope}%
\begin{pgfscope}%
\pgfsetbuttcap%
\pgfsetroundjoin%
\definecolor{currentfill}{rgb}{0.000000,0.000000,0.000000}%
\pgfsetfillcolor{currentfill}%
\pgfsetlinewidth{0.803000pt}%
\definecolor{currentstroke}{rgb}{0.000000,0.000000,0.000000}%
\pgfsetstrokecolor{currentstroke}%
\pgfsetdash{}{0pt}%
\pgfsys@defobject{currentmarker}{\pgfqpoint{0.000000in}{-0.048611in}}{\pgfqpoint{0.000000in}{0.000000in}}{%
\pgfpathmoveto{\pgfqpoint{0.000000in}{0.000000in}}%
\pgfpathlineto{\pgfqpoint{0.000000in}{-0.048611in}}%
\pgfusepath{stroke,fill}%
}%
\begin{pgfscope}%
\pgfsys@transformshift{0.747164in}{0.541249in}%
\pgfsys@useobject{currentmarker}{}%
\end{pgfscope}%
\end{pgfscope}%
\begin{pgfscope}%
\definecolor{textcolor}{rgb}{0.000000,0.000000,0.000000}%
\pgfsetstrokecolor{textcolor}%
\pgfsetfillcolor{textcolor}%
\pgftext[x=0.778414in, y=0.379791in, left, base,rotate=90.000000]{\color{textcolor}\rmfamily\fontsize{9.000000}{10.800000}\selectfont 1}%
\end{pgfscope}%
\begin{pgfscope}%
\pgfsetbuttcap%
\pgfsetroundjoin%
\definecolor{currentfill}{rgb}{0.000000,0.000000,0.000000}%
\pgfsetfillcolor{currentfill}%
\pgfsetlinewidth{0.803000pt}%
\definecolor{currentstroke}{rgb}{0.000000,0.000000,0.000000}%
\pgfsetstrokecolor{currentstroke}%
\pgfsetdash{}{0pt}%
\pgfsys@defobject{currentmarker}{\pgfqpoint{0.000000in}{-0.048611in}}{\pgfqpoint{0.000000in}{0.000000in}}{%
\pgfpathmoveto{\pgfqpoint{0.000000in}{0.000000in}}%
\pgfpathlineto{\pgfqpoint{0.000000in}{-0.048611in}}%
\pgfusepath{stroke,fill}%
}%
\begin{pgfscope}%
\pgfsys@transformshift{1.186770in}{0.541249in}%
\pgfsys@useobject{currentmarker}{}%
\end{pgfscope}%
\end{pgfscope}%
\begin{pgfscope}%
\definecolor{textcolor}{rgb}{0.000000,0.000000,0.000000}%
\pgfsetstrokecolor{textcolor}%
\pgfsetfillcolor{textcolor}%
\pgftext[x=1.218020in, y=0.379791in, left, base,rotate=90.000000]{\color{textcolor}\rmfamily\fontsize{9.000000}{10.800000}\selectfont 2}%
\end{pgfscope}%
\begin{pgfscope}%
\pgfsetbuttcap%
\pgfsetroundjoin%
\definecolor{currentfill}{rgb}{0.000000,0.000000,0.000000}%
\pgfsetfillcolor{currentfill}%
\pgfsetlinewidth{0.803000pt}%
\definecolor{currentstroke}{rgb}{0.000000,0.000000,0.000000}%
\pgfsetstrokecolor{currentstroke}%
\pgfsetdash{}{0pt}%
\pgfsys@defobject{currentmarker}{\pgfqpoint{0.000000in}{-0.048611in}}{\pgfqpoint{0.000000in}{0.000000in}}{%
\pgfpathmoveto{\pgfqpoint{0.000000in}{0.000000in}}%
\pgfpathlineto{\pgfqpoint{0.000000in}{-0.048611in}}%
\pgfusepath{stroke,fill}%
}%
\begin{pgfscope}%
\pgfsys@transformshift{1.626377in}{0.541249in}%
\pgfsys@useobject{currentmarker}{}%
\end{pgfscope}%
\end{pgfscope}%
\begin{pgfscope}%
\definecolor{textcolor}{rgb}{0.000000,0.000000,0.000000}%
\pgfsetstrokecolor{textcolor}%
\pgfsetfillcolor{textcolor}%
\pgftext[x=1.657627in, y=0.379791in, left, base,rotate=90.000000]{\color{textcolor}\rmfamily\fontsize{9.000000}{10.800000}\selectfont 4}%
\end{pgfscope}%
\begin{pgfscope}%
\pgfsetbuttcap%
\pgfsetroundjoin%
\definecolor{currentfill}{rgb}{0.000000,0.000000,0.000000}%
\pgfsetfillcolor{currentfill}%
\pgfsetlinewidth{0.803000pt}%
\definecolor{currentstroke}{rgb}{0.000000,0.000000,0.000000}%
\pgfsetstrokecolor{currentstroke}%
\pgfsetdash{}{0pt}%
\pgfsys@defobject{currentmarker}{\pgfqpoint{0.000000in}{-0.048611in}}{\pgfqpoint{0.000000in}{0.000000in}}{%
\pgfpathmoveto{\pgfqpoint{0.000000in}{0.000000in}}%
\pgfpathlineto{\pgfqpoint{0.000000in}{-0.048611in}}%
\pgfusepath{stroke,fill}%
}%
\begin{pgfscope}%
\pgfsys@transformshift{2.065983in}{0.541249in}%
\pgfsys@useobject{currentmarker}{}%
\end{pgfscope}%
\end{pgfscope}%
\begin{pgfscope}%
\definecolor{textcolor}{rgb}{0.000000,0.000000,0.000000}%
\pgfsetstrokecolor{textcolor}%
\pgfsetfillcolor{textcolor}%
\pgftext[x=2.097233in, y=0.379791in, left, base,rotate=90.000000]{\color{textcolor}\rmfamily\fontsize{9.000000}{10.800000}\selectfont 8}%
\end{pgfscope}%
\begin{pgfscope}%
\pgfsetbuttcap%
\pgfsetroundjoin%
\definecolor{currentfill}{rgb}{0.000000,0.000000,0.000000}%
\pgfsetfillcolor{currentfill}%
\pgfsetlinewidth{0.803000pt}%
\definecolor{currentstroke}{rgb}{0.000000,0.000000,0.000000}%
\pgfsetstrokecolor{currentstroke}%
\pgfsetdash{}{0pt}%
\pgfsys@defobject{currentmarker}{\pgfqpoint{0.000000in}{-0.048611in}}{\pgfqpoint{0.000000in}{0.000000in}}{%
\pgfpathmoveto{\pgfqpoint{0.000000in}{0.000000in}}%
\pgfpathlineto{\pgfqpoint{0.000000in}{-0.048611in}}%
\pgfusepath{stroke,fill}%
}%
\begin{pgfscope}%
\pgfsys@transformshift{2.505590in}{0.541249in}%
\pgfsys@useobject{currentmarker}{}%
\end{pgfscope}%
\end{pgfscope}%
\begin{pgfscope}%
\definecolor{textcolor}{rgb}{0.000000,0.000000,0.000000}%
\pgfsetstrokecolor{textcolor}%
\pgfsetfillcolor{textcolor}%
\pgftext[x=2.536840in, y=0.315556in, left, base,rotate=90.000000]{\color{textcolor}\rmfamily\fontsize{9.000000}{10.800000}\selectfont 16}%
\end{pgfscope}%
\begin{pgfscope}%
\pgfsetbuttcap%
\pgfsetroundjoin%
\definecolor{currentfill}{rgb}{0.000000,0.000000,0.000000}%
\pgfsetfillcolor{currentfill}%
\pgfsetlinewidth{0.803000pt}%
\definecolor{currentstroke}{rgb}{0.000000,0.000000,0.000000}%
\pgfsetstrokecolor{currentstroke}%
\pgfsetdash{}{0pt}%
\pgfsys@defobject{currentmarker}{\pgfqpoint{0.000000in}{-0.048611in}}{\pgfqpoint{0.000000in}{0.000000in}}{%
\pgfpathmoveto{\pgfqpoint{0.000000in}{0.000000in}}%
\pgfpathlineto{\pgfqpoint{0.000000in}{-0.048611in}}%
\pgfusepath{stroke,fill}%
}%
\begin{pgfscope}%
\pgfsys@transformshift{2.945197in}{0.541249in}%
\pgfsys@useobject{currentmarker}{}%
\end{pgfscope}%
\end{pgfscope}%
\begin{pgfscope}%
\definecolor{textcolor}{rgb}{0.000000,0.000000,0.000000}%
\pgfsetstrokecolor{textcolor}%
\pgfsetfillcolor{textcolor}%
\pgftext[x=2.976447in, y=0.315556in, left, base,rotate=90.000000]{\color{textcolor}\rmfamily\fontsize{9.000000}{10.800000}\selectfont 32}%
\end{pgfscope}%
\begin{pgfscope}%
\definecolor{textcolor}{rgb}{0.000000,0.000000,0.000000}%
\pgfsetstrokecolor{textcolor}%
\pgfsetfillcolor{textcolor}%
\pgftext[x=1.846180in,y=0.260000in,,top]{\color{textcolor}\rmfamily\fontsize{9.000000}{10.800000}\selectfont \(\displaystyle \alpha(/255)\)}%
\end{pgfscope}%
\begin{pgfscope}%
\pgfpathrectangle{\pgfqpoint{0.527360in}{0.541249in}}{\pgfqpoint{2.637640in}{1.923751in}}%
\pgfusepath{clip}%
\pgfsetrectcap%
\pgfsetroundjoin%
\pgfsetlinewidth{0.803000pt}%
\definecolor{currentstroke}{rgb}{0.690196,0.690196,0.690196}%
\pgfsetstrokecolor{currentstroke}%
\pgfsetdash{}{0pt}%
\pgfpathmoveto{\pgfqpoint{0.527360in}{0.541249in}}%
\pgfpathlineto{\pgfqpoint{3.165000in}{0.541249in}}%
\pgfusepath{stroke}%
\end{pgfscope}%
\begin{pgfscope}%
\pgfsetbuttcap%
\pgfsetroundjoin%
\definecolor{currentfill}{rgb}{0.000000,0.000000,0.000000}%
\pgfsetfillcolor{currentfill}%
\pgfsetlinewidth{0.803000pt}%
\definecolor{currentstroke}{rgb}{0.000000,0.000000,0.000000}%
\pgfsetstrokecolor{currentstroke}%
\pgfsetdash{}{0pt}%
\pgfsys@defobject{currentmarker}{\pgfqpoint{-0.048611in}{0.000000in}}{\pgfqpoint{-0.000000in}{0.000000in}}{%
\pgfpathmoveto{\pgfqpoint{-0.000000in}{0.000000in}}%
\pgfpathlineto{\pgfqpoint{-0.048611in}{0.000000in}}%
\pgfusepath{stroke,fill}%
}%
\begin{pgfscope}%
\pgfsys@transformshift{0.527360in}{0.541249in}%
\pgfsys@useobject{currentmarker}{}%
\end{pgfscope}%
\end{pgfscope}%
\begin{pgfscope}%
\definecolor{textcolor}{rgb}{0.000000,0.000000,0.000000}%
\pgfsetstrokecolor{textcolor}%
\pgfsetfillcolor{textcolor}%
\pgftext[x=0.365902in, y=0.497846in, left, base]{\color{textcolor}\rmfamily\fontsize{9.000000}{10.800000}\selectfont \(\displaystyle {0}\)}%
\end{pgfscope}%
\begin{pgfscope}%
\pgfpathrectangle{\pgfqpoint{0.527360in}{0.541249in}}{\pgfqpoint{2.637640in}{1.923751in}}%
\pgfusepath{clip}%
\pgfsetrectcap%
\pgfsetroundjoin%
\pgfsetlinewidth{0.803000pt}%
\definecolor{currentstroke}{rgb}{0.690196,0.690196,0.690196}%
\pgfsetstrokecolor{currentstroke}%
\pgfsetdash{}{0pt}%
\pgfpathmoveto{\pgfqpoint{0.527360in}{0.861874in}}%
\pgfpathlineto{\pgfqpoint{3.165000in}{0.861874in}}%
\pgfusepath{stroke}%
\end{pgfscope}%
\begin{pgfscope}%
\pgfsetbuttcap%
\pgfsetroundjoin%
\definecolor{currentfill}{rgb}{0.000000,0.000000,0.000000}%
\pgfsetfillcolor{currentfill}%
\pgfsetlinewidth{0.803000pt}%
\definecolor{currentstroke}{rgb}{0.000000,0.000000,0.000000}%
\pgfsetstrokecolor{currentstroke}%
\pgfsetdash{}{0pt}%
\pgfsys@defobject{currentmarker}{\pgfqpoint{-0.048611in}{0.000000in}}{\pgfqpoint{-0.000000in}{0.000000in}}{%
\pgfpathmoveto{\pgfqpoint{-0.000000in}{0.000000in}}%
\pgfpathlineto{\pgfqpoint{-0.048611in}{0.000000in}}%
\pgfusepath{stroke,fill}%
}%
\begin{pgfscope}%
\pgfsys@transformshift{0.527360in}{0.861874in}%
\pgfsys@useobject{currentmarker}{}%
\end{pgfscope}%
\end{pgfscope}%
\begin{pgfscope}%
\definecolor{textcolor}{rgb}{0.000000,0.000000,0.000000}%
\pgfsetstrokecolor{textcolor}%
\pgfsetfillcolor{textcolor}%
\pgftext[x=0.301667in, y=0.818472in, left, base]{\color{textcolor}\rmfamily\fontsize{9.000000}{10.800000}\selectfont \(\displaystyle {10}\)}%
\end{pgfscope}%
\begin{pgfscope}%
\pgfpathrectangle{\pgfqpoint{0.527360in}{0.541249in}}{\pgfqpoint{2.637640in}{1.923751in}}%
\pgfusepath{clip}%
\pgfsetrectcap%
\pgfsetroundjoin%
\pgfsetlinewidth{0.803000pt}%
\definecolor{currentstroke}{rgb}{0.690196,0.690196,0.690196}%
\pgfsetstrokecolor{currentstroke}%
\pgfsetdash{}{0pt}%
\pgfpathmoveto{\pgfqpoint{0.527360in}{1.182499in}}%
\pgfpathlineto{\pgfqpoint{3.165000in}{1.182499in}}%
\pgfusepath{stroke}%
\end{pgfscope}%
\begin{pgfscope}%
\pgfsetbuttcap%
\pgfsetroundjoin%
\definecolor{currentfill}{rgb}{0.000000,0.000000,0.000000}%
\pgfsetfillcolor{currentfill}%
\pgfsetlinewidth{0.803000pt}%
\definecolor{currentstroke}{rgb}{0.000000,0.000000,0.000000}%
\pgfsetstrokecolor{currentstroke}%
\pgfsetdash{}{0pt}%
\pgfsys@defobject{currentmarker}{\pgfqpoint{-0.048611in}{0.000000in}}{\pgfqpoint{-0.000000in}{0.000000in}}{%
\pgfpathmoveto{\pgfqpoint{-0.000000in}{0.000000in}}%
\pgfpathlineto{\pgfqpoint{-0.048611in}{0.000000in}}%
\pgfusepath{stroke,fill}%
}%
\begin{pgfscope}%
\pgfsys@transformshift{0.527360in}{1.182499in}%
\pgfsys@useobject{currentmarker}{}%
\end{pgfscope}%
\end{pgfscope}%
\begin{pgfscope}%
\definecolor{textcolor}{rgb}{0.000000,0.000000,0.000000}%
\pgfsetstrokecolor{textcolor}%
\pgfsetfillcolor{textcolor}%
\pgftext[x=0.301667in, y=1.139097in, left, base]{\color{textcolor}\rmfamily\fontsize{9.000000}{10.800000}\selectfont \(\displaystyle {20}\)}%
\end{pgfscope}%
\begin{pgfscope}%
\pgfpathrectangle{\pgfqpoint{0.527360in}{0.541249in}}{\pgfqpoint{2.637640in}{1.923751in}}%
\pgfusepath{clip}%
\pgfsetrectcap%
\pgfsetroundjoin%
\pgfsetlinewidth{0.803000pt}%
\definecolor{currentstroke}{rgb}{0.690196,0.690196,0.690196}%
\pgfsetstrokecolor{currentstroke}%
\pgfsetdash{}{0pt}%
\pgfpathmoveto{\pgfqpoint{0.527360in}{1.503125in}}%
\pgfpathlineto{\pgfqpoint{3.165000in}{1.503125in}}%
\pgfusepath{stroke}%
\end{pgfscope}%
\begin{pgfscope}%
\pgfsetbuttcap%
\pgfsetroundjoin%
\definecolor{currentfill}{rgb}{0.000000,0.000000,0.000000}%
\pgfsetfillcolor{currentfill}%
\pgfsetlinewidth{0.803000pt}%
\definecolor{currentstroke}{rgb}{0.000000,0.000000,0.000000}%
\pgfsetstrokecolor{currentstroke}%
\pgfsetdash{}{0pt}%
\pgfsys@defobject{currentmarker}{\pgfqpoint{-0.048611in}{0.000000in}}{\pgfqpoint{-0.000000in}{0.000000in}}{%
\pgfpathmoveto{\pgfqpoint{-0.000000in}{0.000000in}}%
\pgfpathlineto{\pgfqpoint{-0.048611in}{0.000000in}}%
\pgfusepath{stroke,fill}%
}%
\begin{pgfscope}%
\pgfsys@transformshift{0.527360in}{1.503125in}%
\pgfsys@useobject{currentmarker}{}%
\end{pgfscope}%
\end{pgfscope}%
\begin{pgfscope}%
\definecolor{textcolor}{rgb}{0.000000,0.000000,0.000000}%
\pgfsetstrokecolor{textcolor}%
\pgfsetfillcolor{textcolor}%
\pgftext[x=0.301667in, y=1.459722in, left, base]{\color{textcolor}\rmfamily\fontsize{9.000000}{10.800000}\selectfont \(\displaystyle {30}\)}%
\end{pgfscope}%
\begin{pgfscope}%
\pgfpathrectangle{\pgfqpoint{0.527360in}{0.541249in}}{\pgfqpoint{2.637640in}{1.923751in}}%
\pgfusepath{clip}%
\pgfsetrectcap%
\pgfsetroundjoin%
\pgfsetlinewidth{0.803000pt}%
\definecolor{currentstroke}{rgb}{0.690196,0.690196,0.690196}%
\pgfsetstrokecolor{currentstroke}%
\pgfsetdash{}{0pt}%
\pgfpathmoveto{\pgfqpoint{0.527360in}{1.823750in}}%
\pgfpathlineto{\pgfqpoint{3.165000in}{1.823750in}}%
\pgfusepath{stroke}%
\end{pgfscope}%
\begin{pgfscope}%
\pgfsetbuttcap%
\pgfsetroundjoin%
\definecolor{currentfill}{rgb}{0.000000,0.000000,0.000000}%
\pgfsetfillcolor{currentfill}%
\pgfsetlinewidth{0.803000pt}%
\definecolor{currentstroke}{rgb}{0.000000,0.000000,0.000000}%
\pgfsetstrokecolor{currentstroke}%
\pgfsetdash{}{0pt}%
\pgfsys@defobject{currentmarker}{\pgfqpoint{-0.048611in}{0.000000in}}{\pgfqpoint{-0.000000in}{0.000000in}}{%
\pgfpathmoveto{\pgfqpoint{-0.000000in}{0.000000in}}%
\pgfpathlineto{\pgfqpoint{-0.048611in}{0.000000in}}%
\pgfusepath{stroke,fill}%
}%
\begin{pgfscope}%
\pgfsys@transformshift{0.527360in}{1.823750in}%
\pgfsys@useobject{currentmarker}{}%
\end{pgfscope}%
\end{pgfscope}%
\begin{pgfscope}%
\definecolor{textcolor}{rgb}{0.000000,0.000000,0.000000}%
\pgfsetstrokecolor{textcolor}%
\pgfsetfillcolor{textcolor}%
\pgftext[x=0.301667in, y=1.780347in, left, base]{\color{textcolor}\rmfamily\fontsize{9.000000}{10.800000}\selectfont \(\displaystyle {40}\)}%
\end{pgfscope}%
\begin{pgfscope}%
\pgfpathrectangle{\pgfqpoint{0.527360in}{0.541249in}}{\pgfqpoint{2.637640in}{1.923751in}}%
\pgfusepath{clip}%
\pgfsetrectcap%
\pgfsetroundjoin%
\pgfsetlinewidth{0.803000pt}%
\definecolor{currentstroke}{rgb}{0.690196,0.690196,0.690196}%
\pgfsetstrokecolor{currentstroke}%
\pgfsetdash{}{0pt}%
\pgfpathmoveto{\pgfqpoint{0.527360in}{2.144375in}}%
\pgfpathlineto{\pgfqpoint{3.165000in}{2.144375in}}%
\pgfusepath{stroke}%
\end{pgfscope}%
\begin{pgfscope}%
\pgfsetbuttcap%
\pgfsetroundjoin%
\definecolor{currentfill}{rgb}{0.000000,0.000000,0.000000}%
\pgfsetfillcolor{currentfill}%
\pgfsetlinewidth{0.803000pt}%
\definecolor{currentstroke}{rgb}{0.000000,0.000000,0.000000}%
\pgfsetstrokecolor{currentstroke}%
\pgfsetdash{}{0pt}%
\pgfsys@defobject{currentmarker}{\pgfqpoint{-0.048611in}{0.000000in}}{\pgfqpoint{-0.000000in}{0.000000in}}{%
\pgfpathmoveto{\pgfqpoint{-0.000000in}{0.000000in}}%
\pgfpathlineto{\pgfqpoint{-0.048611in}{0.000000in}}%
\pgfusepath{stroke,fill}%
}%
\begin{pgfscope}%
\pgfsys@transformshift{0.527360in}{2.144375in}%
\pgfsys@useobject{currentmarker}{}%
\end{pgfscope}%
\end{pgfscope}%
\begin{pgfscope}%
\definecolor{textcolor}{rgb}{0.000000,0.000000,0.000000}%
\pgfsetstrokecolor{textcolor}%
\pgfsetfillcolor{textcolor}%
\pgftext[x=0.301667in, y=2.100972in, left, base]{\color{textcolor}\rmfamily\fontsize{9.000000}{10.800000}\selectfont \(\displaystyle {50}\)}%
\end{pgfscope}%
\begin{pgfscope}%
\pgfpathrectangle{\pgfqpoint{0.527360in}{0.541249in}}{\pgfqpoint{2.637640in}{1.923751in}}%
\pgfusepath{clip}%
\pgfsetrectcap%
\pgfsetroundjoin%
\pgfsetlinewidth{0.803000pt}%
\definecolor{currentstroke}{rgb}{0.690196,0.690196,0.690196}%
\pgfsetstrokecolor{currentstroke}%
\pgfsetdash{}{0pt}%
\pgfpathmoveto{\pgfqpoint{0.527360in}{2.465000in}}%
\pgfpathlineto{\pgfqpoint{3.165000in}{2.465000in}}%
\pgfusepath{stroke}%
\end{pgfscope}%
\begin{pgfscope}%
\pgfsetbuttcap%
\pgfsetroundjoin%
\definecolor{currentfill}{rgb}{0.000000,0.000000,0.000000}%
\pgfsetfillcolor{currentfill}%
\pgfsetlinewidth{0.803000pt}%
\definecolor{currentstroke}{rgb}{0.000000,0.000000,0.000000}%
\pgfsetstrokecolor{currentstroke}%
\pgfsetdash{}{0pt}%
\pgfsys@defobject{currentmarker}{\pgfqpoint{-0.048611in}{0.000000in}}{\pgfqpoint{-0.000000in}{0.000000in}}{%
\pgfpathmoveto{\pgfqpoint{-0.000000in}{0.000000in}}%
\pgfpathlineto{\pgfqpoint{-0.048611in}{0.000000in}}%
\pgfusepath{stroke,fill}%
}%
\begin{pgfscope}%
\pgfsys@transformshift{0.527360in}{2.465000in}%
\pgfsys@useobject{currentmarker}{}%
\end{pgfscope}%
\end{pgfscope}%
\begin{pgfscope}%
\definecolor{textcolor}{rgb}{0.000000,0.000000,0.000000}%
\pgfsetstrokecolor{textcolor}%
\pgfsetfillcolor{textcolor}%
\pgftext[x=0.301667in, y=2.421597in, left, base]{\color{textcolor}\rmfamily\fontsize{9.000000}{10.800000}\selectfont \(\displaystyle {60}\)}%
\end{pgfscope}%
\begin{pgfscope}%
\definecolor{textcolor}{rgb}{0.000000,0.000000,0.000000}%
\pgfsetstrokecolor{textcolor}%
\pgfsetfillcolor{textcolor}%
\pgftext[x=0.246111in,y=1.503125in,,bottom,rotate=90.000000]{\color{textcolor}\rmfamily\fontsize{9.000000}{10.800000}\selectfont PSNR}%
\end{pgfscope}%
\begin{pgfscope}%
\pgfpathrectangle{\pgfqpoint{0.527360in}{0.541249in}}{\pgfqpoint{2.637640in}{1.923751in}}%
\pgfusepath{clip}%
\pgfsetbuttcap%
\pgfsetmiterjoin%
\definecolor{currentfill}{rgb}{0.121569,0.466667,0.705882}%
\pgfsetfillcolor{currentfill}%
\pgfsetlinewidth{0.000000pt}%
\definecolor{currentstroke}{rgb}{0.000000,0.000000,0.000000}%
\pgfsetstrokecolor{currentstroke}%
\pgfsetstrokeopacity{0.000000}%
\pgfsetdash{}{0pt}%
\pgfpathmoveto{\pgfqpoint{0.637262in}{0.541249in}}%
\pgfpathlineto{\pgfqpoint{0.710530in}{0.541249in}}%
\pgfpathlineto{\pgfqpoint{0.710530in}{1.118374in}}%
\pgfpathlineto{\pgfqpoint{0.637262in}{1.118374in}}%
\pgfpathlineto{\pgfqpoint{0.637262in}{0.541249in}}%
\pgfpathclose%
\pgfusepath{fill}%
\end{pgfscope}%
\begin{pgfscope}%
\pgfpathrectangle{\pgfqpoint{0.527360in}{0.541249in}}{\pgfqpoint{2.637640in}{1.923751in}}%
\pgfusepath{clip}%
\pgfsetbuttcap%
\pgfsetmiterjoin%
\definecolor{currentfill}{rgb}{0.121569,0.466667,0.705882}%
\pgfsetfillcolor{currentfill}%
\pgfsetlinewidth{0.000000pt}%
\definecolor{currentstroke}{rgb}{0.000000,0.000000,0.000000}%
\pgfsetstrokecolor{currentstroke}%
\pgfsetstrokeopacity{0.000000}%
\pgfsetdash{}{0pt}%
\pgfpathmoveto{\pgfqpoint{1.076869in}{0.541249in}}%
\pgfpathlineto{\pgfqpoint{1.150136in}{0.541249in}}%
\pgfpathlineto{\pgfqpoint{1.150136in}{0.964474in}}%
\pgfpathlineto{\pgfqpoint{1.076869in}{0.964474in}}%
\pgfpathlineto{\pgfqpoint{1.076869in}{0.541249in}}%
\pgfpathclose%
\pgfusepath{fill}%
\end{pgfscope}%
\begin{pgfscope}%
\pgfpathrectangle{\pgfqpoint{0.527360in}{0.541249in}}{\pgfqpoint{2.637640in}{1.923751in}}%
\pgfusepath{clip}%
\pgfsetbuttcap%
\pgfsetmiterjoin%
\definecolor{currentfill}{rgb}{0.121569,0.466667,0.705882}%
\pgfsetfillcolor{currentfill}%
\pgfsetlinewidth{0.000000pt}%
\definecolor{currentstroke}{rgb}{0.000000,0.000000,0.000000}%
\pgfsetstrokecolor{currentstroke}%
\pgfsetstrokeopacity{0.000000}%
\pgfsetdash{}{0pt}%
\pgfpathmoveto{\pgfqpoint{1.516475in}{0.541249in}}%
\pgfpathlineto{\pgfqpoint{1.589743in}{0.541249in}}%
\pgfpathlineto{\pgfqpoint{1.589743in}{0.877906in}}%
\pgfpathlineto{\pgfqpoint{1.516475in}{0.877906in}}%
\pgfpathlineto{\pgfqpoint{1.516475in}{0.541249in}}%
\pgfpathclose%
\pgfusepath{fill}%
\end{pgfscope}%
\begin{pgfscope}%
\pgfpathrectangle{\pgfqpoint{0.527360in}{0.541249in}}{\pgfqpoint{2.637640in}{1.923751in}}%
\pgfusepath{clip}%
\pgfsetbuttcap%
\pgfsetmiterjoin%
\definecolor{currentfill}{rgb}{0.121569,0.466667,0.705882}%
\pgfsetfillcolor{currentfill}%
\pgfsetlinewidth{0.000000pt}%
\definecolor{currentstroke}{rgb}{0.000000,0.000000,0.000000}%
\pgfsetstrokecolor{currentstroke}%
\pgfsetstrokeopacity{0.000000}%
\pgfsetdash{}{0pt}%
\pgfpathmoveto{\pgfqpoint{1.956082in}{0.541249in}}%
\pgfpathlineto{\pgfqpoint{2.029350in}{0.541249in}}%
\pgfpathlineto{\pgfqpoint{2.029350in}{0.855462in}}%
\pgfpathlineto{\pgfqpoint{1.956082in}{0.855462in}}%
\pgfpathlineto{\pgfqpoint{1.956082in}{0.541249in}}%
\pgfpathclose%
\pgfusepath{fill}%
\end{pgfscope}%
\begin{pgfscope}%
\pgfpathrectangle{\pgfqpoint{0.527360in}{0.541249in}}{\pgfqpoint{2.637640in}{1.923751in}}%
\pgfusepath{clip}%
\pgfsetbuttcap%
\pgfsetmiterjoin%
\definecolor{currentfill}{rgb}{0.121569,0.466667,0.705882}%
\pgfsetfillcolor{currentfill}%
\pgfsetlinewidth{0.000000pt}%
\definecolor{currentstroke}{rgb}{0.000000,0.000000,0.000000}%
\pgfsetstrokecolor{currentstroke}%
\pgfsetstrokeopacity{0.000000}%
\pgfsetdash{}{0pt}%
\pgfpathmoveto{\pgfqpoint{2.395688in}{0.541249in}}%
\pgfpathlineto{\pgfqpoint{2.468956in}{0.541249in}}%
\pgfpathlineto{\pgfqpoint{2.468956in}{0.820193in}}%
\pgfpathlineto{\pgfqpoint{2.395688in}{0.820193in}}%
\pgfpathlineto{\pgfqpoint{2.395688in}{0.541249in}}%
\pgfpathclose%
\pgfusepath{fill}%
\end{pgfscope}%
\begin{pgfscope}%
\pgfpathrectangle{\pgfqpoint{0.527360in}{0.541249in}}{\pgfqpoint{2.637640in}{1.923751in}}%
\pgfusepath{clip}%
\pgfsetbuttcap%
\pgfsetmiterjoin%
\definecolor{currentfill}{rgb}{0.121569,0.466667,0.705882}%
\pgfsetfillcolor{currentfill}%
\pgfsetlinewidth{0.000000pt}%
\definecolor{currentstroke}{rgb}{0.000000,0.000000,0.000000}%
\pgfsetstrokecolor{currentstroke}%
\pgfsetstrokeopacity{0.000000}%
\pgfsetdash{}{0pt}%
\pgfpathmoveto{\pgfqpoint{2.835295in}{0.541249in}}%
\pgfpathlineto{\pgfqpoint{2.908563in}{0.541249in}}%
\pgfpathlineto{\pgfqpoint{2.908563in}{0.794543in}}%
\pgfpathlineto{\pgfqpoint{2.835295in}{0.794543in}}%
\pgfpathlineto{\pgfqpoint{2.835295in}{0.541249in}}%
\pgfpathclose%
\pgfusepath{fill}%
\end{pgfscope}%
\begin{pgfscope}%
\pgfpathrectangle{\pgfqpoint{0.527360in}{0.541249in}}{\pgfqpoint{2.637640in}{1.923751in}}%
\pgfusepath{clip}%
\pgfsetbuttcap%
\pgfsetmiterjoin%
\definecolor{currentfill}{rgb}{1.000000,0.498039,0.054902}%
\pgfsetfillcolor{currentfill}%
\pgfsetlinewidth{0.000000pt}%
\definecolor{currentstroke}{rgb}{0.000000,0.000000,0.000000}%
\pgfsetstrokecolor{currentstroke}%
\pgfsetstrokeopacity{0.000000}%
\pgfsetdash{}{0pt}%
\pgfpathmoveto{\pgfqpoint{0.710530in}{0.541249in}}%
\pgfpathlineto{\pgfqpoint{0.783797in}{0.541249in}}%
\pgfpathlineto{\pgfqpoint{0.783797in}{1.503125in}}%
\pgfpathlineto{\pgfqpoint{0.710530in}{1.503125in}}%
\pgfpathlineto{\pgfqpoint{0.710530in}{0.541249in}}%
\pgfpathclose%
\pgfusepath{fill}%
\end{pgfscope}%
\begin{pgfscope}%
\pgfpathrectangle{\pgfqpoint{0.527360in}{0.541249in}}{\pgfqpoint{2.637640in}{1.923751in}}%
\pgfusepath{clip}%
\pgfsetbuttcap%
\pgfsetmiterjoin%
\definecolor{currentfill}{rgb}{1.000000,0.498039,0.054902}%
\pgfsetfillcolor{currentfill}%
\pgfsetlinewidth{0.000000pt}%
\definecolor{currentstroke}{rgb}{0.000000,0.000000,0.000000}%
\pgfsetstrokecolor{currentstroke}%
\pgfsetstrokeopacity{0.000000}%
\pgfsetdash{}{0pt}%
\pgfpathmoveto{\pgfqpoint{1.150136in}{0.541249in}}%
\pgfpathlineto{\pgfqpoint{1.223404in}{0.541249in}}%
\pgfpathlineto{\pgfqpoint{1.223404in}{1.310750in}}%
\pgfpathlineto{\pgfqpoint{1.150136in}{1.310750in}}%
\pgfpathlineto{\pgfqpoint{1.150136in}{0.541249in}}%
\pgfpathclose%
\pgfusepath{fill}%
\end{pgfscope}%
\begin{pgfscope}%
\pgfpathrectangle{\pgfqpoint{0.527360in}{0.541249in}}{\pgfqpoint{2.637640in}{1.923751in}}%
\pgfusepath{clip}%
\pgfsetbuttcap%
\pgfsetmiterjoin%
\definecolor{currentfill}{rgb}{1.000000,0.498039,0.054902}%
\pgfsetfillcolor{currentfill}%
\pgfsetlinewidth{0.000000pt}%
\definecolor{currentstroke}{rgb}{0.000000,0.000000,0.000000}%
\pgfsetstrokecolor{currentstroke}%
\pgfsetstrokeopacity{0.000000}%
\pgfsetdash{}{0pt}%
\pgfpathmoveto{\pgfqpoint{1.589743in}{0.541249in}}%
\pgfpathlineto{\pgfqpoint{1.663011in}{0.541249in}}%
\pgfpathlineto{\pgfqpoint{1.663011in}{1.180896in}}%
\pgfpathlineto{\pgfqpoint{1.589743in}{1.180896in}}%
\pgfpathlineto{\pgfqpoint{1.589743in}{0.541249in}}%
\pgfpathclose%
\pgfusepath{fill}%
\end{pgfscope}%
\begin{pgfscope}%
\pgfpathrectangle{\pgfqpoint{0.527360in}{0.541249in}}{\pgfqpoint{2.637640in}{1.923751in}}%
\pgfusepath{clip}%
\pgfsetbuttcap%
\pgfsetmiterjoin%
\definecolor{currentfill}{rgb}{1.000000,0.498039,0.054902}%
\pgfsetfillcolor{currentfill}%
\pgfsetlinewidth{0.000000pt}%
\definecolor{currentstroke}{rgb}{0.000000,0.000000,0.000000}%
\pgfsetstrokecolor{currentstroke}%
\pgfsetstrokeopacity{0.000000}%
\pgfsetdash{}{0pt}%
\pgfpathmoveto{\pgfqpoint{2.029350in}{0.541249in}}%
\pgfpathlineto{\pgfqpoint{2.102617in}{0.541249in}}%
\pgfpathlineto{\pgfqpoint{2.102617in}{1.079899in}}%
\pgfpathlineto{\pgfqpoint{2.029350in}{1.079899in}}%
\pgfpathlineto{\pgfqpoint{2.029350in}{0.541249in}}%
\pgfpathclose%
\pgfusepath{fill}%
\end{pgfscope}%
\begin{pgfscope}%
\pgfpathrectangle{\pgfqpoint{0.527360in}{0.541249in}}{\pgfqpoint{2.637640in}{1.923751in}}%
\pgfusepath{clip}%
\pgfsetbuttcap%
\pgfsetmiterjoin%
\definecolor{currentfill}{rgb}{1.000000,0.498039,0.054902}%
\pgfsetfillcolor{currentfill}%
\pgfsetlinewidth{0.000000pt}%
\definecolor{currentstroke}{rgb}{0.000000,0.000000,0.000000}%
\pgfsetstrokecolor{currentstroke}%
\pgfsetstrokeopacity{0.000000}%
\pgfsetdash{}{0pt}%
\pgfpathmoveto{\pgfqpoint{2.468956in}{0.541249in}}%
\pgfpathlineto{\pgfqpoint{2.542224in}{0.541249in}}%
\pgfpathlineto{\pgfqpoint{2.542224in}{1.022187in}}%
\pgfpathlineto{\pgfqpoint{2.468956in}{1.022187in}}%
\pgfpathlineto{\pgfqpoint{2.468956in}{0.541249in}}%
\pgfpathclose%
\pgfusepath{fill}%
\end{pgfscope}%
\begin{pgfscope}%
\pgfpathrectangle{\pgfqpoint{0.527360in}{0.541249in}}{\pgfqpoint{2.637640in}{1.923751in}}%
\pgfusepath{clip}%
\pgfsetbuttcap%
\pgfsetmiterjoin%
\definecolor{currentfill}{rgb}{1.000000,0.498039,0.054902}%
\pgfsetfillcolor{currentfill}%
\pgfsetlinewidth{0.000000pt}%
\definecolor{currentstroke}{rgb}{0.000000,0.000000,0.000000}%
\pgfsetstrokecolor{currentstroke}%
\pgfsetstrokeopacity{0.000000}%
\pgfsetdash{}{0pt}%
\pgfpathmoveto{\pgfqpoint{2.908563in}{0.541249in}}%
\pgfpathlineto{\pgfqpoint{2.981831in}{0.541249in}}%
\pgfpathlineto{\pgfqpoint{2.981831in}{0.980506in}}%
\pgfpathlineto{\pgfqpoint{2.908563in}{0.980506in}}%
\pgfpathlineto{\pgfqpoint{2.908563in}{0.541249in}}%
\pgfpathclose%
\pgfusepath{fill}%
\end{pgfscope}%
\begin{pgfscope}%
\pgfpathrectangle{\pgfqpoint{0.527360in}{0.541249in}}{\pgfqpoint{2.637640in}{1.923751in}}%
\pgfusepath{clip}%
\pgfsetbuttcap%
\pgfsetmiterjoin%
\definecolor{currentfill}{rgb}{0.172549,0.627451,0.172549}%
\pgfsetfillcolor{currentfill}%
\pgfsetlinewidth{0.000000pt}%
\definecolor{currentstroke}{rgb}{0.000000,0.000000,0.000000}%
\pgfsetstrokecolor{currentstroke}%
\pgfsetstrokeopacity{0.000000}%
\pgfsetdash{}{0pt}%
\pgfpathmoveto{\pgfqpoint{0.783797in}{0.541249in}}%
\pgfpathlineto{\pgfqpoint{0.857065in}{0.541249in}}%
\pgfpathlineto{\pgfqpoint{0.857065in}{1.415915in}}%
\pgfpathlineto{\pgfqpoint{0.783797in}{1.415915in}}%
\pgfpathlineto{\pgfqpoint{0.783797in}{0.541249in}}%
\pgfpathclose%
\pgfusepath{fill}%
\end{pgfscope}%
\begin{pgfscope}%
\pgfpathrectangle{\pgfqpoint{0.527360in}{0.541249in}}{\pgfqpoint{2.637640in}{1.923751in}}%
\pgfusepath{clip}%
\pgfsetbuttcap%
\pgfsetmiterjoin%
\definecolor{currentfill}{rgb}{0.172549,0.627451,0.172549}%
\pgfsetfillcolor{currentfill}%
\pgfsetlinewidth{0.000000pt}%
\definecolor{currentstroke}{rgb}{0.000000,0.000000,0.000000}%
\pgfsetstrokecolor{currentstroke}%
\pgfsetstrokeopacity{0.000000}%
\pgfsetdash{}{0pt}%
\pgfpathmoveto{\pgfqpoint{1.223404in}{0.541249in}}%
\pgfpathlineto{\pgfqpoint{1.296672in}{0.541249in}}%
\pgfpathlineto{\pgfqpoint{1.296672in}{1.265541in}}%
\pgfpathlineto{\pgfqpoint{1.223404in}{1.265541in}}%
\pgfpathlineto{\pgfqpoint{1.223404in}{0.541249in}}%
\pgfpathclose%
\pgfusepath{fill}%
\end{pgfscope}%
\begin{pgfscope}%
\pgfpathrectangle{\pgfqpoint{0.527360in}{0.541249in}}{\pgfqpoint{2.637640in}{1.923751in}}%
\pgfusepath{clip}%
\pgfsetbuttcap%
\pgfsetmiterjoin%
\definecolor{currentfill}{rgb}{0.172549,0.627451,0.172549}%
\pgfsetfillcolor{currentfill}%
\pgfsetlinewidth{0.000000pt}%
\definecolor{currentstroke}{rgb}{0.000000,0.000000,0.000000}%
\pgfsetstrokecolor{currentstroke}%
\pgfsetstrokeopacity{0.000000}%
\pgfsetdash{}{0pt}%
\pgfpathmoveto{\pgfqpoint{1.663011in}{0.541249in}}%
\pgfpathlineto{\pgfqpoint{1.736278in}{0.541249in}}%
\pgfpathlineto{\pgfqpoint{1.736278in}{1.177690in}}%
\pgfpathlineto{\pgfqpoint{1.663011in}{1.177690in}}%
\pgfpathlineto{\pgfqpoint{1.663011in}{0.541249in}}%
\pgfpathclose%
\pgfusepath{fill}%
\end{pgfscope}%
\begin{pgfscope}%
\pgfpathrectangle{\pgfqpoint{0.527360in}{0.541249in}}{\pgfqpoint{2.637640in}{1.923751in}}%
\pgfusepath{clip}%
\pgfsetbuttcap%
\pgfsetmiterjoin%
\definecolor{currentfill}{rgb}{0.172549,0.627451,0.172549}%
\pgfsetfillcolor{currentfill}%
\pgfsetlinewidth{0.000000pt}%
\definecolor{currentstroke}{rgb}{0.000000,0.000000,0.000000}%
\pgfsetstrokecolor{currentstroke}%
\pgfsetstrokeopacity{0.000000}%
\pgfsetdash{}{0pt}%
\pgfpathmoveto{\pgfqpoint{2.102617in}{0.541249in}}%
\pgfpathlineto{\pgfqpoint{2.175885in}{0.541249in}}%
\pgfpathlineto{\pgfqpoint{2.175885in}{1.123504in}}%
\pgfpathlineto{\pgfqpoint{2.102617in}{1.123504in}}%
\pgfpathlineto{\pgfqpoint{2.102617in}{0.541249in}}%
\pgfpathclose%
\pgfusepath{fill}%
\end{pgfscope}%
\begin{pgfscope}%
\pgfpathrectangle{\pgfqpoint{0.527360in}{0.541249in}}{\pgfqpoint{2.637640in}{1.923751in}}%
\pgfusepath{clip}%
\pgfsetbuttcap%
\pgfsetmiterjoin%
\definecolor{currentfill}{rgb}{0.172549,0.627451,0.172549}%
\pgfsetfillcolor{currentfill}%
\pgfsetlinewidth{0.000000pt}%
\definecolor{currentstroke}{rgb}{0.000000,0.000000,0.000000}%
\pgfsetstrokecolor{currentstroke}%
\pgfsetstrokeopacity{0.000000}%
\pgfsetdash{}{0pt}%
\pgfpathmoveto{\pgfqpoint{2.542224in}{0.541249in}}%
\pgfpathlineto{\pgfqpoint{2.615492in}{0.541249in}}%
\pgfpathlineto{\pgfqpoint{2.615492in}{1.070281in}}%
\pgfpathlineto{\pgfqpoint{2.542224in}{1.070281in}}%
\pgfpathlineto{\pgfqpoint{2.542224in}{0.541249in}}%
\pgfpathclose%
\pgfusepath{fill}%
\end{pgfscope}%
\begin{pgfscope}%
\pgfpathrectangle{\pgfqpoint{0.527360in}{0.541249in}}{\pgfqpoint{2.637640in}{1.923751in}}%
\pgfusepath{clip}%
\pgfsetbuttcap%
\pgfsetmiterjoin%
\definecolor{currentfill}{rgb}{0.172549,0.627451,0.172549}%
\pgfsetfillcolor{currentfill}%
\pgfsetlinewidth{0.000000pt}%
\definecolor{currentstroke}{rgb}{0.000000,0.000000,0.000000}%
\pgfsetstrokecolor{currentstroke}%
\pgfsetstrokeopacity{0.000000}%
\pgfsetdash{}{0pt}%
\pgfpathmoveto{\pgfqpoint{2.981831in}{0.541249in}}%
\pgfpathlineto{\pgfqpoint{3.055098in}{0.541249in}}%
\pgfpathlineto{\pgfqpoint{3.055098in}{0.962230in}}%
\pgfpathlineto{\pgfqpoint{2.981831in}{0.962230in}}%
\pgfpathlineto{\pgfqpoint{2.981831in}{0.541249in}}%
\pgfpathclose%
\pgfusepath{fill}%
\end{pgfscope}%
\begin{pgfscope}%
\pgfsetrectcap%
\pgfsetmiterjoin%
\pgfsetlinewidth{0.803000pt}%
\definecolor{currentstroke}{rgb}{0.000000,0.000000,0.000000}%
\pgfsetstrokecolor{currentstroke}%
\pgfsetdash{}{0pt}%
\pgfpathmoveto{\pgfqpoint{0.527360in}{0.541249in}}%
\pgfpathlineto{\pgfqpoint{0.527360in}{2.465000in}}%
\pgfusepath{stroke}%
\end{pgfscope}%
\begin{pgfscope}%
\pgfsetrectcap%
\pgfsetmiterjoin%
\pgfsetlinewidth{0.803000pt}%
\definecolor{currentstroke}{rgb}{0.000000,0.000000,0.000000}%
\pgfsetstrokecolor{currentstroke}%
\pgfsetdash{}{0pt}%
\pgfpathmoveto{\pgfqpoint{3.165000in}{0.541249in}}%
\pgfpathlineto{\pgfqpoint{3.165000in}{2.465000in}}%
\pgfusepath{stroke}%
\end{pgfscope}%
\begin{pgfscope}%
\pgfsetrectcap%
\pgfsetmiterjoin%
\pgfsetlinewidth{0.803000pt}%
\definecolor{currentstroke}{rgb}{0.000000,0.000000,0.000000}%
\pgfsetstrokecolor{currentstroke}%
\pgfsetdash{}{0pt}%
\pgfpathmoveto{\pgfqpoint{0.527360in}{0.541249in}}%
\pgfpathlineto{\pgfqpoint{3.165000in}{0.541249in}}%
\pgfusepath{stroke}%
\end{pgfscope}%
\begin{pgfscope}%
\pgfsetrectcap%
\pgfsetmiterjoin%
\pgfsetlinewidth{0.803000pt}%
\definecolor{currentstroke}{rgb}{0.000000,0.000000,0.000000}%
\pgfsetstrokecolor{currentstroke}%
\pgfsetdash{}{0pt}%
\pgfpathmoveto{\pgfqpoint{0.527360in}{2.465000in}}%
\pgfpathlineto{\pgfqpoint{3.165000in}{2.465000in}}%
\pgfusepath{stroke}%
\end{pgfscope}%
\begin{pgfscope}%
\pgfsetbuttcap%
\pgfsetmiterjoin%
\definecolor{currentfill}{rgb}{1.000000,1.000000,1.000000}%
\pgfsetfillcolor{currentfill}%
\pgfsetfillopacity{0.800000}%
\pgfsetlinewidth{1.003750pt}%
\definecolor{currentstroke}{rgb}{0.800000,0.800000,0.800000}%
\pgfsetstrokecolor{currentstroke}%
\pgfsetstrokeopacity{0.800000}%
\pgfsetdash{}{0pt}%
\pgfpathmoveto{\pgfqpoint{2.134237in}{1.842083in}}%
\pgfpathlineto{\pgfqpoint{3.077500in}{1.842083in}}%
\pgfpathquadraticcurveto{\pgfqpoint{3.102500in}{1.842083in}}{\pgfqpoint{3.102500in}{1.867083in}}%
\pgfpathlineto{\pgfqpoint{3.102500in}{2.377500in}}%
\pgfpathquadraticcurveto{\pgfqpoint{3.102500in}{2.402500in}}{\pgfqpoint{3.077500in}{2.402500in}}%
\pgfpathlineto{\pgfqpoint{2.134237in}{2.402500in}}%
\pgfpathquadraticcurveto{\pgfqpoint{2.109237in}{2.402500in}}{\pgfqpoint{2.109237in}{2.377500in}}%
\pgfpathlineto{\pgfqpoint{2.109237in}{1.867083in}}%
\pgfpathquadraticcurveto{\pgfqpoint{2.109237in}{1.842083in}}{\pgfqpoint{2.134237in}{1.842083in}}%
\pgfpathlineto{\pgfqpoint{2.134237in}{1.842083in}}%
\pgfpathclose%
\pgfusepath{stroke,fill}%
\end{pgfscope}%
\begin{pgfscope}%
\pgfsetbuttcap%
\pgfsetmiterjoin%
\definecolor{currentfill}{rgb}{0.121569,0.466667,0.705882}%
\pgfsetfillcolor{currentfill}%
\pgfsetlinewidth{0.000000pt}%
\definecolor{currentstroke}{rgb}{0.000000,0.000000,0.000000}%
\pgfsetstrokecolor{currentstroke}%
\pgfsetstrokeopacity{0.000000}%
\pgfsetdash{}{0pt}%
\pgfpathmoveto{\pgfqpoint{2.159237in}{2.265000in}}%
\pgfpathlineto{\pgfqpoint{2.409237in}{2.265000in}}%
\pgfpathlineto{\pgfqpoint{2.409237in}{2.352500in}}%
\pgfpathlineto{\pgfqpoint{2.159237in}{2.352500in}}%
\pgfpathlineto{\pgfqpoint{2.159237in}{2.265000in}}%
\pgfpathclose%
\pgfusepath{fill}%
\end{pgfscope}%
\begin{pgfscope}%
\definecolor{textcolor}{rgb}{0.000000,0.000000,0.000000}%
\pgfsetstrokecolor{textcolor}%
\pgfsetfillcolor{textcolor}%
\pgftext[x=2.509237in,y=2.265000in,left,base]{\color{textcolor}\rmfamily\fontsize{9.000000}{10.800000}\selectfont ESRGAN}%
\end{pgfscope}%
\begin{pgfscope}%
\pgfsetbuttcap%
\pgfsetmiterjoin%
\definecolor{currentfill}{rgb}{1.000000,0.498039,0.054902}%
\pgfsetfillcolor{currentfill}%
\pgfsetlinewidth{0.000000pt}%
\definecolor{currentstroke}{rgb}{0.000000,0.000000,0.000000}%
\pgfsetstrokecolor{currentstroke}%
\pgfsetstrokeopacity{0.000000}%
\pgfsetdash{}{0pt}%
\pgfpathmoveto{\pgfqpoint{2.159237in}{2.090694in}}%
\pgfpathlineto{\pgfqpoint{2.409237in}{2.090694in}}%
\pgfpathlineto{\pgfqpoint{2.409237in}{2.178194in}}%
\pgfpathlineto{\pgfqpoint{2.159237in}{2.178194in}}%
\pgfpathlineto{\pgfqpoint{2.159237in}{2.090694in}}%
\pgfpathclose%
\pgfusepath{fill}%
\end{pgfscope}%
\begin{pgfscope}%
\definecolor{textcolor}{rgb}{0.000000,0.000000,0.000000}%
\pgfsetstrokecolor{textcolor}%
\pgfsetfillcolor{textcolor}%
\pgftext[x=2.509237in,y=2.090694in,left,base]{\color{textcolor}\rmfamily\fontsize{9.000000}{10.800000}\selectfont EDSR}%
\end{pgfscope}%
\begin{pgfscope}%
\pgfsetbuttcap%
\pgfsetmiterjoin%
\definecolor{currentfill}{rgb}{0.172549,0.627451,0.172549}%
\pgfsetfillcolor{currentfill}%
\pgfsetlinewidth{0.000000pt}%
\definecolor{currentstroke}{rgb}{0.000000,0.000000,0.000000}%
\pgfsetstrokecolor{currentstroke}%
\pgfsetstrokeopacity{0.000000}%
\pgfsetdash{}{0pt}%
\pgfpathmoveto{\pgfqpoint{2.159237in}{1.916389in}}%
\pgfpathlineto{\pgfqpoint{2.409237in}{1.916389in}}%
\pgfpathlineto{\pgfqpoint{2.409237in}{2.003889in}}%
\pgfpathlineto{\pgfqpoint{2.159237in}{2.003889in}}%
\pgfpathlineto{\pgfqpoint{2.159237in}{1.916389in}}%
\pgfpathclose%
\pgfusepath{fill}%
\end{pgfscope}%
\begin{pgfscope}%
\definecolor{textcolor}{rgb}{0.000000,0.000000,0.000000}%
\pgfsetstrokecolor{textcolor}%
\pgfsetfillcolor{textcolor}%
\pgftext[x=2.509237in,y=1.916389in,left,base]{\color{textcolor}\rmfamily\fontsize{9.000000}{10.800000}\selectfont SRFlow}%
\end{pgfscope}%
\end{pgfpicture}%
\makeatother%
\endgroup%

%% file: latex/figures/inp_set5.pgf
\begingroup%
\makeatletter%
\begin{pgfpicture}%
\pgfpathrectangle{\pgfpointorigin}{\pgfqpoint{3.300000in}{2.600000in}}%
\pgfusepath{use as bounding box, clip}%
\begin{pgfscope}%
\pgfsetbuttcap%
\pgfsetmiterjoin%
\definecolor{currentfill}{rgb}{1.000000,1.000000,1.000000}%
\pgfsetfillcolor{currentfill}%
\pgfsetlinewidth{0.000000pt}%
\definecolor{currentstroke}{rgb}{1.000000,1.000000,1.000000}%
\pgfsetstrokecolor{currentstroke}%
\pgfsetdash{}{0pt}%
\pgfpathmoveto{\pgfqpoint{0.000000in}{0.000000in}}%
\pgfpathlineto{\pgfqpoint{3.300000in}{0.000000in}}%
\pgfpathlineto{\pgfqpoint{3.300000in}{2.600000in}}%
\pgfpathlineto{\pgfqpoint{0.000000in}{2.600000in}}%
\pgfpathlineto{\pgfqpoint{0.000000in}{0.000000in}}%
\pgfpathclose%
\pgfusepath{fill}%
\end{pgfscope}%
\begin{pgfscope}%
\pgfsetbuttcap%
\pgfsetmiterjoin%
\definecolor{currentfill}{rgb}{1.000000,1.000000,1.000000}%
\pgfsetfillcolor{currentfill}%
\pgfsetlinewidth{0.000000pt}%
\definecolor{currentstroke}{rgb}{0.000000,0.000000,0.000000}%
\pgfsetstrokecolor{currentstroke}%
\pgfsetstrokeopacity{0.000000}%
\pgfsetdash{}{0pt}%
\pgfpathmoveto{\pgfqpoint{0.527360in}{0.541249in}}%
\pgfpathlineto{\pgfqpoint{3.165000in}{0.541249in}}%
\pgfpathlineto{\pgfqpoint{3.165000in}{2.465000in}}%
\pgfpathlineto{\pgfqpoint{0.527360in}{2.465000in}}%
\pgfpathlineto{\pgfqpoint{0.527360in}{0.541249in}}%
\pgfpathclose%
\pgfusepath{fill}%
\end{pgfscope}%
\begin{pgfscope}%
\pgfsetbuttcap%
\pgfsetroundjoin%
\definecolor{currentfill}{rgb}{0.000000,0.000000,0.000000}%
\pgfsetfillcolor{currentfill}%
\pgfsetlinewidth{0.803000pt}%
\definecolor{currentstroke}{rgb}{0.000000,0.000000,0.000000}%
\pgfsetstrokecolor{currentstroke}%
\pgfsetdash{}{0pt}%
\pgfsys@defobject{currentmarker}{\pgfqpoint{0.000000in}{-0.048611in}}{\pgfqpoint{0.000000in}{0.000000in}}{%
\pgfpathmoveto{\pgfqpoint{0.000000in}{0.000000in}}%
\pgfpathlineto{\pgfqpoint{0.000000in}{-0.048611in}}%
\pgfusepath{stroke,fill}%
}%
\begin{pgfscope}%
\pgfsys@transformshift{0.747164in}{0.541249in}%
\pgfsys@useobject{currentmarker}{}%
\end{pgfscope}%
\end{pgfscope}%
\begin{pgfscope}%
\definecolor{textcolor}{rgb}{0.000000,0.000000,0.000000}%
\pgfsetstrokecolor{textcolor}%
\pgfsetfillcolor{textcolor}%
\pgftext[x=0.778414in, y=0.379791in, left, base,rotate=90.000000]{\color{textcolor}\rmfamily\fontsize{9.000000}{10.800000}\selectfont 1}%
\end{pgfscope}%
\begin{pgfscope}%
\pgfsetbuttcap%
\pgfsetroundjoin%
\definecolor{currentfill}{rgb}{0.000000,0.000000,0.000000}%
\pgfsetfillcolor{currentfill}%
\pgfsetlinewidth{0.803000pt}%
\definecolor{currentstroke}{rgb}{0.000000,0.000000,0.000000}%
\pgfsetstrokecolor{currentstroke}%
\pgfsetdash{}{0pt}%
\pgfsys@defobject{currentmarker}{\pgfqpoint{0.000000in}{-0.048611in}}{\pgfqpoint{0.000000in}{0.000000in}}{%
\pgfpathmoveto{\pgfqpoint{0.000000in}{0.000000in}}%
\pgfpathlineto{\pgfqpoint{0.000000in}{-0.048611in}}%
\pgfusepath{stroke,fill}%
}%
\begin{pgfscope}%
\pgfsys@transformshift{1.186770in}{0.541249in}%
\pgfsys@useobject{currentmarker}{}%
\end{pgfscope}%
\end{pgfscope}%
\begin{pgfscope}%
\definecolor{textcolor}{rgb}{0.000000,0.000000,0.000000}%
\pgfsetstrokecolor{textcolor}%
\pgfsetfillcolor{textcolor}%
\pgftext[x=1.218020in, y=0.379791in, left, base,rotate=90.000000]{\color{textcolor}\rmfamily\fontsize{9.000000}{10.800000}\selectfont 2}%
\end{pgfscope}%
\begin{pgfscope}%
\pgfsetbuttcap%
\pgfsetroundjoin%
\definecolor{currentfill}{rgb}{0.000000,0.000000,0.000000}%
\pgfsetfillcolor{currentfill}%
\pgfsetlinewidth{0.803000pt}%
\definecolor{currentstroke}{rgb}{0.000000,0.000000,0.000000}%
\pgfsetstrokecolor{currentstroke}%
\pgfsetdash{}{0pt}%
\pgfsys@defobject{currentmarker}{\pgfqpoint{0.000000in}{-0.048611in}}{\pgfqpoint{0.000000in}{0.000000in}}{%
\pgfpathmoveto{\pgfqpoint{0.000000in}{0.000000in}}%
\pgfpathlineto{\pgfqpoint{0.000000in}{-0.048611in}}%
\pgfusepath{stroke,fill}%
}%
\begin{pgfscope}%
\pgfsys@transformshift{1.626377in}{0.541249in}%
\pgfsys@useobject{currentmarker}{}%
\end{pgfscope}%
\end{pgfscope}%
\begin{pgfscope}%
\definecolor{textcolor}{rgb}{0.000000,0.000000,0.000000}%
\pgfsetstrokecolor{textcolor}%
\pgfsetfillcolor{textcolor}%
\pgftext[x=1.657627in, y=0.379791in, left, base,rotate=90.000000]{\color{textcolor}\rmfamily\fontsize{9.000000}{10.800000}\selectfont 4}%
\end{pgfscope}%
\begin{pgfscope}%
\pgfsetbuttcap%
\pgfsetroundjoin%
\definecolor{currentfill}{rgb}{0.000000,0.000000,0.000000}%
\pgfsetfillcolor{currentfill}%
\pgfsetlinewidth{0.803000pt}%
\definecolor{currentstroke}{rgb}{0.000000,0.000000,0.000000}%
\pgfsetstrokecolor{currentstroke}%
\pgfsetdash{}{0pt}%
\pgfsys@defobject{currentmarker}{\pgfqpoint{0.000000in}{-0.048611in}}{\pgfqpoint{0.000000in}{0.000000in}}{%
\pgfpathmoveto{\pgfqpoint{0.000000in}{0.000000in}}%
\pgfpathlineto{\pgfqpoint{0.000000in}{-0.048611in}}%
\pgfusepath{stroke,fill}%
}%
\begin{pgfscope}%
\pgfsys@transformshift{2.065983in}{0.541249in}%
\pgfsys@useobject{currentmarker}{}%
\end{pgfscope}%
\end{pgfscope}%
\begin{pgfscope}%
\definecolor{textcolor}{rgb}{0.000000,0.000000,0.000000}%
\pgfsetstrokecolor{textcolor}%
\pgfsetfillcolor{textcolor}%
\pgftext[x=2.097233in, y=0.379791in, left, base,rotate=90.000000]{\color{textcolor}\rmfamily\fontsize{9.000000}{10.800000}\selectfont 8}%
\end{pgfscope}%
\begin{pgfscope}%
\pgfsetbuttcap%
\pgfsetroundjoin%
\definecolor{currentfill}{rgb}{0.000000,0.000000,0.000000}%
\pgfsetfillcolor{currentfill}%
\pgfsetlinewidth{0.803000pt}%
\definecolor{currentstroke}{rgb}{0.000000,0.000000,0.000000}%
\pgfsetstrokecolor{currentstroke}%
\pgfsetdash{}{0pt}%
\pgfsys@defobject{currentmarker}{\pgfqpoint{0.000000in}{-0.048611in}}{\pgfqpoint{0.000000in}{0.000000in}}{%
\pgfpathmoveto{\pgfqpoint{0.000000in}{0.000000in}}%
\pgfpathlineto{\pgfqpoint{0.000000in}{-0.048611in}}%
\pgfusepath{stroke,fill}%
}%
\begin{pgfscope}%
\pgfsys@transformshift{2.505590in}{0.541249in}%
\pgfsys@useobject{currentmarker}{}%
\end{pgfscope}%
\end{pgfscope}%
\begin{pgfscope}%
\definecolor{textcolor}{rgb}{0.000000,0.000000,0.000000}%
\pgfsetstrokecolor{textcolor}%
\pgfsetfillcolor{textcolor}%
\pgftext[x=2.536840in, y=0.315556in, left, base,rotate=90.000000]{\color{textcolor}\rmfamily\fontsize{9.000000}{10.800000}\selectfont 16}%
\end{pgfscope}%
\begin{pgfscope}%
\pgfsetbuttcap%
\pgfsetroundjoin%
\definecolor{currentfill}{rgb}{0.000000,0.000000,0.000000}%
\pgfsetfillcolor{currentfill}%
\pgfsetlinewidth{0.803000pt}%
\definecolor{currentstroke}{rgb}{0.000000,0.000000,0.000000}%
\pgfsetstrokecolor{currentstroke}%
\pgfsetdash{}{0pt}%
\pgfsys@defobject{currentmarker}{\pgfqpoint{0.000000in}{-0.048611in}}{\pgfqpoint{0.000000in}{0.000000in}}{%
\pgfpathmoveto{\pgfqpoint{0.000000in}{0.000000in}}%
\pgfpathlineto{\pgfqpoint{0.000000in}{-0.048611in}}%
\pgfusepath{stroke,fill}%
}%
\begin{pgfscope}%
\pgfsys@transformshift{2.945197in}{0.541249in}%
\pgfsys@useobject{currentmarker}{}%
\end{pgfscope}%
\end{pgfscope}%
\begin{pgfscope}%
\definecolor{textcolor}{rgb}{0.000000,0.000000,0.000000}%
\pgfsetstrokecolor{textcolor}%
\pgfsetfillcolor{textcolor}%
\pgftext[x=2.976447in, y=0.315556in, left, base,rotate=90.000000]{\color{textcolor}\rmfamily\fontsize{9.000000}{10.800000}\selectfont 32}%
\end{pgfscope}%
\begin{pgfscope}%
\definecolor{textcolor}{rgb}{0.000000,0.000000,0.000000}%
\pgfsetstrokecolor{textcolor}%
\pgfsetfillcolor{textcolor}%
\pgftext[x=1.846180in,y=0.260000in,,top]{\color{textcolor}\rmfamily\fontsize{9.000000}{10.800000}\selectfont \(\displaystyle \alpha(/255)\)}%
\end{pgfscope}%
\begin{pgfscope}%
\pgfpathrectangle{\pgfqpoint{0.527360in}{0.541249in}}{\pgfqpoint{2.637640in}{1.923751in}}%
\pgfusepath{clip}%
\pgfsetrectcap%
\pgfsetroundjoin%
\pgfsetlinewidth{0.803000pt}%
\definecolor{currentstroke}{rgb}{0.690196,0.690196,0.690196}%
\pgfsetstrokecolor{currentstroke}%
\pgfsetdash{}{0pt}%
\pgfpathmoveto{\pgfqpoint{0.527360in}{0.541249in}}%
\pgfpathlineto{\pgfqpoint{3.165000in}{0.541249in}}%
\pgfusepath{stroke}%
\end{pgfscope}%
\begin{pgfscope}%
\pgfsetbuttcap%
\pgfsetroundjoin%
\definecolor{currentfill}{rgb}{0.000000,0.000000,0.000000}%
\pgfsetfillcolor{currentfill}%
\pgfsetlinewidth{0.803000pt}%
\definecolor{currentstroke}{rgb}{0.000000,0.000000,0.000000}%
\pgfsetstrokecolor{currentstroke}%
\pgfsetdash{}{0pt}%
\pgfsys@defobject{currentmarker}{\pgfqpoint{-0.048611in}{0.000000in}}{\pgfqpoint{-0.000000in}{0.000000in}}{%
\pgfpathmoveto{\pgfqpoint{-0.000000in}{0.000000in}}%
\pgfpathlineto{\pgfqpoint{-0.048611in}{0.000000in}}%
\pgfusepath{stroke,fill}%
}%
\begin{pgfscope}%
\pgfsys@transformshift{0.527360in}{0.541249in}%
\pgfsys@useobject{currentmarker}{}%
\end{pgfscope}%
\end{pgfscope}%
\begin{pgfscope}%
\definecolor{textcolor}{rgb}{0.000000,0.000000,0.000000}%
\pgfsetstrokecolor{textcolor}%
\pgfsetfillcolor{textcolor}%
\pgftext[x=0.365902in, y=0.497846in, left, base]{\color{textcolor}\rmfamily\fontsize{9.000000}{10.800000}\selectfont \(\displaystyle {0}\)}%
\end{pgfscope}%
\begin{pgfscope}%
\pgfpathrectangle{\pgfqpoint{0.527360in}{0.541249in}}{\pgfqpoint{2.637640in}{1.923751in}}%
\pgfusepath{clip}%
\pgfsetrectcap%
\pgfsetroundjoin%
\pgfsetlinewidth{0.803000pt}%
\definecolor{currentstroke}{rgb}{0.690196,0.690196,0.690196}%
\pgfsetstrokecolor{currentstroke}%
\pgfsetdash{}{0pt}%
\pgfpathmoveto{\pgfqpoint{0.527360in}{0.861874in}}%
\pgfpathlineto{\pgfqpoint{3.165000in}{0.861874in}}%
\pgfusepath{stroke}%
\end{pgfscope}%
\begin{pgfscope}%
\pgfsetbuttcap%
\pgfsetroundjoin%
\definecolor{currentfill}{rgb}{0.000000,0.000000,0.000000}%
\pgfsetfillcolor{currentfill}%
\pgfsetlinewidth{0.803000pt}%
\definecolor{currentstroke}{rgb}{0.000000,0.000000,0.000000}%
\pgfsetstrokecolor{currentstroke}%
\pgfsetdash{}{0pt}%
\pgfsys@defobject{currentmarker}{\pgfqpoint{-0.048611in}{0.000000in}}{\pgfqpoint{-0.000000in}{0.000000in}}{%
\pgfpathmoveto{\pgfqpoint{-0.000000in}{0.000000in}}%
\pgfpathlineto{\pgfqpoint{-0.048611in}{0.000000in}}%
\pgfusepath{stroke,fill}%
}%
\begin{pgfscope}%
\pgfsys@transformshift{0.527360in}{0.861874in}%
\pgfsys@useobject{currentmarker}{}%
\end{pgfscope}%
\end{pgfscope}%
\begin{pgfscope}%
\definecolor{textcolor}{rgb}{0.000000,0.000000,0.000000}%
\pgfsetstrokecolor{textcolor}%
\pgfsetfillcolor{textcolor}%
\pgftext[x=0.301667in, y=0.818472in, left, base]{\color{textcolor}\rmfamily\fontsize{9.000000}{10.800000}\selectfont \(\displaystyle {10}\)}%
\end{pgfscope}%
\begin{pgfscope}%
\pgfpathrectangle{\pgfqpoint{0.527360in}{0.541249in}}{\pgfqpoint{2.637640in}{1.923751in}}%
\pgfusepath{clip}%
\pgfsetrectcap%
\pgfsetroundjoin%
\pgfsetlinewidth{0.803000pt}%
\definecolor{currentstroke}{rgb}{0.690196,0.690196,0.690196}%
\pgfsetstrokecolor{currentstroke}%
\pgfsetdash{}{0pt}%
\pgfpathmoveto{\pgfqpoint{0.527360in}{1.182499in}}%
\pgfpathlineto{\pgfqpoint{3.165000in}{1.182499in}}%
\pgfusepath{stroke}%
\end{pgfscope}%
\begin{pgfscope}%
\pgfsetbuttcap%
\pgfsetroundjoin%
\definecolor{currentfill}{rgb}{0.000000,0.000000,0.000000}%
\pgfsetfillcolor{currentfill}%
\pgfsetlinewidth{0.803000pt}%
\definecolor{currentstroke}{rgb}{0.000000,0.000000,0.000000}%
\pgfsetstrokecolor{currentstroke}%
\pgfsetdash{}{0pt}%
\pgfsys@defobject{currentmarker}{\pgfqpoint{-0.048611in}{0.000000in}}{\pgfqpoint{-0.000000in}{0.000000in}}{%
\pgfpathmoveto{\pgfqpoint{-0.000000in}{0.000000in}}%
\pgfpathlineto{\pgfqpoint{-0.048611in}{0.000000in}}%
\pgfusepath{stroke,fill}%
}%
\begin{pgfscope}%
\pgfsys@transformshift{0.527360in}{1.182499in}%
\pgfsys@useobject{currentmarker}{}%
\end{pgfscope}%
\end{pgfscope}%
\begin{pgfscope}%
\definecolor{textcolor}{rgb}{0.000000,0.000000,0.000000}%
\pgfsetstrokecolor{textcolor}%
\pgfsetfillcolor{textcolor}%
\pgftext[x=0.301667in, y=1.139097in, left, base]{\color{textcolor}\rmfamily\fontsize{9.000000}{10.800000}\selectfont \(\displaystyle {20}\)}%
\end{pgfscope}%
\begin{pgfscope}%
\pgfpathrectangle{\pgfqpoint{0.527360in}{0.541249in}}{\pgfqpoint{2.637640in}{1.923751in}}%
\pgfusepath{clip}%
\pgfsetrectcap%
\pgfsetroundjoin%
\pgfsetlinewidth{0.803000pt}%
\definecolor{currentstroke}{rgb}{0.690196,0.690196,0.690196}%
\pgfsetstrokecolor{currentstroke}%
\pgfsetdash{}{0pt}%
\pgfpathmoveto{\pgfqpoint{0.527360in}{1.503125in}}%
\pgfpathlineto{\pgfqpoint{3.165000in}{1.503125in}}%
\pgfusepath{stroke}%
\end{pgfscope}%
\begin{pgfscope}%
\pgfsetbuttcap%
\pgfsetroundjoin%
\definecolor{currentfill}{rgb}{0.000000,0.000000,0.000000}%
\pgfsetfillcolor{currentfill}%
\pgfsetlinewidth{0.803000pt}%
\definecolor{currentstroke}{rgb}{0.000000,0.000000,0.000000}%
\pgfsetstrokecolor{currentstroke}%
\pgfsetdash{}{0pt}%
\pgfsys@defobject{currentmarker}{\pgfqpoint{-0.048611in}{0.000000in}}{\pgfqpoint{-0.000000in}{0.000000in}}{%
\pgfpathmoveto{\pgfqpoint{-0.000000in}{0.000000in}}%
\pgfpathlineto{\pgfqpoint{-0.048611in}{0.000000in}}%
\pgfusepath{stroke,fill}%
}%
\begin{pgfscope}%
\pgfsys@transformshift{0.527360in}{1.503125in}%
\pgfsys@useobject{currentmarker}{}%
\end{pgfscope}%
\end{pgfscope}%
\begin{pgfscope}%
\definecolor{textcolor}{rgb}{0.000000,0.000000,0.000000}%
\pgfsetstrokecolor{textcolor}%
\pgfsetfillcolor{textcolor}%
\pgftext[x=0.301667in, y=1.459722in, left, base]{\color{textcolor}\rmfamily\fontsize{9.000000}{10.800000}\selectfont \(\displaystyle {30}\)}%
\end{pgfscope}%
\begin{pgfscope}%
\pgfpathrectangle{\pgfqpoint{0.527360in}{0.541249in}}{\pgfqpoint{2.637640in}{1.923751in}}%
\pgfusepath{clip}%
\pgfsetrectcap%
\pgfsetroundjoin%
\pgfsetlinewidth{0.803000pt}%
\definecolor{currentstroke}{rgb}{0.690196,0.690196,0.690196}%
\pgfsetstrokecolor{currentstroke}%
\pgfsetdash{}{0pt}%
\pgfpathmoveto{\pgfqpoint{0.527360in}{1.823750in}}%
\pgfpathlineto{\pgfqpoint{3.165000in}{1.823750in}}%
\pgfusepath{stroke}%
\end{pgfscope}%
\begin{pgfscope}%
\pgfsetbuttcap%
\pgfsetroundjoin%
\definecolor{currentfill}{rgb}{0.000000,0.000000,0.000000}%
\pgfsetfillcolor{currentfill}%
\pgfsetlinewidth{0.803000pt}%
\definecolor{currentstroke}{rgb}{0.000000,0.000000,0.000000}%
\pgfsetstrokecolor{currentstroke}%
\pgfsetdash{}{0pt}%
\pgfsys@defobject{currentmarker}{\pgfqpoint{-0.048611in}{0.000000in}}{\pgfqpoint{-0.000000in}{0.000000in}}{%
\pgfpathmoveto{\pgfqpoint{-0.000000in}{0.000000in}}%
\pgfpathlineto{\pgfqpoint{-0.048611in}{0.000000in}}%
\pgfusepath{stroke,fill}%
}%
\begin{pgfscope}%
\pgfsys@transformshift{0.527360in}{1.823750in}%
\pgfsys@useobject{currentmarker}{}%
\end{pgfscope}%
\end{pgfscope}%
\begin{pgfscope}%
\definecolor{textcolor}{rgb}{0.000000,0.000000,0.000000}%
\pgfsetstrokecolor{textcolor}%
\pgfsetfillcolor{textcolor}%
\pgftext[x=0.301667in, y=1.780347in, left, base]{\color{textcolor}\rmfamily\fontsize{9.000000}{10.800000}\selectfont \(\displaystyle {40}\)}%
\end{pgfscope}%
\begin{pgfscope}%
\pgfpathrectangle{\pgfqpoint{0.527360in}{0.541249in}}{\pgfqpoint{2.637640in}{1.923751in}}%
\pgfusepath{clip}%
\pgfsetrectcap%
\pgfsetroundjoin%
\pgfsetlinewidth{0.803000pt}%
\definecolor{currentstroke}{rgb}{0.690196,0.690196,0.690196}%
\pgfsetstrokecolor{currentstroke}%
\pgfsetdash{}{0pt}%
\pgfpathmoveto{\pgfqpoint{0.527360in}{2.144375in}}%
\pgfpathlineto{\pgfqpoint{3.165000in}{2.144375in}}%
\pgfusepath{stroke}%
\end{pgfscope}%
\begin{pgfscope}%
\pgfsetbuttcap%
\pgfsetroundjoin%
\definecolor{currentfill}{rgb}{0.000000,0.000000,0.000000}%
\pgfsetfillcolor{currentfill}%
\pgfsetlinewidth{0.803000pt}%
\definecolor{currentstroke}{rgb}{0.000000,0.000000,0.000000}%
\pgfsetstrokecolor{currentstroke}%
\pgfsetdash{}{0pt}%
\pgfsys@defobject{currentmarker}{\pgfqpoint{-0.048611in}{0.000000in}}{\pgfqpoint{-0.000000in}{0.000000in}}{%
\pgfpathmoveto{\pgfqpoint{-0.000000in}{0.000000in}}%
\pgfpathlineto{\pgfqpoint{-0.048611in}{0.000000in}}%
\pgfusepath{stroke,fill}%
}%
\begin{pgfscope}%
\pgfsys@transformshift{0.527360in}{2.144375in}%
\pgfsys@useobject{currentmarker}{}%
\end{pgfscope}%
\end{pgfscope}%
\begin{pgfscope}%
\definecolor{textcolor}{rgb}{0.000000,0.000000,0.000000}%
\pgfsetstrokecolor{textcolor}%
\pgfsetfillcolor{textcolor}%
\pgftext[x=0.301667in, y=2.100972in, left, base]{\color{textcolor}\rmfamily\fontsize{9.000000}{10.800000}\selectfont \(\displaystyle {50}\)}%
\end{pgfscope}%
\begin{pgfscope}%
\pgfpathrectangle{\pgfqpoint{0.527360in}{0.541249in}}{\pgfqpoint{2.637640in}{1.923751in}}%
\pgfusepath{clip}%
\pgfsetrectcap%
\pgfsetroundjoin%
\pgfsetlinewidth{0.803000pt}%
\definecolor{currentstroke}{rgb}{0.690196,0.690196,0.690196}%
\pgfsetstrokecolor{currentstroke}%
\pgfsetdash{}{0pt}%
\pgfpathmoveto{\pgfqpoint{0.527360in}{2.465000in}}%
\pgfpathlineto{\pgfqpoint{3.165000in}{2.465000in}}%
\pgfusepath{stroke}%
\end{pgfscope}%
\begin{pgfscope}%
\pgfsetbuttcap%
\pgfsetroundjoin%
\definecolor{currentfill}{rgb}{0.000000,0.000000,0.000000}%
\pgfsetfillcolor{currentfill}%
\pgfsetlinewidth{0.803000pt}%
\definecolor{currentstroke}{rgb}{0.000000,0.000000,0.000000}%
\pgfsetstrokecolor{currentstroke}%
\pgfsetdash{}{0pt}%
\pgfsys@defobject{currentmarker}{\pgfqpoint{-0.048611in}{0.000000in}}{\pgfqpoint{-0.000000in}{0.000000in}}{%
\pgfpathmoveto{\pgfqpoint{-0.000000in}{0.000000in}}%
\pgfpathlineto{\pgfqpoint{-0.048611in}{0.000000in}}%
\pgfusepath{stroke,fill}%
}%
\begin{pgfscope}%
\pgfsys@transformshift{0.527360in}{2.465000in}%
\pgfsys@useobject{currentmarker}{}%
\end{pgfscope}%
\end{pgfscope}%
\begin{pgfscope}%
\definecolor{textcolor}{rgb}{0.000000,0.000000,0.000000}%
\pgfsetstrokecolor{textcolor}%
\pgfsetfillcolor{textcolor}%
\pgftext[x=0.301667in, y=2.421597in, left, base]{\color{textcolor}\rmfamily\fontsize{9.000000}{10.800000}\selectfont \(\displaystyle {60}\)}%
\end{pgfscope}%
\begin{pgfscope}%
\definecolor{textcolor}{rgb}{0.000000,0.000000,0.000000}%
\pgfsetstrokecolor{textcolor}%
\pgfsetfillcolor{textcolor}%
\pgftext[x=0.246111in,y=1.503125in,,bottom,rotate=90.000000]{\color{textcolor}\rmfamily\fontsize{9.000000}{10.800000}\selectfont PSNR}%
\end{pgfscope}%
\begin{pgfscope}%
\pgfpathrectangle{\pgfqpoint{0.527360in}{0.541249in}}{\pgfqpoint{2.637640in}{1.923751in}}%
\pgfusepath{clip}%
\pgfsetbuttcap%
\pgfsetmiterjoin%
\definecolor{currentfill}{rgb}{0.121569,0.466667,0.705882}%
\pgfsetfillcolor{currentfill}%
\pgfsetlinewidth{0.000000pt}%
\definecolor{currentstroke}{rgb}{0.000000,0.000000,0.000000}%
\pgfsetstrokecolor{currentstroke}%
\pgfsetstrokeopacity{0.000000}%
\pgfsetdash{}{0pt}%
\pgfpathmoveto{\pgfqpoint{0.637262in}{0.541249in}}%
\pgfpathlineto{\pgfqpoint{0.710530in}{0.541249in}}%
\pgfpathlineto{\pgfqpoint{0.710530in}{2.160406in}}%
\pgfpathlineto{\pgfqpoint{0.637262in}{2.160406in}}%
\pgfpathlineto{\pgfqpoint{0.637262in}{0.541249in}}%
\pgfpathclose%
\pgfusepath{fill}%
\end{pgfscope}%
\begin{pgfscope}%
\pgfpathrectangle{\pgfqpoint{0.527360in}{0.541249in}}{\pgfqpoint{2.637640in}{1.923751in}}%
\pgfusepath{clip}%
\pgfsetbuttcap%
\pgfsetmiterjoin%
\definecolor{currentfill}{rgb}{0.121569,0.466667,0.705882}%
\pgfsetfillcolor{currentfill}%
\pgfsetlinewidth{0.000000pt}%
\definecolor{currentstroke}{rgb}{0.000000,0.000000,0.000000}%
\pgfsetstrokecolor{currentstroke}%
\pgfsetstrokeopacity{0.000000}%
\pgfsetdash{}{0pt}%
\pgfpathmoveto{\pgfqpoint{1.076869in}{0.541249in}}%
\pgfpathlineto{\pgfqpoint{1.150136in}{0.541249in}}%
\pgfpathlineto{\pgfqpoint{1.150136in}{2.096281in}}%
\pgfpathlineto{\pgfqpoint{1.076869in}{2.096281in}}%
\pgfpathlineto{\pgfqpoint{1.076869in}{0.541249in}}%
\pgfpathclose%
\pgfusepath{fill}%
\end{pgfscope}%
\begin{pgfscope}%
\pgfpathrectangle{\pgfqpoint{0.527360in}{0.541249in}}{\pgfqpoint{2.637640in}{1.923751in}}%
\pgfusepath{clip}%
\pgfsetbuttcap%
\pgfsetmiterjoin%
\definecolor{currentfill}{rgb}{0.121569,0.466667,0.705882}%
\pgfsetfillcolor{currentfill}%
\pgfsetlinewidth{0.000000pt}%
\definecolor{currentstroke}{rgb}{0.000000,0.000000,0.000000}%
\pgfsetstrokecolor{currentstroke}%
\pgfsetstrokeopacity{0.000000}%
\pgfsetdash{}{0pt}%
\pgfpathmoveto{\pgfqpoint{1.516475in}{0.541249in}}%
\pgfpathlineto{\pgfqpoint{1.589743in}{0.541249in}}%
\pgfpathlineto{\pgfqpoint{1.589743in}{1.968031in}}%
\pgfpathlineto{\pgfqpoint{1.516475in}{1.968031in}}%
\pgfpathlineto{\pgfqpoint{1.516475in}{0.541249in}}%
\pgfpathclose%
\pgfusepath{fill}%
\end{pgfscope}%
\begin{pgfscope}%
\pgfpathrectangle{\pgfqpoint{0.527360in}{0.541249in}}{\pgfqpoint{2.637640in}{1.923751in}}%
\pgfusepath{clip}%
\pgfsetbuttcap%
\pgfsetmiterjoin%
\definecolor{currentfill}{rgb}{0.121569,0.466667,0.705882}%
\pgfsetfillcolor{currentfill}%
\pgfsetlinewidth{0.000000pt}%
\definecolor{currentstroke}{rgb}{0.000000,0.000000,0.000000}%
\pgfsetstrokecolor{currentstroke}%
\pgfsetstrokeopacity{0.000000}%
\pgfsetdash{}{0pt}%
\pgfpathmoveto{\pgfqpoint{1.956082in}{0.541249in}}%
\pgfpathlineto{\pgfqpoint{2.029350in}{0.541249in}}%
\pgfpathlineto{\pgfqpoint{2.029350in}{1.825353in}}%
\pgfpathlineto{\pgfqpoint{1.956082in}{1.825353in}}%
\pgfpathlineto{\pgfqpoint{1.956082in}{0.541249in}}%
\pgfpathclose%
\pgfusepath{fill}%
\end{pgfscope}%
\begin{pgfscope}%
\pgfpathrectangle{\pgfqpoint{0.527360in}{0.541249in}}{\pgfqpoint{2.637640in}{1.923751in}}%
\pgfusepath{clip}%
\pgfsetbuttcap%
\pgfsetmiterjoin%
\definecolor{currentfill}{rgb}{0.121569,0.466667,0.705882}%
\pgfsetfillcolor{currentfill}%
\pgfsetlinewidth{0.000000pt}%
\definecolor{currentstroke}{rgb}{0.000000,0.000000,0.000000}%
\pgfsetstrokecolor{currentstroke}%
\pgfsetstrokeopacity{0.000000}%
\pgfsetdash{}{0pt}%
\pgfpathmoveto{\pgfqpoint{2.395688in}{0.541249in}}%
\pgfpathlineto{\pgfqpoint{2.468956in}{0.541249in}}%
\pgfpathlineto{\pgfqpoint{2.468956in}{1.727562in}}%
\pgfpathlineto{\pgfqpoint{2.395688in}{1.727562in}}%
\pgfpathlineto{\pgfqpoint{2.395688in}{0.541249in}}%
\pgfpathclose%
\pgfusepath{fill}%
\end{pgfscope}%
\begin{pgfscope}%
\pgfpathrectangle{\pgfqpoint{0.527360in}{0.541249in}}{\pgfqpoint{2.637640in}{1.923751in}}%
\pgfusepath{clip}%
\pgfsetbuttcap%
\pgfsetmiterjoin%
\definecolor{currentfill}{rgb}{0.121569,0.466667,0.705882}%
\pgfsetfillcolor{currentfill}%
\pgfsetlinewidth{0.000000pt}%
\definecolor{currentstroke}{rgb}{0.000000,0.000000,0.000000}%
\pgfsetstrokecolor{currentstroke}%
\pgfsetstrokeopacity{0.000000}%
\pgfsetdash{}{0pt}%
\pgfpathmoveto{\pgfqpoint{2.835295in}{0.541249in}}%
\pgfpathlineto{\pgfqpoint{2.908563in}{0.541249in}}%
\pgfpathlineto{\pgfqpoint{2.908563in}{1.573662in}}%
\pgfpathlineto{\pgfqpoint{2.835295in}{1.573662in}}%
\pgfpathlineto{\pgfqpoint{2.835295in}{0.541249in}}%
\pgfpathclose%
\pgfusepath{fill}%
\end{pgfscope}%
\begin{pgfscope}%
\pgfpathrectangle{\pgfqpoint{0.527360in}{0.541249in}}{\pgfqpoint{2.637640in}{1.923751in}}%
\pgfusepath{clip}%
\pgfsetbuttcap%
\pgfsetmiterjoin%
\definecolor{currentfill}{rgb}{1.000000,0.498039,0.054902}%
\pgfsetfillcolor{currentfill}%
\pgfsetlinewidth{0.000000pt}%
\definecolor{currentstroke}{rgb}{0.000000,0.000000,0.000000}%
\pgfsetstrokecolor{currentstroke}%
\pgfsetstrokeopacity{0.000000}%
\pgfsetdash{}{0pt}%
\pgfpathmoveto{\pgfqpoint{0.710530in}{0.541249in}}%
\pgfpathlineto{\pgfqpoint{0.783797in}{0.541249in}}%
\pgfpathlineto{\pgfqpoint{0.783797in}{2.182850in}}%
\pgfpathlineto{\pgfqpoint{0.710530in}{2.182850in}}%
\pgfpathlineto{\pgfqpoint{0.710530in}{0.541249in}}%
\pgfpathclose%
\pgfusepath{fill}%
\end{pgfscope}%
\begin{pgfscope}%
\pgfpathrectangle{\pgfqpoint{0.527360in}{0.541249in}}{\pgfqpoint{2.637640in}{1.923751in}}%
\pgfusepath{clip}%
\pgfsetbuttcap%
\pgfsetmiterjoin%
\definecolor{currentfill}{rgb}{1.000000,0.498039,0.054902}%
\pgfsetfillcolor{currentfill}%
\pgfsetlinewidth{0.000000pt}%
\definecolor{currentstroke}{rgb}{0.000000,0.000000,0.000000}%
\pgfsetstrokecolor{currentstroke}%
\pgfsetstrokeopacity{0.000000}%
\pgfsetdash{}{0pt}%
\pgfpathmoveto{\pgfqpoint{1.150136in}{0.541249in}}%
\pgfpathlineto{\pgfqpoint{1.223404in}{0.541249in}}%
\pgfpathlineto{\pgfqpoint{1.223404in}{2.112312in}}%
\pgfpathlineto{\pgfqpoint{1.150136in}{2.112312in}}%
\pgfpathlineto{\pgfqpoint{1.150136in}{0.541249in}}%
\pgfpathclose%
\pgfusepath{fill}%
\end{pgfscope}%
\begin{pgfscope}%
\pgfpathrectangle{\pgfqpoint{0.527360in}{0.541249in}}{\pgfqpoint{2.637640in}{1.923751in}}%
\pgfusepath{clip}%
\pgfsetbuttcap%
\pgfsetmiterjoin%
\definecolor{currentfill}{rgb}{1.000000,0.498039,0.054902}%
\pgfsetfillcolor{currentfill}%
\pgfsetlinewidth{0.000000pt}%
\definecolor{currentstroke}{rgb}{0.000000,0.000000,0.000000}%
\pgfsetstrokecolor{currentstroke}%
\pgfsetstrokeopacity{0.000000}%
\pgfsetdash{}{0pt}%
\pgfpathmoveto{\pgfqpoint{1.589743in}{0.541249in}}%
\pgfpathlineto{\pgfqpoint{1.663011in}{0.541249in}}%
\pgfpathlineto{\pgfqpoint{1.663011in}{1.984062in}}%
\pgfpathlineto{\pgfqpoint{1.589743in}{1.984062in}}%
\pgfpathlineto{\pgfqpoint{1.589743in}{0.541249in}}%
\pgfpathclose%
\pgfusepath{fill}%
\end{pgfscope}%
\begin{pgfscope}%
\pgfpathrectangle{\pgfqpoint{0.527360in}{0.541249in}}{\pgfqpoint{2.637640in}{1.923751in}}%
\pgfusepath{clip}%
\pgfsetbuttcap%
\pgfsetmiterjoin%
\definecolor{currentfill}{rgb}{1.000000,0.498039,0.054902}%
\pgfsetfillcolor{currentfill}%
\pgfsetlinewidth{0.000000pt}%
\definecolor{currentstroke}{rgb}{0.000000,0.000000,0.000000}%
\pgfsetstrokecolor{currentstroke}%
\pgfsetstrokeopacity{0.000000}%
\pgfsetdash{}{0pt}%
\pgfpathmoveto{\pgfqpoint{2.029350in}{0.541249in}}%
\pgfpathlineto{\pgfqpoint{2.102617in}{0.541249in}}%
\pgfpathlineto{\pgfqpoint{2.102617in}{1.849400in}}%
\pgfpathlineto{\pgfqpoint{2.029350in}{1.849400in}}%
\pgfpathlineto{\pgfqpoint{2.029350in}{0.541249in}}%
\pgfpathclose%
\pgfusepath{fill}%
\end{pgfscope}%
\begin{pgfscope}%
\pgfpathrectangle{\pgfqpoint{0.527360in}{0.541249in}}{\pgfqpoint{2.637640in}{1.923751in}}%
\pgfusepath{clip}%
\pgfsetbuttcap%
\pgfsetmiterjoin%
\definecolor{currentfill}{rgb}{1.000000,0.498039,0.054902}%
\pgfsetfillcolor{currentfill}%
\pgfsetlinewidth{0.000000pt}%
\definecolor{currentstroke}{rgb}{0.000000,0.000000,0.000000}%
\pgfsetstrokecolor{currentstroke}%
\pgfsetstrokeopacity{0.000000}%
\pgfsetdash{}{0pt}%
\pgfpathmoveto{\pgfqpoint{2.468956in}{0.541249in}}%
\pgfpathlineto{\pgfqpoint{2.542224in}{0.541249in}}%
\pgfpathlineto{\pgfqpoint{2.542224in}{1.759625in}}%
\pgfpathlineto{\pgfqpoint{2.468956in}{1.759625in}}%
\pgfpathlineto{\pgfqpoint{2.468956in}{0.541249in}}%
\pgfpathclose%
\pgfusepath{fill}%
\end{pgfscope}%
\begin{pgfscope}%
\pgfpathrectangle{\pgfqpoint{0.527360in}{0.541249in}}{\pgfqpoint{2.637640in}{1.923751in}}%
\pgfusepath{clip}%
\pgfsetbuttcap%
\pgfsetmiterjoin%
\definecolor{currentfill}{rgb}{1.000000,0.498039,0.054902}%
\pgfsetfillcolor{currentfill}%
\pgfsetlinewidth{0.000000pt}%
\definecolor{currentstroke}{rgb}{0.000000,0.000000,0.000000}%
\pgfsetstrokecolor{currentstroke}%
\pgfsetstrokeopacity{0.000000}%
\pgfsetdash{}{0pt}%
\pgfpathmoveto{\pgfqpoint{2.908563in}{0.541249in}}%
\pgfpathlineto{\pgfqpoint{2.981831in}{0.541249in}}%
\pgfpathlineto{\pgfqpoint{2.981831in}{1.599312in}}%
\pgfpathlineto{\pgfqpoint{2.908563in}{1.599312in}}%
\pgfpathlineto{\pgfqpoint{2.908563in}{0.541249in}}%
\pgfpathclose%
\pgfusepath{fill}%
\end{pgfscope}%
\begin{pgfscope}%
\pgfpathrectangle{\pgfqpoint{0.527360in}{0.541249in}}{\pgfqpoint{2.637640in}{1.923751in}}%
\pgfusepath{clip}%
\pgfsetbuttcap%
\pgfsetmiterjoin%
\definecolor{currentfill}{rgb}{0.172549,0.627451,0.172549}%
\pgfsetfillcolor{currentfill}%
\pgfsetlinewidth{0.000000pt}%
\definecolor{currentstroke}{rgb}{0.000000,0.000000,0.000000}%
\pgfsetstrokecolor{currentstroke}%
\pgfsetstrokeopacity{0.000000}%
\pgfsetdash{}{0pt}%
\pgfpathmoveto{\pgfqpoint{0.783797in}{0.541249in}}%
\pgfpathlineto{\pgfqpoint{0.857065in}{0.541249in}}%
\pgfpathlineto{\pgfqpoint{0.857065in}{2.165536in}}%
\pgfpathlineto{\pgfqpoint{0.783797in}{2.165536in}}%
\pgfpathlineto{\pgfqpoint{0.783797in}{0.541249in}}%
\pgfpathclose%
\pgfusepath{fill}%
\end{pgfscope}%
\begin{pgfscope}%
\pgfpathrectangle{\pgfqpoint{0.527360in}{0.541249in}}{\pgfqpoint{2.637640in}{1.923751in}}%
\pgfusepath{clip}%
\pgfsetbuttcap%
\pgfsetmiterjoin%
\definecolor{currentfill}{rgb}{0.172549,0.627451,0.172549}%
\pgfsetfillcolor{currentfill}%
\pgfsetlinewidth{0.000000pt}%
\definecolor{currentstroke}{rgb}{0.000000,0.000000,0.000000}%
\pgfsetstrokecolor{currentstroke}%
\pgfsetstrokeopacity{0.000000}%
\pgfsetdash{}{0pt}%
\pgfpathmoveto{\pgfqpoint{1.223404in}{0.541249in}}%
\pgfpathlineto{\pgfqpoint{1.296672in}{0.541249in}}%
\pgfpathlineto{\pgfqpoint{1.296672in}{2.011315in}}%
\pgfpathlineto{\pgfqpoint{1.223404in}{2.011315in}}%
\pgfpathlineto{\pgfqpoint{1.223404in}{0.541249in}}%
\pgfpathclose%
\pgfusepath{fill}%
\end{pgfscope}%
\begin{pgfscope}%
\pgfpathrectangle{\pgfqpoint{0.527360in}{0.541249in}}{\pgfqpoint{2.637640in}{1.923751in}}%
\pgfusepath{clip}%
\pgfsetbuttcap%
\pgfsetmiterjoin%
\definecolor{currentfill}{rgb}{0.172549,0.627451,0.172549}%
\pgfsetfillcolor{currentfill}%
\pgfsetlinewidth{0.000000pt}%
\definecolor{currentstroke}{rgb}{0.000000,0.000000,0.000000}%
\pgfsetstrokecolor{currentstroke}%
\pgfsetstrokeopacity{0.000000}%
\pgfsetdash{}{0pt}%
\pgfpathmoveto{\pgfqpoint{1.663011in}{0.541249in}}%
\pgfpathlineto{\pgfqpoint{1.736278in}{0.541249in}}%
\pgfpathlineto{\pgfqpoint{1.736278in}{1.833048in}}%
\pgfpathlineto{\pgfqpoint{1.663011in}{1.833048in}}%
\pgfpathlineto{\pgfqpoint{1.663011in}{0.541249in}}%
\pgfpathclose%
\pgfusepath{fill}%
\end{pgfscope}%
\begin{pgfscope}%
\pgfpathrectangle{\pgfqpoint{0.527360in}{0.541249in}}{\pgfqpoint{2.637640in}{1.923751in}}%
\pgfusepath{clip}%
\pgfsetbuttcap%
\pgfsetmiterjoin%
\definecolor{currentfill}{rgb}{0.172549,0.627451,0.172549}%
\pgfsetfillcolor{currentfill}%
\pgfsetlinewidth{0.000000pt}%
\definecolor{currentstroke}{rgb}{0.000000,0.000000,0.000000}%
\pgfsetstrokecolor{currentstroke}%
\pgfsetstrokeopacity{0.000000}%
\pgfsetdash{}{0pt}%
\pgfpathmoveto{\pgfqpoint{2.102617in}{0.541249in}}%
\pgfpathlineto{\pgfqpoint{2.175885in}{0.541249in}}%
\pgfpathlineto{\pgfqpoint{2.175885in}{1.643238in}}%
\pgfpathlineto{\pgfqpoint{2.102617in}{1.643238in}}%
\pgfpathlineto{\pgfqpoint{2.102617in}{0.541249in}}%
\pgfpathclose%
\pgfusepath{fill}%
\end{pgfscope}%
\begin{pgfscope}%
\pgfpathrectangle{\pgfqpoint{0.527360in}{0.541249in}}{\pgfqpoint{2.637640in}{1.923751in}}%
\pgfusepath{clip}%
\pgfsetbuttcap%
\pgfsetmiterjoin%
\definecolor{currentfill}{rgb}{0.172549,0.627451,0.172549}%
\pgfsetfillcolor{currentfill}%
\pgfsetlinewidth{0.000000pt}%
\definecolor{currentstroke}{rgb}{0.000000,0.000000,0.000000}%
\pgfsetstrokecolor{currentstroke}%
\pgfsetstrokeopacity{0.000000}%
\pgfsetdash{}{0pt}%
\pgfpathmoveto{\pgfqpoint{2.542224in}{0.541249in}}%
\pgfpathlineto{\pgfqpoint{2.615492in}{0.541249in}}%
\pgfpathlineto{\pgfqpoint{2.615492in}{1.461443in}}%
\pgfpathlineto{\pgfqpoint{2.542224in}{1.461443in}}%
\pgfpathlineto{\pgfqpoint{2.542224in}{0.541249in}}%
\pgfpathclose%
\pgfusepath{fill}%
\end{pgfscope}%
\begin{pgfscope}%
\pgfpathrectangle{\pgfqpoint{0.527360in}{0.541249in}}{\pgfqpoint{2.637640in}{1.923751in}}%
\pgfusepath{clip}%
\pgfsetbuttcap%
\pgfsetmiterjoin%
\definecolor{currentfill}{rgb}{0.172549,0.627451,0.172549}%
\pgfsetfillcolor{currentfill}%
\pgfsetlinewidth{0.000000pt}%
\definecolor{currentstroke}{rgb}{0.000000,0.000000,0.000000}%
\pgfsetstrokecolor{currentstroke}%
\pgfsetstrokeopacity{0.000000}%
\pgfsetdash{}{0pt}%
\pgfpathmoveto{\pgfqpoint{2.981831in}{0.541249in}}%
\pgfpathlineto{\pgfqpoint{3.055098in}{0.541249in}}%
\pgfpathlineto{\pgfqpoint{3.055098in}{1.284779in}}%
\pgfpathlineto{\pgfqpoint{2.981831in}{1.284779in}}%
\pgfpathlineto{\pgfqpoint{2.981831in}{0.541249in}}%
\pgfpathclose%
\pgfusepath{fill}%
\end{pgfscope}%
\begin{pgfscope}%
\pgfsetrectcap%
\pgfsetmiterjoin%
\pgfsetlinewidth{0.803000pt}%
\definecolor{currentstroke}{rgb}{0.000000,0.000000,0.000000}%
\pgfsetstrokecolor{currentstroke}%
\pgfsetdash{}{0pt}%
\pgfpathmoveto{\pgfqpoint{0.527360in}{0.541249in}}%
\pgfpathlineto{\pgfqpoint{0.527360in}{2.465000in}}%
\pgfusepath{stroke}%
\end{pgfscope}%
\begin{pgfscope}%
\pgfsetrectcap%
\pgfsetmiterjoin%
\pgfsetlinewidth{0.803000pt}%
\definecolor{currentstroke}{rgb}{0.000000,0.000000,0.000000}%
\pgfsetstrokecolor{currentstroke}%
\pgfsetdash{}{0pt}%
\pgfpathmoveto{\pgfqpoint{3.165000in}{0.541249in}}%
\pgfpathlineto{\pgfqpoint{3.165000in}{2.465000in}}%
\pgfusepath{stroke}%
\end{pgfscope}%
\begin{pgfscope}%
\pgfsetrectcap%
\pgfsetmiterjoin%
\pgfsetlinewidth{0.803000pt}%
\definecolor{currentstroke}{rgb}{0.000000,0.000000,0.000000}%
\pgfsetstrokecolor{currentstroke}%
\pgfsetdash{}{0pt}%
\pgfpathmoveto{\pgfqpoint{0.527360in}{0.541249in}}%
\pgfpathlineto{\pgfqpoint{3.165000in}{0.541249in}}%
\pgfusepath{stroke}%
\end{pgfscope}%
\begin{pgfscope}%
\pgfsetrectcap%
\pgfsetmiterjoin%
\pgfsetlinewidth{0.803000pt}%
\definecolor{currentstroke}{rgb}{0.000000,0.000000,0.000000}%
\pgfsetstrokecolor{currentstroke}%
\pgfsetdash{}{0pt}%
\pgfpathmoveto{\pgfqpoint{0.527360in}{2.465000in}}%
\pgfpathlineto{\pgfqpoint{3.165000in}{2.465000in}}%
\pgfusepath{stroke}%
\end{pgfscope}%
\begin{pgfscope}%
\pgfsetbuttcap%
\pgfsetmiterjoin%
\definecolor{currentfill}{rgb}{1.000000,1.000000,1.000000}%
\pgfsetfillcolor{currentfill}%
\pgfsetfillopacity{0.800000}%
\pgfsetlinewidth{1.003750pt}%
\definecolor{currentstroke}{rgb}{0.800000,0.800000,0.800000}%
\pgfsetstrokecolor{currentstroke}%
\pgfsetstrokeopacity{0.800000}%
\pgfsetdash{}{0pt}%
\pgfpathmoveto{\pgfqpoint{2.134237in}{1.842083in}}%
\pgfpathlineto{\pgfqpoint{3.077500in}{1.842083in}}%
\pgfpathquadraticcurveto{\pgfqpoint{3.102500in}{1.842083in}}{\pgfqpoint{3.102500in}{1.867083in}}%
\pgfpathlineto{\pgfqpoint{3.102500in}{2.377500in}}%
\pgfpathquadraticcurveto{\pgfqpoint{3.102500in}{2.402500in}}{\pgfqpoint{3.077500in}{2.402500in}}%
\pgfpathlineto{\pgfqpoint{2.134237in}{2.402500in}}%
\pgfpathquadraticcurveto{\pgfqpoint{2.109237in}{2.402500in}}{\pgfqpoint{2.109237in}{2.377500in}}%
\pgfpathlineto{\pgfqpoint{2.109237in}{1.867083in}}%
\pgfpathquadraticcurveto{\pgfqpoint{2.109237in}{1.842083in}}{\pgfqpoint{2.134237in}{1.842083in}}%
\pgfpathlineto{\pgfqpoint{2.134237in}{1.842083in}}%
\pgfpathclose%
\pgfusepath{stroke,fill}%
\end{pgfscope}%
\begin{pgfscope}%
\pgfsetbuttcap%
\pgfsetmiterjoin%
\definecolor{currentfill}{rgb}{0.121569,0.466667,0.705882}%
\pgfsetfillcolor{currentfill}%
\pgfsetlinewidth{0.000000pt}%
\definecolor{currentstroke}{rgb}{0.000000,0.000000,0.000000}%
\pgfsetstrokecolor{currentstroke}%
\pgfsetstrokeopacity{0.000000}%
\pgfsetdash{}{0pt}%
\pgfpathmoveto{\pgfqpoint{2.159237in}{2.265000in}}%
\pgfpathlineto{\pgfqpoint{2.409237in}{2.265000in}}%
\pgfpathlineto{\pgfqpoint{2.409237in}{2.352500in}}%
\pgfpathlineto{\pgfqpoint{2.159237in}{2.352500in}}%
\pgfpathlineto{\pgfqpoint{2.159237in}{2.265000in}}%
\pgfpathclose%
\pgfusepath{fill}%
\end{pgfscope}%
\begin{pgfscope}%
\definecolor{textcolor}{rgb}{0.000000,0.000000,0.000000}%
\pgfsetstrokecolor{textcolor}%
\pgfsetfillcolor{textcolor}%
\pgftext[x=2.509237in,y=2.265000in,left,base]{\color{textcolor}\rmfamily\fontsize{9.000000}{10.800000}\selectfont ESRGAN}%
\end{pgfscope}%
\begin{pgfscope}%
\pgfsetbuttcap%
\pgfsetmiterjoin%
\definecolor{currentfill}{rgb}{1.000000,0.498039,0.054902}%
\pgfsetfillcolor{currentfill}%
\pgfsetlinewidth{0.000000pt}%
\definecolor{currentstroke}{rgb}{0.000000,0.000000,0.000000}%
\pgfsetstrokecolor{currentstroke}%
\pgfsetstrokeopacity{0.000000}%
\pgfsetdash{}{0pt}%
\pgfpathmoveto{\pgfqpoint{2.159237in}{2.090694in}}%
\pgfpathlineto{\pgfqpoint{2.409237in}{2.090694in}}%
\pgfpathlineto{\pgfqpoint{2.409237in}{2.178194in}}%
\pgfpathlineto{\pgfqpoint{2.159237in}{2.178194in}}%
\pgfpathlineto{\pgfqpoint{2.159237in}{2.090694in}}%
\pgfpathclose%
\pgfusepath{fill}%
\end{pgfscope}%
\begin{pgfscope}%
\definecolor{textcolor}{rgb}{0.000000,0.000000,0.000000}%
\pgfsetstrokecolor{textcolor}%
\pgfsetfillcolor{textcolor}%
\pgftext[x=2.509237in,y=2.090694in,left,base]{\color{textcolor}\rmfamily\fontsize{9.000000}{10.800000}\selectfont EDSR}%
\end{pgfscope}%
\begin{pgfscope}%
\pgfsetbuttcap%
\pgfsetmiterjoin%
\definecolor{currentfill}{rgb}{0.172549,0.627451,0.172549}%
\pgfsetfillcolor{currentfill}%
\pgfsetlinewidth{0.000000pt}%
\definecolor{currentstroke}{rgb}{0.000000,0.000000,0.000000}%
\pgfsetstrokecolor{currentstroke}%
\pgfsetstrokeopacity{0.000000}%
\pgfsetdash{}{0pt}%
\pgfpathmoveto{\pgfqpoint{2.159237in}{1.916389in}}%
\pgfpathlineto{\pgfqpoint{2.409237in}{1.916389in}}%
\pgfpathlineto{\pgfqpoint{2.409237in}{2.003889in}}%
\pgfpathlineto{\pgfqpoint{2.159237in}{2.003889in}}%
\pgfpathlineto{\pgfqpoint{2.159237in}{1.916389in}}%
\pgfpathclose%
\pgfusepath{fill}%
\end{pgfscope}%
\begin{pgfscope}%
\definecolor{textcolor}{rgb}{0.000000,0.000000,0.000000}%
\pgfsetstrokecolor{textcolor}%
\pgfsetfillcolor{textcolor}%
\pgftext[x=2.509237in,y=1.916389in,left,base]{\color{textcolor}\rmfamily\fontsize{9.000000}{10.800000}\selectfont SRFlow}%
\end{pgfscope}%
\end{pgfpicture}%
\makeatother%
\endgroup%

%% file: latex/figures/adv_set14.pgf
\begingroup%
\makeatletter%
\begin{pgfpicture}%
\pgfpathrectangle{\pgfpointorigin}{\pgfqpoint{3.300000in}{2.600000in}}%
\pgfusepath{use as bounding box, clip}%
\begin{pgfscope}%
\pgfsetbuttcap%
\pgfsetmiterjoin%
\definecolor{currentfill}{rgb}{1.000000,1.000000,1.000000}%
\pgfsetfillcolor{currentfill}%
\pgfsetlinewidth{0.000000pt}%
\definecolor{currentstroke}{rgb}{1.000000,1.000000,1.000000}%
\pgfsetstrokecolor{currentstroke}%
\pgfsetdash{}{0pt}%
\pgfpathmoveto{\pgfqpoint{0.000000in}{0.000000in}}%
\pgfpathlineto{\pgfqpoint{3.300000in}{0.000000in}}%
\pgfpathlineto{\pgfqpoint{3.300000in}{2.600000in}}%
\pgfpathlineto{\pgfqpoint{0.000000in}{2.600000in}}%
\pgfpathlineto{\pgfqpoint{0.000000in}{0.000000in}}%
\pgfpathclose%
\pgfusepath{fill}%
\end{pgfscope}%
\begin{pgfscope}%
\pgfsetbuttcap%
\pgfsetmiterjoin%
\definecolor{currentfill}{rgb}{1.000000,1.000000,1.000000}%
\pgfsetfillcolor{currentfill}%
\pgfsetlinewidth{0.000000pt}%
\definecolor{currentstroke}{rgb}{0.000000,0.000000,0.000000}%
\pgfsetstrokecolor{currentstroke}%
\pgfsetstrokeopacity{0.000000}%
\pgfsetdash{}{0pt}%
\pgfpathmoveto{\pgfqpoint{0.527360in}{0.541249in}}%
\pgfpathlineto{\pgfqpoint{3.165000in}{0.541249in}}%
\pgfpathlineto{\pgfqpoint{3.165000in}{2.465000in}}%
\pgfpathlineto{\pgfqpoint{0.527360in}{2.465000in}}%
\pgfpathlineto{\pgfqpoint{0.527360in}{0.541249in}}%
\pgfpathclose%
\pgfusepath{fill}%
\end{pgfscope}%
\begin{pgfscope}%
\pgfsetbuttcap%
\pgfsetroundjoin%
\definecolor{currentfill}{rgb}{0.000000,0.000000,0.000000}%
\pgfsetfillcolor{currentfill}%
\pgfsetlinewidth{0.803000pt}%
\definecolor{currentstroke}{rgb}{0.000000,0.000000,0.000000}%
\pgfsetstrokecolor{currentstroke}%
\pgfsetdash{}{0pt}%
\pgfsys@defobject{currentmarker}{\pgfqpoint{0.000000in}{-0.048611in}}{\pgfqpoint{0.000000in}{0.000000in}}{%
\pgfpathmoveto{\pgfqpoint{0.000000in}{0.000000in}}%
\pgfpathlineto{\pgfqpoint{0.000000in}{-0.048611in}}%
\pgfusepath{stroke,fill}%
}%
\begin{pgfscope}%
\pgfsys@transformshift{0.747164in}{0.541249in}%
\pgfsys@useobject{currentmarker}{}%
\end{pgfscope}%
\end{pgfscope}%
\begin{pgfscope}%
\definecolor{textcolor}{rgb}{0.000000,0.000000,0.000000}%
\pgfsetstrokecolor{textcolor}%
\pgfsetfillcolor{textcolor}%
\pgftext[x=0.778414in, y=0.379791in, left, base,rotate=90.000000]{\color{textcolor}\rmfamily\fontsize{9.000000}{10.800000}\selectfont 1}%
\end{pgfscope}%
\begin{pgfscope}%
\pgfsetbuttcap%
\pgfsetroundjoin%
\definecolor{currentfill}{rgb}{0.000000,0.000000,0.000000}%
\pgfsetfillcolor{currentfill}%
\pgfsetlinewidth{0.803000pt}%
\definecolor{currentstroke}{rgb}{0.000000,0.000000,0.000000}%
\pgfsetstrokecolor{currentstroke}%
\pgfsetdash{}{0pt}%
\pgfsys@defobject{currentmarker}{\pgfqpoint{0.000000in}{-0.048611in}}{\pgfqpoint{0.000000in}{0.000000in}}{%
\pgfpathmoveto{\pgfqpoint{0.000000in}{0.000000in}}%
\pgfpathlineto{\pgfqpoint{0.000000in}{-0.048611in}}%
\pgfusepath{stroke,fill}%
}%
\begin{pgfscope}%
\pgfsys@transformshift{1.186770in}{0.541249in}%
\pgfsys@useobject{currentmarker}{}%
\end{pgfscope}%
\end{pgfscope}%
\begin{pgfscope}%
\definecolor{textcolor}{rgb}{0.000000,0.000000,0.000000}%
\pgfsetstrokecolor{textcolor}%
\pgfsetfillcolor{textcolor}%
\pgftext[x=1.218020in, y=0.379791in, left, base,rotate=90.000000]{\color{textcolor}\rmfamily\fontsize{9.000000}{10.800000}\selectfont 2}%
\end{pgfscope}%
\begin{pgfscope}%
\pgfsetbuttcap%
\pgfsetroundjoin%
\definecolor{currentfill}{rgb}{0.000000,0.000000,0.000000}%
\pgfsetfillcolor{currentfill}%
\pgfsetlinewidth{0.803000pt}%
\definecolor{currentstroke}{rgb}{0.000000,0.000000,0.000000}%
\pgfsetstrokecolor{currentstroke}%
\pgfsetdash{}{0pt}%
\pgfsys@defobject{currentmarker}{\pgfqpoint{0.000000in}{-0.048611in}}{\pgfqpoint{0.000000in}{0.000000in}}{%
\pgfpathmoveto{\pgfqpoint{0.000000in}{0.000000in}}%
\pgfpathlineto{\pgfqpoint{0.000000in}{-0.048611in}}%
\pgfusepath{stroke,fill}%
}%
\begin{pgfscope}%
\pgfsys@transformshift{1.626377in}{0.541249in}%
\pgfsys@useobject{currentmarker}{}%
\end{pgfscope}%
\end{pgfscope}%
\begin{pgfscope}%
\definecolor{textcolor}{rgb}{0.000000,0.000000,0.000000}%
\pgfsetstrokecolor{textcolor}%
\pgfsetfillcolor{textcolor}%
\pgftext[x=1.657627in, y=0.379791in, left, base,rotate=90.000000]{\color{textcolor}\rmfamily\fontsize{9.000000}{10.800000}\selectfont 4}%
\end{pgfscope}%
\begin{pgfscope}%
\pgfsetbuttcap%
\pgfsetroundjoin%
\definecolor{currentfill}{rgb}{0.000000,0.000000,0.000000}%
\pgfsetfillcolor{currentfill}%
\pgfsetlinewidth{0.803000pt}%
\definecolor{currentstroke}{rgb}{0.000000,0.000000,0.000000}%
\pgfsetstrokecolor{currentstroke}%
\pgfsetdash{}{0pt}%
\pgfsys@defobject{currentmarker}{\pgfqpoint{0.000000in}{-0.048611in}}{\pgfqpoint{0.000000in}{0.000000in}}{%
\pgfpathmoveto{\pgfqpoint{0.000000in}{0.000000in}}%
\pgfpathlineto{\pgfqpoint{0.000000in}{-0.048611in}}%
\pgfusepath{stroke,fill}%
}%
\begin{pgfscope}%
\pgfsys@transformshift{2.065983in}{0.541249in}%
\pgfsys@useobject{currentmarker}{}%
\end{pgfscope}%
\end{pgfscope}%
\begin{pgfscope}%
\definecolor{textcolor}{rgb}{0.000000,0.000000,0.000000}%
\pgfsetstrokecolor{textcolor}%
\pgfsetfillcolor{textcolor}%
\pgftext[x=2.097233in, y=0.379791in, left, base,rotate=90.000000]{\color{textcolor}\rmfamily\fontsize{9.000000}{10.800000}\selectfont 8}%
\end{pgfscope}%
\begin{pgfscope}%
\pgfsetbuttcap%
\pgfsetroundjoin%
\definecolor{currentfill}{rgb}{0.000000,0.000000,0.000000}%
\pgfsetfillcolor{currentfill}%
\pgfsetlinewidth{0.803000pt}%
\definecolor{currentstroke}{rgb}{0.000000,0.000000,0.000000}%
\pgfsetstrokecolor{currentstroke}%
\pgfsetdash{}{0pt}%
\pgfsys@defobject{currentmarker}{\pgfqpoint{0.000000in}{-0.048611in}}{\pgfqpoint{0.000000in}{0.000000in}}{%
\pgfpathmoveto{\pgfqpoint{0.000000in}{0.000000in}}%
\pgfpathlineto{\pgfqpoint{0.000000in}{-0.048611in}}%
\pgfusepath{stroke,fill}%
}%
\begin{pgfscope}%
\pgfsys@transformshift{2.505590in}{0.541249in}%
\pgfsys@useobject{currentmarker}{}%
\end{pgfscope}%
\end{pgfscope}%
\begin{pgfscope}%
\definecolor{textcolor}{rgb}{0.000000,0.000000,0.000000}%
\pgfsetstrokecolor{textcolor}%
\pgfsetfillcolor{textcolor}%
\pgftext[x=2.536840in, y=0.315556in, left, base,rotate=90.000000]{\color{textcolor}\rmfamily\fontsize{9.000000}{10.800000}\selectfont 16}%
\end{pgfscope}%
\begin{pgfscope}%
\pgfsetbuttcap%
\pgfsetroundjoin%
\definecolor{currentfill}{rgb}{0.000000,0.000000,0.000000}%
\pgfsetfillcolor{currentfill}%
\pgfsetlinewidth{0.803000pt}%
\definecolor{currentstroke}{rgb}{0.000000,0.000000,0.000000}%
\pgfsetstrokecolor{currentstroke}%
\pgfsetdash{}{0pt}%
\pgfsys@defobject{currentmarker}{\pgfqpoint{0.000000in}{-0.048611in}}{\pgfqpoint{0.000000in}{0.000000in}}{%
\pgfpathmoveto{\pgfqpoint{0.000000in}{0.000000in}}%
\pgfpathlineto{\pgfqpoint{0.000000in}{-0.048611in}}%
\pgfusepath{stroke,fill}%
}%
\begin{pgfscope}%
\pgfsys@transformshift{2.945197in}{0.541249in}%
\pgfsys@useobject{currentmarker}{}%
\end{pgfscope}%
\end{pgfscope}%
\begin{pgfscope}%
\definecolor{textcolor}{rgb}{0.000000,0.000000,0.000000}%
\pgfsetstrokecolor{textcolor}%
\pgfsetfillcolor{textcolor}%
\pgftext[x=2.976447in, y=0.315556in, left, base,rotate=90.000000]{\color{textcolor}\rmfamily\fontsize{9.000000}{10.800000}\selectfont 32}%
\end{pgfscope}%
\begin{pgfscope}%
\definecolor{textcolor}{rgb}{0.000000,0.000000,0.000000}%
\pgfsetstrokecolor{textcolor}%
\pgfsetfillcolor{textcolor}%
\pgftext[x=1.846180in,y=0.260000in,,top]{\color{textcolor}\rmfamily\fontsize{9.000000}{10.800000}\selectfont \(\displaystyle \alpha(/255)\)}%
\end{pgfscope}%
\begin{pgfscope}%
\pgfpathrectangle{\pgfqpoint{0.527360in}{0.541249in}}{\pgfqpoint{2.637640in}{1.923751in}}%
\pgfusepath{clip}%
\pgfsetrectcap%
\pgfsetroundjoin%
\pgfsetlinewidth{0.803000pt}%
\definecolor{currentstroke}{rgb}{0.690196,0.690196,0.690196}%
\pgfsetstrokecolor{currentstroke}%
\pgfsetdash{}{0pt}%
\pgfpathmoveto{\pgfqpoint{0.527360in}{0.541249in}}%
\pgfpathlineto{\pgfqpoint{3.165000in}{0.541249in}}%
\pgfusepath{stroke}%
\end{pgfscope}%
\begin{pgfscope}%
\pgfsetbuttcap%
\pgfsetroundjoin%
\definecolor{currentfill}{rgb}{0.000000,0.000000,0.000000}%
\pgfsetfillcolor{currentfill}%
\pgfsetlinewidth{0.803000pt}%
\definecolor{currentstroke}{rgb}{0.000000,0.000000,0.000000}%
\pgfsetstrokecolor{currentstroke}%
\pgfsetdash{}{0pt}%
\pgfsys@defobject{currentmarker}{\pgfqpoint{-0.048611in}{0.000000in}}{\pgfqpoint{-0.000000in}{0.000000in}}{%
\pgfpathmoveto{\pgfqpoint{-0.000000in}{0.000000in}}%
\pgfpathlineto{\pgfqpoint{-0.048611in}{0.000000in}}%
\pgfusepath{stroke,fill}%
}%
\begin{pgfscope}%
\pgfsys@transformshift{0.527360in}{0.541249in}%
\pgfsys@useobject{currentmarker}{}%
\end{pgfscope}%
\end{pgfscope}%
\begin{pgfscope}%
\definecolor{textcolor}{rgb}{0.000000,0.000000,0.000000}%
\pgfsetstrokecolor{textcolor}%
\pgfsetfillcolor{textcolor}%
\pgftext[x=0.365902in, y=0.497846in, left, base]{\color{textcolor}\rmfamily\fontsize{9.000000}{10.800000}\selectfont \(\displaystyle {0}\)}%
\end{pgfscope}%
\begin{pgfscope}%
\pgfpathrectangle{\pgfqpoint{0.527360in}{0.541249in}}{\pgfqpoint{2.637640in}{1.923751in}}%
\pgfusepath{clip}%
\pgfsetrectcap%
\pgfsetroundjoin%
\pgfsetlinewidth{0.803000pt}%
\definecolor{currentstroke}{rgb}{0.690196,0.690196,0.690196}%
\pgfsetstrokecolor{currentstroke}%
\pgfsetdash{}{0pt}%
\pgfpathmoveto{\pgfqpoint{0.527360in}{0.861874in}}%
\pgfpathlineto{\pgfqpoint{3.165000in}{0.861874in}}%
\pgfusepath{stroke}%
\end{pgfscope}%
\begin{pgfscope}%
\pgfsetbuttcap%
\pgfsetroundjoin%
\definecolor{currentfill}{rgb}{0.000000,0.000000,0.000000}%
\pgfsetfillcolor{currentfill}%
\pgfsetlinewidth{0.803000pt}%
\definecolor{currentstroke}{rgb}{0.000000,0.000000,0.000000}%
\pgfsetstrokecolor{currentstroke}%
\pgfsetdash{}{0pt}%
\pgfsys@defobject{currentmarker}{\pgfqpoint{-0.048611in}{0.000000in}}{\pgfqpoint{-0.000000in}{0.000000in}}{%
\pgfpathmoveto{\pgfqpoint{-0.000000in}{0.000000in}}%
\pgfpathlineto{\pgfqpoint{-0.048611in}{0.000000in}}%
\pgfusepath{stroke,fill}%
}%
\begin{pgfscope}%
\pgfsys@transformshift{0.527360in}{0.861874in}%
\pgfsys@useobject{currentmarker}{}%
\end{pgfscope}%
\end{pgfscope}%
\begin{pgfscope}%
\definecolor{textcolor}{rgb}{0.000000,0.000000,0.000000}%
\pgfsetstrokecolor{textcolor}%
\pgfsetfillcolor{textcolor}%
\pgftext[x=0.301667in, y=0.818472in, left, base]{\color{textcolor}\rmfamily\fontsize{9.000000}{10.800000}\selectfont \(\displaystyle {10}\)}%
\end{pgfscope}%
\begin{pgfscope}%
\pgfpathrectangle{\pgfqpoint{0.527360in}{0.541249in}}{\pgfqpoint{2.637640in}{1.923751in}}%
\pgfusepath{clip}%
\pgfsetrectcap%
\pgfsetroundjoin%
\pgfsetlinewidth{0.803000pt}%
\definecolor{currentstroke}{rgb}{0.690196,0.690196,0.690196}%
\pgfsetstrokecolor{currentstroke}%
\pgfsetdash{}{0pt}%
\pgfpathmoveto{\pgfqpoint{0.527360in}{1.182499in}}%
\pgfpathlineto{\pgfqpoint{3.165000in}{1.182499in}}%
\pgfusepath{stroke}%
\end{pgfscope}%
\begin{pgfscope}%
\pgfsetbuttcap%
\pgfsetroundjoin%
\definecolor{currentfill}{rgb}{0.000000,0.000000,0.000000}%
\pgfsetfillcolor{currentfill}%
\pgfsetlinewidth{0.803000pt}%
\definecolor{currentstroke}{rgb}{0.000000,0.000000,0.000000}%
\pgfsetstrokecolor{currentstroke}%
\pgfsetdash{}{0pt}%
\pgfsys@defobject{currentmarker}{\pgfqpoint{-0.048611in}{0.000000in}}{\pgfqpoint{-0.000000in}{0.000000in}}{%
\pgfpathmoveto{\pgfqpoint{-0.000000in}{0.000000in}}%
\pgfpathlineto{\pgfqpoint{-0.048611in}{0.000000in}}%
\pgfusepath{stroke,fill}%
}%
\begin{pgfscope}%
\pgfsys@transformshift{0.527360in}{1.182499in}%
\pgfsys@useobject{currentmarker}{}%
\end{pgfscope}%
\end{pgfscope}%
\begin{pgfscope}%
\definecolor{textcolor}{rgb}{0.000000,0.000000,0.000000}%
\pgfsetstrokecolor{textcolor}%
\pgfsetfillcolor{textcolor}%
\pgftext[x=0.301667in, y=1.139097in, left, base]{\color{textcolor}\rmfamily\fontsize{9.000000}{10.800000}\selectfont \(\displaystyle {20}\)}%
\end{pgfscope}%
\begin{pgfscope}%
\pgfpathrectangle{\pgfqpoint{0.527360in}{0.541249in}}{\pgfqpoint{2.637640in}{1.923751in}}%
\pgfusepath{clip}%
\pgfsetrectcap%
\pgfsetroundjoin%
\pgfsetlinewidth{0.803000pt}%
\definecolor{currentstroke}{rgb}{0.690196,0.690196,0.690196}%
\pgfsetstrokecolor{currentstroke}%
\pgfsetdash{}{0pt}%
\pgfpathmoveto{\pgfqpoint{0.527360in}{1.503125in}}%
\pgfpathlineto{\pgfqpoint{3.165000in}{1.503125in}}%
\pgfusepath{stroke}%
\end{pgfscope}%
\begin{pgfscope}%
\pgfsetbuttcap%
\pgfsetroundjoin%
\definecolor{currentfill}{rgb}{0.000000,0.000000,0.000000}%
\pgfsetfillcolor{currentfill}%
\pgfsetlinewidth{0.803000pt}%
\definecolor{currentstroke}{rgb}{0.000000,0.000000,0.000000}%
\pgfsetstrokecolor{currentstroke}%
\pgfsetdash{}{0pt}%
\pgfsys@defobject{currentmarker}{\pgfqpoint{-0.048611in}{0.000000in}}{\pgfqpoint{-0.000000in}{0.000000in}}{%
\pgfpathmoveto{\pgfqpoint{-0.000000in}{0.000000in}}%
\pgfpathlineto{\pgfqpoint{-0.048611in}{0.000000in}}%
\pgfusepath{stroke,fill}%
}%
\begin{pgfscope}%
\pgfsys@transformshift{0.527360in}{1.503125in}%
\pgfsys@useobject{currentmarker}{}%
\end{pgfscope}%
\end{pgfscope}%
\begin{pgfscope}%
\definecolor{textcolor}{rgb}{0.000000,0.000000,0.000000}%
\pgfsetstrokecolor{textcolor}%
\pgfsetfillcolor{textcolor}%
\pgftext[x=0.301667in, y=1.459722in, left, base]{\color{textcolor}\rmfamily\fontsize{9.000000}{10.800000}\selectfont \(\displaystyle {30}\)}%
\end{pgfscope}%
\begin{pgfscope}%
\pgfpathrectangle{\pgfqpoint{0.527360in}{0.541249in}}{\pgfqpoint{2.637640in}{1.923751in}}%
\pgfusepath{clip}%
\pgfsetrectcap%
\pgfsetroundjoin%
\pgfsetlinewidth{0.803000pt}%
\definecolor{currentstroke}{rgb}{0.690196,0.690196,0.690196}%
\pgfsetstrokecolor{currentstroke}%
\pgfsetdash{}{0pt}%
\pgfpathmoveto{\pgfqpoint{0.527360in}{1.823750in}}%
\pgfpathlineto{\pgfqpoint{3.165000in}{1.823750in}}%
\pgfusepath{stroke}%
\end{pgfscope}%
\begin{pgfscope}%
\pgfsetbuttcap%
\pgfsetroundjoin%
\definecolor{currentfill}{rgb}{0.000000,0.000000,0.000000}%
\pgfsetfillcolor{currentfill}%
\pgfsetlinewidth{0.803000pt}%
\definecolor{currentstroke}{rgb}{0.000000,0.000000,0.000000}%
\pgfsetstrokecolor{currentstroke}%
\pgfsetdash{}{0pt}%
\pgfsys@defobject{currentmarker}{\pgfqpoint{-0.048611in}{0.000000in}}{\pgfqpoint{-0.000000in}{0.000000in}}{%
\pgfpathmoveto{\pgfqpoint{-0.000000in}{0.000000in}}%
\pgfpathlineto{\pgfqpoint{-0.048611in}{0.000000in}}%
\pgfusepath{stroke,fill}%
}%
\begin{pgfscope}%
\pgfsys@transformshift{0.527360in}{1.823750in}%
\pgfsys@useobject{currentmarker}{}%
\end{pgfscope}%
\end{pgfscope}%
\begin{pgfscope}%
\definecolor{textcolor}{rgb}{0.000000,0.000000,0.000000}%
\pgfsetstrokecolor{textcolor}%
\pgfsetfillcolor{textcolor}%
\pgftext[x=0.301667in, y=1.780347in, left, base]{\color{textcolor}\rmfamily\fontsize{9.000000}{10.800000}\selectfont \(\displaystyle {40}\)}%
\end{pgfscope}%
\begin{pgfscope}%
\pgfpathrectangle{\pgfqpoint{0.527360in}{0.541249in}}{\pgfqpoint{2.637640in}{1.923751in}}%
\pgfusepath{clip}%
\pgfsetrectcap%
\pgfsetroundjoin%
\pgfsetlinewidth{0.803000pt}%
\definecolor{currentstroke}{rgb}{0.690196,0.690196,0.690196}%
\pgfsetstrokecolor{currentstroke}%
\pgfsetdash{}{0pt}%
\pgfpathmoveto{\pgfqpoint{0.527360in}{2.144375in}}%
\pgfpathlineto{\pgfqpoint{3.165000in}{2.144375in}}%
\pgfusepath{stroke}%
\end{pgfscope}%
\begin{pgfscope}%
\pgfsetbuttcap%
\pgfsetroundjoin%
\definecolor{currentfill}{rgb}{0.000000,0.000000,0.000000}%
\pgfsetfillcolor{currentfill}%
\pgfsetlinewidth{0.803000pt}%
\definecolor{currentstroke}{rgb}{0.000000,0.000000,0.000000}%
\pgfsetstrokecolor{currentstroke}%
\pgfsetdash{}{0pt}%
\pgfsys@defobject{currentmarker}{\pgfqpoint{-0.048611in}{0.000000in}}{\pgfqpoint{-0.000000in}{0.000000in}}{%
\pgfpathmoveto{\pgfqpoint{-0.000000in}{0.000000in}}%
\pgfpathlineto{\pgfqpoint{-0.048611in}{0.000000in}}%
\pgfusepath{stroke,fill}%
}%
\begin{pgfscope}%
\pgfsys@transformshift{0.527360in}{2.144375in}%
\pgfsys@useobject{currentmarker}{}%
\end{pgfscope}%
\end{pgfscope}%
\begin{pgfscope}%
\definecolor{textcolor}{rgb}{0.000000,0.000000,0.000000}%
\pgfsetstrokecolor{textcolor}%
\pgfsetfillcolor{textcolor}%
\pgftext[x=0.301667in, y=2.100972in, left, base]{\color{textcolor}\rmfamily\fontsize{9.000000}{10.800000}\selectfont \(\displaystyle {50}\)}%
\end{pgfscope}%
\begin{pgfscope}%
\pgfpathrectangle{\pgfqpoint{0.527360in}{0.541249in}}{\pgfqpoint{2.637640in}{1.923751in}}%
\pgfusepath{clip}%
\pgfsetrectcap%
\pgfsetroundjoin%
\pgfsetlinewidth{0.803000pt}%
\definecolor{currentstroke}{rgb}{0.690196,0.690196,0.690196}%
\pgfsetstrokecolor{currentstroke}%
\pgfsetdash{}{0pt}%
\pgfpathmoveto{\pgfqpoint{0.527360in}{2.465000in}}%
\pgfpathlineto{\pgfqpoint{3.165000in}{2.465000in}}%
\pgfusepath{stroke}%
\end{pgfscope}%
\begin{pgfscope}%
\pgfsetbuttcap%
\pgfsetroundjoin%
\definecolor{currentfill}{rgb}{0.000000,0.000000,0.000000}%
\pgfsetfillcolor{currentfill}%
\pgfsetlinewidth{0.803000pt}%
\definecolor{currentstroke}{rgb}{0.000000,0.000000,0.000000}%
\pgfsetstrokecolor{currentstroke}%
\pgfsetdash{}{0pt}%
\pgfsys@defobject{currentmarker}{\pgfqpoint{-0.048611in}{0.000000in}}{\pgfqpoint{-0.000000in}{0.000000in}}{%
\pgfpathmoveto{\pgfqpoint{-0.000000in}{0.000000in}}%
\pgfpathlineto{\pgfqpoint{-0.048611in}{0.000000in}}%
\pgfusepath{stroke,fill}%
}%
\begin{pgfscope}%
\pgfsys@transformshift{0.527360in}{2.465000in}%
\pgfsys@useobject{currentmarker}{}%
\end{pgfscope}%
\end{pgfscope}%
\begin{pgfscope}%
\definecolor{textcolor}{rgb}{0.000000,0.000000,0.000000}%
\pgfsetstrokecolor{textcolor}%
\pgfsetfillcolor{textcolor}%
\pgftext[x=0.301667in, y=2.421597in, left, base]{\color{textcolor}\rmfamily\fontsize{9.000000}{10.800000}\selectfont \(\displaystyle {60}\)}%
\end{pgfscope}%
\begin{pgfscope}%
\definecolor{textcolor}{rgb}{0.000000,0.000000,0.000000}%
\pgfsetstrokecolor{textcolor}%
\pgfsetfillcolor{textcolor}%
\pgftext[x=0.246111in,y=1.503125in,,bottom,rotate=90.000000]{\color{textcolor}\rmfamily\fontsize{9.000000}{10.800000}\selectfont PSNR}%
\end{pgfscope}%
\begin{pgfscope}%
\pgfpathrectangle{\pgfqpoint{0.527360in}{0.541249in}}{\pgfqpoint{2.637640in}{1.923751in}}%
\pgfusepath{clip}%
\pgfsetbuttcap%
\pgfsetmiterjoin%
\definecolor{currentfill}{rgb}{0.121569,0.466667,0.705882}%
\pgfsetfillcolor{currentfill}%
\pgfsetlinewidth{0.000000pt}%
\definecolor{currentstroke}{rgb}{0.000000,0.000000,0.000000}%
\pgfsetstrokecolor{currentstroke}%
\pgfsetstrokeopacity{0.000000}%
\pgfsetdash{}{0pt}%
\pgfpathmoveto{\pgfqpoint{0.637262in}{0.541249in}}%
\pgfpathlineto{\pgfqpoint{0.710530in}{0.541249in}}%
\pgfpathlineto{\pgfqpoint{0.710530in}{1.022187in}}%
\pgfpathlineto{\pgfqpoint{0.637262in}{1.022187in}}%
\pgfpathlineto{\pgfqpoint{0.637262in}{0.541249in}}%
\pgfpathclose%
\pgfusepath{fill}%
\end{pgfscope}%
\begin{pgfscope}%
\pgfpathrectangle{\pgfqpoint{0.527360in}{0.541249in}}{\pgfqpoint{2.637640in}{1.923751in}}%
\pgfusepath{clip}%
\pgfsetbuttcap%
\pgfsetmiterjoin%
\definecolor{currentfill}{rgb}{0.121569,0.466667,0.705882}%
\pgfsetfillcolor{currentfill}%
\pgfsetlinewidth{0.000000pt}%
\definecolor{currentstroke}{rgb}{0.000000,0.000000,0.000000}%
\pgfsetstrokecolor{currentstroke}%
\pgfsetstrokeopacity{0.000000}%
\pgfsetdash{}{0pt}%
\pgfpathmoveto{\pgfqpoint{1.076869in}{0.541249in}}%
\pgfpathlineto{\pgfqpoint{1.150136in}{0.541249in}}%
\pgfpathlineto{\pgfqpoint{1.150136in}{0.925999in}}%
\pgfpathlineto{\pgfqpoint{1.076869in}{0.925999in}}%
\pgfpathlineto{\pgfqpoint{1.076869in}{0.541249in}}%
\pgfpathclose%
\pgfusepath{fill}%
\end{pgfscope}%
\begin{pgfscope}%
\pgfpathrectangle{\pgfqpoint{0.527360in}{0.541249in}}{\pgfqpoint{2.637640in}{1.923751in}}%
\pgfusepath{clip}%
\pgfsetbuttcap%
\pgfsetmiterjoin%
\definecolor{currentfill}{rgb}{0.121569,0.466667,0.705882}%
\pgfsetfillcolor{currentfill}%
\pgfsetlinewidth{0.000000pt}%
\definecolor{currentstroke}{rgb}{0.000000,0.000000,0.000000}%
\pgfsetstrokecolor{currentstroke}%
\pgfsetstrokeopacity{0.000000}%
\pgfsetdash{}{0pt}%
\pgfpathmoveto{\pgfqpoint{1.516475in}{0.541249in}}%
\pgfpathlineto{\pgfqpoint{1.589743in}{0.541249in}}%
\pgfpathlineto{\pgfqpoint{1.589743in}{0.865081in}}%
\pgfpathlineto{\pgfqpoint{1.516475in}{0.865081in}}%
\pgfpathlineto{\pgfqpoint{1.516475in}{0.541249in}}%
\pgfpathclose%
\pgfusepath{fill}%
\end{pgfscope}%
\begin{pgfscope}%
\pgfpathrectangle{\pgfqpoint{0.527360in}{0.541249in}}{\pgfqpoint{2.637640in}{1.923751in}}%
\pgfusepath{clip}%
\pgfsetbuttcap%
\pgfsetmiterjoin%
\definecolor{currentfill}{rgb}{0.121569,0.466667,0.705882}%
\pgfsetfillcolor{currentfill}%
\pgfsetlinewidth{0.000000pt}%
\definecolor{currentstroke}{rgb}{0.000000,0.000000,0.000000}%
\pgfsetstrokecolor{currentstroke}%
\pgfsetstrokeopacity{0.000000}%
\pgfsetdash{}{0pt}%
\pgfpathmoveto{\pgfqpoint{1.956082in}{0.541249in}}%
\pgfpathlineto{\pgfqpoint{2.029350in}{0.541249in}}%
\pgfpathlineto{\pgfqpoint{2.029350in}{0.845843in}}%
\pgfpathlineto{\pgfqpoint{1.956082in}{0.845843in}}%
\pgfpathlineto{\pgfqpoint{1.956082in}{0.541249in}}%
\pgfpathclose%
\pgfusepath{fill}%
\end{pgfscope}%
\begin{pgfscope}%
\pgfpathrectangle{\pgfqpoint{0.527360in}{0.541249in}}{\pgfqpoint{2.637640in}{1.923751in}}%
\pgfusepath{clip}%
\pgfsetbuttcap%
\pgfsetmiterjoin%
\definecolor{currentfill}{rgb}{0.121569,0.466667,0.705882}%
\pgfsetfillcolor{currentfill}%
\pgfsetlinewidth{0.000000pt}%
\definecolor{currentstroke}{rgb}{0.000000,0.000000,0.000000}%
\pgfsetstrokecolor{currentstroke}%
\pgfsetstrokeopacity{0.000000}%
\pgfsetdash{}{0pt}%
\pgfpathmoveto{\pgfqpoint{2.395688in}{0.541249in}}%
\pgfpathlineto{\pgfqpoint{2.468956in}{0.541249in}}%
\pgfpathlineto{\pgfqpoint{2.468956in}{0.820193in}}%
\pgfpathlineto{\pgfqpoint{2.395688in}{0.820193in}}%
\pgfpathlineto{\pgfqpoint{2.395688in}{0.541249in}}%
\pgfpathclose%
\pgfusepath{fill}%
\end{pgfscope}%
\begin{pgfscope}%
\pgfpathrectangle{\pgfqpoint{0.527360in}{0.541249in}}{\pgfqpoint{2.637640in}{1.923751in}}%
\pgfusepath{clip}%
\pgfsetbuttcap%
\pgfsetmiterjoin%
\definecolor{currentfill}{rgb}{0.121569,0.466667,0.705882}%
\pgfsetfillcolor{currentfill}%
\pgfsetlinewidth{0.000000pt}%
\definecolor{currentstroke}{rgb}{0.000000,0.000000,0.000000}%
\pgfsetstrokecolor{currentstroke}%
\pgfsetstrokeopacity{0.000000}%
\pgfsetdash{}{0pt}%
\pgfpathmoveto{\pgfqpoint{2.835295in}{0.541249in}}%
\pgfpathlineto{\pgfqpoint{2.908563in}{0.541249in}}%
\pgfpathlineto{\pgfqpoint{2.908563in}{0.772099in}}%
\pgfpathlineto{\pgfqpoint{2.835295in}{0.772099in}}%
\pgfpathlineto{\pgfqpoint{2.835295in}{0.541249in}}%
\pgfpathclose%
\pgfusepath{fill}%
\end{pgfscope}%
\begin{pgfscope}%
\pgfpathrectangle{\pgfqpoint{0.527360in}{0.541249in}}{\pgfqpoint{2.637640in}{1.923751in}}%
\pgfusepath{clip}%
\pgfsetbuttcap%
\pgfsetmiterjoin%
\definecolor{currentfill}{rgb}{1.000000,0.498039,0.054902}%
\pgfsetfillcolor{currentfill}%
\pgfsetlinewidth{0.000000pt}%
\definecolor{currentstroke}{rgb}{0.000000,0.000000,0.000000}%
\pgfsetstrokecolor{currentstroke}%
\pgfsetstrokeopacity{0.000000}%
\pgfsetdash{}{0pt}%
\pgfpathmoveto{\pgfqpoint{0.710530in}{0.541249in}}%
\pgfpathlineto{\pgfqpoint{0.783797in}{0.541249in}}%
\pgfpathlineto{\pgfqpoint{0.783797in}{1.406937in}}%
\pgfpathlineto{\pgfqpoint{0.710530in}{1.406937in}}%
\pgfpathlineto{\pgfqpoint{0.710530in}{0.541249in}}%
\pgfpathclose%
\pgfusepath{fill}%
\end{pgfscope}%
\begin{pgfscope}%
\pgfpathrectangle{\pgfqpoint{0.527360in}{0.541249in}}{\pgfqpoint{2.637640in}{1.923751in}}%
\pgfusepath{clip}%
\pgfsetbuttcap%
\pgfsetmiterjoin%
\definecolor{currentfill}{rgb}{1.000000,0.498039,0.054902}%
\pgfsetfillcolor{currentfill}%
\pgfsetlinewidth{0.000000pt}%
\definecolor{currentstroke}{rgb}{0.000000,0.000000,0.000000}%
\pgfsetstrokecolor{currentstroke}%
\pgfsetstrokeopacity{0.000000}%
\pgfsetdash{}{0pt}%
\pgfpathmoveto{\pgfqpoint{1.150136in}{0.541249in}}%
\pgfpathlineto{\pgfqpoint{1.223404in}{0.541249in}}%
\pgfpathlineto{\pgfqpoint{1.223404in}{1.246624in}}%
\pgfpathlineto{\pgfqpoint{1.150136in}{1.246624in}}%
\pgfpathlineto{\pgfqpoint{1.150136in}{0.541249in}}%
\pgfpathclose%
\pgfusepath{fill}%
\end{pgfscope}%
\begin{pgfscope}%
\pgfpathrectangle{\pgfqpoint{0.527360in}{0.541249in}}{\pgfqpoint{2.637640in}{1.923751in}}%
\pgfusepath{clip}%
\pgfsetbuttcap%
\pgfsetmiterjoin%
\definecolor{currentfill}{rgb}{1.000000,0.498039,0.054902}%
\pgfsetfillcolor{currentfill}%
\pgfsetlinewidth{0.000000pt}%
\definecolor{currentstroke}{rgb}{0.000000,0.000000,0.000000}%
\pgfsetstrokecolor{currentstroke}%
\pgfsetstrokeopacity{0.000000}%
\pgfsetdash{}{0pt}%
\pgfpathmoveto{\pgfqpoint{1.589743in}{0.541249in}}%
\pgfpathlineto{\pgfqpoint{1.663011in}{0.541249in}}%
\pgfpathlineto{\pgfqpoint{1.663011in}{1.127993in}}%
\pgfpathlineto{\pgfqpoint{1.589743in}{1.127993in}}%
\pgfpathlineto{\pgfqpoint{1.589743in}{0.541249in}}%
\pgfpathclose%
\pgfusepath{fill}%
\end{pgfscope}%
\begin{pgfscope}%
\pgfpathrectangle{\pgfqpoint{0.527360in}{0.541249in}}{\pgfqpoint{2.637640in}{1.923751in}}%
\pgfusepath{clip}%
\pgfsetbuttcap%
\pgfsetmiterjoin%
\definecolor{currentfill}{rgb}{1.000000,0.498039,0.054902}%
\pgfsetfillcolor{currentfill}%
\pgfsetlinewidth{0.000000pt}%
\definecolor{currentstroke}{rgb}{0.000000,0.000000,0.000000}%
\pgfsetstrokecolor{currentstroke}%
\pgfsetstrokeopacity{0.000000}%
\pgfsetdash{}{0pt}%
\pgfpathmoveto{\pgfqpoint{2.029350in}{0.541249in}}%
\pgfpathlineto{\pgfqpoint{2.102617in}{0.541249in}}%
\pgfpathlineto{\pgfqpoint{2.102617in}{1.054249in}}%
\pgfpathlineto{\pgfqpoint{2.029350in}{1.054249in}}%
\pgfpathlineto{\pgfqpoint{2.029350in}{0.541249in}}%
\pgfpathclose%
\pgfusepath{fill}%
\end{pgfscope}%
\begin{pgfscope}%
\pgfpathrectangle{\pgfqpoint{0.527360in}{0.541249in}}{\pgfqpoint{2.637640in}{1.923751in}}%
\pgfusepath{clip}%
\pgfsetbuttcap%
\pgfsetmiterjoin%
\definecolor{currentfill}{rgb}{1.000000,0.498039,0.054902}%
\pgfsetfillcolor{currentfill}%
\pgfsetlinewidth{0.000000pt}%
\definecolor{currentstroke}{rgb}{0.000000,0.000000,0.000000}%
\pgfsetstrokecolor{currentstroke}%
\pgfsetstrokeopacity{0.000000}%
\pgfsetdash{}{0pt}%
\pgfpathmoveto{\pgfqpoint{2.468956in}{0.541249in}}%
\pgfpathlineto{\pgfqpoint{2.542224in}{0.541249in}}%
\pgfpathlineto{\pgfqpoint{2.542224in}{1.006156in}}%
\pgfpathlineto{\pgfqpoint{2.468956in}{1.006156in}}%
\pgfpathlineto{\pgfqpoint{2.468956in}{0.541249in}}%
\pgfpathclose%
\pgfusepath{fill}%
\end{pgfscope}%
\begin{pgfscope}%
\pgfpathrectangle{\pgfqpoint{0.527360in}{0.541249in}}{\pgfqpoint{2.637640in}{1.923751in}}%
\pgfusepath{clip}%
\pgfsetbuttcap%
\pgfsetmiterjoin%
\definecolor{currentfill}{rgb}{1.000000,0.498039,0.054902}%
\pgfsetfillcolor{currentfill}%
\pgfsetlinewidth{0.000000pt}%
\definecolor{currentstroke}{rgb}{0.000000,0.000000,0.000000}%
\pgfsetstrokecolor{currentstroke}%
\pgfsetstrokeopacity{0.000000}%
\pgfsetdash{}{0pt}%
\pgfpathmoveto{\pgfqpoint{2.908563in}{0.541249in}}%
\pgfpathlineto{\pgfqpoint{2.981831in}{0.541249in}}%
\pgfpathlineto{\pgfqpoint{2.981831in}{0.958062in}}%
\pgfpathlineto{\pgfqpoint{2.908563in}{0.958062in}}%
\pgfpathlineto{\pgfqpoint{2.908563in}{0.541249in}}%
\pgfpathclose%
\pgfusepath{fill}%
\end{pgfscope}%
\begin{pgfscope}%
\pgfpathrectangle{\pgfqpoint{0.527360in}{0.541249in}}{\pgfqpoint{2.637640in}{1.923751in}}%
\pgfusepath{clip}%
\pgfsetbuttcap%
\pgfsetmiterjoin%
\definecolor{currentfill}{rgb}{0.172549,0.627451,0.172549}%
\pgfsetfillcolor{currentfill}%
\pgfsetlinewidth{0.000000pt}%
\definecolor{currentstroke}{rgb}{0.000000,0.000000,0.000000}%
\pgfsetstrokecolor{currentstroke}%
\pgfsetstrokeopacity{0.000000}%
\pgfsetdash{}{0pt}%
\pgfpathmoveto{\pgfqpoint{0.783797in}{0.541249in}}%
\pgfpathlineto{\pgfqpoint{0.857065in}{0.541249in}}%
\pgfpathlineto{\pgfqpoint{0.857065in}{1.340023in}}%
\pgfpathlineto{\pgfqpoint{0.783797in}{1.340023in}}%
\pgfpathlineto{\pgfqpoint{0.783797in}{0.541249in}}%
\pgfpathclose%
\pgfusepath{fill}%
\end{pgfscope}%
\begin{pgfscope}%
\pgfpathrectangle{\pgfqpoint{0.527360in}{0.541249in}}{\pgfqpoint{2.637640in}{1.923751in}}%
\pgfusepath{clip}%
\pgfsetbuttcap%
\pgfsetmiterjoin%
\definecolor{currentfill}{rgb}{0.172549,0.627451,0.172549}%
\pgfsetfillcolor{currentfill}%
\pgfsetlinewidth{0.000000pt}%
\definecolor{currentstroke}{rgb}{0.000000,0.000000,0.000000}%
\pgfsetstrokecolor{currentstroke}%
\pgfsetstrokeopacity{0.000000}%
\pgfsetdash{}{0pt}%
\pgfpathmoveto{\pgfqpoint{1.223404in}{0.541249in}}%
\pgfpathlineto{\pgfqpoint{1.296672in}{0.541249in}}%
\pgfpathlineto{\pgfqpoint{1.296672in}{1.230593in}}%
\pgfpathlineto{\pgfqpoint{1.223404in}{1.230593in}}%
\pgfpathlineto{\pgfqpoint{1.223404in}{0.541249in}}%
\pgfpathclose%
\pgfusepath{fill}%
\end{pgfscope}%
\begin{pgfscope}%
\pgfpathrectangle{\pgfqpoint{0.527360in}{0.541249in}}{\pgfqpoint{2.637640in}{1.923751in}}%
\pgfusepath{clip}%
\pgfsetbuttcap%
\pgfsetmiterjoin%
\definecolor{currentfill}{rgb}{0.172549,0.627451,0.172549}%
\pgfsetfillcolor{currentfill}%
\pgfsetlinewidth{0.000000pt}%
\definecolor{currentstroke}{rgb}{0.000000,0.000000,0.000000}%
\pgfsetstrokecolor{currentstroke}%
\pgfsetstrokeopacity{0.000000}%
\pgfsetdash{}{0pt}%
\pgfpathmoveto{\pgfqpoint{1.663011in}{0.541249in}}%
\pgfpathlineto{\pgfqpoint{1.736278in}{0.541249in}}%
\pgfpathlineto{\pgfqpoint{1.736278in}{1.161659in}}%
\pgfpathlineto{\pgfqpoint{1.663011in}{1.161659in}}%
\pgfpathlineto{\pgfqpoint{1.663011in}{0.541249in}}%
\pgfpathclose%
\pgfusepath{fill}%
\end{pgfscope}%
\begin{pgfscope}%
\pgfpathrectangle{\pgfqpoint{0.527360in}{0.541249in}}{\pgfqpoint{2.637640in}{1.923751in}}%
\pgfusepath{clip}%
\pgfsetbuttcap%
\pgfsetmiterjoin%
\definecolor{currentfill}{rgb}{0.172549,0.627451,0.172549}%
\pgfsetfillcolor{currentfill}%
\pgfsetlinewidth{0.000000pt}%
\definecolor{currentstroke}{rgb}{0.000000,0.000000,0.000000}%
\pgfsetstrokecolor{currentstroke}%
\pgfsetstrokeopacity{0.000000}%
\pgfsetdash{}{0pt}%
\pgfpathmoveto{\pgfqpoint{2.102617in}{0.541249in}}%
\pgfpathlineto{\pgfqpoint{2.175885in}{0.541249in}}%
\pgfpathlineto{\pgfqpoint{2.175885in}{1.109076in}}%
\pgfpathlineto{\pgfqpoint{2.102617in}{1.109076in}}%
\pgfpathlineto{\pgfqpoint{2.102617in}{0.541249in}}%
\pgfpathclose%
\pgfusepath{fill}%
\end{pgfscope}%
\begin{pgfscope}%
\pgfpathrectangle{\pgfqpoint{0.527360in}{0.541249in}}{\pgfqpoint{2.637640in}{1.923751in}}%
\pgfusepath{clip}%
\pgfsetbuttcap%
\pgfsetmiterjoin%
\definecolor{currentfill}{rgb}{0.172549,0.627451,0.172549}%
\pgfsetfillcolor{currentfill}%
\pgfsetlinewidth{0.000000pt}%
\definecolor{currentstroke}{rgb}{0.000000,0.000000,0.000000}%
\pgfsetstrokecolor{currentstroke}%
\pgfsetstrokeopacity{0.000000}%
\pgfsetdash{}{0pt}%
\pgfpathmoveto{\pgfqpoint{2.542224in}{0.541249in}}%
\pgfpathlineto{\pgfqpoint{2.615492in}{0.541249in}}%
\pgfpathlineto{\pgfqpoint{2.615492in}{1.067716in}}%
\pgfpathlineto{\pgfqpoint{2.542224in}{1.067716in}}%
\pgfpathlineto{\pgfqpoint{2.542224in}{0.541249in}}%
\pgfpathclose%
\pgfusepath{fill}%
\end{pgfscope}%
\begin{pgfscope}%
\pgfpathrectangle{\pgfqpoint{0.527360in}{0.541249in}}{\pgfqpoint{2.637640in}{1.923751in}}%
\pgfusepath{clip}%
\pgfsetbuttcap%
\pgfsetmiterjoin%
\definecolor{currentfill}{rgb}{0.172549,0.627451,0.172549}%
\pgfsetfillcolor{currentfill}%
\pgfsetlinewidth{0.000000pt}%
\definecolor{currentstroke}{rgb}{0.000000,0.000000,0.000000}%
\pgfsetstrokecolor{currentstroke}%
\pgfsetstrokeopacity{0.000000}%
\pgfsetdash{}{0pt}%
\pgfpathmoveto{\pgfqpoint{2.981831in}{0.541249in}}%
\pgfpathlineto{\pgfqpoint{3.055098in}{0.541249in}}%
\pgfpathlineto{\pgfqpoint{3.055098in}{0.979864in}}%
\pgfpathlineto{\pgfqpoint{2.981831in}{0.979864in}}%
\pgfpathlineto{\pgfqpoint{2.981831in}{0.541249in}}%
\pgfpathclose%
\pgfusepath{fill}%
\end{pgfscope}%
\begin{pgfscope}%
\pgfsetrectcap%
\pgfsetmiterjoin%
\pgfsetlinewidth{0.803000pt}%
\definecolor{currentstroke}{rgb}{0.000000,0.000000,0.000000}%
\pgfsetstrokecolor{currentstroke}%
\pgfsetdash{}{0pt}%
\pgfpathmoveto{\pgfqpoint{0.527360in}{0.541249in}}%
\pgfpathlineto{\pgfqpoint{0.527360in}{2.465000in}}%
\pgfusepath{stroke}%
\end{pgfscope}%
\begin{pgfscope}%
\pgfsetrectcap%
\pgfsetmiterjoin%
\pgfsetlinewidth{0.803000pt}%
\definecolor{currentstroke}{rgb}{0.000000,0.000000,0.000000}%
\pgfsetstrokecolor{currentstroke}%
\pgfsetdash{}{0pt}%
\pgfpathmoveto{\pgfqpoint{3.165000in}{0.541249in}}%
\pgfpathlineto{\pgfqpoint{3.165000in}{2.465000in}}%
\pgfusepath{stroke}%
\end{pgfscope}%
\begin{pgfscope}%
\pgfsetrectcap%
\pgfsetmiterjoin%
\pgfsetlinewidth{0.803000pt}%
\definecolor{currentstroke}{rgb}{0.000000,0.000000,0.000000}%
\pgfsetstrokecolor{currentstroke}%
\pgfsetdash{}{0pt}%
\pgfpathmoveto{\pgfqpoint{0.527360in}{0.541249in}}%
\pgfpathlineto{\pgfqpoint{3.165000in}{0.541249in}}%
\pgfusepath{stroke}%
\end{pgfscope}%
\begin{pgfscope}%
\pgfsetrectcap%
\pgfsetmiterjoin%
\pgfsetlinewidth{0.803000pt}%
\definecolor{currentstroke}{rgb}{0.000000,0.000000,0.000000}%
\pgfsetstrokecolor{currentstroke}%
\pgfsetdash{}{0pt}%
\pgfpathmoveto{\pgfqpoint{0.527360in}{2.465000in}}%
\pgfpathlineto{\pgfqpoint{3.165000in}{2.465000in}}%
\pgfusepath{stroke}%
\end{pgfscope}%
\begin{pgfscope}%
\pgfsetbuttcap%
\pgfsetmiterjoin%
\definecolor{currentfill}{rgb}{1.000000,1.000000,1.000000}%
\pgfsetfillcolor{currentfill}%
\pgfsetfillopacity{0.800000}%
\pgfsetlinewidth{1.003750pt}%
\definecolor{currentstroke}{rgb}{0.800000,0.800000,0.800000}%
\pgfsetstrokecolor{currentstroke}%
\pgfsetstrokeopacity{0.800000}%
\pgfsetdash{}{0pt}%
\pgfpathmoveto{\pgfqpoint{2.134237in}{1.842083in}}%
\pgfpathlineto{\pgfqpoint{3.077500in}{1.842083in}}%
\pgfpathquadraticcurveto{\pgfqpoint{3.102500in}{1.842083in}}{\pgfqpoint{3.102500in}{1.867083in}}%
\pgfpathlineto{\pgfqpoint{3.102500in}{2.377500in}}%
\pgfpathquadraticcurveto{\pgfqpoint{3.102500in}{2.402500in}}{\pgfqpoint{3.077500in}{2.402500in}}%
\pgfpathlineto{\pgfqpoint{2.134237in}{2.402500in}}%
\pgfpathquadraticcurveto{\pgfqpoint{2.109237in}{2.402500in}}{\pgfqpoint{2.109237in}{2.377500in}}%
\pgfpathlineto{\pgfqpoint{2.109237in}{1.867083in}}%
\pgfpathquadraticcurveto{\pgfqpoint{2.109237in}{1.842083in}}{\pgfqpoint{2.134237in}{1.842083in}}%
\pgfpathlineto{\pgfqpoint{2.134237in}{1.842083in}}%
\pgfpathclose%
\pgfusepath{stroke,fill}%
\end{pgfscope}%
\begin{pgfscope}%
\pgfsetbuttcap%
\pgfsetmiterjoin%
\definecolor{currentfill}{rgb}{0.121569,0.466667,0.705882}%
\pgfsetfillcolor{currentfill}%
\pgfsetlinewidth{0.000000pt}%
\definecolor{currentstroke}{rgb}{0.000000,0.000000,0.000000}%
\pgfsetstrokecolor{currentstroke}%
\pgfsetstrokeopacity{0.000000}%
\pgfsetdash{}{0pt}%
\pgfpathmoveto{\pgfqpoint{2.159237in}{2.265000in}}%
\pgfpathlineto{\pgfqpoint{2.409237in}{2.265000in}}%
\pgfpathlineto{\pgfqpoint{2.409237in}{2.352500in}}%
\pgfpathlineto{\pgfqpoint{2.159237in}{2.352500in}}%
\pgfpathlineto{\pgfqpoint{2.159237in}{2.265000in}}%
\pgfpathclose%
\pgfusepath{fill}%
\end{pgfscope}%
\begin{pgfscope}%
\definecolor{textcolor}{rgb}{0.000000,0.000000,0.000000}%
\pgfsetstrokecolor{textcolor}%
\pgfsetfillcolor{textcolor}%
\pgftext[x=2.509237in,y=2.265000in,left,base]{\color{textcolor}\rmfamily\fontsize{9.000000}{10.800000}\selectfont ESRGAN}%
\end{pgfscope}%
\begin{pgfscope}%
\pgfsetbuttcap%
\pgfsetmiterjoin%
\definecolor{currentfill}{rgb}{1.000000,0.498039,0.054902}%
\pgfsetfillcolor{currentfill}%
\pgfsetlinewidth{0.000000pt}%
\definecolor{currentstroke}{rgb}{0.000000,0.000000,0.000000}%
\pgfsetstrokecolor{currentstroke}%
\pgfsetstrokeopacity{0.000000}%
\pgfsetdash{}{0pt}%
\pgfpathmoveto{\pgfqpoint{2.159237in}{2.090694in}}%
\pgfpathlineto{\pgfqpoint{2.409237in}{2.090694in}}%
\pgfpathlineto{\pgfqpoint{2.409237in}{2.178194in}}%
\pgfpathlineto{\pgfqpoint{2.159237in}{2.178194in}}%
\pgfpathlineto{\pgfqpoint{2.159237in}{2.090694in}}%
\pgfpathclose%
\pgfusepath{fill}%
\end{pgfscope}%
\begin{pgfscope}%
\definecolor{textcolor}{rgb}{0.000000,0.000000,0.000000}%
\pgfsetstrokecolor{textcolor}%
\pgfsetfillcolor{textcolor}%
\pgftext[x=2.509237in,y=2.090694in,left,base]{\color{textcolor}\rmfamily\fontsize{9.000000}{10.800000}\selectfont EDSR}%
\end{pgfscope}%
\begin{pgfscope}%
\pgfsetbuttcap%
\pgfsetmiterjoin%
\definecolor{currentfill}{rgb}{0.172549,0.627451,0.172549}%
\pgfsetfillcolor{currentfill}%
\pgfsetlinewidth{0.000000pt}%
\definecolor{currentstroke}{rgb}{0.000000,0.000000,0.000000}%
\pgfsetstrokecolor{currentstroke}%
\pgfsetstrokeopacity{0.000000}%
\pgfsetdash{}{0pt}%
\pgfpathmoveto{\pgfqpoint{2.159237in}{1.916389in}}%
\pgfpathlineto{\pgfqpoint{2.409237in}{1.916389in}}%
\pgfpathlineto{\pgfqpoint{2.409237in}{2.003889in}}%
\pgfpathlineto{\pgfqpoint{2.159237in}{2.003889in}}%
\pgfpathlineto{\pgfqpoint{2.159237in}{1.916389in}}%
\pgfpathclose%
\pgfusepath{fill}%
\end{pgfscope}%
\begin{pgfscope}%
\definecolor{textcolor}{rgb}{0.000000,0.000000,0.000000}%
\pgfsetstrokecolor{textcolor}%
\pgfsetfillcolor{textcolor}%
\pgftext[x=2.509237in,y=1.916389in,left,base]{\color{textcolor}\rmfamily\fontsize{9.000000}{10.800000}\selectfont SRFlow}%
\end{pgfscope}%
\end{pgfpicture}%
\makeatother%
\endgroup%

%% file: latex/figures/inp_set14.pgf
\begingroup%
\makeatletter%
\begin{pgfpicture}%
\pgfpathrectangle{\pgfpointorigin}{\pgfqpoint{3.300000in}{2.600000in}}%
\pgfusepath{use as bounding box, clip}%
\begin{pgfscope}%
\pgfsetbuttcap%
\pgfsetmiterjoin%
\definecolor{currentfill}{rgb}{1.000000,1.000000,1.000000}%
\pgfsetfillcolor{currentfill}%
\pgfsetlinewidth{0.000000pt}%
\definecolor{currentstroke}{rgb}{1.000000,1.000000,1.000000}%
\pgfsetstrokecolor{currentstroke}%
\pgfsetdash{}{0pt}%
\pgfpathmoveto{\pgfqpoint{0.000000in}{0.000000in}}%
\pgfpathlineto{\pgfqpoint{3.300000in}{0.000000in}}%
\pgfpathlineto{\pgfqpoint{3.300000in}{2.600000in}}%
\pgfpathlineto{\pgfqpoint{0.000000in}{2.600000in}}%
\pgfpathlineto{\pgfqpoint{0.000000in}{0.000000in}}%
\pgfpathclose%
\pgfusepath{fill}%
\end{pgfscope}%
\begin{pgfscope}%
\pgfsetbuttcap%
\pgfsetmiterjoin%
\definecolor{currentfill}{rgb}{1.000000,1.000000,1.000000}%
\pgfsetfillcolor{currentfill}%
\pgfsetlinewidth{0.000000pt}%
\definecolor{currentstroke}{rgb}{0.000000,0.000000,0.000000}%
\pgfsetstrokecolor{currentstroke}%
\pgfsetstrokeopacity{0.000000}%
\pgfsetdash{}{0pt}%
\pgfpathmoveto{\pgfqpoint{0.527360in}{0.541249in}}%
\pgfpathlineto{\pgfqpoint{3.165000in}{0.541249in}}%
\pgfpathlineto{\pgfqpoint{3.165000in}{2.465000in}}%
\pgfpathlineto{\pgfqpoint{0.527360in}{2.465000in}}%
\pgfpathlineto{\pgfqpoint{0.527360in}{0.541249in}}%
\pgfpathclose%
\pgfusepath{fill}%
\end{pgfscope}%
\begin{pgfscope}%
\pgfsetbuttcap%
\pgfsetroundjoin%
\definecolor{currentfill}{rgb}{0.000000,0.000000,0.000000}%
\pgfsetfillcolor{currentfill}%
\pgfsetlinewidth{0.803000pt}%
\definecolor{currentstroke}{rgb}{0.000000,0.000000,0.000000}%
\pgfsetstrokecolor{currentstroke}%
\pgfsetdash{}{0pt}%
\pgfsys@defobject{currentmarker}{\pgfqpoint{0.000000in}{-0.048611in}}{\pgfqpoint{0.000000in}{0.000000in}}{%
\pgfpathmoveto{\pgfqpoint{0.000000in}{0.000000in}}%
\pgfpathlineto{\pgfqpoint{0.000000in}{-0.048611in}}%
\pgfusepath{stroke,fill}%
}%
\begin{pgfscope}%
\pgfsys@transformshift{0.747164in}{0.541249in}%
\pgfsys@useobject{currentmarker}{}%
\end{pgfscope}%
\end{pgfscope}%
\begin{pgfscope}%
\definecolor{textcolor}{rgb}{0.000000,0.000000,0.000000}%
\pgfsetstrokecolor{textcolor}%
\pgfsetfillcolor{textcolor}%
\pgftext[x=0.778414in, y=0.379791in, left, base,rotate=90.000000]{\color{textcolor}\rmfamily\fontsize{9.000000}{10.800000}\selectfont 1}%
\end{pgfscope}%
\begin{pgfscope}%
\pgfsetbuttcap%
\pgfsetroundjoin%
\definecolor{currentfill}{rgb}{0.000000,0.000000,0.000000}%
\pgfsetfillcolor{currentfill}%
\pgfsetlinewidth{0.803000pt}%
\definecolor{currentstroke}{rgb}{0.000000,0.000000,0.000000}%
\pgfsetstrokecolor{currentstroke}%
\pgfsetdash{}{0pt}%
\pgfsys@defobject{currentmarker}{\pgfqpoint{0.000000in}{-0.048611in}}{\pgfqpoint{0.000000in}{0.000000in}}{%
\pgfpathmoveto{\pgfqpoint{0.000000in}{0.000000in}}%
\pgfpathlineto{\pgfqpoint{0.000000in}{-0.048611in}}%
\pgfusepath{stroke,fill}%
}%
\begin{pgfscope}%
\pgfsys@transformshift{1.186770in}{0.541249in}%
\pgfsys@useobject{currentmarker}{}%
\end{pgfscope}%
\end{pgfscope}%
\begin{pgfscope}%
\definecolor{textcolor}{rgb}{0.000000,0.000000,0.000000}%
\pgfsetstrokecolor{textcolor}%
\pgfsetfillcolor{textcolor}%
\pgftext[x=1.218020in, y=0.379791in, left, base,rotate=90.000000]{\color{textcolor}\rmfamily\fontsize{9.000000}{10.800000}\selectfont 2}%
\end{pgfscope}%
\begin{pgfscope}%
\pgfsetbuttcap%
\pgfsetroundjoin%
\definecolor{currentfill}{rgb}{0.000000,0.000000,0.000000}%
\pgfsetfillcolor{currentfill}%
\pgfsetlinewidth{0.803000pt}%
\definecolor{currentstroke}{rgb}{0.000000,0.000000,0.000000}%
\pgfsetstrokecolor{currentstroke}%
\pgfsetdash{}{0pt}%
\pgfsys@defobject{currentmarker}{\pgfqpoint{0.000000in}{-0.048611in}}{\pgfqpoint{0.000000in}{0.000000in}}{%
\pgfpathmoveto{\pgfqpoint{0.000000in}{0.000000in}}%
\pgfpathlineto{\pgfqpoint{0.000000in}{-0.048611in}}%
\pgfusepath{stroke,fill}%
}%
\begin{pgfscope}%
\pgfsys@transformshift{1.626377in}{0.541249in}%
\pgfsys@useobject{currentmarker}{}%
\end{pgfscope}%
\end{pgfscope}%
\begin{pgfscope}%
\definecolor{textcolor}{rgb}{0.000000,0.000000,0.000000}%
\pgfsetstrokecolor{textcolor}%
\pgfsetfillcolor{textcolor}%
\pgftext[x=1.657627in, y=0.379791in, left, base,rotate=90.000000]{\color{textcolor}\rmfamily\fontsize{9.000000}{10.800000}\selectfont 4}%
\end{pgfscope}%
\begin{pgfscope}%
\pgfsetbuttcap%
\pgfsetroundjoin%
\definecolor{currentfill}{rgb}{0.000000,0.000000,0.000000}%
\pgfsetfillcolor{currentfill}%
\pgfsetlinewidth{0.803000pt}%
\definecolor{currentstroke}{rgb}{0.000000,0.000000,0.000000}%
\pgfsetstrokecolor{currentstroke}%
\pgfsetdash{}{0pt}%
\pgfsys@defobject{currentmarker}{\pgfqpoint{0.000000in}{-0.048611in}}{\pgfqpoint{0.000000in}{0.000000in}}{%
\pgfpathmoveto{\pgfqpoint{0.000000in}{0.000000in}}%
\pgfpathlineto{\pgfqpoint{0.000000in}{-0.048611in}}%
\pgfusepath{stroke,fill}%
}%
\begin{pgfscope}%
\pgfsys@transformshift{2.065983in}{0.541249in}%
\pgfsys@useobject{currentmarker}{}%
\end{pgfscope}%
\end{pgfscope}%
\begin{pgfscope}%
\definecolor{textcolor}{rgb}{0.000000,0.000000,0.000000}%
\pgfsetstrokecolor{textcolor}%
\pgfsetfillcolor{textcolor}%
\pgftext[x=2.097233in, y=0.379791in, left, base,rotate=90.000000]{\color{textcolor}\rmfamily\fontsize{9.000000}{10.800000}\selectfont 8}%
\end{pgfscope}%
\begin{pgfscope}%
\pgfsetbuttcap%
\pgfsetroundjoin%
\definecolor{currentfill}{rgb}{0.000000,0.000000,0.000000}%
\pgfsetfillcolor{currentfill}%
\pgfsetlinewidth{0.803000pt}%
\definecolor{currentstroke}{rgb}{0.000000,0.000000,0.000000}%
\pgfsetstrokecolor{currentstroke}%
\pgfsetdash{}{0pt}%
\pgfsys@defobject{currentmarker}{\pgfqpoint{0.000000in}{-0.048611in}}{\pgfqpoint{0.000000in}{0.000000in}}{%
\pgfpathmoveto{\pgfqpoint{0.000000in}{0.000000in}}%
\pgfpathlineto{\pgfqpoint{0.000000in}{-0.048611in}}%
\pgfusepath{stroke,fill}%
}%
\begin{pgfscope}%
\pgfsys@transformshift{2.505590in}{0.541249in}%
\pgfsys@useobject{currentmarker}{}%
\end{pgfscope}%
\end{pgfscope}%
\begin{pgfscope}%
\definecolor{textcolor}{rgb}{0.000000,0.000000,0.000000}%
\pgfsetstrokecolor{textcolor}%
\pgfsetfillcolor{textcolor}%
\pgftext[x=2.536840in, y=0.315556in, left, base,rotate=90.000000]{\color{textcolor}\rmfamily\fontsize{9.000000}{10.800000}\selectfont 16}%
\end{pgfscope}%
\begin{pgfscope}%
\pgfsetbuttcap%
\pgfsetroundjoin%
\definecolor{currentfill}{rgb}{0.000000,0.000000,0.000000}%
\pgfsetfillcolor{currentfill}%
\pgfsetlinewidth{0.803000pt}%
\definecolor{currentstroke}{rgb}{0.000000,0.000000,0.000000}%
\pgfsetstrokecolor{currentstroke}%
\pgfsetdash{}{0pt}%
\pgfsys@defobject{currentmarker}{\pgfqpoint{0.000000in}{-0.048611in}}{\pgfqpoint{0.000000in}{0.000000in}}{%
\pgfpathmoveto{\pgfqpoint{0.000000in}{0.000000in}}%
\pgfpathlineto{\pgfqpoint{0.000000in}{-0.048611in}}%
\pgfusepath{stroke,fill}%
}%
\begin{pgfscope}%
\pgfsys@transformshift{2.945197in}{0.541249in}%
\pgfsys@useobject{currentmarker}{}%
\end{pgfscope}%
\end{pgfscope}%
\begin{pgfscope}%
\definecolor{textcolor}{rgb}{0.000000,0.000000,0.000000}%
\pgfsetstrokecolor{textcolor}%
\pgfsetfillcolor{textcolor}%
\pgftext[x=2.976447in, y=0.315556in, left, base,rotate=90.000000]{\color{textcolor}\rmfamily\fontsize{9.000000}{10.800000}\selectfont 32}%
\end{pgfscope}%
\begin{pgfscope}%
\definecolor{textcolor}{rgb}{0.000000,0.000000,0.000000}%
\pgfsetstrokecolor{textcolor}%
\pgfsetfillcolor{textcolor}%
\pgftext[x=1.846180in,y=0.260000in,,top]{\color{textcolor}\rmfamily\fontsize{9.000000}{10.800000}\selectfont \(\displaystyle \alpha(/255)\)}%
\end{pgfscope}%
\begin{pgfscope}%
\pgfpathrectangle{\pgfqpoint{0.527360in}{0.541249in}}{\pgfqpoint{2.637640in}{1.923751in}}%
\pgfusepath{clip}%
\pgfsetrectcap%
\pgfsetroundjoin%
\pgfsetlinewidth{0.803000pt}%
\definecolor{currentstroke}{rgb}{0.690196,0.690196,0.690196}%
\pgfsetstrokecolor{currentstroke}%
\pgfsetdash{}{0pt}%
\pgfpathmoveto{\pgfqpoint{0.527360in}{0.541249in}}%
\pgfpathlineto{\pgfqpoint{3.165000in}{0.541249in}}%
\pgfusepath{stroke}%
\end{pgfscope}%
\begin{pgfscope}%
\pgfsetbuttcap%
\pgfsetroundjoin%
\definecolor{currentfill}{rgb}{0.000000,0.000000,0.000000}%
\pgfsetfillcolor{currentfill}%
\pgfsetlinewidth{0.803000pt}%
\definecolor{currentstroke}{rgb}{0.000000,0.000000,0.000000}%
\pgfsetstrokecolor{currentstroke}%
\pgfsetdash{}{0pt}%
\pgfsys@defobject{currentmarker}{\pgfqpoint{-0.048611in}{0.000000in}}{\pgfqpoint{-0.000000in}{0.000000in}}{%
\pgfpathmoveto{\pgfqpoint{-0.000000in}{0.000000in}}%
\pgfpathlineto{\pgfqpoint{-0.048611in}{0.000000in}}%
\pgfusepath{stroke,fill}%
}%
\begin{pgfscope}%
\pgfsys@transformshift{0.527360in}{0.541249in}%
\pgfsys@useobject{currentmarker}{}%
\end{pgfscope}%
\end{pgfscope}%
\begin{pgfscope}%
\definecolor{textcolor}{rgb}{0.000000,0.000000,0.000000}%
\pgfsetstrokecolor{textcolor}%
\pgfsetfillcolor{textcolor}%
\pgftext[x=0.365902in, y=0.497846in, left, base]{\color{textcolor}\rmfamily\fontsize{9.000000}{10.800000}\selectfont \(\displaystyle {0}\)}%
\end{pgfscope}%
\begin{pgfscope}%
\pgfpathrectangle{\pgfqpoint{0.527360in}{0.541249in}}{\pgfqpoint{2.637640in}{1.923751in}}%
\pgfusepath{clip}%
\pgfsetrectcap%
\pgfsetroundjoin%
\pgfsetlinewidth{0.803000pt}%
\definecolor{currentstroke}{rgb}{0.690196,0.690196,0.690196}%
\pgfsetstrokecolor{currentstroke}%
\pgfsetdash{}{0pt}%
\pgfpathmoveto{\pgfqpoint{0.527360in}{0.861874in}}%
\pgfpathlineto{\pgfqpoint{3.165000in}{0.861874in}}%
\pgfusepath{stroke}%
\end{pgfscope}%
\begin{pgfscope}%
\pgfsetbuttcap%
\pgfsetroundjoin%
\definecolor{currentfill}{rgb}{0.000000,0.000000,0.000000}%
\pgfsetfillcolor{currentfill}%
\pgfsetlinewidth{0.803000pt}%
\definecolor{currentstroke}{rgb}{0.000000,0.000000,0.000000}%
\pgfsetstrokecolor{currentstroke}%
\pgfsetdash{}{0pt}%
\pgfsys@defobject{currentmarker}{\pgfqpoint{-0.048611in}{0.000000in}}{\pgfqpoint{-0.000000in}{0.000000in}}{%
\pgfpathmoveto{\pgfqpoint{-0.000000in}{0.000000in}}%
\pgfpathlineto{\pgfqpoint{-0.048611in}{0.000000in}}%
\pgfusepath{stroke,fill}%
}%
\begin{pgfscope}%
\pgfsys@transformshift{0.527360in}{0.861874in}%
\pgfsys@useobject{currentmarker}{}%
\end{pgfscope}%
\end{pgfscope}%
\begin{pgfscope}%
\definecolor{textcolor}{rgb}{0.000000,0.000000,0.000000}%
\pgfsetstrokecolor{textcolor}%
\pgfsetfillcolor{textcolor}%
\pgftext[x=0.301667in, y=0.818472in, left, base]{\color{textcolor}\rmfamily\fontsize{9.000000}{10.800000}\selectfont \(\displaystyle {10}\)}%
\end{pgfscope}%
\begin{pgfscope}%
\pgfpathrectangle{\pgfqpoint{0.527360in}{0.541249in}}{\pgfqpoint{2.637640in}{1.923751in}}%
\pgfusepath{clip}%
\pgfsetrectcap%
\pgfsetroundjoin%
\pgfsetlinewidth{0.803000pt}%
\definecolor{currentstroke}{rgb}{0.690196,0.690196,0.690196}%
\pgfsetstrokecolor{currentstroke}%
\pgfsetdash{}{0pt}%
\pgfpathmoveto{\pgfqpoint{0.527360in}{1.182499in}}%
\pgfpathlineto{\pgfqpoint{3.165000in}{1.182499in}}%
\pgfusepath{stroke}%
\end{pgfscope}%
\begin{pgfscope}%
\pgfsetbuttcap%
\pgfsetroundjoin%
\definecolor{currentfill}{rgb}{0.000000,0.000000,0.000000}%
\pgfsetfillcolor{currentfill}%
\pgfsetlinewidth{0.803000pt}%
\definecolor{currentstroke}{rgb}{0.000000,0.000000,0.000000}%
\pgfsetstrokecolor{currentstroke}%
\pgfsetdash{}{0pt}%
\pgfsys@defobject{currentmarker}{\pgfqpoint{-0.048611in}{0.000000in}}{\pgfqpoint{-0.000000in}{0.000000in}}{%
\pgfpathmoveto{\pgfqpoint{-0.000000in}{0.000000in}}%
\pgfpathlineto{\pgfqpoint{-0.048611in}{0.000000in}}%
\pgfusepath{stroke,fill}%
}%
\begin{pgfscope}%
\pgfsys@transformshift{0.527360in}{1.182499in}%
\pgfsys@useobject{currentmarker}{}%
\end{pgfscope}%
\end{pgfscope}%
\begin{pgfscope}%
\definecolor{textcolor}{rgb}{0.000000,0.000000,0.000000}%
\pgfsetstrokecolor{textcolor}%
\pgfsetfillcolor{textcolor}%
\pgftext[x=0.301667in, y=1.139097in, left, base]{\color{textcolor}\rmfamily\fontsize{9.000000}{10.800000}\selectfont \(\displaystyle {20}\)}%
\end{pgfscope}%
\begin{pgfscope}%
\pgfpathrectangle{\pgfqpoint{0.527360in}{0.541249in}}{\pgfqpoint{2.637640in}{1.923751in}}%
\pgfusepath{clip}%
\pgfsetrectcap%
\pgfsetroundjoin%
\pgfsetlinewidth{0.803000pt}%
\definecolor{currentstroke}{rgb}{0.690196,0.690196,0.690196}%
\pgfsetstrokecolor{currentstroke}%
\pgfsetdash{}{0pt}%
\pgfpathmoveto{\pgfqpoint{0.527360in}{1.503125in}}%
\pgfpathlineto{\pgfqpoint{3.165000in}{1.503125in}}%
\pgfusepath{stroke}%
\end{pgfscope}%
\begin{pgfscope}%
\pgfsetbuttcap%
\pgfsetroundjoin%
\definecolor{currentfill}{rgb}{0.000000,0.000000,0.000000}%
\pgfsetfillcolor{currentfill}%
\pgfsetlinewidth{0.803000pt}%
\definecolor{currentstroke}{rgb}{0.000000,0.000000,0.000000}%
\pgfsetstrokecolor{currentstroke}%
\pgfsetdash{}{0pt}%
\pgfsys@defobject{currentmarker}{\pgfqpoint{-0.048611in}{0.000000in}}{\pgfqpoint{-0.000000in}{0.000000in}}{%
\pgfpathmoveto{\pgfqpoint{-0.000000in}{0.000000in}}%
\pgfpathlineto{\pgfqpoint{-0.048611in}{0.000000in}}%
\pgfusepath{stroke,fill}%
}%
\begin{pgfscope}%
\pgfsys@transformshift{0.527360in}{1.503125in}%
\pgfsys@useobject{currentmarker}{}%
\end{pgfscope}%
\end{pgfscope}%
\begin{pgfscope}%
\definecolor{textcolor}{rgb}{0.000000,0.000000,0.000000}%
\pgfsetstrokecolor{textcolor}%
\pgfsetfillcolor{textcolor}%
\pgftext[x=0.301667in, y=1.459722in, left, base]{\color{textcolor}\rmfamily\fontsize{9.000000}{10.800000}\selectfont \(\displaystyle {30}\)}%
\end{pgfscope}%
\begin{pgfscope}%
\pgfpathrectangle{\pgfqpoint{0.527360in}{0.541249in}}{\pgfqpoint{2.637640in}{1.923751in}}%
\pgfusepath{clip}%
\pgfsetrectcap%
\pgfsetroundjoin%
\pgfsetlinewidth{0.803000pt}%
\definecolor{currentstroke}{rgb}{0.690196,0.690196,0.690196}%
\pgfsetstrokecolor{currentstroke}%
\pgfsetdash{}{0pt}%
\pgfpathmoveto{\pgfqpoint{0.527360in}{1.823750in}}%
\pgfpathlineto{\pgfqpoint{3.165000in}{1.823750in}}%
\pgfusepath{stroke}%
\end{pgfscope}%
\begin{pgfscope}%
\pgfsetbuttcap%
\pgfsetroundjoin%
\definecolor{currentfill}{rgb}{0.000000,0.000000,0.000000}%
\pgfsetfillcolor{currentfill}%
\pgfsetlinewidth{0.803000pt}%
\definecolor{currentstroke}{rgb}{0.000000,0.000000,0.000000}%
\pgfsetstrokecolor{currentstroke}%
\pgfsetdash{}{0pt}%
\pgfsys@defobject{currentmarker}{\pgfqpoint{-0.048611in}{0.000000in}}{\pgfqpoint{-0.000000in}{0.000000in}}{%
\pgfpathmoveto{\pgfqpoint{-0.000000in}{0.000000in}}%
\pgfpathlineto{\pgfqpoint{-0.048611in}{0.000000in}}%
\pgfusepath{stroke,fill}%
}%
\begin{pgfscope}%
\pgfsys@transformshift{0.527360in}{1.823750in}%
\pgfsys@useobject{currentmarker}{}%
\end{pgfscope}%
\end{pgfscope}%
\begin{pgfscope}%
\definecolor{textcolor}{rgb}{0.000000,0.000000,0.000000}%
\pgfsetstrokecolor{textcolor}%
\pgfsetfillcolor{textcolor}%
\pgftext[x=0.301667in, y=1.780347in, left, base]{\color{textcolor}\rmfamily\fontsize{9.000000}{10.800000}\selectfont \(\displaystyle {40}\)}%
\end{pgfscope}%
\begin{pgfscope}%
\pgfpathrectangle{\pgfqpoint{0.527360in}{0.541249in}}{\pgfqpoint{2.637640in}{1.923751in}}%
\pgfusepath{clip}%
\pgfsetrectcap%
\pgfsetroundjoin%
\pgfsetlinewidth{0.803000pt}%
\definecolor{currentstroke}{rgb}{0.690196,0.690196,0.690196}%
\pgfsetstrokecolor{currentstroke}%
\pgfsetdash{}{0pt}%
\pgfpathmoveto{\pgfqpoint{0.527360in}{2.144375in}}%
\pgfpathlineto{\pgfqpoint{3.165000in}{2.144375in}}%
\pgfusepath{stroke}%
\end{pgfscope}%
\begin{pgfscope}%
\pgfsetbuttcap%
\pgfsetroundjoin%
\definecolor{currentfill}{rgb}{0.000000,0.000000,0.000000}%
\pgfsetfillcolor{currentfill}%
\pgfsetlinewidth{0.803000pt}%
\definecolor{currentstroke}{rgb}{0.000000,0.000000,0.000000}%
\pgfsetstrokecolor{currentstroke}%
\pgfsetdash{}{0pt}%
\pgfsys@defobject{currentmarker}{\pgfqpoint{-0.048611in}{0.000000in}}{\pgfqpoint{-0.000000in}{0.000000in}}{%
\pgfpathmoveto{\pgfqpoint{-0.000000in}{0.000000in}}%
\pgfpathlineto{\pgfqpoint{-0.048611in}{0.000000in}}%
\pgfusepath{stroke,fill}%
}%
\begin{pgfscope}%
\pgfsys@transformshift{0.527360in}{2.144375in}%
\pgfsys@useobject{currentmarker}{}%
\end{pgfscope}%
\end{pgfscope}%
\begin{pgfscope}%
\definecolor{textcolor}{rgb}{0.000000,0.000000,0.000000}%
\pgfsetstrokecolor{textcolor}%
\pgfsetfillcolor{textcolor}%
\pgftext[x=0.301667in, y=2.100972in, left, base]{\color{textcolor}\rmfamily\fontsize{9.000000}{10.800000}\selectfont \(\displaystyle {50}\)}%
\end{pgfscope}%
\begin{pgfscope}%
\pgfpathrectangle{\pgfqpoint{0.527360in}{0.541249in}}{\pgfqpoint{2.637640in}{1.923751in}}%
\pgfusepath{clip}%
\pgfsetrectcap%
\pgfsetroundjoin%
\pgfsetlinewidth{0.803000pt}%
\definecolor{currentstroke}{rgb}{0.690196,0.690196,0.690196}%
\pgfsetstrokecolor{currentstroke}%
\pgfsetdash{}{0pt}%
\pgfpathmoveto{\pgfqpoint{0.527360in}{2.465000in}}%
\pgfpathlineto{\pgfqpoint{3.165000in}{2.465000in}}%
\pgfusepath{stroke}%
\end{pgfscope}%
\begin{pgfscope}%
\pgfsetbuttcap%
\pgfsetroundjoin%
\definecolor{currentfill}{rgb}{0.000000,0.000000,0.000000}%
\pgfsetfillcolor{currentfill}%
\pgfsetlinewidth{0.803000pt}%
\definecolor{currentstroke}{rgb}{0.000000,0.000000,0.000000}%
\pgfsetstrokecolor{currentstroke}%
\pgfsetdash{}{0pt}%
\pgfsys@defobject{currentmarker}{\pgfqpoint{-0.048611in}{0.000000in}}{\pgfqpoint{-0.000000in}{0.000000in}}{%
\pgfpathmoveto{\pgfqpoint{-0.000000in}{0.000000in}}%
\pgfpathlineto{\pgfqpoint{-0.048611in}{0.000000in}}%
\pgfusepath{stroke,fill}%
}%
\begin{pgfscope}%
\pgfsys@transformshift{0.527360in}{2.465000in}%
\pgfsys@useobject{currentmarker}{}%
\end{pgfscope}%
\end{pgfscope}%
\begin{pgfscope}%
\definecolor{textcolor}{rgb}{0.000000,0.000000,0.000000}%
\pgfsetstrokecolor{textcolor}%
\pgfsetfillcolor{textcolor}%
\pgftext[x=0.301667in, y=2.421597in, left, base]{\color{textcolor}\rmfamily\fontsize{9.000000}{10.800000}\selectfont \(\displaystyle {60}\)}%
\end{pgfscope}%
\begin{pgfscope}%
\definecolor{textcolor}{rgb}{0.000000,0.000000,0.000000}%
\pgfsetstrokecolor{textcolor}%
\pgfsetfillcolor{textcolor}%
\pgftext[x=0.246111in,y=1.503125in,,bottom,rotate=90.000000]{\color{textcolor}\rmfamily\fontsize{9.000000}{10.800000}\selectfont PSNR}%
\end{pgfscope}%
\begin{pgfscope}%
\pgfpathrectangle{\pgfqpoint{0.527360in}{0.541249in}}{\pgfqpoint{2.637640in}{1.923751in}}%
\pgfusepath{clip}%
\pgfsetbuttcap%
\pgfsetmiterjoin%
\definecolor{currentfill}{rgb}{0.121569,0.466667,0.705882}%
\pgfsetfillcolor{currentfill}%
\pgfsetlinewidth{0.000000pt}%
\definecolor{currentstroke}{rgb}{0.000000,0.000000,0.000000}%
\pgfsetstrokecolor{currentstroke}%
\pgfsetstrokeopacity{0.000000}%
\pgfsetdash{}{0pt}%
\pgfpathmoveto{\pgfqpoint{0.637262in}{0.541249in}}%
\pgfpathlineto{\pgfqpoint{0.710530in}{0.541249in}}%
\pgfpathlineto{\pgfqpoint{0.710530in}{2.176437in}}%
\pgfpathlineto{\pgfqpoint{0.637262in}{2.176437in}}%
\pgfpathlineto{\pgfqpoint{0.637262in}{0.541249in}}%
\pgfpathclose%
\pgfusepath{fill}%
\end{pgfscope}%
\begin{pgfscope}%
\pgfpathrectangle{\pgfqpoint{0.527360in}{0.541249in}}{\pgfqpoint{2.637640in}{1.923751in}}%
\pgfusepath{clip}%
\pgfsetbuttcap%
\pgfsetmiterjoin%
\definecolor{currentfill}{rgb}{0.121569,0.466667,0.705882}%
\pgfsetfillcolor{currentfill}%
\pgfsetlinewidth{0.000000pt}%
\definecolor{currentstroke}{rgb}{0.000000,0.000000,0.000000}%
\pgfsetstrokecolor{currentstroke}%
\pgfsetstrokeopacity{0.000000}%
\pgfsetdash{}{0pt}%
\pgfpathmoveto{\pgfqpoint{1.076869in}{0.541249in}}%
\pgfpathlineto{\pgfqpoint{1.150136in}{0.541249in}}%
\pgfpathlineto{\pgfqpoint{1.150136in}{2.118725in}}%
\pgfpathlineto{\pgfqpoint{1.076869in}{2.118725in}}%
\pgfpathlineto{\pgfqpoint{1.076869in}{0.541249in}}%
\pgfpathclose%
\pgfusepath{fill}%
\end{pgfscope}%
\begin{pgfscope}%
\pgfpathrectangle{\pgfqpoint{0.527360in}{0.541249in}}{\pgfqpoint{2.637640in}{1.923751in}}%
\pgfusepath{clip}%
\pgfsetbuttcap%
\pgfsetmiterjoin%
\definecolor{currentfill}{rgb}{0.121569,0.466667,0.705882}%
\pgfsetfillcolor{currentfill}%
\pgfsetlinewidth{0.000000pt}%
\definecolor{currentstroke}{rgb}{0.000000,0.000000,0.000000}%
\pgfsetstrokecolor{currentstroke}%
\pgfsetstrokeopacity{0.000000}%
\pgfsetdash{}{0pt}%
\pgfpathmoveto{\pgfqpoint{1.516475in}{0.541249in}}%
\pgfpathlineto{\pgfqpoint{1.589743in}{0.541249in}}%
\pgfpathlineto{\pgfqpoint{1.589743in}{1.990475in}}%
\pgfpathlineto{\pgfqpoint{1.516475in}{1.990475in}}%
\pgfpathlineto{\pgfqpoint{1.516475in}{0.541249in}}%
\pgfpathclose%
\pgfusepath{fill}%
\end{pgfscope}%
\begin{pgfscope}%
\pgfpathrectangle{\pgfqpoint{0.527360in}{0.541249in}}{\pgfqpoint{2.637640in}{1.923751in}}%
\pgfusepath{clip}%
\pgfsetbuttcap%
\pgfsetmiterjoin%
\definecolor{currentfill}{rgb}{0.121569,0.466667,0.705882}%
\pgfsetfillcolor{currentfill}%
\pgfsetlinewidth{0.000000pt}%
\definecolor{currentstroke}{rgb}{0.000000,0.000000,0.000000}%
\pgfsetstrokecolor{currentstroke}%
\pgfsetstrokeopacity{0.000000}%
\pgfsetdash{}{0pt}%
\pgfpathmoveto{\pgfqpoint{1.956082in}{0.541249in}}%
\pgfpathlineto{\pgfqpoint{2.029350in}{0.541249in}}%
\pgfpathlineto{\pgfqpoint{2.029350in}{1.855812in}}%
\pgfpathlineto{\pgfqpoint{1.956082in}{1.855812in}}%
\pgfpathlineto{\pgfqpoint{1.956082in}{0.541249in}}%
\pgfpathclose%
\pgfusepath{fill}%
\end{pgfscope}%
\begin{pgfscope}%
\pgfpathrectangle{\pgfqpoint{0.527360in}{0.541249in}}{\pgfqpoint{2.637640in}{1.923751in}}%
\pgfusepath{clip}%
\pgfsetbuttcap%
\pgfsetmiterjoin%
\definecolor{currentfill}{rgb}{0.121569,0.466667,0.705882}%
\pgfsetfillcolor{currentfill}%
\pgfsetlinewidth{0.000000pt}%
\definecolor{currentstroke}{rgb}{0.000000,0.000000,0.000000}%
\pgfsetstrokecolor{currentstroke}%
\pgfsetstrokeopacity{0.000000}%
\pgfsetdash{}{0pt}%
\pgfpathmoveto{\pgfqpoint{2.395688in}{0.541249in}}%
\pgfpathlineto{\pgfqpoint{2.468956in}{0.541249in}}%
\pgfpathlineto{\pgfqpoint{2.468956in}{1.766037in}}%
\pgfpathlineto{\pgfqpoint{2.395688in}{1.766037in}}%
\pgfpathlineto{\pgfqpoint{2.395688in}{0.541249in}}%
\pgfpathclose%
\pgfusepath{fill}%
\end{pgfscope}%
\begin{pgfscope}%
\pgfpathrectangle{\pgfqpoint{0.527360in}{0.541249in}}{\pgfqpoint{2.637640in}{1.923751in}}%
\pgfusepath{clip}%
\pgfsetbuttcap%
\pgfsetmiterjoin%
\definecolor{currentfill}{rgb}{0.121569,0.466667,0.705882}%
\pgfsetfillcolor{currentfill}%
\pgfsetlinewidth{0.000000pt}%
\definecolor{currentstroke}{rgb}{0.000000,0.000000,0.000000}%
\pgfsetstrokecolor{currentstroke}%
\pgfsetstrokeopacity{0.000000}%
\pgfsetdash{}{0pt}%
\pgfpathmoveto{\pgfqpoint{2.835295in}{0.541249in}}%
\pgfpathlineto{\pgfqpoint{2.908563in}{0.541249in}}%
\pgfpathlineto{\pgfqpoint{2.908563in}{1.599312in}}%
\pgfpathlineto{\pgfqpoint{2.835295in}{1.599312in}}%
\pgfpathlineto{\pgfqpoint{2.835295in}{0.541249in}}%
\pgfpathclose%
\pgfusepath{fill}%
\end{pgfscope}%
\begin{pgfscope}%
\pgfpathrectangle{\pgfqpoint{0.527360in}{0.541249in}}{\pgfqpoint{2.637640in}{1.923751in}}%
\pgfusepath{clip}%
\pgfsetbuttcap%
\pgfsetmiterjoin%
\definecolor{currentfill}{rgb}{1.000000,0.498039,0.054902}%
\pgfsetfillcolor{currentfill}%
\pgfsetlinewidth{0.000000pt}%
\definecolor{currentstroke}{rgb}{0.000000,0.000000,0.000000}%
\pgfsetstrokecolor{currentstroke}%
\pgfsetstrokeopacity{0.000000}%
\pgfsetdash{}{0pt}%
\pgfpathmoveto{\pgfqpoint{0.710530in}{0.541249in}}%
\pgfpathlineto{\pgfqpoint{0.783797in}{0.541249in}}%
\pgfpathlineto{\pgfqpoint{0.783797in}{2.208500in}}%
\pgfpathlineto{\pgfqpoint{0.710530in}{2.208500in}}%
\pgfpathlineto{\pgfqpoint{0.710530in}{0.541249in}}%
\pgfpathclose%
\pgfusepath{fill}%
\end{pgfscope}%
\begin{pgfscope}%
\pgfpathrectangle{\pgfqpoint{0.527360in}{0.541249in}}{\pgfqpoint{2.637640in}{1.923751in}}%
\pgfusepath{clip}%
\pgfsetbuttcap%
\pgfsetmiterjoin%
\definecolor{currentfill}{rgb}{1.000000,0.498039,0.054902}%
\pgfsetfillcolor{currentfill}%
\pgfsetlinewidth{0.000000pt}%
\definecolor{currentstroke}{rgb}{0.000000,0.000000,0.000000}%
\pgfsetstrokecolor{currentstroke}%
\pgfsetstrokeopacity{0.000000}%
\pgfsetdash{}{0pt}%
\pgfpathmoveto{\pgfqpoint{1.150136in}{0.541249in}}%
\pgfpathlineto{\pgfqpoint{1.223404in}{0.541249in}}%
\pgfpathlineto{\pgfqpoint{1.223404in}{2.142772in}}%
\pgfpathlineto{\pgfqpoint{1.150136in}{2.142772in}}%
\pgfpathlineto{\pgfqpoint{1.150136in}{0.541249in}}%
\pgfpathclose%
\pgfusepath{fill}%
\end{pgfscope}%
\begin{pgfscope}%
\pgfpathrectangle{\pgfqpoint{0.527360in}{0.541249in}}{\pgfqpoint{2.637640in}{1.923751in}}%
\pgfusepath{clip}%
\pgfsetbuttcap%
\pgfsetmiterjoin%
\definecolor{currentfill}{rgb}{1.000000,0.498039,0.054902}%
\pgfsetfillcolor{currentfill}%
\pgfsetlinewidth{0.000000pt}%
\definecolor{currentstroke}{rgb}{0.000000,0.000000,0.000000}%
\pgfsetstrokecolor{currentstroke}%
\pgfsetstrokeopacity{0.000000}%
\pgfsetdash{}{0pt}%
\pgfpathmoveto{\pgfqpoint{1.589743in}{0.541249in}}%
\pgfpathlineto{\pgfqpoint{1.663011in}{0.541249in}}%
\pgfpathlineto{\pgfqpoint{1.663011in}{2.032156in}}%
\pgfpathlineto{\pgfqpoint{1.589743in}{2.032156in}}%
\pgfpathlineto{\pgfqpoint{1.589743in}{0.541249in}}%
\pgfpathclose%
\pgfusepath{fill}%
\end{pgfscope}%
\begin{pgfscope}%
\pgfpathrectangle{\pgfqpoint{0.527360in}{0.541249in}}{\pgfqpoint{2.637640in}{1.923751in}}%
\pgfusepath{clip}%
\pgfsetbuttcap%
\pgfsetmiterjoin%
\definecolor{currentfill}{rgb}{1.000000,0.498039,0.054902}%
\pgfsetfillcolor{currentfill}%
\pgfsetlinewidth{0.000000pt}%
\definecolor{currentstroke}{rgb}{0.000000,0.000000,0.000000}%
\pgfsetstrokecolor{currentstroke}%
\pgfsetstrokeopacity{0.000000}%
\pgfsetdash{}{0pt}%
\pgfpathmoveto{\pgfqpoint{2.029350in}{0.541249in}}%
\pgfpathlineto{\pgfqpoint{2.102617in}{0.541249in}}%
\pgfpathlineto{\pgfqpoint{2.102617in}{1.919937in}}%
\pgfpathlineto{\pgfqpoint{2.029350in}{1.919937in}}%
\pgfpathlineto{\pgfqpoint{2.029350in}{0.541249in}}%
\pgfpathclose%
\pgfusepath{fill}%
\end{pgfscope}%
\begin{pgfscope}%
\pgfpathrectangle{\pgfqpoint{0.527360in}{0.541249in}}{\pgfqpoint{2.637640in}{1.923751in}}%
\pgfusepath{clip}%
\pgfsetbuttcap%
\pgfsetmiterjoin%
\definecolor{currentfill}{rgb}{1.000000,0.498039,0.054902}%
\pgfsetfillcolor{currentfill}%
\pgfsetlinewidth{0.000000pt}%
\definecolor{currentstroke}{rgb}{0.000000,0.000000,0.000000}%
\pgfsetstrokecolor{currentstroke}%
\pgfsetstrokeopacity{0.000000}%
\pgfsetdash{}{0pt}%
\pgfpathmoveto{\pgfqpoint{2.468956in}{0.541249in}}%
\pgfpathlineto{\pgfqpoint{2.542224in}{0.541249in}}%
\pgfpathlineto{\pgfqpoint{2.542224in}{1.801306in}}%
\pgfpathlineto{\pgfqpoint{2.468956in}{1.801306in}}%
\pgfpathlineto{\pgfqpoint{2.468956in}{0.541249in}}%
\pgfpathclose%
\pgfusepath{fill}%
\end{pgfscope}%
\begin{pgfscope}%
\pgfpathrectangle{\pgfqpoint{0.527360in}{0.541249in}}{\pgfqpoint{2.637640in}{1.923751in}}%
\pgfusepath{clip}%
\pgfsetbuttcap%
\pgfsetmiterjoin%
\definecolor{currentfill}{rgb}{1.000000,0.498039,0.054902}%
\pgfsetfillcolor{currentfill}%
\pgfsetlinewidth{0.000000pt}%
\definecolor{currentstroke}{rgb}{0.000000,0.000000,0.000000}%
\pgfsetstrokecolor{currentstroke}%
\pgfsetstrokeopacity{0.000000}%
\pgfsetdash{}{0pt}%
\pgfpathmoveto{\pgfqpoint{2.908563in}{0.541249in}}%
\pgfpathlineto{\pgfqpoint{2.981831in}{0.541249in}}%
\pgfpathlineto{\pgfqpoint{2.981831in}{1.647406in}}%
\pgfpathlineto{\pgfqpoint{2.908563in}{1.647406in}}%
\pgfpathlineto{\pgfqpoint{2.908563in}{0.541249in}}%
\pgfpathclose%
\pgfusepath{fill}%
\end{pgfscope}%
\begin{pgfscope}%
\pgfpathrectangle{\pgfqpoint{0.527360in}{0.541249in}}{\pgfqpoint{2.637640in}{1.923751in}}%
\pgfusepath{clip}%
\pgfsetbuttcap%
\pgfsetmiterjoin%
\definecolor{currentfill}{rgb}{0.172549,0.627451,0.172549}%
\pgfsetfillcolor{currentfill}%
\pgfsetlinewidth{0.000000pt}%
\definecolor{currentstroke}{rgb}{0.000000,0.000000,0.000000}%
\pgfsetstrokecolor{currentstroke}%
\pgfsetstrokeopacity{0.000000}%
\pgfsetdash{}{0pt}%
\pgfpathmoveto{\pgfqpoint{0.783797in}{0.541249in}}%
\pgfpathlineto{\pgfqpoint{0.857065in}{0.541249in}}%
\pgfpathlineto{\pgfqpoint{0.857065in}{2.167780in}}%
\pgfpathlineto{\pgfqpoint{0.783797in}{2.167780in}}%
\pgfpathlineto{\pgfqpoint{0.783797in}{0.541249in}}%
\pgfpathclose%
\pgfusepath{fill}%
\end{pgfscope}%
\begin{pgfscope}%
\pgfpathrectangle{\pgfqpoint{0.527360in}{0.541249in}}{\pgfqpoint{2.637640in}{1.923751in}}%
\pgfusepath{clip}%
\pgfsetbuttcap%
\pgfsetmiterjoin%
\definecolor{currentfill}{rgb}{0.172549,0.627451,0.172549}%
\pgfsetfillcolor{currentfill}%
\pgfsetlinewidth{0.000000pt}%
\definecolor{currentstroke}{rgb}{0.000000,0.000000,0.000000}%
\pgfsetstrokecolor{currentstroke}%
\pgfsetstrokeopacity{0.000000}%
\pgfsetdash{}{0pt}%
\pgfpathmoveto{\pgfqpoint{1.223404in}{0.541249in}}%
\pgfpathlineto{\pgfqpoint{1.296672in}{0.541249in}}%
\pgfpathlineto{\pgfqpoint{1.296672in}{2.016125in}}%
\pgfpathlineto{\pgfqpoint{1.223404in}{2.016125in}}%
\pgfpathlineto{\pgfqpoint{1.223404in}{0.541249in}}%
\pgfpathclose%
\pgfusepath{fill}%
\end{pgfscope}%
\begin{pgfscope}%
\pgfpathrectangle{\pgfqpoint{0.527360in}{0.541249in}}{\pgfqpoint{2.637640in}{1.923751in}}%
\pgfusepath{clip}%
\pgfsetbuttcap%
\pgfsetmiterjoin%
\definecolor{currentfill}{rgb}{0.172549,0.627451,0.172549}%
\pgfsetfillcolor{currentfill}%
\pgfsetlinewidth{0.000000pt}%
\definecolor{currentstroke}{rgb}{0.000000,0.000000,0.000000}%
\pgfsetstrokecolor{currentstroke}%
\pgfsetstrokeopacity{0.000000}%
\pgfsetdash{}{0pt}%
\pgfpathmoveto{\pgfqpoint{1.663011in}{0.541249in}}%
\pgfpathlineto{\pgfqpoint{1.736278in}{0.541249in}}%
\pgfpathlineto{\pgfqpoint{1.736278in}{1.835613in}}%
\pgfpathlineto{\pgfqpoint{1.663011in}{1.835613in}}%
\pgfpathlineto{\pgfqpoint{1.663011in}{0.541249in}}%
\pgfpathclose%
\pgfusepath{fill}%
\end{pgfscope}%
\begin{pgfscope}%
\pgfpathrectangle{\pgfqpoint{0.527360in}{0.541249in}}{\pgfqpoint{2.637640in}{1.923751in}}%
\pgfusepath{clip}%
\pgfsetbuttcap%
\pgfsetmiterjoin%
\definecolor{currentfill}{rgb}{0.172549,0.627451,0.172549}%
\pgfsetfillcolor{currentfill}%
\pgfsetlinewidth{0.000000pt}%
\definecolor{currentstroke}{rgb}{0.000000,0.000000,0.000000}%
\pgfsetstrokecolor{currentstroke}%
\pgfsetstrokeopacity{0.000000}%
\pgfsetdash{}{0pt}%
\pgfpathmoveto{\pgfqpoint{2.102617in}{0.541249in}}%
\pgfpathlineto{\pgfqpoint{2.175885in}{0.541249in}}%
\pgfpathlineto{\pgfqpoint{2.175885in}{1.643558in}}%
\pgfpathlineto{\pgfqpoint{2.102617in}{1.643558in}}%
\pgfpathlineto{\pgfqpoint{2.102617in}{0.541249in}}%
\pgfpathclose%
\pgfusepath{fill}%
\end{pgfscope}%
\begin{pgfscope}%
\pgfpathrectangle{\pgfqpoint{0.527360in}{0.541249in}}{\pgfqpoint{2.637640in}{1.923751in}}%
\pgfusepath{clip}%
\pgfsetbuttcap%
\pgfsetmiterjoin%
\definecolor{currentfill}{rgb}{0.172549,0.627451,0.172549}%
\pgfsetfillcolor{currentfill}%
\pgfsetlinewidth{0.000000pt}%
\definecolor{currentstroke}{rgb}{0.000000,0.000000,0.000000}%
\pgfsetstrokecolor{currentstroke}%
\pgfsetstrokeopacity{0.000000}%
\pgfsetdash{}{0pt}%
\pgfpathmoveto{\pgfqpoint{2.542224in}{0.541249in}}%
\pgfpathlineto{\pgfqpoint{2.615492in}{0.541249in}}%
\pgfpathlineto{\pgfqpoint{2.615492in}{1.459520in}}%
\pgfpathlineto{\pgfqpoint{2.542224in}{1.459520in}}%
\pgfpathlineto{\pgfqpoint{2.542224in}{0.541249in}}%
\pgfpathclose%
\pgfusepath{fill}%
\end{pgfscope}%
\begin{pgfscope}%
\pgfpathrectangle{\pgfqpoint{0.527360in}{0.541249in}}{\pgfqpoint{2.637640in}{1.923751in}}%
\pgfusepath{clip}%
\pgfsetbuttcap%
\pgfsetmiterjoin%
\definecolor{currentfill}{rgb}{0.172549,0.627451,0.172549}%
\pgfsetfillcolor{currentfill}%
\pgfsetlinewidth{0.000000pt}%
\definecolor{currentstroke}{rgb}{0.000000,0.000000,0.000000}%
\pgfsetstrokecolor{currentstroke}%
\pgfsetstrokeopacity{0.000000}%
\pgfsetdash{}{0pt}%
\pgfpathmoveto{\pgfqpoint{2.981831in}{0.541249in}}%
\pgfpathlineto{\pgfqpoint{3.055098in}{0.541249in}}%
\pgfpathlineto{\pgfqpoint{3.055098in}{1.278687in}}%
\pgfpathlineto{\pgfqpoint{2.981831in}{1.278687in}}%
\pgfpathlineto{\pgfqpoint{2.981831in}{0.541249in}}%
\pgfpathclose%
\pgfusepath{fill}%
\end{pgfscope}%
\begin{pgfscope}%
\pgfsetrectcap%
\pgfsetmiterjoin%
\pgfsetlinewidth{0.803000pt}%
\definecolor{currentstroke}{rgb}{0.000000,0.000000,0.000000}%
\pgfsetstrokecolor{currentstroke}%
\pgfsetdash{}{0pt}%
\pgfpathmoveto{\pgfqpoint{0.527360in}{0.541249in}}%
\pgfpathlineto{\pgfqpoint{0.527360in}{2.465000in}}%
\pgfusepath{stroke}%
\end{pgfscope}%
\begin{pgfscope}%
\pgfsetrectcap%
\pgfsetmiterjoin%
\pgfsetlinewidth{0.803000pt}%
\definecolor{currentstroke}{rgb}{0.000000,0.000000,0.000000}%
\pgfsetstrokecolor{currentstroke}%
\pgfsetdash{}{0pt}%
\pgfpathmoveto{\pgfqpoint{3.165000in}{0.541249in}}%
\pgfpathlineto{\pgfqpoint{3.165000in}{2.465000in}}%
\pgfusepath{stroke}%
\end{pgfscope}%
\begin{pgfscope}%
\pgfsetrectcap%
\pgfsetmiterjoin%
\pgfsetlinewidth{0.803000pt}%
\definecolor{currentstroke}{rgb}{0.000000,0.000000,0.000000}%
\pgfsetstrokecolor{currentstroke}%
\pgfsetdash{}{0pt}%
\pgfpathmoveto{\pgfqpoint{0.527360in}{0.541249in}}%
\pgfpathlineto{\pgfqpoint{3.165000in}{0.541249in}}%
\pgfusepath{stroke}%
\end{pgfscope}%
\begin{pgfscope}%
\pgfsetrectcap%
\pgfsetmiterjoin%
\pgfsetlinewidth{0.803000pt}%
\definecolor{currentstroke}{rgb}{0.000000,0.000000,0.000000}%
\pgfsetstrokecolor{currentstroke}%
\pgfsetdash{}{0pt}%
\pgfpathmoveto{\pgfqpoint{0.527360in}{2.465000in}}%
\pgfpathlineto{\pgfqpoint{3.165000in}{2.465000in}}%
\pgfusepath{stroke}%
\end{pgfscope}%
\begin{pgfscope}%
\pgfsetbuttcap%
\pgfsetmiterjoin%
\definecolor{currentfill}{rgb}{1.000000,1.000000,1.000000}%
\pgfsetfillcolor{currentfill}%
\pgfsetfillopacity{0.800000}%
\pgfsetlinewidth{1.003750pt}%
\definecolor{currentstroke}{rgb}{0.800000,0.800000,0.800000}%
\pgfsetstrokecolor{currentstroke}%
\pgfsetstrokeopacity{0.800000}%
\pgfsetdash{}{0pt}%
\pgfpathmoveto{\pgfqpoint{2.134237in}{1.842083in}}%
\pgfpathlineto{\pgfqpoint{3.077500in}{1.842083in}}%
\pgfpathquadraticcurveto{\pgfqpoint{3.102500in}{1.842083in}}{\pgfqpoint{3.102500in}{1.867083in}}%
\pgfpathlineto{\pgfqpoint{3.102500in}{2.377500in}}%
\pgfpathquadraticcurveto{\pgfqpoint{3.102500in}{2.402500in}}{\pgfqpoint{3.077500in}{2.402500in}}%
\pgfpathlineto{\pgfqpoint{2.134237in}{2.402500in}}%
\pgfpathquadraticcurveto{\pgfqpoint{2.109237in}{2.402500in}}{\pgfqpoint{2.109237in}{2.377500in}}%
\pgfpathlineto{\pgfqpoint{2.109237in}{1.867083in}}%
\pgfpathquadraticcurveto{\pgfqpoint{2.109237in}{1.842083in}}{\pgfqpoint{2.134237in}{1.842083in}}%
\pgfpathlineto{\pgfqpoint{2.134237in}{1.842083in}}%
\pgfpathclose%
\pgfusepath{stroke,fill}%
\end{pgfscope}%
\begin{pgfscope}%
\pgfsetbuttcap%
\pgfsetmiterjoin%
\definecolor{currentfill}{rgb}{0.121569,0.466667,0.705882}%
\pgfsetfillcolor{currentfill}%
\pgfsetlinewidth{0.000000pt}%
\definecolor{currentstroke}{rgb}{0.000000,0.000000,0.000000}%
\pgfsetstrokecolor{currentstroke}%
\pgfsetstrokeopacity{0.000000}%
\pgfsetdash{}{0pt}%
\pgfpathmoveto{\pgfqpoint{2.159237in}{2.265000in}}%
\pgfpathlineto{\pgfqpoint{2.409237in}{2.265000in}}%
\pgfpathlineto{\pgfqpoint{2.409237in}{2.352500in}}%
\pgfpathlineto{\pgfqpoint{2.159237in}{2.352500in}}%
\pgfpathlineto{\pgfqpoint{2.159237in}{2.265000in}}%
\pgfpathclose%
\pgfusepath{fill}%
\end{pgfscope}%
\begin{pgfscope}%
\definecolor{textcolor}{rgb}{0.000000,0.000000,0.000000}%
\pgfsetstrokecolor{textcolor}%
\pgfsetfillcolor{textcolor}%
\pgftext[x=2.509237in,y=2.265000in,left,base]{\color{textcolor}\rmfamily\fontsize{9.000000}{10.800000}\selectfont ESRGAN}%
\end{pgfscope}%
\begin{pgfscope}%
\pgfsetbuttcap%
\pgfsetmiterjoin%
\definecolor{currentfill}{rgb}{1.000000,0.498039,0.054902}%
\pgfsetfillcolor{currentfill}%
\pgfsetlinewidth{0.000000pt}%
\definecolor{currentstroke}{rgb}{0.000000,0.000000,0.000000}%
\pgfsetstrokecolor{currentstroke}%
\pgfsetstrokeopacity{0.000000}%
\pgfsetdash{}{0pt}%
\pgfpathmoveto{\pgfqpoint{2.159237in}{2.090694in}}%
\pgfpathlineto{\pgfqpoint{2.409237in}{2.090694in}}%
\pgfpathlineto{\pgfqpoint{2.409237in}{2.178194in}}%
\pgfpathlineto{\pgfqpoint{2.159237in}{2.178194in}}%
\pgfpathlineto{\pgfqpoint{2.159237in}{2.090694in}}%
\pgfpathclose%
\pgfusepath{fill}%
\end{pgfscope}%
\begin{pgfscope}%
\definecolor{textcolor}{rgb}{0.000000,0.000000,0.000000}%
\pgfsetstrokecolor{textcolor}%
\pgfsetfillcolor{textcolor}%
\pgftext[x=2.509237in,y=2.090694in,left,base]{\color{textcolor}\rmfamily\fontsize{9.000000}{10.800000}\selectfont EDSR}%
\end{pgfscope}%
\begin{pgfscope}%
\pgfsetbuttcap%
\pgfsetmiterjoin%
\definecolor{currentfill}{rgb}{0.172549,0.627451,0.172549}%
\pgfsetfillcolor{currentfill}%
\pgfsetlinewidth{0.000000pt}%
\definecolor{currentstroke}{rgb}{0.000000,0.000000,0.000000}%
\pgfsetstrokecolor{currentstroke}%
\pgfsetstrokeopacity{0.000000}%
\pgfsetdash{}{0pt}%
\pgfpathmoveto{\pgfqpoint{2.159237in}{1.916389in}}%
\pgfpathlineto{\pgfqpoint{2.409237in}{1.916389in}}%
\pgfpathlineto{\pgfqpoint{2.409237in}{2.003889in}}%
\pgfpathlineto{\pgfqpoint{2.159237in}{2.003889in}}%
\pgfpathlineto{\pgfqpoint{2.159237in}{1.916389in}}%
\pgfpathclose%
\pgfusepath{fill}%
\end{pgfscope}%
\begin{pgfscope}%
\definecolor{textcolor}{rgb}{0.000000,0.000000,0.000000}%
\pgfsetstrokecolor{textcolor}%
\pgfsetfillcolor{textcolor}%
\pgftext[x=2.509237in,y=1.916389in,left,base]{\color{textcolor}\rmfamily\fontsize{9.000000}{10.800000}\selectfont SRFlow}%
\end{pgfscope}%
\end{pgfpicture}%
\makeatother%
\endgroup%

%% file: latex/figures/adv_bsd.pgf
\begingroup%
\makeatletter%
\begin{pgfpicture}%
\pgfpathrectangle{\pgfpointorigin}{\pgfqpoint{3.300000in}{2.600000in}}%
\pgfusepath{use as bounding box, clip}%
\begin{pgfscope}%
\pgfsetbuttcap%
\pgfsetmiterjoin%
\definecolor{currentfill}{rgb}{1.000000,1.000000,1.000000}%
\pgfsetfillcolor{currentfill}%
\pgfsetlinewidth{0.000000pt}%
\definecolor{currentstroke}{rgb}{1.000000,1.000000,1.000000}%
\pgfsetstrokecolor{currentstroke}%
\pgfsetdash{}{0pt}%
\pgfpathmoveto{\pgfqpoint{0.000000in}{0.000000in}}%
\pgfpathlineto{\pgfqpoint{3.300000in}{0.000000in}}%
\pgfpathlineto{\pgfqpoint{3.300000in}{2.600000in}}%
\pgfpathlineto{\pgfqpoint{0.000000in}{2.600000in}}%
\pgfpathlineto{\pgfqpoint{0.000000in}{0.000000in}}%
\pgfpathclose%
\pgfusepath{fill}%
\end{pgfscope}%
\begin{pgfscope}%
\pgfsetbuttcap%
\pgfsetmiterjoin%
\definecolor{currentfill}{rgb}{1.000000,1.000000,1.000000}%
\pgfsetfillcolor{currentfill}%
\pgfsetlinewidth{0.000000pt}%
\definecolor{currentstroke}{rgb}{0.000000,0.000000,0.000000}%
\pgfsetstrokecolor{currentstroke}%
\pgfsetstrokeopacity{0.000000}%
\pgfsetdash{}{0pt}%
\pgfpathmoveto{\pgfqpoint{0.527360in}{0.541249in}}%
\pgfpathlineto{\pgfqpoint{3.165000in}{0.541249in}}%
\pgfpathlineto{\pgfqpoint{3.165000in}{2.465000in}}%
\pgfpathlineto{\pgfqpoint{0.527360in}{2.465000in}}%
\pgfpathlineto{\pgfqpoint{0.527360in}{0.541249in}}%
\pgfpathclose%
\pgfusepath{fill}%
\end{pgfscope}%
\begin{pgfscope}%
\pgfsetbuttcap%
\pgfsetroundjoin%
\definecolor{currentfill}{rgb}{0.000000,0.000000,0.000000}%
\pgfsetfillcolor{currentfill}%
\pgfsetlinewidth{0.803000pt}%
\definecolor{currentstroke}{rgb}{0.000000,0.000000,0.000000}%
\pgfsetstrokecolor{currentstroke}%
\pgfsetdash{}{0pt}%
\pgfsys@defobject{currentmarker}{\pgfqpoint{0.000000in}{-0.048611in}}{\pgfqpoint{0.000000in}{0.000000in}}{%
\pgfpathmoveto{\pgfqpoint{0.000000in}{0.000000in}}%
\pgfpathlineto{\pgfqpoint{0.000000in}{-0.048611in}}%
\pgfusepath{stroke,fill}%
}%
\begin{pgfscope}%
\pgfsys@transformshift{0.747164in}{0.541249in}%
\pgfsys@useobject{currentmarker}{}%
\end{pgfscope}%
\end{pgfscope}%
\begin{pgfscope}%
\definecolor{textcolor}{rgb}{0.000000,0.000000,0.000000}%
\pgfsetstrokecolor{textcolor}%
\pgfsetfillcolor{textcolor}%
\pgftext[x=0.778414in, y=0.379791in, left, base,rotate=90.000000]{\color{textcolor}\rmfamily\fontsize{9.000000}{10.800000}\selectfont 1}%
\end{pgfscope}%
\begin{pgfscope}%
\pgfsetbuttcap%
\pgfsetroundjoin%
\definecolor{currentfill}{rgb}{0.000000,0.000000,0.000000}%
\pgfsetfillcolor{currentfill}%
\pgfsetlinewidth{0.803000pt}%
\definecolor{currentstroke}{rgb}{0.000000,0.000000,0.000000}%
\pgfsetstrokecolor{currentstroke}%
\pgfsetdash{}{0pt}%
\pgfsys@defobject{currentmarker}{\pgfqpoint{0.000000in}{-0.048611in}}{\pgfqpoint{0.000000in}{0.000000in}}{%
\pgfpathmoveto{\pgfqpoint{0.000000in}{0.000000in}}%
\pgfpathlineto{\pgfqpoint{0.000000in}{-0.048611in}}%
\pgfusepath{stroke,fill}%
}%
\begin{pgfscope}%
\pgfsys@transformshift{1.186770in}{0.541249in}%
\pgfsys@useobject{currentmarker}{}%
\end{pgfscope}%
\end{pgfscope}%
\begin{pgfscope}%
\definecolor{textcolor}{rgb}{0.000000,0.000000,0.000000}%
\pgfsetstrokecolor{textcolor}%
\pgfsetfillcolor{textcolor}%
\pgftext[x=1.218020in, y=0.379791in, left, base,rotate=90.000000]{\color{textcolor}\rmfamily\fontsize{9.000000}{10.800000}\selectfont 2}%
\end{pgfscope}%
\begin{pgfscope}%
\pgfsetbuttcap%
\pgfsetroundjoin%
\definecolor{currentfill}{rgb}{0.000000,0.000000,0.000000}%
\pgfsetfillcolor{currentfill}%
\pgfsetlinewidth{0.803000pt}%
\definecolor{currentstroke}{rgb}{0.000000,0.000000,0.000000}%
\pgfsetstrokecolor{currentstroke}%
\pgfsetdash{}{0pt}%
\pgfsys@defobject{currentmarker}{\pgfqpoint{0.000000in}{-0.048611in}}{\pgfqpoint{0.000000in}{0.000000in}}{%
\pgfpathmoveto{\pgfqpoint{0.000000in}{0.000000in}}%
\pgfpathlineto{\pgfqpoint{0.000000in}{-0.048611in}}%
\pgfusepath{stroke,fill}%
}%
\begin{pgfscope}%
\pgfsys@transformshift{1.626377in}{0.541249in}%
\pgfsys@useobject{currentmarker}{}%
\end{pgfscope}%
\end{pgfscope}%
\begin{pgfscope}%
\definecolor{textcolor}{rgb}{0.000000,0.000000,0.000000}%
\pgfsetstrokecolor{textcolor}%
\pgfsetfillcolor{textcolor}%
\pgftext[x=1.657627in, y=0.379791in, left, base,rotate=90.000000]{\color{textcolor}\rmfamily\fontsize{9.000000}{10.800000}\selectfont 4}%
\end{pgfscope}%
\begin{pgfscope}%
\pgfsetbuttcap%
\pgfsetroundjoin%
\definecolor{currentfill}{rgb}{0.000000,0.000000,0.000000}%
\pgfsetfillcolor{currentfill}%
\pgfsetlinewidth{0.803000pt}%
\definecolor{currentstroke}{rgb}{0.000000,0.000000,0.000000}%
\pgfsetstrokecolor{currentstroke}%
\pgfsetdash{}{0pt}%
\pgfsys@defobject{currentmarker}{\pgfqpoint{0.000000in}{-0.048611in}}{\pgfqpoint{0.000000in}{0.000000in}}{%
\pgfpathmoveto{\pgfqpoint{0.000000in}{0.000000in}}%
\pgfpathlineto{\pgfqpoint{0.000000in}{-0.048611in}}%
\pgfusepath{stroke,fill}%
}%
\begin{pgfscope}%
\pgfsys@transformshift{2.065983in}{0.541249in}%
\pgfsys@useobject{currentmarker}{}%
\end{pgfscope}%
\end{pgfscope}%
\begin{pgfscope}%
\definecolor{textcolor}{rgb}{0.000000,0.000000,0.000000}%
\pgfsetstrokecolor{textcolor}%
\pgfsetfillcolor{textcolor}%
\pgftext[x=2.097233in, y=0.379791in, left, base,rotate=90.000000]{\color{textcolor}\rmfamily\fontsize{9.000000}{10.800000}\selectfont 8}%
\end{pgfscope}%
\begin{pgfscope}%
\pgfsetbuttcap%
\pgfsetroundjoin%
\definecolor{currentfill}{rgb}{0.000000,0.000000,0.000000}%
\pgfsetfillcolor{currentfill}%
\pgfsetlinewidth{0.803000pt}%
\definecolor{currentstroke}{rgb}{0.000000,0.000000,0.000000}%
\pgfsetstrokecolor{currentstroke}%
\pgfsetdash{}{0pt}%
\pgfsys@defobject{currentmarker}{\pgfqpoint{0.000000in}{-0.048611in}}{\pgfqpoint{0.000000in}{0.000000in}}{%
\pgfpathmoveto{\pgfqpoint{0.000000in}{0.000000in}}%
\pgfpathlineto{\pgfqpoint{0.000000in}{-0.048611in}}%
\pgfusepath{stroke,fill}%
}%
\begin{pgfscope}%
\pgfsys@transformshift{2.505590in}{0.541249in}%
\pgfsys@useobject{currentmarker}{}%
\end{pgfscope}%
\end{pgfscope}%
\begin{pgfscope}%
\definecolor{textcolor}{rgb}{0.000000,0.000000,0.000000}%
\pgfsetstrokecolor{textcolor}%
\pgfsetfillcolor{textcolor}%
\pgftext[x=2.536840in, y=0.315556in, left, base,rotate=90.000000]{\color{textcolor}\rmfamily\fontsize{9.000000}{10.800000}\selectfont 16}%
\end{pgfscope}%
\begin{pgfscope}%
\pgfsetbuttcap%
\pgfsetroundjoin%
\definecolor{currentfill}{rgb}{0.000000,0.000000,0.000000}%
\pgfsetfillcolor{currentfill}%
\pgfsetlinewidth{0.803000pt}%
\definecolor{currentstroke}{rgb}{0.000000,0.000000,0.000000}%
\pgfsetstrokecolor{currentstroke}%
\pgfsetdash{}{0pt}%
\pgfsys@defobject{currentmarker}{\pgfqpoint{0.000000in}{-0.048611in}}{\pgfqpoint{0.000000in}{0.000000in}}{%
\pgfpathmoveto{\pgfqpoint{0.000000in}{0.000000in}}%
\pgfpathlineto{\pgfqpoint{0.000000in}{-0.048611in}}%
\pgfusepath{stroke,fill}%
}%
\begin{pgfscope}%
\pgfsys@transformshift{2.945197in}{0.541249in}%
\pgfsys@useobject{currentmarker}{}%
\end{pgfscope}%
\end{pgfscope}%
\begin{pgfscope}%
\definecolor{textcolor}{rgb}{0.000000,0.000000,0.000000}%
\pgfsetstrokecolor{textcolor}%
\pgfsetfillcolor{textcolor}%
\pgftext[x=2.976447in, y=0.315556in, left, base,rotate=90.000000]{\color{textcolor}\rmfamily\fontsize{9.000000}{10.800000}\selectfont 32}%
\end{pgfscope}%
\begin{pgfscope}%
\definecolor{textcolor}{rgb}{0.000000,0.000000,0.000000}%
\pgfsetstrokecolor{textcolor}%
\pgfsetfillcolor{textcolor}%
\pgftext[x=1.846180in,y=0.260000in,,top]{\color{textcolor}\rmfamily\fontsize{9.000000}{10.800000}\selectfont \(\displaystyle \alpha(/255)\)}%
\end{pgfscope}%
\begin{pgfscope}%
\pgfpathrectangle{\pgfqpoint{0.527360in}{0.541249in}}{\pgfqpoint{2.637640in}{1.923751in}}%
\pgfusepath{clip}%
\pgfsetrectcap%
\pgfsetroundjoin%
\pgfsetlinewidth{0.803000pt}%
\definecolor{currentstroke}{rgb}{0.690196,0.690196,0.690196}%
\pgfsetstrokecolor{currentstroke}%
\pgfsetdash{}{0pt}%
\pgfpathmoveto{\pgfqpoint{0.527360in}{0.541249in}}%
\pgfpathlineto{\pgfqpoint{3.165000in}{0.541249in}}%
\pgfusepath{stroke}%
\end{pgfscope}%
\begin{pgfscope}%
\pgfsetbuttcap%
\pgfsetroundjoin%
\definecolor{currentfill}{rgb}{0.000000,0.000000,0.000000}%
\pgfsetfillcolor{currentfill}%
\pgfsetlinewidth{0.803000pt}%
\definecolor{currentstroke}{rgb}{0.000000,0.000000,0.000000}%
\pgfsetstrokecolor{currentstroke}%
\pgfsetdash{}{0pt}%
\pgfsys@defobject{currentmarker}{\pgfqpoint{-0.048611in}{0.000000in}}{\pgfqpoint{-0.000000in}{0.000000in}}{%
\pgfpathmoveto{\pgfqpoint{-0.000000in}{0.000000in}}%
\pgfpathlineto{\pgfqpoint{-0.048611in}{0.000000in}}%
\pgfusepath{stroke,fill}%
}%
\begin{pgfscope}%
\pgfsys@transformshift{0.527360in}{0.541249in}%
\pgfsys@useobject{currentmarker}{}%
\end{pgfscope}%
\end{pgfscope}%
\begin{pgfscope}%
\definecolor{textcolor}{rgb}{0.000000,0.000000,0.000000}%
\pgfsetstrokecolor{textcolor}%
\pgfsetfillcolor{textcolor}%
\pgftext[x=0.365902in, y=0.497846in, left, base]{\color{textcolor}\rmfamily\fontsize{9.000000}{10.800000}\selectfont \(\displaystyle {0}\)}%
\end{pgfscope}%
\begin{pgfscope}%
\pgfpathrectangle{\pgfqpoint{0.527360in}{0.541249in}}{\pgfqpoint{2.637640in}{1.923751in}}%
\pgfusepath{clip}%
\pgfsetrectcap%
\pgfsetroundjoin%
\pgfsetlinewidth{0.803000pt}%
\definecolor{currentstroke}{rgb}{0.690196,0.690196,0.690196}%
\pgfsetstrokecolor{currentstroke}%
\pgfsetdash{}{0pt}%
\pgfpathmoveto{\pgfqpoint{0.527360in}{0.861874in}}%
\pgfpathlineto{\pgfqpoint{3.165000in}{0.861874in}}%
\pgfusepath{stroke}%
\end{pgfscope}%
\begin{pgfscope}%
\pgfsetbuttcap%
\pgfsetroundjoin%
\definecolor{currentfill}{rgb}{0.000000,0.000000,0.000000}%
\pgfsetfillcolor{currentfill}%
\pgfsetlinewidth{0.803000pt}%
\definecolor{currentstroke}{rgb}{0.000000,0.000000,0.000000}%
\pgfsetstrokecolor{currentstroke}%
\pgfsetdash{}{0pt}%
\pgfsys@defobject{currentmarker}{\pgfqpoint{-0.048611in}{0.000000in}}{\pgfqpoint{-0.000000in}{0.000000in}}{%
\pgfpathmoveto{\pgfqpoint{-0.000000in}{0.000000in}}%
\pgfpathlineto{\pgfqpoint{-0.048611in}{0.000000in}}%
\pgfusepath{stroke,fill}%
}%
\begin{pgfscope}%
\pgfsys@transformshift{0.527360in}{0.861874in}%
\pgfsys@useobject{currentmarker}{}%
\end{pgfscope}%
\end{pgfscope}%
\begin{pgfscope}%
\definecolor{textcolor}{rgb}{0.000000,0.000000,0.000000}%
\pgfsetstrokecolor{textcolor}%
\pgfsetfillcolor{textcolor}%
\pgftext[x=0.301667in, y=0.818472in, left, base]{\color{textcolor}\rmfamily\fontsize{9.000000}{10.800000}\selectfont \(\displaystyle {10}\)}%
\end{pgfscope}%
\begin{pgfscope}%
\pgfpathrectangle{\pgfqpoint{0.527360in}{0.541249in}}{\pgfqpoint{2.637640in}{1.923751in}}%
\pgfusepath{clip}%
\pgfsetrectcap%
\pgfsetroundjoin%
\pgfsetlinewidth{0.803000pt}%
\definecolor{currentstroke}{rgb}{0.690196,0.690196,0.690196}%
\pgfsetstrokecolor{currentstroke}%
\pgfsetdash{}{0pt}%
\pgfpathmoveto{\pgfqpoint{0.527360in}{1.182499in}}%
\pgfpathlineto{\pgfqpoint{3.165000in}{1.182499in}}%
\pgfusepath{stroke}%
\end{pgfscope}%
\begin{pgfscope}%
\pgfsetbuttcap%
\pgfsetroundjoin%
\definecolor{currentfill}{rgb}{0.000000,0.000000,0.000000}%
\pgfsetfillcolor{currentfill}%
\pgfsetlinewidth{0.803000pt}%
\definecolor{currentstroke}{rgb}{0.000000,0.000000,0.000000}%
\pgfsetstrokecolor{currentstroke}%
\pgfsetdash{}{0pt}%
\pgfsys@defobject{currentmarker}{\pgfqpoint{-0.048611in}{0.000000in}}{\pgfqpoint{-0.000000in}{0.000000in}}{%
\pgfpathmoveto{\pgfqpoint{-0.000000in}{0.000000in}}%
\pgfpathlineto{\pgfqpoint{-0.048611in}{0.000000in}}%
\pgfusepath{stroke,fill}%
}%
\begin{pgfscope}%
\pgfsys@transformshift{0.527360in}{1.182499in}%
\pgfsys@useobject{currentmarker}{}%
\end{pgfscope}%
\end{pgfscope}%
\begin{pgfscope}%
\definecolor{textcolor}{rgb}{0.000000,0.000000,0.000000}%
\pgfsetstrokecolor{textcolor}%
\pgfsetfillcolor{textcolor}%
\pgftext[x=0.301667in, y=1.139097in, left, base]{\color{textcolor}\rmfamily\fontsize{9.000000}{10.800000}\selectfont \(\displaystyle {20}\)}%
\end{pgfscope}%
\begin{pgfscope}%
\pgfpathrectangle{\pgfqpoint{0.527360in}{0.541249in}}{\pgfqpoint{2.637640in}{1.923751in}}%
\pgfusepath{clip}%
\pgfsetrectcap%
\pgfsetroundjoin%
\pgfsetlinewidth{0.803000pt}%
\definecolor{currentstroke}{rgb}{0.690196,0.690196,0.690196}%
\pgfsetstrokecolor{currentstroke}%
\pgfsetdash{}{0pt}%
\pgfpathmoveto{\pgfqpoint{0.527360in}{1.503125in}}%
\pgfpathlineto{\pgfqpoint{3.165000in}{1.503125in}}%
\pgfusepath{stroke}%
\end{pgfscope}%
\begin{pgfscope}%
\pgfsetbuttcap%
\pgfsetroundjoin%
\definecolor{currentfill}{rgb}{0.000000,0.000000,0.000000}%
\pgfsetfillcolor{currentfill}%
\pgfsetlinewidth{0.803000pt}%
\definecolor{currentstroke}{rgb}{0.000000,0.000000,0.000000}%
\pgfsetstrokecolor{currentstroke}%
\pgfsetdash{}{0pt}%
\pgfsys@defobject{currentmarker}{\pgfqpoint{-0.048611in}{0.000000in}}{\pgfqpoint{-0.000000in}{0.000000in}}{%
\pgfpathmoveto{\pgfqpoint{-0.000000in}{0.000000in}}%
\pgfpathlineto{\pgfqpoint{-0.048611in}{0.000000in}}%
\pgfusepath{stroke,fill}%
}%
\begin{pgfscope}%
\pgfsys@transformshift{0.527360in}{1.503125in}%
\pgfsys@useobject{currentmarker}{}%
\end{pgfscope}%
\end{pgfscope}%
\begin{pgfscope}%
\definecolor{textcolor}{rgb}{0.000000,0.000000,0.000000}%
\pgfsetstrokecolor{textcolor}%
\pgfsetfillcolor{textcolor}%
\pgftext[x=0.301667in, y=1.459722in, left, base]{\color{textcolor}\rmfamily\fontsize{9.000000}{10.800000}\selectfont \(\displaystyle {30}\)}%
\end{pgfscope}%
\begin{pgfscope}%
\pgfpathrectangle{\pgfqpoint{0.527360in}{0.541249in}}{\pgfqpoint{2.637640in}{1.923751in}}%
\pgfusepath{clip}%
\pgfsetrectcap%
\pgfsetroundjoin%
\pgfsetlinewidth{0.803000pt}%
\definecolor{currentstroke}{rgb}{0.690196,0.690196,0.690196}%
\pgfsetstrokecolor{currentstroke}%
\pgfsetdash{}{0pt}%
\pgfpathmoveto{\pgfqpoint{0.527360in}{1.823750in}}%
\pgfpathlineto{\pgfqpoint{3.165000in}{1.823750in}}%
\pgfusepath{stroke}%
\end{pgfscope}%
\begin{pgfscope}%
\pgfsetbuttcap%
\pgfsetroundjoin%
\definecolor{currentfill}{rgb}{0.000000,0.000000,0.000000}%
\pgfsetfillcolor{currentfill}%
\pgfsetlinewidth{0.803000pt}%
\definecolor{currentstroke}{rgb}{0.000000,0.000000,0.000000}%
\pgfsetstrokecolor{currentstroke}%
\pgfsetdash{}{0pt}%
\pgfsys@defobject{currentmarker}{\pgfqpoint{-0.048611in}{0.000000in}}{\pgfqpoint{-0.000000in}{0.000000in}}{%
\pgfpathmoveto{\pgfqpoint{-0.000000in}{0.000000in}}%
\pgfpathlineto{\pgfqpoint{-0.048611in}{0.000000in}}%
\pgfusepath{stroke,fill}%
}%
\begin{pgfscope}%
\pgfsys@transformshift{0.527360in}{1.823750in}%
\pgfsys@useobject{currentmarker}{}%
\end{pgfscope}%
\end{pgfscope}%
\begin{pgfscope}%
\definecolor{textcolor}{rgb}{0.000000,0.000000,0.000000}%
\pgfsetstrokecolor{textcolor}%
\pgfsetfillcolor{textcolor}%
\pgftext[x=0.301667in, y=1.780347in, left, base]{\color{textcolor}\rmfamily\fontsize{9.000000}{10.800000}\selectfont \(\displaystyle {40}\)}%
\end{pgfscope}%
\begin{pgfscope}%
\pgfpathrectangle{\pgfqpoint{0.527360in}{0.541249in}}{\pgfqpoint{2.637640in}{1.923751in}}%
\pgfusepath{clip}%
\pgfsetrectcap%
\pgfsetroundjoin%
\pgfsetlinewidth{0.803000pt}%
\definecolor{currentstroke}{rgb}{0.690196,0.690196,0.690196}%
\pgfsetstrokecolor{currentstroke}%
\pgfsetdash{}{0pt}%
\pgfpathmoveto{\pgfqpoint{0.527360in}{2.144375in}}%
\pgfpathlineto{\pgfqpoint{3.165000in}{2.144375in}}%
\pgfusepath{stroke}%
\end{pgfscope}%
\begin{pgfscope}%
\pgfsetbuttcap%
\pgfsetroundjoin%
\definecolor{currentfill}{rgb}{0.000000,0.000000,0.000000}%
\pgfsetfillcolor{currentfill}%
\pgfsetlinewidth{0.803000pt}%
\definecolor{currentstroke}{rgb}{0.000000,0.000000,0.000000}%
\pgfsetstrokecolor{currentstroke}%
\pgfsetdash{}{0pt}%
\pgfsys@defobject{currentmarker}{\pgfqpoint{-0.048611in}{0.000000in}}{\pgfqpoint{-0.000000in}{0.000000in}}{%
\pgfpathmoveto{\pgfqpoint{-0.000000in}{0.000000in}}%
\pgfpathlineto{\pgfqpoint{-0.048611in}{0.000000in}}%
\pgfusepath{stroke,fill}%
}%
\begin{pgfscope}%
\pgfsys@transformshift{0.527360in}{2.144375in}%
\pgfsys@useobject{currentmarker}{}%
\end{pgfscope}%
\end{pgfscope}%
\begin{pgfscope}%
\definecolor{textcolor}{rgb}{0.000000,0.000000,0.000000}%
\pgfsetstrokecolor{textcolor}%
\pgfsetfillcolor{textcolor}%
\pgftext[x=0.301667in, y=2.100972in, left, base]{\color{textcolor}\rmfamily\fontsize{9.000000}{10.800000}\selectfont \(\displaystyle {50}\)}%
\end{pgfscope}%
\begin{pgfscope}%
\pgfpathrectangle{\pgfqpoint{0.527360in}{0.541249in}}{\pgfqpoint{2.637640in}{1.923751in}}%
\pgfusepath{clip}%
\pgfsetrectcap%
\pgfsetroundjoin%
\pgfsetlinewidth{0.803000pt}%
\definecolor{currentstroke}{rgb}{0.690196,0.690196,0.690196}%
\pgfsetstrokecolor{currentstroke}%
\pgfsetdash{}{0pt}%
\pgfpathmoveto{\pgfqpoint{0.527360in}{2.465000in}}%
\pgfpathlineto{\pgfqpoint{3.165000in}{2.465000in}}%
\pgfusepath{stroke}%
\end{pgfscope}%
\begin{pgfscope}%
\pgfsetbuttcap%
\pgfsetroundjoin%
\definecolor{currentfill}{rgb}{0.000000,0.000000,0.000000}%
\pgfsetfillcolor{currentfill}%
\pgfsetlinewidth{0.803000pt}%
\definecolor{currentstroke}{rgb}{0.000000,0.000000,0.000000}%
\pgfsetstrokecolor{currentstroke}%
\pgfsetdash{}{0pt}%
\pgfsys@defobject{currentmarker}{\pgfqpoint{-0.048611in}{0.000000in}}{\pgfqpoint{-0.000000in}{0.000000in}}{%
\pgfpathmoveto{\pgfqpoint{-0.000000in}{0.000000in}}%
\pgfpathlineto{\pgfqpoint{-0.048611in}{0.000000in}}%
\pgfusepath{stroke,fill}%
}%
\begin{pgfscope}%
\pgfsys@transformshift{0.527360in}{2.465000in}%
\pgfsys@useobject{currentmarker}{}%
\end{pgfscope}%
\end{pgfscope}%
\begin{pgfscope}%
\definecolor{textcolor}{rgb}{0.000000,0.000000,0.000000}%
\pgfsetstrokecolor{textcolor}%
\pgfsetfillcolor{textcolor}%
\pgftext[x=0.301667in, y=2.421597in, left, base]{\color{textcolor}\rmfamily\fontsize{9.000000}{10.800000}\selectfont \(\displaystyle {60}\)}%
\end{pgfscope}%
\begin{pgfscope}%
\definecolor{textcolor}{rgb}{0.000000,0.000000,0.000000}%
\pgfsetstrokecolor{textcolor}%
\pgfsetfillcolor{textcolor}%
\pgftext[x=0.246111in,y=1.503125in,,bottom,rotate=90.000000]{\color{textcolor}\rmfamily\fontsize{9.000000}{10.800000}\selectfont PSNR}%
\end{pgfscope}%
\begin{pgfscope}%
\pgfpathrectangle{\pgfqpoint{0.527360in}{0.541249in}}{\pgfqpoint{2.637640in}{1.923751in}}%
\pgfusepath{clip}%
\pgfsetbuttcap%
\pgfsetmiterjoin%
\definecolor{currentfill}{rgb}{0.121569,0.466667,0.705882}%
\pgfsetfillcolor{currentfill}%
\pgfsetlinewidth{0.000000pt}%
\definecolor{currentstroke}{rgb}{0.000000,0.000000,0.000000}%
\pgfsetstrokecolor{currentstroke}%
\pgfsetstrokeopacity{0.000000}%
\pgfsetdash{}{0pt}%
\pgfpathmoveto{\pgfqpoint{0.637262in}{0.541249in}}%
\pgfpathlineto{\pgfqpoint{0.710530in}{0.541249in}}%
\pgfpathlineto{\pgfqpoint{0.710530in}{1.018981in}}%
\pgfpathlineto{\pgfqpoint{0.637262in}{1.018981in}}%
\pgfpathlineto{\pgfqpoint{0.637262in}{0.541249in}}%
\pgfpathclose%
\pgfusepath{fill}%
\end{pgfscope}%
\begin{pgfscope}%
\pgfpathrectangle{\pgfqpoint{0.527360in}{0.541249in}}{\pgfqpoint{2.637640in}{1.923751in}}%
\pgfusepath{clip}%
\pgfsetbuttcap%
\pgfsetmiterjoin%
\definecolor{currentfill}{rgb}{0.121569,0.466667,0.705882}%
\pgfsetfillcolor{currentfill}%
\pgfsetlinewidth{0.000000pt}%
\definecolor{currentstroke}{rgb}{0.000000,0.000000,0.000000}%
\pgfsetstrokecolor{currentstroke}%
\pgfsetstrokeopacity{0.000000}%
\pgfsetdash{}{0pt}%
\pgfpathmoveto{\pgfqpoint{1.076869in}{0.541249in}}%
\pgfpathlineto{\pgfqpoint{1.150136in}{0.541249in}}%
\pgfpathlineto{\pgfqpoint{1.150136in}{0.925999in}}%
\pgfpathlineto{\pgfqpoint{1.076869in}{0.925999in}}%
\pgfpathlineto{\pgfqpoint{1.076869in}{0.541249in}}%
\pgfpathclose%
\pgfusepath{fill}%
\end{pgfscope}%
\begin{pgfscope}%
\pgfpathrectangle{\pgfqpoint{0.527360in}{0.541249in}}{\pgfqpoint{2.637640in}{1.923751in}}%
\pgfusepath{clip}%
\pgfsetbuttcap%
\pgfsetmiterjoin%
\definecolor{currentfill}{rgb}{0.121569,0.466667,0.705882}%
\pgfsetfillcolor{currentfill}%
\pgfsetlinewidth{0.000000pt}%
\definecolor{currentstroke}{rgb}{0.000000,0.000000,0.000000}%
\pgfsetstrokecolor{currentstroke}%
\pgfsetstrokeopacity{0.000000}%
\pgfsetdash{}{0pt}%
\pgfpathmoveto{\pgfqpoint{1.516475in}{0.541249in}}%
\pgfpathlineto{\pgfqpoint{1.589743in}{0.541249in}}%
\pgfpathlineto{\pgfqpoint{1.589743in}{0.868287in}}%
\pgfpathlineto{\pgfqpoint{1.516475in}{0.868287in}}%
\pgfpathlineto{\pgfqpoint{1.516475in}{0.541249in}}%
\pgfpathclose%
\pgfusepath{fill}%
\end{pgfscope}%
\begin{pgfscope}%
\pgfpathrectangle{\pgfqpoint{0.527360in}{0.541249in}}{\pgfqpoint{2.637640in}{1.923751in}}%
\pgfusepath{clip}%
\pgfsetbuttcap%
\pgfsetmiterjoin%
\definecolor{currentfill}{rgb}{0.121569,0.466667,0.705882}%
\pgfsetfillcolor{currentfill}%
\pgfsetlinewidth{0.000000pt}%
\definecolor{currentstroke}{rgb}{0.000000,0.000000,0.000000}%
\pgfsetstrokecolor{currentstroke}%
\pgfsetstrokeopacity{0.000000}%
\pgfsetdash{}{0pt}%
\pgfpathmoveto{\pgfqpoint{1.956082in}{0.541249in}}%
\pgfpathlineto{\pgfqpoint{2.029350in}{0.541249in}}%
\pgfpathlineto{\pgfqpoint{2.029350in}{0.855462in}}%
\pgfpathlineto{\pgfqpoint{1.956082in}{0.855462in}}%
\pgfpathlineto{\pgfqpoint{1.956082in}{0.541249in}}%
\pgfpathclose%
\pgfusepath{fill}%
\end{pgfscope}%
\begin{pgfscope}%
\pgfpathrectangle{\pgfqpoint{0.527360in}{0.541249in}}{\pgfqpoint{2.637640in}{1.923751in}}%
\pgfusepath{clip}%
\pgfsetbuttcap%
\pgfsetmiterjoin%
\definecolor{currentfill}{rgb}{0.121569,0.466667,0.705882}%
\pgfsetfillcolor{currentfill}%
\pgfsetlinewidth{0.000000pt}%
\definecolor{currentstroke}{rgb}{0.000000,0.000000,0.000000}%
\pgfsetstrokecolor{currentstroke}%
\pgfsetstrokeopacity{0.000000}%
\pgfsetdash{}{0pt}%
\pgfpathmoveto{\pgfqpoint{2.395688in}{0.541249in}}%
\pgfpathlineto{\pgfqpoint{2.468956in}{0.541249in}}%
\pgfpathlineto{\pgfqpoint{2.468956in}{0.829812in}}%
\pgfpathlineto{\pgfqpoint{2.395688in}{0.829812in}}%
\pgfpathlineto{\pgfqpoint{2.395688in}{0.541249in}}%
\pgfpathclose%
\pgfusepath{fill}%
\end{pgfscope}%
\begin{pgfscope}%
\pgfpathrectangle{\pgfqpoint{0.527360in}{0.541249in}}{\pgfqpoint{2.637640in}{1.923751in}}%
\pgfusepath{clip}%
\pgfsetbuttcap%
\pgfsetmiterjoin%
\definecolor{currentfill}{rgb}{0.121569,0.466667,0.705882}%
\pgfsetfillcolor{currentfill}%
\pgfsetlinewidth{0.000000pt}%
\definecolor{currentstroke}{rgb}{0.000000,0.000000,0.000000}%
\pgfsetstrokecolor{currentstroke}%
\pgfsetstrokeopacity{0.000000}%
\pgfsetdash{}{0pt}%
\pgfpathmoveto{\pgfqpoint{2.835295in}{0.541249in}}%
\pgfpathlineto{\pgfqpoint{2.908563in}{0.541249in}}%
\pgfpathlineto{\pgfqpoint{2.908563in}{0.807368in}}%
\pgfpathlineto{\pgfqpoint{2.835295in}{0.807368in}}%
\pgfpathlineto{\pgfqpoint{2.835295in}{0.541249in}}%
\pgfpathclose%
\pgfusepath{fill}%
\end{pgfscope}%
\begin{pgfscope}%
\pgfpathrectangle{\pgfqpoint{0.527360in}{0.541249in}}{\pgfqpoint{2.637640in}{1.923751in}}%
\pgfusepath{clip}%
\pgfsetbuttcap%
\pgfsetmiterjoin%
\definecolor{currentfill}{rgb}{1.000000,0.498039,0.054902}%
\pgfsetfillcolor{currentfill}%
\pgfsetlinewidth{0.000000pt}%
\definecolor{currentstroke}{rgb}{0.000000,0.000000,0.000000}%
\pgfsetstrokecolor{currentstroke}%
\pgfsetstrokeopacity{0.000000}%
\pgfsetdash{}{0pt}%
\pgfpathmoveto{\pgfqpoint{0.710530in}{0.541249in}}%
\pgfpathlineto{\pgfqpoint{0.783797in}{0.541249in}}%
\pgfpathlineto{\pgfqpoint{0.783797in}{1.422968in}}%
\pgfpathlineto{\pgfqpoint{0.710530in}{1.422968in}}%
\pgfpathlineto{\pgfqpoint{0.710530in}{0.541249in}}%
\pgfpathclose%
\pgfusepath{fill}%
\end{pgfscope}%
\begin{pgfscope}%
\pgfpathrectangle{\pgfqpoint{0.527360in}{0.541249in}}{\pgfqpoint{2.637640in}{1.923751in}}%
\pgfusepath{clip}%
\pgfsetbuttcap%
\pgfsetmiterjoin%
\definecolor{currentfill}{rgb}{1.000000,0.498039,0.054902}%
\pgfsetfillcolor{currentfill}%
\pgfsetlinewidth{0.000000pt}%
\definecolor{currentstroke}{rgb}{0.000000,0.000000,0.000000}%
\pgfsetstrokecolor{currentstroke}%
\pgfsetstrokeopacity{0.000000}%
\pgfsetdash{}{0pt}%
\pgfpathmoveto{\pgfqpoint{1.150136in}{0.541249in}}%
\pgfpathlineto{\pgfqpoint{1.223404in}{0.541249in}}%
\pgfpathlineto{\pgfqpoint{1.223404in}{1.230593in}}%
\pgfpathlineto{\pgfqpoint{1.150136in}{1.230593in}}%
\pgfpathlineto{\pgfqpoint{1.150136in}{0.541249in}}%
\pgfpathclose%
\pgfusepath{fill}%
\end{pgfscope}%
\begin{pgfscope}%
\pgfpathrectangle{\pgfqpoint{0.527360in}{0.541249in}}{\pgfqpoint{2.637640in}{1.923751in}}%
\pgfusepath{clip}%
\pgfsetbuttcap%
\pgfsetmiterjoin%
\definecolor{currentfill}{rgb}{1.000000,0.498039,0.054902}%
\pgfsetfillcolor{currentfill}%
\pgfsetlinewidth{0.000000pt}%
\definecolor{currentstroke}{rgb}{0.000000,0.000000,0.000000}%
\pgfsetstrokecolor{currentstroke}%
\pgfsetstrokeopacity{0.000000}%
\pgfsetdash{}{0pt}%
\pgfpathmoveto{\pgfqpoint{1.589743in}{0.541249in}}%
\pgfpathlineto{\pgfqpoint{1.663011in}{0.541249in}}%
\pgfpathlineto{\pgfqpoint{1.663011in}{1.118374in}}%
\pgfpathlineto{\pgfqpoint{1.589743in}{1.118374in}}%
\pgfpathlineto{\pgfqpoint{1.589743in}{0.541249in}}%
\pgfpathclose%
\pgfusepath{fill}%
\end{pgfscope}%
\begin{pgfscope}%
\pgfpathrectangle{\pgfqpoint{0.527360in}{0.541249in}}{\pgfqpoint{2.637640in}{1.923751in}}%
\pgfusepath{clip}%
\pgfsetbuttcap%
\pgfsetmiterjoin%
\definecolor{currentfill}{rgb}{1.000000,0.498039,0.054902}%
\pgfsetfillcolor{currentfill}%
\pgfsetlinewidth{0.000000pt}%
\definecolor{currentstroke}{rgb}{0.000000,0.000000,0.000000}%
\pgfsetstrokecolor{currentstroke}%
\pgfsetstrokeopacity{0.000000}%
\pgfsetdash{}{0pt}%
\pgfpathmoveto{\pgfqpoint{2.029350in}{0.541249in}}%
\pgfpathlineto{\pgfqpoint{2.102617in}{0.541249in}}%
\pgfpathlineto{\pgfqpoint{2.102617in}{1.057456in}}%
\pgfpathlineto{\pgfqpoint{2.029350in}{1.057456in}}%
\pgfpathlineto{\pgfqpoint{2.029350in}{0.541249in}}%
\pgfpathclose%
\pgfusepath{fill}%
\end{pgfscope}%
\begin{pgfscope}%
\pgfpathrectangle{\pgfqpoint{0.527360in}{0.541249in}}{\pgfqpoint{2.637640in}{1.923751in}}%
\pgfusepath{clip}%
\pgfsetbuttcap%
\pgfsetmiterjoin%
\definecolor{currentfill}{rgb}{1.000000,0.498039,0.054902}%
\pgfsetfillcolor{currentfill}%
\pgfsetlinewidth{0.000000pt}%
\definecolor{currentstroke}{rgb}{0.000000,0.000000,0.000000}%
\pgfsetstrokecolor{currentstroke}%
\pgfsetstrokeopacity{0.000000}%
\pgfsetdash{}{0pt}%
\pgfpathmoveto{\pgfqpoint{2.468956in}{0.541249in}}%
\pgfpathlineto{\pgfqpoint{2.542224in}{0.541249in}}%
\pgfpathlineto{\pgfqpoint{2.542224in}{1.015774in}}%
\pgfpathlineto{\pgfqpoint{2.468956in}{1.015774in}}%
\pgfpathlineto{\pgfqpoint{2.468956in}{0.541249in}}%
\pgfpathclose%
\pgfusepath{fill}%
\end{pgfscope}%
\begin{pgfscope}%
\pgfpathrectangle{\pgfqpoint{0.527360in}{0.541249in}}{\pgfqpoint{2.637640in}{1.923751in}}%
\pgfusepath{clip}%
\pgfsetbuttcap%
\pgfsetmiterjoin%
\definecolor{currentfill}{rgb}{1.000000,0.498039,0.054902}%
\pgfsetfillcolor{currentfill}%
\pgfsetlinewidth{0.000000pt}%
\definecolor{currentstroke}{rgb}{0.000000,0.000000,0.000000}%
\pgfsetstrokecolor{currentstroke}%
\pgfsetstrokeopacity{0.000000}%
\pgfsetdash{}{0pt}%
\pgfpathmoveto{\pgfqpoint{2.908563in}{0.541249in}}%
\pgfpathlineto{\pgfqpoint{2.981831in}{0.541249in}}%
\pgfpathlineto{\pgfqpoint{2.981831in}{0.958062in}}%
\pgfpathlineto{\pgfqpoint{2.908563in}{0.958062in}}%
\pgfpathlineto{\pgfqpoint{2.908563in}{0.541249in}}%
\pgfpathclose%
\pgfusepath{fill}%
\end{pgfscope}%
\begin{pgfscope}%
\pgfpathrectangle{\pgfqpoint{0.527360in}{0.541249in}}{\pgfqpoint{2.637640in}{1.923751in}}%
\pgfusepath{clip}%
\pgfsetbuttcap%
\pgfsetmiterjoin%
\definecolor{currentfill}{rgb}{0.172549,0.627451,0.172549}%
\pgfsetfillcolor{currentfill}%
\pgfsetlinewidth{0.000000pt}%
\definecolor{currentstroke}{rgb}{0.000000,0.000000,0.000000}%
\pgfsetstrokecolor{currentstroke}%
\pgfsetstrokeopacity{0.000000}%
\pgfsetdash{}{0pt}%
\pgfpathmoveto{\pgfqpoint{0.783797in}{0.541249in}}%
\pgfpathlineto{\pgfqpoint{0.857065in}{0.541249in}}%
\pgfpathlineto{\pgfqpoint{0.857065in}{1.287985in}}%
\pgfpathlineto{\pgfqpoint{0.783797in}{1.287985in}}%
\pgfpathlineto{\pgfqpoint{0.783797in}{0.541249in}}%
\pgfpathclose%
\pgfusepath{fill}%
\end{pgfscope}%
\begin{pgfscope}%
\pgfpathrectangle{\pgfqpoint{0.527360in}{0.541249in}}{\pgfqpoint{2.637640in}{1.923751in}}%
\pgfusepath{clip}%
\pgfsetbuttcap%
\pgfsetmiterjoin%
\definecolor{currentfill}{rgb}{0.172549,0.627451,0.172549}%
\pgfsetfillcolor{currentfill}%
\pgfsetlinewidth{0.000000pt}%
\definecolor{currentstroke}{rgb}{0.000000,0.000000,0.000000}%
\pgfsetstrokecolor{currentstroke}%
\pgfsetstrokeopacity{0.000000}%
\pgfsetdash{}{0pt}%
\pgfpathmoveto{\pgfqpoint{1.223404in}{0.541249in}}%
\pgfpathlineto{\pgfqpoint{1.296672in}{0.541249in}}%
\pgfpathlineto{\pgfqpoint{1.296672in}{1.215524in}}%
\pgfpathlineto{\pgfqpoint{1.223404in}{1.215524in}}%
\pgfpathlineto{\pgfqpoint{1.223404in}{0.541249in}}%
\pgfpathclose%
\pgfusepath{fill}%
\end{pgfscope}%
\begin{pgfscope}%
\pgfpathrectangle{\pgfqpoint{0.527360in}{0.541249in}}{\pgfqpoint{2.637640in}{1.923751in}}%
\pgfusepath{clip}%
\pgfsetbuttcap%
\pgfsetmiterjoin%
\definecolor{currentfill}{rgb}{0.172549,0.627451,0.172549}%
\pgfsetfillcolor{currentfill}%
\pgfsetlinewidth{0.000000pt}%
\definecolor{currentstroke}{rgb}{0.000000,0.000000,0.000000}%
\pgfsetstrokecolor{currentstroke}%
\pgfsetstrokeopacity{0.000000}%
\pgfsetdash{}{0pt}%
\pgfpathmoveto{\pgfqpoint{1.663011in}{0.541249in}}%
\pgfpathlineto{\pgfqpoint{1.736278in}{0.541249in}}%
\pgfpathlineto{\pgfqpoint{1.736278in}{1.154605in}}%
\pgfpathlineto{\pgfqpoint{1.663011in}{1.154605in}}%
\pgfpathlineto{\pgfqpoint{1.663011in}{0.541249in}}%
\pgfpathclose%
\pgfusepath{fill}%
\end{pgfscope}%
\begin{pgfscope}%
\pgfpathrectangle{\pgfqpoint{0.527360in}{0.541249in}}{\pgfqpoint{2.637640in}{1.923751in}}%
\pgfusepath{clip}%
\pgfsetbuttcap%
\pgfsetmiterjoin%
\definecolor{currentfill}{rgb}{0.172549,0.627451,0.172549}%
\pgfsetfillcolor{currentfill}%
\pgfsetlinewidth{0.000000pt}%
\definecolor{currentstroke}{rgb}{0.000000,0.000000,0.000000}%
\pgfsetstrokecolor{currentstroke}%
\pgfsetstrokeopacity{0.000000}%
\pgfsetdash{}{0pt}%
\pgfpathmoveto{\pgfqpoint{2.102617in}{0.541249in}}%
\pgfpathlineto{\pgfqpoint{2.175885in}{0.541249in}}%
\pgfpathlineto{\pgfqpoint{2.175885in}{1.095931in}}%
\pgfpathlineto{\pgfqpoint{2.102617in}{1.095931in}}%
\pgfpathlineto{\pgfqpoint{2.102617in}{0.541249in}}%
\pgfpathclose%
\pgfusepath{fill}%
\end{pgfscope}%
\begin{pgfscope}%
\pgfpathrectangle{\pgfqpoint{0.527360in}{0.541249in}}{\pgfqpoint{2.637640in}{1.923751in}}%
\pgfusepath{clip}%
\pgfsetbuttcap%
\pgfsetmiterjoin%
\definecolor{currentfill}{rgb}{0.172549,0.627451,0.172549}%
\pgfsetfillcolor{currentfill}%
\pgfsetlinewidth{0.000000pt}%
\definecolor{currentstroke}{rgb}{0.000000,0.000000,0.000000}%
\pgfsetstrokecolor{currentstroke}%
\pgfsetstrokeopacity{0.000000}%
\pgfsetdash{}{0pt}%
\pgfpathmoveto{\pgfqpoint{2.542224in}{0.541249in}}%
\pgfpathlineto{\pgfqpoint{2.615492in}{0.541249in}}%
\pgfpathlineto{\pgfqpoint{2.615492in}{1.046554in}}%
\pgfpathlineto{\pgfqpoint{2.542224in}{1.046554in}}%
\pgfpathlineto{\pgfqpoint{2.542224in}{0.541249in}}%
\pgfpathclose%
\pgfusepath{fill}%
\end{pgfscope}%
\begin{pgfscope}%
\pgfpathrectangle{\pgfqpoint{0.527360in}{0.541249in}}{\pgfqpoint{2.637640in}{1.923751in}}%
\pgfusepath{clip}%
\pgfsetbuttcap%
\pgfsetmiterjoin%
\definecolor{currentfill}{rgb}{0.172549,0.627451,0.172549}%
\pgfsetfillcolor{currentfill}%
\pgfsetlinewidth{0.000000pt}%
\definecolor{currentstroke}{rgb}{0.000000,0.000000,0.000000}%
\pgfsetstrokecolor{currentstroke}%
\pgfsetstrokeopacity{0.000000}%
\pgfsetdash{}{0pt}%
\pgfpathmoveto{\pgfqpoint{2.981831in}{0.541249in}}%
\pgfpathlineto{\pgfqpoint{3.055098in}{0.541249in}}%
\pgfpathlineto{\pgfqpoint{3.055098in}{0.964474in}}%
\pgfpathlineto{\pgfqpoint{2.981831in}{0.964474in}}%
\pgfpathlineto{\pgfqpoint{2.981831in}{0.541249in}}%
\pgfpathclose%
\pgfusepath{fill}%
\end{pgfscope}%
\begin{pgfscope}%
\pgfsetrectcap%
\pgfsetmiterjoin%
\pgfsetlinewidth{0.803000pt}%
\definecolor{currentstroke}{rgb}{0.000000,0.000000,0.000000}%
\pgfsetstrokecolor{currentstroke}%
\pgfsetdash{}{0pt}%
\pgfpathmoveto{\pgfqpoint{0.527360in}{0.541249in}}%
\pgfpathlineto{\pgfqpoint{0.527360in}{2.465000in}}%
\pgfusepath{stroke}%
\end{pgfscope}%
\begin{pgfscope}%
\pgfsetrectcap%
\pgfsetmiterjoin%
\pgfsetlinewidth{0.803000pt}%
\definecolor{currentstroke}{rgb}{0.000000,0.000000,0.000000}%
\pgfsetstrokecolor{currentstroke}%
\pgfsetdash{}{0pt}%
\pgfpathmoveto{\pgfqpoint{3.165000in}{0.541249in}}%
\pgfpathlineto{\pgfqpoint{3.165000in}{2.465000in}}%
\pgfusepath{stroke}%
\end{pgfscope}%
\begin{pgfscope}%
\pgfsetrectcap%
\pgfsetmiterjoin%
\pgfsetlinewidth{0.803000pt}%
\definecolor{currentstroke}{rgb}{0.000000,0.000000,0.000000}%
\pgfsetstrokecolor{currentstroke}%
\pgfsetdash{}{0pt}%
\pgfpathmoveto{\pgfqpoint{0.527360in}{0.541249in}}%
\pgfpathlineto{\pgfqpoint{3.165000in}{0.541249in}}%
\pgfusepath{stroke}%
\end{pgfscope}%
\begin{pgfscope}%
\pgfsetrectcap%
\pgfsetmiterjoin%
\pgfsetlinewidth{0.803000pt}%
\definecolor{currentstroke}{rgb}{0.000000,0.000000,0.000000}%
\pgfsetstrokecolor{currentstroke}%
\pgfsetdash{}{0pt}%
\pgfpathmoveto{\pgfqpoint{0.527360in}{2.465000in}}%
\pgfpathlineto{\pgfqpoint{3.165000in}{2.465000in}}%
\pgfusepath{stroke}%
\end{pgfscope}%
\begin{pgfscope}%
\pgfsetbuttcap%
\pgfsetmiterjoin%
\definecolor{currentfill}{rgb}{1.000000,1.000000,1.000000}%
\pgfsetfillcolor{currentfill}%
\pgfsetfillopacity{0.800000}%
\pgfsetlinewidth{1.003750pt}%
\definecolor{currentstroke}{rgb}{0.800000,0.800000,0.800000}%
\pgfsetstrokecolor{currentstroke}%
\pgfsetstrokeopacity{0.800000}%
\pgfsetdash{}{0pt}%
\pgfpathmoveto{\pgfqpoint{2.134237in}{1.842083in}}%
\pgfpathlineto{\pgfqpoint{3.077500in}{1.842083in}}%
\pgfpathquadraticcurveto{\pgfqpoint{3.102500in}{1.842083in}}{\pgfqpoint{3.102500in}{1.867083in}}%
\pgfpathlineto{\pgfqpoint{3.102500in}{2.377500in}}%
\pgfpathquadraticcurveto{\pgfqpoint{3.102500in}{2.402500in}}{\pgfqpoint{3.077500in}{2.402500in}}%
\pgfpathlineto{\pgfqpoint{2.134237in}{2.402500in}}%
\pgfpathquadraticcurveto{\pgfqpoint{2.109237in}{2.402500in}}{\pgfqpoint{2.109237in}{2.377500in}}%
\pgfpathlineto{\pgfqpoint{2.109237in}{1.867083in}}%
\pgfpathquadraticcurveto{\pgfqpoint{2.109237in}{1.842083in}}{\pgfqpoint{2.134237in}{1.842083in}}%
\pgfpathlineto{\pgfqpoint{2.134237in}{1.842083in}}%
\pgfpathclose%
\pgfusepath{stroke,fill}%
\end{pgfscope}%
\begin{pgfscope}%
\pgfsetbuttcap%
\pgfsetmiterjoin%
\definecolor{currentfill}{rgb}{0.121569,0.466667,0.705882}%
\pgfsetfillcolor{currentfill}%
\pgfsetlinewidth{0.000000pt}%
\definecolor{currentstroke}{rgb}{0.000000,0.000000,0.000000}%
\pgfsetstrokecolor{currentstroke}%
\pgfsetstrokeopacity{0.000000}%
\pgfsetdash{}{0pt}%
\pgfpathmoveto{\pgfqpoint{2.159237in}{2.265000in}}%
\pgfpathlineto{\pgfqpoint{2.409237in}{2.265000in}}%
\pgfpathlineto{\pgfqpoint{2.409237in}{2.352500in}}%
\pgfpathlineto{\pgfqpoint{2.159237in}{2.352500in}}%
\pgfpathlineto{\pgfqpoint{2.159237in}{2.265000in}}%
\pgfpathclose%
\pgfusepath{fill}%
\end{pgfscope}%
\begin{pgfscope}%
\definecolor{textcolor}{rgb}{0.000000,0.000000,0.000000}%
\pgfsetstrokecolor{textcolor}%
\pgfsetfillcolor{textcolor}%
\pgftext[x=2.509237in,y=2.265000in,left,base]{\color{textcolor}\rmfamily\fontsize{9.000000}{10.800000}\selectfont ESRGAN}%
\end{pgfscope}%
\begin{pgfscope}%
\pgfsetbuttcap%
\pgfsetmiterjoin%
\definecolor{currentfill}{rgb}{1.000000,0.498039,0.054902}%
\pgfsetfillcolor{currentfill}%
\pgfsetlinewidth{0.000000pt}%
\definecolor{currentstroke}{rgb}{0.000000,0.000000,0.000000}%
\pgfsetstrokecolor{currentstroke}%
\pgfsetstrokeopacity{0.000000}%
\pgfsetdash{}{0pt}%
\pgfpathmoveto{\pgfqpoint{2.159237in}{2.090694in}}%
\pgfpathlineto{\pgfqpoint{2.409237in}{2.090694in}}%
\pgfpathlineto{\pgfqpoint{2.409237in}{2.178194in}}%
\pgfpathlineto{\pgfqpoint{2.159237in}{2.178194in}}%
\pgfpathlineto{\pgfqpoint{2.159237in}{2.090694in}}%
\pgfpathclose%
\pgfusepath{fill}%
\end{pgfscope}%
\begin{pgfscope}%
\definecolor{textcolor}{rgb}{0.000000,0.000000,0.000000}%
\pgfsetstrokecolor{textcolor}%
\pgfsetfillcolor{textcolor}%
\pgftext[x=2.509237in,y=2.090694in,left,base]{\color{textcolor}\rmfamily\fontsize{9.000000}{10.800000}\selectfont EDSR}%
\end{pgfscope}%
\begin{pgfscope}%
\pgfsetbuttcap%
\pgfsetmiterjoin%
\definecolor{currentfill}{rgb}{0.172549,0.627451,0.172549}%
\pgfsetfillcolor{currentfill}%
\pgfsetlinewidth{0.000000pt}%
\definecolor{currentstroke}{rgb}{0.000000,0.000000,0.000000}%
\pgfsetstrokecolor{currentstroke}%
\pgfsetstrokeopacity{0.000000}%
\pgfsetdash{}{0pt}%
\pgfpathmoveto{\pgfqpoint{2.159237in}{1.916389in}}%
\pgfpathlineto{\pgfqpoint{2.409237in}{1.916389in}}%
\pgfpathlineto{\pgfqpoint{2.409237in}{2.003889in}}%
\pgfpathlineto{\pgfqpoint{2.159237in}{2.003889in}}%
\pgfpathlineto{\pgfqpoint{2.159237in}{1.916389in}}%
\pgfpathclose%
\pgfusepath{fill}%
\end{pgfscope}%
\begin{pgfscope}%
\definecolor{textcolor}{rgb}{0.000000,0.000000,0.000000}%
\pgfsetstrokecolor{textcolor}%
\pgfsetfillcolor{textcolor}%
\pgftext[x=2.509237in,y=1.916389in,left,base]{\color{textcolor}\rmfamily\fontsize{9.000000}{10.800000}\selectfont SRFlow}%
\end{pgfscope}%
\end{pgfpicture}%
\makeatother%
\endgroup%

%% file: latex/figures/inp_bsd.pgf
\begingroup%
\makeatletter%
\begin{pgfpicture}%
\pgfpathrectangle{\pgfpointorigin}{\pgfqpoint{3.300000in}{2.600000in}}%
\pgfusepath{use as bounding box, clip}%
\begin{pgfscope}%
\pgfsetbuttcap%
\pgfsetmiterjoin%
\definecolor{currentfill}{rgb}{1.000000,1.000000,1.000000}%
\pgfsetfillcolor{currentfill}%
\pgfsetlinewidth{0.000000pt}%
\definecolor{currentstroke}{rgb}{1.000000,1.000000,1.000000}%
\pgfsetstrokecolor{currentstroke}%
\pgfsetdash{}{0pt}%
\pgfpathmoveto{\pgfqpoint{0.000000in}{0.000000in}}%
\pgfpathlineto{\pgfqpoint{3.300000in}{0.000000in}}%
\pgfpathlineto{\pgfqpoint{3.300000in}{2.600000in}}%
\pgfpathlineto{\pgfqpoint{0.000000in}{2.600000in}}%
\pgfpathlineto{\pgfqpoint{0.000000in}{0.000000in}}%
\pgfpathclose%
\pgfusepath{fill}%
\end{pgfscope}%
\begin{pgfscope}%
\pgfsetbuttcap%
\pgfsetmiterjoin%
\definecolor{currentfill}{rgb}{1.000000,1.000000,1.000000}%
\pgfsetfillcolor{currentfill}%
\pgfsetlinewidth{0.000000pt}%
\definecolor{currentstroke}{rgb}{0.000000,0.000000,0.000000}%
\pgfsetstrokecolor{currentstroke}%
\pgfsetstrokeopacity{0.000000}%
\pgfsetdash{}{0pt}%
\pgfpathmoveto{\pgfqpoint{0.527360in}{0.541249in}}%
\pgfpathlineto{\pgfqpoint{3.165000in}{0.541249in}}%
\pgfpathlineto{\pgfqpoint{3.165000in}{2.465000in}}%
\pgfpathlineto{\pgfqpoint{0.527360in}{2.465000in}}%
\pgfpathlineto{\pgfqpoint{0.527360in}{0.541249in}}%
\pgfpathclose%
\pgfusepath{fill}%
\end{pgfscope}%
\begin{pgfscope}%
\pgfsetbuttcap%
\pgfsetroundjoin%
\definecolor{currentfill}{rgb}{0.000000,0.000000,0.000000}%
\pgfsetfillcolor{currentfill}%
\pgfsetlinewidth{0.803000pt}%
\definecolor{currentstroke}{rgb}{0.000000,0.000000,0.000000}%
\pgfsetstrokecolor{currentstroke}%
\pgfsetdash{}{0pt}%
\pgfsys@defobject{currentmarker}{\pgfqpoint{0.000000in}{-0.048611in}}{\pgfqpoint{0.000000in}{0.000000in}}{%
\pgfpathmoveto{\pgfqpoint{0.000000in}{0.000000in}}%
\pgfpathlineto{\pgfqpoint{0.000000in}{-0.048611in}}%
\pgfusepath{stroke,fill}%
}%
\begin{pgfscope}%
\pgfsys@transformshift{0.747164in}{0.541249in}%
\pgfsys@useobject{currentmarker}{}%
\end{pgfscope}%
\end{pgfscope}%
\begin{pgfscope}%
\definecolor{textcolor}{rgb}{0.000000,0.000000,0.000000}%
\pgfsetstrokecolor{textcolor}%
\pgfsetfillcolor{textcolor}%
\pgftext[x=0.778414in, y=0.379791in, left, base,rotate=90.000000]{\color{textcolor}\rmfamily\fontsize{9.000000}{10.800000}\selectfont 1}%
\end{pgfscope}%
\begin{pgfscope}%
\pgfsetbuttcap%
\pgfsetroundjoin%
\definecolor{currentfill}{rgb}{0.000000,0.000000,0.000000}%
\pgfsetfillcolor{currentfill}%
\pgfsetlinewidth{0.803000pt}%
\definecolor{currentstroke}{rgb}{0.000000,0.000000,0.000000}%
\pgfsetstrokecolor{currentstroke}%
\pgfsetdash{}{0pt}%
\pgfsys@defobject{currentmarker}{\pgfqpoint{0.000000in}{-0.048611in}}{\pgfqpoint{0.000000in}{0.000000in}}{%
\pgfpathmoveto{\pgfqpoint{0.000000in}{0.000000in}}%
\pgfpathlineto{\pgfqpoint{0.000000in}{-0.048611in}}%
\pgfusepath{stroke,fill}%
}%
\begin{pgfscope}%
\pgfsys@transformshift{1.186770in}{0.541249in}%
\pgfsys@useobject{currentmarker}{}%
\end{pgfscope}%
\end{pgfscope}%
\begin{pgfscope}%
\definecolor{textcolor}{rgb}{0.000000,0.000000,0.000000}%
\pgfsetstrokecolor{textcolor}%
\pgfsetfillcolor{textcolor}%
\pgftext[x=1.218020in, y=0.379791in, left, base,rotate=90.000000]{\color{textcolor}\rmfamily\fontsize{9.000000}{10.800000}\selectfont 2}%
\end{pgfscope}%
\begin{pgfscope}%
\pgfsetbuttcap%
\pgfsetroundjoin%
\definecolor{currentfill}{rgb}{0.000000,0.000000,0.000000}%
\pgfsetfillcolor{currentfill}%
\pgfsetlinewidth{0.803000pt}%
\definecolor{currentstroke}{rgb}{0.000000,0.000000,0.000000}%
\pgfsetstrokecolor{currentstroke}%
\pgfsetdash{}{0pt}%
\pgfsys@defobject{currentmarker}{\pgfqpoint{0.000000in}{-0.048611in}}{\pgfqpoint{0.000000in}{0.000000in}}{%
\pgfpathmoveto{\pgfqpoint{0.000000in}{0.000000in}}%
\pgfpathlineto{\pgfqpoint{0.000000in}{-0.048611in}}%
\pgfusepath{stroke,fill}%
}%
\begin{pgfscope}%
\pgfsys@transformshift{1.626377in}{0.541249in}%
\pgfsys@useobject{currentmarker}{}%
\end{pgfscope}%
\end{pgfscope}%
\begin{pgfscope}%
\definecolor{textcolor}{rgb}{0.000000,0.000000,0.000000}%
\pgfsetstrokecolor{textcolor}%
\pgfsetfillcolor{textcolor}%
\pgftext[x=1.657627in, y=0.379791in, left, base,rotate=90.000000]{\color{textcolor}\rmfamily\fontsize{9.000000}{10.800000}\selectfont 4}%
\end{pgfscope}%
\begin{pgfscope}%
\pgfsetbuttcap%
\pgfsetroundjoin%
\definecolor{currentfill}{rgb}{0.000000,0.000000,0.000000}%
\pgfsetfillcolor{currentfill}%
\pgfsetlinewidth{0.803000pt}%
\definecolor{currentstroke}{rgb}{0.000000,0.000000,0.000000}%
\pgfsetstrokecolor{currentstroke}%
\pgfsetdash{}{0pt}%
\pgfsys@defobject{currentmarker}{\pgfqpoint{0.000000in}{-0.048611in}}{\pgfqpoint{0.000000in}{0.000000in}}{%
\pgfpathmoveto{\pgfqpoint{0.000000in}{0.000000in}}%
\pgfpathlineto{\pgfqpoint{0.000000in}{-0.048611in}}%
\pgfusepath{stroke,fill}%
}%
\begin{pgfscope}%
\pgfsys@transformshift{2.065983in}{0.541249in}%
\pgfsys@useobject{currentmarker}{}%
\end{pgfscope}%
\end{pgfscope}%
\begin{pgfscope}%
\definecolor{textcolor}{rgb}{0.000000,0.000000,0.000000}%
\pgfsetstrokecolor{textcolor}%
\pgfsetfillcolor{textcolor}%
\pgftext[x=2.097233in, y=0.379791in, left, base,rotate=90.000000]{\color{textcolor}\rmfamily\fontsize{9.000000}{10.800000}\selectfont 8}%
\end{pgfscope}%
\begin{pgfscope}%
\pgfsetbuttcap%
\pgfsetroundjoin%
\definecolor{currentfill}{rgb}{0.000000,0.000000,0.000000}%
\pgfsetfillcolor{currentfill}%
\pgfsetlinewidth{0.803000pt}%
\definecolor{currentstroke}{rgb}{0.000000,0.000000,0.000000}%
\pgfsetstrokecolor{currentstroke}%
\pgfsetdash{}{0pt}%
\pgfsys@defobject{currentmarker}{\pgfqpoint{0.000000in}{-0.048611in}}{\pgfqpoint{0.000000in}{0.000000in}}{%
\pgfpathmoveto{\pgfqpoint{0.000000in}{0.000000in}}%
\pgfpathlineto{\pgfqpoint{0.000000in}{-0.048611in}}%
\pgfusepath{stroke,fill}%
}%
\begin{pgfscope}%
\pgfsys@transformshift{2.505590in}{0.541249in}%
\pgfsys@useobject{currentmarker}{}%
\end{pgfscope}%
\end{pgfscope}%
\begin{pgfscope}%
\definecolor{textcolor}{rgb}{0.000000,0.000000,0.000000}%
\pgfsetstrokecolor{textcolor}%
\pgfsetfillcolor{textcolor}%
\pgftext[x=2.536840in, y=0.315556in, left, base,rotate=90.000000]{\color{textcolor}\rmfamily\fontsize{9.000000}{10.800000}\selectfont 16}%
\end{pgfscope}%
\begin{pgfscope}%
\pgfsetbuttcap%
\pgfsetroundjoin%
\definecolor{currentfill}{rgb}{0.000000,0.000000,0.000000}%
\pgfsetfillcolor{currentfill}%
\pgfsetlinewidth{0.803000pt}%
\definecolor{currentstroke}{rgb}{0.000000,0.000000,0.000000}%
\pgfsetstrokecolor{currentstroke}%
\pgfsetdash{}{0pt}%
\pgfsys@defobject{currentmarker}{\pgfqpoint{0.000000in}{-0.048611in}}{\pgfqpoint{0.000000in}{0.000000in}}{%
\pgfpathmoveto{\pgfqpoint{0.000000in}{0.000000in}}%
\pgfpathlineto{\pgfqpoint{0.000000in}{-0.048611in}}%
\pgfusepath{stroke,fill}%
}%
\begin{pgfscope}%
\pgfsys@transformshift{2.945197in}{0.541249in}%
\pgfsys@useobject{currentmarker}{}%
\end{pgfscope}%
\end{pgfscope}%
\begin{pgfscope}%
\definecolor{textcolor}{rgb}{0.000000,0.000000,0.000000}%
\pgfsetstrokecolor{textcolor}%
\pgfsetfillcolor{textcolor}%
\pgftext[x=2.976447in, y=0.315556in, left, base,rotate=90.000000]{\color{textcolor}\rmfamily\fontsize{9.000000}{10.800000}\selectfont 32}%
\end{pgfscope}%
\begin{pgfscope}%
\definecolor{textcolor}{rgb}{0.000000,0.000000,0.000000}%
\pgfsetstrokecolor{textcolor}%
\pgfsetfillcolor{textcolor}%
\pgftext[x=1.846180in,y=0.260000in,,top]{\color{textcolor}\rmfamily\fontsize{9.000000}{10.800000}\selectfont \(\displaystyle \alpha(/255)\)}%
\end{pgfscope}%
\begin{pgfscope}%
\pgfpathrectangle{\pgfqpoint{0.527360in}{0.541249in}}{\pgfqpoint{2.637640in}{1.923751in}}%
\pgfusepath{clip}%
\pgfsetrectcap%
\pgfsetroundjoin%
\pgfsetlinewidth{0.803000pt}%
\definecolor{currentstroke}{rgb}{0.690196,0.690196,0.690196}%
\pgfsetstrokecolor{currentstroke}%
\pgfsetdash{}{0pt}%
\pgfpathmoveto{\pgfqpoint{0.527360in}{0.541249in}}%
\pgfpathlineto{\pgfqpoint{3.165000in}{0.541249in}}%
\pgfusepath{stroke}%
\end{pgfscope}%
\begin{pgfscope}%
\pgfsetbuttcap%
\pgfsetroundjoin%
\definecolor{currentfill}{rgb}{0.000000,0.000000,0.000000}%
\pgfsetfillcolor{currentfill}%
\pgfsetlinewidth{0.803000pt}%
\definecolor{currentstroke}{rgb}{0.000000,0.000000,0.000000}%
\pgfsetstrokecolor{currentstroke}%
\pgfsetdash{}{0pt}%
\pgfsys@defobject{currentmarker}{\pgfqpoint{-0.048611in}{0.000000in}}{\pgfqpoint{-0.000000in}{0.000000in}}{%
\pgfpathmoveto{\pgfqpoint{-0.000000in}{0.000000in}}%
\pgfpathlineto{\pgfqpoint{-0.048611in}{0.000000in}}%
\pgfusepath{stroke,fill}%
}%
\begin{pgfscope}%
\pgfsys@transformshift{0.527360in}{0.541249in}%
\pgfsys@useobject{currentmarker}{}%
\end{pgfscope}%
\end{pgfscope}%
\begin{pgfscope}%
\definecolor{textcolor}{rgb}{0.000000,0.000000,0.000000}%
\pgfsetstrokecolor{textcolor}%
\pgfsetfillcolor{textcolor}%
\pgftext[x=0.365902in, y=0.497846in, left, base]{\color{textcolor}\rmfamily\fontsize{9.000000}{10.800000}\selectfont \(\displaystyle {0}\)}%
\end{pgfscope}%
\begin{pgfscope}%
\pgfpathrectangle{\pgfqpoint{0.527360in}{0.541249in}}{\pgfqpoint{2.637640in}{1.923751in}}%
\pgfusepath{clip}%
\pgfsetrectcap%
\pgfsetroundjoin%
\pgfsetlinewidth{0.803000pt}%
\definecolor{currentstroke}{rgb}{0.690196,0.690196,0.690196}%
\pgfsetstrokecolor{currentstroke}%
\pgfsetdash{}{0pt}%
\pgfpathmoveto{\pgfqpoint{0.527360in}{0.861874in}}%
\pgfpathlineto{\pgfqpoint{3.165000in}{0.861874in}}%
\pgfusepath{stroke}%
\end{pgfscope}%
\begin{pgfscope}%
\pgfsetbuttcap%
\pgfsetroundjoin%
\definecolor{currentfill}{rgb}{0.000000,0.000000,0.000000}%
\pgfsetfillcolor{currentfill}%
\pgfsetlinewidth{0.803000pt}%
\definecolor{currentstroke}{rgb}{0.000000,0.000000,0.000000}%
\pgfsetstrokecolor{currentstroke}%
\pgfsetdash{}{0pt}%
\pgfsys@defobject{currentmarker}{\pgfqpoint{-0.048611in}{0.000000in}}{\pgfqpoint{-0.000000in}{0.000000in}}{%
\pgfpathmoveto{\pgfqpoint{-0.000000in}{0.000000in}}%
\pgfpathlineto{\pgfqpoint{-0.048611in}{0.000000in}}%
\pgfusepath{stroke,fill}%
}%
\begin{pgfscope}%
\pgfsys@transformshift{0.527360in}{0.861874in}%
\pgfsys@useobject{currentmarker}{}%
\end{pgfscope}%
\end{pgfscope}%
\begin{pgfscope}%
\definecolor{textcolor}{rgb}{0.000000,0.000000,0.000000}%
\pgfsetstrokecolor{textcolor}%
\pgfsetfillcolor{textcolor}%
\pgftext[x=0.301667in, y=0.818472in, left, base]{\color{textcolor}\rmfamily\fontsize{9.000000}{10.800000}\selectfont \(\displaystyle {10}\)}%
\end{pgfscope}%
\begin{pgfscope}%
\pgfpathrectangle{\pgfqpoint{0.527360in}{0.541249in}}{\pgfqpoint{2.637640in}{1.923751in}}%
\pgfusepath{clip}%
\pgfsetrectcap%
\pgfsetroundjoin%
\pgfsetlinewidth{0.803000pt}%
\definecolor{currentstroke}{rgb}{0.690196,0.690196,0.690196}%
\pgfsetstrokecolor{currentstroke}%
\pgfsetdash{}{0pt}%
\pgfpathmoveto{\pgfqpoint{0.527360in}{1.182499in}}%
\pgfpathlineto{\pgfqpoint{3.165000in}{1.182499in}}%
\pgfusepath{stroke}%
\end{pgfscope}%
\begin{pgfscope}%
\pgfsetbuttcap%
\pgfsetroundjoin%
\definecolor{currentfill}{rgb}{0.000000,0.000000,0.000000}%
\pgfsetfillcolor{currentfill}%
\pgfsetlinewidth{0.803000pt}%
\definecolor{currentstroke}{rgb}{0.000000,0.000000,0.000000}%
\pgfsetstrokecolor{currentstroke}%
\pgfsetdash{}{0pt}%
\pgfsys@defobject{currentmarker}{\pgfqpoint{-0.048611in}{0.000000in}}{\pgfqpoint{-0.000000in}{0.000000in}}{%
\pgfpathmoveto{\pgfqpoint{-0.000000in}{0.000000in}}%
\pgfpathlineto{\pgfqpoint{-0.048611in}{0.000000in}}%
\pgfusepath{stroke,fill}%
}%
\begin{pgfscope}%
\pgfsys@transformshift{0.527360in}{1.182499in}%
\pgfsys@useobject{currentmarker}{}%
\end{pgfscope}%
\end{pgfscope}%
\begin{pgfscope}%
\definecolor{textcolor}{rgb}{0.000000,0.000000,0.000000}%
\pgfsetstrokecolor{textcolor}%
\pgfsetfillcolor{textcolor}%
\pgftext[x=0.301667in, y=1.139097in, left, base]{\color{textcolor}\rmfamily\fontsize{9.000000}{10.800000}\selectfont \(\displaystyle {20}\)}%
\end{pgfscope}%
\begin{pgfscope}%
\pgfpathrectangle{\pgfqpoint{0.527360in}{0.541249in}}{\pgfqpoint{2.637640in}{1.923751in}}%
\pgfusepath{clip}%
\pgfsetrectcap%
\pgfsetroundjoin%
\pgfsetlinewidth{0.803000pt}%
\definecolor{currentstroke}{rgb}{0.690196,0.690196,0.690196}%
\pgfsetstrokecolor{currentstroke}%
\pgfsetdash{}{0pt}%
\pgfpathmoveto{\pgfqpoint{0.527360in}{1.503125in}}%
\pgfpathlineto{\pgfqpoint{3.165000in}{1.503125in}}%
\pgfusepath{stroke}%
\end{pgfscope}%
\begin{pgfscope}%
\pgfsetbuttcap%
\pgfsetroundjoin%
\definecolor{currentfill}{rgb}{0.000000,0.000000,0.000000}%
\pgfsetfillcolor{currentfill}%
\pgfsetlinewidth{0.803000pt}%
\definecolor{currentstroke}{rgb}{0.000000,0.000000,0.000000}%
\pgfsetstrokecolor{currentstroke}%
\pgfsetdash{}{0pt}%
\pgfsys@defobject{currentmarker}{\pgfqpoint{-0.048611in}{0.000000in}}{\pgfqpoint{-0.000000in}{0.000000in}}{%
\pgfpathmoveto{\pgfqpoint{-0.000000in}{0.000000in}}%
\pgfpathlineto{\pgfqpoint{-0.048611in}{0.000000in}}%
\pgfusepath{stroke,fill}%
}%
\begin{pgfscope}%
\pgfsys@transformshift{0.527360in}{1.503125in}%
\pgfsys@useobject{currentmarker}{}%
\end{pgfscope}%
\end{pgfscope}%
\begin{pgfscope}%
\definecolor{textcolor}{rgb}{0.000000,0.000000,0.000000}%
\pgfsetstrokecolor{textcolor}%
\pgfsetfillcolor{textcolor}%
\pgftext[x=0.301667in, y=1.459722in, left, base]{\color{textcolor}\rmfamily\fontsize{9.000000}{10.800000}\selectfont \(\displaystyle {30}\)}%
\end{pgfscope}%
\begin{pgfscope}%
\pgfpathrectangle{\pgfqpoint{0.527360in}{0.541249in}}{\pgfqpoint{2.637640in}{1.923751in}}%
\pgfusepath{clip}%
\pgfsetrectcap%
\pgfsetroundjoin%
\pgfsetlinewidth{0.803000pt}%
\definecolor{currentstroke}{rgb}{0.690196,0.690196,0.690196}%
\pgfsetstrokecolor{currentstroke}%
\pgfsetdash{}{0pt}%
\pgfpathmoveto{\pgfqpoint{0.527360in}{1.823750in}}%
\pgfpathlineto{\pgfqpoint{3.165000in}{1.823750in}}%
\pgfusepath{stroke}%
\end{pgfscope}%
\begin{pgfscope}%
\pgfsetbuttcap%
\pgfsetroundjoin%
\definecolor{currentfill}{rgb}{0.000000,0.000000,0.000000}%
\pgfsetfillcolor{currentfill}%
\pgfsetlinewidth{0.803000pt}%
\definecolor{currentstroke}{rgb}{0.000000,0.000000,0.000000}%
\pgfsetstrokecolor{currentstroke}%
\pgfsetdash{}{0pt}%
\pgfsys@defobject{currentmarker}{\pgfqpoint{-0.048611in}{0.000000in}}{\pgfqpoint{-0.000000in}{0.000000in}}{%
\pgfpathmoveto{\pgfqpoint{-0.000000in}{0.000000in}}%
\pgfpathlineto{\pgfqpoint{-0.048611in}{0.000000in}}%
\pgfusepath{stroke,fill}%
}%
\begin{pgfscope}%
\pgfsys@transformshift{0.527360in}{1.823750in}%
\pgfsys@useobject{currentmarker}{}%
\end{pgfscope}%
\end{pgfscope}%
\begin{pgfscope}%
\definecolor{textcolor}{rgb}{0.000000,0.000000,0.000000}%
\pgfsetstrokecolor{textcolor}%
\pgfsetfillcolor{textcolor}%
\pgftext[x=0.301667in, y=1.780347in, left, base]{\color{textcolor}\rmfamily\fontsize{9.000000}{10.800000}\selectfont \(\displaystyle {40}\)}%
\end{pgfscope}%
\begin{pgfscope}%
\pgfpathrectangle{\pgfqpoint{0.527360in}{0.541249in}}{\pgfqpoint{2.637640in}{1.923751in}}%
\pgfusepath{clip}%
\pgfsetrectcap%
\pgfsetroundjoin%
\pgfsetlinewidth{0.803000pt}%
\definecolor{currentstroke}{rgb}{0.690196,0.690196,0.690196}%
\pgfsetstrokecolor{currentstroke}%
\pgfsetdash{}{0pt}%
\pgfpathmoveto{\pgfqpoint{0.527360in}{2.144375in}}%
\pgfpathlineto{\pgfqpoint{3.165000in}{2.144375in}}%
\pgfusepath{stroke}%
\end{pgfscope}%
\begin{pgfscope}%
\pgfsetbuttcap%
\pgfsetroundjoin%
\definecolor{currentfill}{rgb}{0.000000,0.000000,0.000000}%
\pgfsetfillcolor{currentfill}%
\pgfsetlinewidth{0.803000pt}%
\definecolor{currentstroke}{rgb}{0.000000,0.000000,0.000000}%
\pgfsetstrokecolor{currentstroke}%
\pgfsetdash{}{0pt}%
\pgfsys@defobject{currentmarker}{\pgfqpoint{-0.048611in}{0.000000in}}{\pgfqpoint{-0.000000in}{0.000000in}}{%
\pgfpathmoveto{\pgfqpoint{-0.000000in}{0.000000in}}%
\pgfpathlineto{\pgfqpoint{-0.048611in}{0.000000in}}%
\pgfusepath{stroke,fill}%
}%
\begin{pgfscope}%
\pgfsys@transformshift{0.527360in}{2.144375in}%
\pgfsys@useobject{currentmarker}{}%
\end{pgfscope}%
\end{pgfscope}%
\begin{pgfscope}%
\definecolor{textcolor}{rgb}{0.000000,0.000000,0.000000}%
\pgfsetstrokecolor{textcolor}%
\pgfsetfillcolor{textcolor}%
\pgftext[x=0.301667in, y=2.100972in, left, base]{\color{textcolor}\rmfamily\fontsize{9.000000}{10.800000}\selectfont \(\displaystyle {50}\)}%
\end{pgfscope}%
\begin{pgfscope}%
\pgfpathrectangle{\pgfqpoint{0.527360in}{0.541249in}}{\pgfqpoint{2.637640in}{1.923751in}}%
\pgfusepath{clip}%
\pgfsetrectcap%
\pgfsetroundjoin%
\pgfsetlinewidth{0.803000pt}%
\definecolor{currentstroke}{rgb}{0.690196,0.690196,0.690196}%
\pgfsetstrokecolor{currentstroke}%
\pgfsetdash{}{0pt}%
\pgfpathmoveto{\pgfqpoint{0.527360in}{2.465000in}}%
\pgfpathlineto{\pgfqpoint{3.165000in}{2.465000in}}%
\pgfusepath{stroke}%
\end{pgfscope}%
\begin{pgfscope}%
\pgfsetbuttcap%
\pgfsetroundjoin%
\definecolor{currentfill}{rgb}{0.000000,0.000000,0.000000}%
\pgfsetfillcolor{currentfill}%
\pgfsetlinewidth{0.803000pt}%
\definecolor{currentstroke}{rgb}{0.000000,0.000000,0.000000}%
\pgfsetstrokecolor{currentstroke}%
\pgfsetdash{}{0pt}%
\pgfsys@defobject{currentmarker}{\pgfqpoint{-0.048611in}{0.000000in}}{\pgfqpoint{-0.000000in}{0.000000in}}{%
\pgfpathmoveto{\pgfqpoint{-0.000000in}{0.000000in}}%
\pgfpathlineto{\pgfqpoint{-0.048611in}{0.000000in}}%
\pgfusepath{stroke,fill}%
}%
\begin{pgfscope}%
\pgfsys@transformshift{0.527360in}{2.465000in}%
\pgfsys@useobject{currentmarker}{}%
\end{pgfscope}%
\end{pgfscope}%
\begin{pgfscope}%
\definecolor{textcolor}{rgb}{0.000000,0.000000,0.000000}%
\pgfsetstrokecolor{textcolor}%
\pgfsetfillcolor{textcolor}%
\pgftext[x=0.301667in, y=2.421597in, left, base]{\color{textcolor}\rmfamily\fontsize{9.000000}{10.800000}\selectfont \(\displaystyle {60}\)}%
\end{pgfscope}%
\begin{pgfscope}%
\definecolor{textcolor}{rgb}{0.000000,0.000000,0.000000}%
\pgfsetstrokecolor{textcolor}%
\pgfsetfillcolor{textcolor}%
\pgftext[x=0.246111in,y=1.503125in,,bottom,rotate=90.000000]{\color{textcolor}\rmfamily\fontsize{9.000000}{10.800000}\selectfont PSNR}%
\end{pgfscope}%
\begin{pgfscope}%
\pgfpathrectangle{\pgfqpoint{0.527360in}{0.541249in}}{\pgfqpoint{2.637640in}{1.923751in}}%
\pgfusepath{clip}%
\pgfsetbuttcap%
\pgfsetmiterjoin%
\definecolor{currentfill}{rgb}{0.121569,0.466667,0.705882}%
\pgfsetfillcolor{currentfill}%
\pgfsetlinewidth{0.000000pt}%
\definecolor{currentstroke}{rgb}{0.000000,0.000000,0.000000}%
\pgfsetstrokecolor{currentstroke}%
\pgfsetstrokeopacity{0.000000}%
\pgfsetdash{}{0pt}%
\pgfpathmoveto{\pgfqpoint{0.637262in}{0.541249in}}%
\pgfpathlineto{\pgfqpoint{0.710530in}{0.541249in}}%
\pgfpathlineto{\pgfqpoint{0.710530in}{2.186056in}}%
\pgfpathlineto{\pgfqpoint{0.637262in}{2.186056in}}%
\pgfpathlineto{\pgfqpoint{0.637262in}{0.541249in}}%
\pgfpathclose%
\pgfusepath{fill}%
\end{pgfscope}%
\begin{pgfscope}%
\pgfpathrectangle{\pgfqpoint{0.527360in}{0.541249in}}{\pgfqpoint{2.637640in}{1.923751in}}%
\pgfusepath{clip}%
\pgfsetbuttcap%
\pgfsetmiterjoin%
\definecolor{currentfill}{rgb}{0.121569,0.466667,0.705882}%
\pgfsetfillcolor{currentfill}%
\pgfsetlinewidth{0.000000pt}%
\definecolor{currentstroke}{rgb}{0.000000,0.000000,0.000000}%
\pgfsetstrokecolor{currentstroke}%
\pgfsetstrokeopacity{0.000000}%
\pgfsetdash{}{0pt}%
\pgfpathmoveto{\pgfqpoint{1.076869in}{0.541249in}}%
\pgfpathlineto{\pgfqpoint{1.150136in}{0.541249in}}%
\pgfpathlineto{\pgfqpoint{1.150136in}{2.142772in}}%
\pgfpathlineto{\pgfqpoint{1.076869in}{2.142772in}}%
\pgfpathlineto{\pgfqpoint{1.076869in}{0.541249in}}%
\pgfpathclose%
\pgfusepath{fill}%
\end{pgfscope}%
\begin{pgfscope}%
\pgfpathrectangle{\pgfqpoint{0.527360in}{0.541249in}}{\pgfqpoint{2.637640in}{1.923751in}}%
\pgfusepath{clip}%
\pgfsetbuttcap%
\pgfsetmiterjoin%
\definecolor{currentfill}{rgb}{0.121569,0.466667,0.705882}%
\pgfsetfillcolor{currentfill}%
\pgfsetlinewidth{0.000000pt}%
\definecolor{currentstroke}{rgb}{0.000000,0.000000,0.000000}%
\pgfsetstrokecolor{currentstroke}%
\pgfsetstrokeopacity{0.000000}%
\pgfsetdash{}{0pt}%
\pgfpathmoveto{\pgfqpoint{1.516475in}{0.541249in}}%
\pgfpathlineto{\pgfqpoint{1.589743in}{0.541249in}}%
\pgfpathlineto{\pgfqpoint{1.589743in}{2.016125in}}%
\pgfpathlineto{\pgfqpoint{1.516475in}{2.016125in}}%
\pgfpathlineto{\pgfqpoint{1.516475in}{0.541249in}}%
\pgfpathclose%
\pgfusepath{fill}%
\end{pgfscope}%
\begin{pgfscope}%
\pgfpathrectangle{\pgfqpoint{0.527360in}{0.541249in}}{\pgfqpoint{2.637640in}{1.923751in}}%
\pgfusepath{clip}%
\pgfsetbuttcap%
\pgfsetmiterjoin%
\definecolor{currentfill}{rgb}{0.121569,0.466667,0.705882}%
\pgfsetfillcolor{currentfill}%
\pgfsetlinewidth{0.000000pt}%
\definecolor{currentstroke}{rgb}{0.000000,0.000000,0.000000}%
\pgfsetstrokecolor{currentstroke}%
\pgfsetstrokeopacity{0.000000}%
\pgfsetdash{}{0pt}%
\pgfpathmoveto{\pgfqpoint{1.956082in}{0.541249in}}%
\pgfpathlineto{\pgfqpoint{2.029350in}{0.541249in}}%
\pgfpathlineto{\pgfqpoint{2.029350in}{1.929556in}}%
\pgfpathlineto{\pgfqpoint{1.956082in}{1.929556in}}%
\pgfpathlineto{\pgfqpoint{1.956082in}{0.541249in}}%
\pgfpathclose%
\pgfusepath{fill}%
\end{pgfscope}%
\begin{pgfscope}%
\pgfpathrectangle{\pgfqpoint{0.527360in}{0.541249in}}{\pgfqpoint{2.637640in}{1.923751in}}%
\pgfusepath{clip}%
\pgfsetbuttcap%
\pgfsetmiterjoin%
\definecolor{currentfill}{rgb}{0.121569,0.466667,0.705882}%
\pgfsetfillcolor{currentfill}%
\pgfsetlinewidth{0.000000pt}%
\definecolor{currentstroke}{rgb}{0.000000,0.000000,0.000000}%
\pgfsetstrokecolor{currentstroke}%
\pgfsetstrokeopacity{0.000000}%
\pgfsetdash{}{0pt}%
\pgfpathmoveto{\pgfqpoint{2.395688in}{0.541249in}}%
\pgfpathlineto{\pgfqpoint{2.468956in}{0.541249in}}%
\pgfpathlineto{\pgfqpoint{2.468956in}{1.794893in}}%
\pgfpathlineto{\pgfqpoint{2.395688in}{1.794893in}}%
\pgfpathlineto{\pgfqpoint{2.395688in}{0.541249in}}%
\pgfpathclose%
\pgfusepath{fill}%
\end{pgfscope}%
\begin{pgfscope}%
\pgfpathrectangle{\pgfqpoint{0.527360in}{0.541249in}}{\pgfqpoint{2.637640in}{1.923751in}}%
\pgfusepath{clip}%
\pgfsetbuttcap%
\pgfsetmiterjoin%
\definecolor{currentfill}{rgb}{0.121569,0.466667,0.705882}%
\pgfsetfillcolor{currentfill}%
\pgfsetlinewidth{0.000000pt}%
\definecolor{currentstroke}{rgb}{0.000000,0.000000,0.000000}%
\pgfsetstrokecolor{currentstroke}%
\pgfsetstrokeopacity{0.000000}%
\pgfsetdash{}{0pt}%
\pgfpathmoveto{\pgfqpoint{2.835295in}{0.541249in}}%
\pgfpathlineto{\pgfqpoint{2.908563in}{0.541249in}}%
\pgfpathlineto{\pgfqpoint{2.908563in}{1.657025in}}%
\pgfpathlineto{\pgfqpoint{2.835295in}{1.657025in}}%
\pgfpathlineto{\pgfqpoint{2.835295in}{0.541249in}}%
\pgfpathclose%
\pgfusepath{fill}%
\end{pgfscope}%
\begin{pgfscope}%
\pgfpathrectangle{\pgfqpoint{0.527360in}{0.541249in}}{\pgfqpoint{2.637640in}{1.923751in}}%
\pgfusepath{clip}%
\pgfsetbuttcap%
\pgfsetmiterjoin%
\definecolor{currentfill}{rgb}{1.000000,0.498039,0.054902}%
\pgfsetfillcolor{currentfill}%
\pgfsetlinewidth{0.000000pt}%
\definecolor{currentstroke}{rgb}{0.000000,0.000000,0.000000}%
\pgfsetstrokecolor{currentstroke}%
\pgfsetstrokeopacity{0.000000}%
\pgfsetdash{}{0pt}%
\pgfpathmoveto{\pgfqpoint{0.710530in}{0.541249in}}%
\pgfpathlineto{\pgfqpoint{0.783797in}{0.541249in}}%
\pgfpathlineto{\pgfqpoint{0.783797in}{2.208500in}}%
\pgfpathlineto{\pgfqpoint{0.710530in}{2.208500in}}%
\pgfpathlineto{\pgfqpoint{0.710530in}{0.541249in}}%
\pgfpathclose%
\pgfusepath{fill}%
\end{pgfscope}%
\begin{pgfscope}%
\pgfpathrectangle{\pgfqpoint{0.527360in}{0.541249in}}{\pgfqpoint{2.637640in}{1.923751in}}%
\pgfusepath{clip}%
\pgfsetbuttcap%
\pgfsetmiterjoin%
\definecolor{currentfill}{rgb}{1.000000,0.498039,0.054902}%
\pgfsetfillcolor{currentfill}%
\pgfsetlinewidth{0.000000pt}%
\definecolor{currentstroke}{rgb}{0.000000,0.000000,0.000000}%
\pgfsetstrokecolor{currentstroke}%
\pgfsetstrokeopacity{0.000000}%
\pgfsetdash{}{0pt}%
\pgfpathmoveto{\pgfqpoint{1.150136in}{0.541249in}}%
\pgfpathlineto{\pgfqpoint{1.223404in}{0.541249in}}%
\pgfpathlineto{\pgfqpoint{1.223404in}{2.147581in}}%
\pgfpathlineto{\pgfqpoint{1.150136in}{2.147581in}}%
\pgfpathlineto{\pgfqpoint{1.150136in}{0.541249in}}%
\pgfpathclose%
\pgfusepath{fill}%
\end{pgfscope}%
\begin{pgfscope}%
\pgfpathrectangle{\pgfqpoint{0.527360in}{0.541249in}}{\pgfqpoint{2.637640in}{1.923751in}}%
\pgfusepath{clip}%
\pgfsetbuttcap%
\pgfsetmiterjoin%
\definecolor{currentfill}{rgb}{1.000000,0.498039,0.054902}%
\pgfsetfillcolor{currentfill}%
\pgfsetlinewidth{0.000000pt}%
\definecolor{currentstroke}{rgb}{0.000000,0.000000,0.000000}%
\pgfsetstrokecolor{currentstroke}%
\pgfsetstrokeopacity{0.000000}%
\pgfsetdash{}{0pt}%
\pgfpathmoveto{\pgfqpoint{1.589743in}{0.541249in}}%
\pgfpathlineto{\pgfqpoint{1.663011in}{0.541249in}}%
\pgfpathlineto{\pgfqpoint{1.663011in}{2.038569in}}%
\pgfpathlineto{\pgfqpoint{1.589743in}{2.038569in}}%
\pgfpathlineto{\pgfqpoint{1.589743in}{0.541249in}}%
\pgfpathclose%
\pgfusepath{fill}%
\end{pgfscope}%
\begin{pgfscope}%
\pgfpathrectangle{\pgfqpoint{0.527360in}{0.541249in}}{\pgfqpoint{2.637640in}{1.923751in}}%
\pgfusepath{clip}%
\pgfsetbuttcap%
\pgfsetmiterjoin%
\definecolor{currentfill}{rgb}{1.000000,0.498039,0.054902}%
\pgfsetfillcolor{currentfill}%
\pgfsetlinewidth{0.000000pt}%
\definecolor{currentstroke}{rgb}{0.000000,0.000000,0.000000}%
\pgfsetstrokecolor{currentstroke}%
\pgfsetstrokeopacity{0.000000}%
\pgfsetdash{}{0pt}%
\pgfpathmoveto{\pgfqpoint{2.029350in}{0.541249in}}%
\pgfpathlineto{\pgfqpoint{2.102617in}{0.541249in}}%
\pgfpathlineto{\pgfqpoint{2.102617in}{1.952000in}}%
\pgfpathlineto{\pgfqpoint{2.029350in}{1.952000in}}%
\pgfpathlineto{\pgfqpoint{2.029350in}{0.541249in}}%
\pgfpathclose%
\pgfusepath{fill}%
\end{pgfscope}%
\begin{pgfscope}%
\pgfpathrectangle{\pgfqpoint{0.527360in}{0.541249in}}{\pgfqpoint{2.637640in}{1.923751in}}%
\pgfusepath{clip}%
\pgfsetbuttcap%
\pgfsetmiterjoin%
\definecolor{currentfill}{rgb}{1.000000,0.498039,0.054902}%
\pgfsetfillcolor{currentfill}%
\pgfsetlinewidth{0.000000pt}%
\definecolor{currentstroke}{rgb}{0.000000,0.000000,0.000000}%
\pgfsetstrokecolor{currentstroke}%
\pgfsetstrokeopacity{0.000000}%
\pgfsetdash{}{0pt}%
\pgfpathmoveto{\pgfqpoint{2.468956in}{0.541249in}}%
\pgfpathlineto{\pgfqpoint{2.542224in}{0.541249in}}%
\pgfpathlineto{\pgfqpoint{2.542224in}{1.823750in}}%
\pgfpathlineto{\pgfqpoint{2.468956in}{1.823750in}}%
\pgfpathlineto{\pgfqpoint{2.468956in}{0.541249in}}%
\pgfpathclose%
\pgfusepath{fill}%
\end{pgfscope}%
\begin{pgfscope}%
\pgfpathrectangle{\pgfqpoint{0.527360in}{0.541249in}}{\pgfqpoint{2.637640in}{1.923751in}}%
\pgfusepath{clip}%
\pgfsetbuttcap%
\pgfsetmiterjoin%
\definecolor{currentfill}{rgb}{1.000000,0.498039,0.054902}%
\pgfsetfillcolor{currentfill}%
\pgfsetlinewidth{0.000000pt}%
\definecolor{currentstroke}{rgb}{0.000000,0.000000,0.000000}%
\pgfsetstrokecolor{currentstroke}%
\pgfsetstrokeopacity{0.000000}%
\pgfsetdash{}{0pt}%
\pgfpathmoveto{\pgfqpoint{2.908563in}{0.541249in}}%
\pgfpathlineto{\pgfqpoint{2.981831in}{0.541249in}}%
\pgfpathlineto{\pgfqpoint{2.981831in}{1.663437in}}%
\pgfpathlineto{\pgfqpoint{2.908563in}{1.663437in}}%
\pgfpathlineto{\pgfqpoint{2.908563in}{0.541249in}}%
\pgfpathclose%
\pgfusepath{fill}%
\end{pgfscope}%
\begin{pgfscope}%
\pgfpathrectangle{\pgfqpoint{0.527360in}{0.541249in}}{\pgfqpoint{2.637640in}{1.923751in}}%
\pgfusepath{clip}%
\pgfsetbuttcap%
\pgfsetmiterjoin%
\definecolor{currentfill}{rgb}{0.172549,0.627451,0.172549}%
\pgfsetfillcolor{currentfill}%
\pgfsetlinewidth{0.000000pt}%
\definecolor{currentstroke}{rgb}{0.000000,0.000000,0.000000}%
\pgfsetstrokecolor{currentstroke}%
\pgfsetstrokeopacity{0.000000}%
\pgfsetdash{}{0pt}%
\pgfpathmoveto{\pgfqpoint{0.783797in}{0.541249in}}%
\pgfpathlineto{\pgfqpoint{0.857065in}{0.541249in}}%
\pgfpathlineto{\pgfqpoint{0.857065in}{2.168742in}}%
\pgfpathlineto{\pgfqpoint{0.783797in}{2.168742in}}%
\pgfpathlineto{\pgfqpoint{0.783797in}{0.541249in}}%
\pgfpathclose%
\pgfusepath{fill}%
\end{pgfscope}%
\begin{pgfscope}%
\pgfpathrectangle{\pgfqpoint{0.527360in}{0.541249in}}{\pgfqpoint{2.637640in}{1.923751in}}%
\pgfusepath{clip}%
\pgfsetbuttcap%
\pgfsetmiterjoin%
\definecolor{currentfill}{rgb}{0.172549,0.627451,0.172549}%
\pgfsetfillcolor{currentfill}%
\pgfsetlinewidth{0.000000pt}%
\definecolor{currentstroke}{rgb}{0.000000,0.000000,0.000000}%
\pgfsetstrokecolor{currentstroke}%
\pgfsetstrokeopacity{0.000000}%
\pgfsetdash{}{0pt}%
\pgfpathmoveto{\pgfqpoint{1.223404in}{0.541249in}}%
\pgfpathlineto{\pgfqpoint{1.296672in}{0.541249in}}%
\pgfpathlineto{\pgfqpoint{1.296672in}{2.016125in}}%
\pgfpathlineto{\pgfqpoint{1.223404in}{2.016125in}}%
\pgfpathlineto{\pgfqpoint{1.223404in}{0.541249in}}%
\pgfpathclose%
\pgfusepath{fill}%
\end{pgfscope}%
\begin{pgfscope}%
\pgfpathrectangle{\pgfqpoint{0.527360in}{0.541249in}}{\pgfqpoint{2.637640in}{1.923751in}}%
\pgfusepath{clip}%
\pgfsetbuttcap%
\pgfsetmiterjoin%
\definecolor{currentfill}{rgb}{0.172549,0.627451,0.172549}%
\pgfsetfillcolor{currentfill}%
\pgfsetlinewidth{0.000000pt}%
\definecolor{currentstroke}{rgb}{0.000000,0.000000,0.000000}%
\pgfsetstrokecolor{currentstroke}%
\pgfsetstrokeopacity{0.000000}%
\pgfsetdash{}{0pt}%
\pgfpathmoveto{\pgfqpoint{1.663011in}{0.541249in}}%
\pgfpathlineto{\pgfqpoint{1.736278in}{0.541249in}}%
\pgfpathlineto{\pgfqpoint{1.736278in}{1.832727in}}%
\pgfpathlineto{\pgfqpoint{1.663011in}{1.832727in}}%
\pgfpathlineto{\pgfqpoint{1.663011in}{0.541249in}}%
\pgfpathclose%
\pgfusepath{fill}%
\end{pgfscope}%
\begin{pgfscope}%
\pgfpathrectangle{\pgfqpoint{0.527360in}{0.541249in}}{\pgfqpoint{2.637640in}{1.923751in}}%
\pgfusepath{clip}%
\pgfsetbuttcap%
\pgfsetmiterjoin%
\definecolor{currentfill}{rgb}{0.172549,0.627451,0.172549}%
\pgfsetfillcolor{currentfill}%
\pgfsetlinewidth{0.000000pt}%
\definecolor{currentstroke}{rgb}{0.000000,0.000000,0.000000}%
\pgfsetstrokecolor{currentstroke}%
\pgfsetstrokeopacity{0.000000}%
\pgfsetdash{}{0pt}%
\pgfpathmoveto{\pgfqpoint{2.102617in}{0.541249in}}%
\pgfpathlineto{\pgfqpoint{2.175885in}{0.541249in}}%
\pgfpathlineto{\pgfqpoint{2.175885in}{1.638108in}}%
\pgfpathlineto{\pgfqpoint{2.102617in}{1.638108in}}%
\pgfpathlineto{\pgfqpoint{2.102617in}{0.541249in}}%
\pgfpathclose%
\pgfusepath{fill}%
\end{pgfscope}%
\begin{pgfscope}%
\pgfpathrectangle{\pgfqpoint{0.527360in}{0.541249in}}{\pgfqpoint{2.637640in}{1.923751in}}%
\pgfusepath{clip}%
\pgfsetbuttcap%
\pgfsetmiterjoin%
\definecolor{currentfill}{rgb}{0.172549,0.627451,0.172549}%
\pgfsetfillcolor{currentfill}%
\pgfsetlinewidth{0.000000pt}%
\definecolor{currentstroke}{rgb}{0.000000,0.000000,0.000000}%
\pgfsetstrokecolor{currentstroke}%
\pgfsetstrokeopacity{0.000000}%
\pgfsetdash{}{0pt}%
\pgfpathmoveto{\pgfqpoint{2.542224in}{0.541249in}}%
\pgfpathlineto{\pgfqpoint{2.615492in}{0.541249in}}%
\pgfpathlineto{\pgfqpoint{2.615492in}{1.454069in}}%
\pgfpathlineto{\pgfqpoint{2.542224in}{1.454069in}}%
\pgfpathlineto{\pgfqpoint{2.542224in}{0.541249in}}%
\pgfpathclose%
\pgfusepath{fill}%
\end{pgfscope}%
\begin{pgfscope}%
\pgfpathrectangle{\pgfqpoint{0.527360in}{0.541249in}}{\pgfqpoint{2.637640in}{1.923751in}}%
\pgfusepath{clip}%
\pgfsetbuttcap%
\pgfsetmiterjoin%
\definecolor{currentfill}{rgb}{0.172549,0.627451,0.172549}%
\pgfsetfillcolor{currentfill}%
\pgfsetlinewidth{0.000000pt}%
\definecolor{currentstroke}{rgb}{0.000000,0.000000,0.000000}%
\pgfsetstrokecolor{currentstroke}%
\pgfsetstrokeopacity{0.000000}%
\pgfsetdash{}{0pt}%
\pgfpathmoveto{\pgfqpoint{2.981831in}{0.541249in}}%
\pgfpathlineto{\pgfqpoint{3.055098in}{0.541249in}}%
\pgfpathlineto{\pgfqpoint{3.055098in}{1.275801in}}%
\pgfpathlineto{\pgfqpoint{2.981831in}{1.275801in}}%
\pgfpathlineto{\pgfqpoint{2.981831in}{0.541249in}}%
\pgfpathclose%
\pgfusepath{fill}%
\end{pgfscope}%
\begin{pgfscope}%
\pgfsetrectcap%
\pgfsetmiterjoin%
\pgfsetlinewidth{0.803000pt}%
\definecolor{currentstroke}{rgb}{0.000000,0.000000,0.000000}%
\pgfsetstrokecolor{currentstroke}%
\pgfsetdash{}{0pt}%
\pgfpathmoveto{\pgfqpoint{0.527360in}{0.541249in}}%
\pgfpathlineto{\pgfqpoint{0.527360in}{2.465000in}}%
\pgfusepath{stroke}%
\end{pgfscope}%
\begin{pgfscope}%
\pgfsetrectcap%
\pgfsetmiterjoin%
\pgfsetlinewidth{0.803000pt}%
\definecolor{currentstroke}{rgb}{0.000000,0.000000,0.000000}%
\pgfsetstrokecolor{currentstroke}%
\pgfsetdash{}{0pt}%
\pgfpathmoveto{\pgfqpoint{3.165000in}{0.541249in}}%
\pgfpathlineto{\pgfqpoint{3.165000in}{2.465000in}}%
\pgfusepath{stroke}%
\end{pgfscope}%
\begin{pgfscope}%
\pgfsetrectcap%
\pgfsetmiterjoin%
\pgfsetlinewidth{0.803000pt}%
\definecolor{currentstroke}{rgb}{0.000000,0.000000,0.000000}%
\pgfsetstrokecolor{currentstroke}%
\pgfsetdash{}{0pt}%
\pgfpathmoveto{\pgfqpoint{0.527360in}{0.541249in}}%
\pgfpathlineto{\pgfqpoint{3.165000in}{0.541249in}}%
\pgfusepath{stroke}%
\end{pgfscope}%
\begin{pgfscope}%
\pgfsetrectcap%
\pgfsetmiterjoin%
\pgfsetlinewidth{0.803000pt}%
\definecolor{currentstroke}{rgb}{0.000000,0.000000,0.000000}%
\pgfsetstrokecolor{currentstroke}%
\pgfsetdash{}{0pt}%
\pgfpathmoveto{\pgfqpoint{0.527360in}{2.465000in}}%
\pgfpathlineto{\pgfqpoint{3.165000in}{2.465000in}}%
\pgfusepath{stroke}%
\end{pgfscope}%
\begin{pgfscope}%
\pgfsetbuttcap%
\pgfsetmiterjoin%
\definecolor{currentfill}{rgb}{1.000000,1.000000,1.000000}%
\pgfsetfillcolor{currentfill}%
\pgfsetfillopacity{0.800000}%
\pgfsetlinewidth{1.003750pt}%
\definecolor{currentstroke}{rgb}{0.800000,0.800000,0.800000}%
\pgfsetstrokecolor{currentstroke}%
\pgfsetstrokeopacity{0.800000}%
\pgfsetdash{}{0pt}%
\pgfpathmoveto{\pgfqpoint{2.134237in}{1.842083in}}%
\pgfpathlineto{\pgfqpoint{3.077500in}{1.842083in}}%
\pgfpathquadraticcurveto{\pgfqpoint{3.102500in}{1.842083in}}{\pgfqpoint{3.102500in}{1.867083in}}%
\pgfpathlineto{\pgfqpoint{3.102500in}{2.377500in}}%
\pgfpathquadraticcurveto{\pgfqpoint{3.102500in}{2.402500in}}{\pgfqpoint{3.077500in}{2.402500in}}%
\pgfpathlineto{\pgfqpoint{2.134237in}{2.402500in}}%
\pgfpathquadraticcurveto{\pgfqpoint{2.109237in}{2.402500in}}{\pgfqpoint{2.109237in}{2.377500in}}%
\pgfpathlineto{\pgfqpoint{2.109237in}{1.867083in}}%
\pgfpathquadraticcurveto{\pgfqpoint{2.109237in}{1.842083in}}{\pgfqpoint{2.134237in}{1.842083in}}%
\pgfpathlineto{\pgfqpoint{2.134237in}{1.842083in}}%
\pgfpathclose%
\pgfusepath{stroke,fill}%
\end{pgfscope}%
\begin{pgfscope}%
\pgfsetbuttcap%
\pgfsetmiterjoin%
\definecolor{currentfill}{rgb}{0.121569,0.466667,0.705882}%
\pgfsetfillcolor{currentfill}%
\pgfsetlinewidth{0.000000pt}%
\definecolor{currentstroke}{rgb}{0.000000,0.000000,0.000000}%
\pgfsetstrokecolor{currentstroke}%
\pgfsetstrokeopacity{0.000000}%
\pgfsetdash{}{0pt}%
\pgfpathmoveto{\pgfqpoint{2.159237in}{2.265000in}}%
\pgfpathlineto{\pgfqpoint{2.409237in}{2.265000in}}%
\pgfpathlineto{\pgfqpoint{2.409237in}{2.352500in}}%
\pgfpathlineto{\pgfqpoint{2.159237in}{2.352500in}}%
\pgfpathlineto{\pgfqpoint{2.159237in}{2.265000in}}%
\pgfpathclose%
\pgfusepath{fill}%
\end{pgfscope}%
\begin{pgfscope}%
\definecolor{textcolor}{rgb}{0.000000,0.000000,0.000000}%
\pgfsetstrokecolor{textcolor}%
\pgfsetfillcolor{textcolor}%
\pgftext[x=2.509237in,y=2.265000in,left,base]{\color{textcolor}\rmfamily\fontsize{9.000000}{10.800000}\selectfont ESRGAN}%
\end{pgfscope}%
\begin{pgfscope}%
\pgfsetbuttcap%
\pgfsetmiterjoin%
\definecolor{currentfill}{rgb}{1.000000,0.498039,0.054902}%
\pgfsetfillcolor{currentfill}%
\pgfsetlinewidth{0.000000pt}%
\definecolor{currentstroke}{rgb}{0.000000,0.000000,0.000000}%
\pgfsetstrokecolor{currentstroke}%
\pgfsetstrokeopacity{0.000000}%
\pgfsetdash{}{0pt}%
\pgfpathmoveto{\pgfqpoint{2.159237in}{2.090694in}}%
\pgfpathlineto{\pgfqpoint{2.409237in}{2.090694in}}%
\pgfpathlineto{\pgfqpoint{2.409237in}{2.178194in}}%
\pgfpathlineto{\pgfqpoint{2.159237in}{2.178194in}}%
\pgfpathlineto{\pgfqpoint{2.159237in}{2.090694in}}%
\pgfpathclose%
\pgfusepath{fill}%
\end{pgfscope}%
\begin{pgfscope}%
\definecolor{textcolor}{rgb}{0.000000,0.000000,0.000000}%
\pgfsetstrokecolor{textcolor}%
\pgfsetfillcolor{textcolor}%
\pgftext[x=2.509237in,y=2.090694in,left,base]{\color{textcolor}\rmfamily\fontsize{9.000000}{10.800000}\selectfont EDSR}%
\end{pgfscope}%
\begin{pgfscope}%
\pgfsetbuttcap%
\pgfsetmiterjoin%
\definecolor{currentfill}{rgb}{0.172549,0.627451,0.172549}%
\pgfsetfillcolor{currentfill}%
\pgfsetlinewidth{0.000000pt}%
\definecolor{currentstroke}{rgb}{0.000000,0.000000,0.000000}%
\pgfsetstrokecolor{currentstroke}%
\pgfsetstrokeopacity{0.000000}%
\pgfsetdash{}{0pt}%
\pgfpathmoveto{\pgfqpoint{2.159237in}{1.916389in}}%
\pgfpathlineto{\pgfqpoint{2.409237in}{1.916389in}}%
\pgfpathlineto{\pgfqpoint{2.409237in}{2.003889in}}%
\pgfpathlineto{\pgfqpoint{2.159237in}{2.003889in}}%
\pgfpathlineto{\pgfqpoint{2.159237in}{1.916389in}}%
\pgfpathclose%
\pgfusepath{fill}%
\end{pgfscope}%
\begin{pgfscope}%
\definecolor{textcolor}{rgb}{0.000000,0.000000,0.000000}%
\pgfsetstrokecolor{textcolor}%
\pgfsetfillcolor{textcolor}%
\pgftext[x=2.509237in,y=1.916389in,left,base]{\color{textcolor}\rmfamily\fontsize{9.000000}{10.800000}\selectfont SRFlow}%
\end{pgfscope}%
\end{pgfpicture}%
\makeatother%
\endgroup%